\documentclass[12pt,oneside,noupper,nocenter,bold,letterpaper]{teseir}
\usepackage[dvipsnames]{xcolor}
\usepackage[utf8]{inputenc}
\usepackage{amsmath}
\usepackage{amssymb}
\usepackage{graphicx}
\usepackage{enumitem} 
\usepackage{hyperref} 
\usepackage{float} 
\usepackage{physics} 
\usepackage{subcaption}
\usepackage[USenglish]{babel}
\usepackage{epigraph}
\usepackage{cleveref}
\usepackage{pgf}
\usepackage{bm} 
\usepackage{slashed}
\DeclareMathOperator\arcsinh{arcsinh}
\usepackage{mhchem}
\usepackage[T1]{fontenc}
\usepackage{lmodern}
\usepackage[hmargin=1in,vmargin=1in]{geometry}

\usepackage{multirow,termcal}
\usepackage{setspace}

\newif\ifblackandwhite
\usepackage{etoolbox}
\usepackage{longtable}%

\usepackage{pdflscape}
\usepackage{colortbl}%

\usepackage{booktabs}
\usepackage{makecell}
\ifblackandwhite

\else

\fi

\newcommand{\hiddensubsection}[1]{
  \stepcounter{subsection}
  \subsection*{\arabic{chapter}.\arabic{section}.\arabic{subsection}\hspace{1em}{#1}}
  }

\newcommand{\hiddensubsubsection}[1]{
  \stepcounter{subsubsection}
  \subsubsection*{\arabic{chapter}.\arabic{section}.\arabic{subsection}.\arabic{subsubsection}\hspace{1em}{#1}}
  }

\usepackage{chngcntr}
\counterwithout{footnote}{chapter} 
\begin{document}
\singlespacing
\begin{titlepage}

  \begin{center}
    WASHINGTON UNIVERSITY IN ST.~LOUIS \\
    DEPARTMENT OF PHYSICS \\
    \vspace{1 cm}

    Dissertation Examination Committee:\\
    Mark Alford, Chair\\
    Greg Comer\\
    Bhupal Dev\\
    Wim Dickhoff\\
    Saori Pastore\\

    \vspace{\fill}
    {Transport in Neutron Star Mergers}\\

    by\\
    Steven P. Harris\\

    \vspace{\fill}
    A dissertation presented to\\
    The Graduate School \\
    of Washington University in \\
    partial fullfillment of the \\
    requirements for the degree\\
    of Doctor of Philosophy \\

    \vspace{\fill}
    \vspace{4cm}
    St.~Louis, Missouri\\
    May, 2020
  \end{center}
  
\end{titlepage}
\pagestyle{noheads}
\pagenumbering{roman}
\pagebreak
\tableofcontents
\listoffigures
\doublespacing
\newpage
\pagestyle{noheads}
\phantomsection
\addcontentsline{toc}{chapter}{Acknowledgements}
\begin{center}
{\Huge Acknowledgements}
\end{center}
Completing five years of graduate school is not something I could have done alone, and I want to use this section to thank those who have helped me along the way.  

First, of course, is my thesis adviser Mark Alford.  Despite being department chair and having many other responsibilities, Mark was always available to answer my questions and I always left his office with a better understanding of a given topic than when I arrived.  He has taught me much about breaking up a project into small pieces, which is one of the most valuable skills in research.  In addition, he encouraged me to go to many conferences and took me to Frankfurt for a week to collaborate with several physicists there.  Thank you.

Second, I would like to thank the members of my committee, Wim Dickhoff, Bhupal Dev, Saori Pastore, and Greg Comer for reading and critiquing this thesis.  In particular I would like to thank Bhupal and Wim for serving on my faculty mentoring committee for the past three years, during which they offered many helpful suggestions.

I would like to thank Kuver Sinha, Jeff Fortin, and Pratik Sachdeva with whom I've authored papers during my time in graduate school.  In addition I would like to thank those with whom I have discussed topics in those papers, including Fridolin Weber, Kai Schwenzer, Micaela Oertel, Peter Shternin, Luke Roberts, Armen Sedrakian, Reed Essick, Andreas Reisenegger, Alessandro Drago, J{\"u}rgen Schaffner-Bielich, and Francesc Ferrer.  In particular, I want to thank my fellow group members, past and present, Andreas Windisch, Alex Haber, and Ziyuan Zhang for many fruitful discussions.  

Time spent in my offices, first in Crow Hall and then in Compton Hall, has been enjoyable and that is due to my officemates.  I want to thank all of my friends among the graduate students and postdocs at Wash U.  It has been a pleasure hanging out with you at grad sem, dinners, Cardinals games, and PIB over the past five years.

Most of all, I want to thank my parents for their love and constant support.  Throughout my education, they have done whatever they can to help me and I am forever grateful.  

The week that I spent at the Frankfurt Institute for Advanced Study during the summer of 2019 was very enjoyable and productive.  I would like to thank the members of the institute for their hospitality.  This research in this thesis was partly supported by the U.S. Department of Energy, Office of Science, Office of Nuclear Physics, under Award No.~\#DEFG02-05ER41375.

\begin{flushright}
  Steven Harris
\end{flushright}
\begin{flushleft}
  \textit{Washington University in St.~Louis\\
  May, 2020}
\end{flushleft}
\abstract
{ \centering
  Transport in Neutron Star Mergers \\
  by\\
  Steven P. Harris\\
  Doctor of Philosophy in Physics \\
  Washington University in St.~Louis, \the\year \\
  Professor Mark Alford (Chair) \\
  \vspace{1.0cm}
  }

Neutron star mergers are the only situation in nature in which we find matter compressed to several times nuclear saturation density and temperatures of several tens of MeV.  By observing and numerically simulating neutron star mergers, we can learn about the nature of matter at high temperatures and densities.  Neutron star merger simulations evolve Einstein's equations of general relativity coupled to the equations of relativistic hydrodynamics along with a nuclear equation of state, which describes the neutron star matter.  Many simulations also take into account neutrino transport and electrodynamics.  The purpose of this thesis is to see whether other physical processes, including thermal transport and viscosity, are relevant to neutron star mergers and thus should be included in merger simulations.

After an introduction to the QCD phase diagram, the nuclear equations of state, and neutron star mergers, I discuss three projects related to transport and nuclear matter in neutron star mergers.  The first is the nature of beta equilibrium in the portion of a merger that is transparent to neutrinos.  We calculate the weak interaction (Urca) rates and find that the beta equilibrium condition needs to be modified by adding an additional chemical potential, which changes slightly the particle content in neutrino-transparent beta equilibrium.  Secondly, we calculate the bulk viscosity in neutrino-transparent nuclear matter in conditions encountered in neutron star mergers.  Bulk viscosity arises from a phase lag between the pressure and density in the nuclear matter, which is due to the finite rate of beta equilibration.  When bulk viscosity is sufficiently strong, which happens when the equilibration rate nearly matches the frequency of the density oscillation, it can noticeably dampen the oscillation.  We find that in certain thermodynamic conditions likely encountered in mergers, oscillations in nuclear matter can be damped on timescales on the order of 10 milliseconds, so we conclude that bulk viscosity should be included in merger simulations.  Finally, we study thermal transport due to axions in neutron star mergers.  We conclude that axions are never trapped in mergers, but instead escape, carrying energy away from the merger.  We calculate the cooling time due to the energy carried away by axions and find that within current constraints on the axion-nucleon coupling, axions could cool fluid elements in mergers on timescales which could affect the dynamics of the merger.
\newpage
\pagenumbering{arabic}
\chapter{QCD and nuclear matter}
\pagestyle{myheadings}
\section{QCD and its predecessors}
Despite the Coulomb repulsion of its constituent protons, the nucleus of an atom remains bound.  In 1934, Yukawa proposed that there is an attractive force, called the strong force or the nuclear force, that keeps the neutrons and protons together in the nucleus despite the repulsive electromagnetic interaction.  The strong interaction must be very short-ranged, as it is not observed in everyday life (unlike the long-ranged forces of electromagnetism or gravity).  The first serious model of the strong interaction was proposed by Yukawa in 1934 \cite{Yukawa:1935xg}.  He suggested that the nucleons interact by exchanging a meson, generating an attractive force.  As the nuclear force was observed to act on distances of the size of the nucleus (a few femtometers, or Fermi), the mass of the mediator meson was expected\footnote{The general idea here is that two nucleons in the nucleus will interact by producing a virtual pion, which has energy $\Delta E$ and lives for time $\Delta t$.  This virtual pion production is allowed by the energy-time uncertainty principle provided that $\Delta E_{\pi}\Delta t \sim \hbar$.  The energy $\Delta E_{\pi}$ to create the pion is the pion rest mass-energy, $m_{\pi}c^2$, and the pion has range $r=c\Delta t$.  Putting this all together, $\Delta E_{\pi}\Delta t = (m_{\pi}c^2)(r/c) \sim \hbar$, or $m_{\pi}c^2 \sim \hbar c/ r$ and so if the range of the pion-mediated strong interaction is 1 Fermi, then $m_{\pi}c^2 \sim 200 \text{ MeV}$ \cite{Ryder:1985wq}.  See Ref.~\cite{2016iqm..book.....G} for a discussion of the use of the energy-time uncertainty principle in this context.} to be around 100 MeV.  In contrast, electromagnetism is a long-range force, because it is mediated by the photon, which is massless \cite{Griffiths:2008zz}.  One decade later, the exchanged meson for the strong force was determined to be the pion, with mass around 140 MeV \cite{Marshak:1947zz,Lattes:1947mw,Lattes:1947mx}.  By the middle of the twentieth century, the theory of the strong interaction involved protons and neutrons as fundamental particles, which interact by exchanging pions \cite{Griffiths:2008zz}.

Beginning in the late 1940s, a bevy of new particles was discovered, including mesons like the kaon, eta, phi, omega, and rho and baryons like the lambda, sigma, and xi.  Surprisingly, these particles could be organized into patterns based on their electric charge and strangeness, called the baryon octet, meson octet, and the baryon decuplet \cite{Griffiths:2008zz}.  In 1964, Gell-Mann and Zweig independently proposed that all baryons and mesons were made of fundamental particles called quarks, which explained the octet and decuplet patterns \cite{GellMann:1964nj,Zweig:1981pd,Zweig:1964jf}.  Baryons are made up of three quarks (antibaryons are made of three anti-quarks), while mesons are made of a quark and an anti-quark.  Within this model, there are three ``flavors'' of quarks, called up, down, and strange.  Soon after, Greenberg, Nambu, and Han \cite{Greenberg:1964pe,Han:1965pf} proposed that quarks have an additional degree of freedom, later called color \cite{Lichtenberg:1978pc,GellMann:1981ph}.  There are three colors, red, green, and blue, and all hadrons (particles interacting via the strong interaction) are color-neutral.  Color explains why hadrons must be either baryons, anti-baryons, or mesons.  Later, it was determined that there are actually six flavors of quarks (the additional flavors being charm, bottom, and top), but these three flavors are much heavier than the original three \cite{Griffiths:2008zz}.  

The modern theory of the strong interaction, called quantum chromodynamics (QCD), is a Yang-Mills theory \cite{Yang:1954ek} with gauge group\footnote{See Ref.~\cite{Zee:2016fuk} for a review of Lie groups.} SU(3), an idea first proposed in \cite{Fritzsch:1972jv,Fritzsch:1973pi}.  A Yang-Mills theory describes fermionic matter fields interacting via a non-abelian gauge field.  In the case of QCD, the matter fields are quark fields $q_f = (q_R,q_B,q_G)^T$ of flavor $f\in \{d,u,s,c,b,t\}$ and the gauge theory is SU(3), which has 8 gauge bosons, called gluons $A^{\mu}_a$ where $a\in\{1,2,3,4,5,6,7,8\}$ \cite{Maggiore:2005qv}.  

The QCD Lagrangian is\footnote{Throughout this thesis, I use natural units where $\hbar \equiv 1$, $c \equiv 1$, and $k_B \equiv 1$, which implies that $\hbar c = (6.58\times 10^{-16} \text{ eV s})\times (3\times 10^8\text{ m/s}) = 197 \text{ MeV fm}.$  Also it implies that $1 \text{ MeV} = 1.16\times 10^{10} \text{ K}$.  In natural units, all dimensions are expressible in powers of energy units, typically MeV.  Thus, energy, mass, momentum, frequency, and temperature are measured in MeV, while distance and time are measured in $\text{ MeV}^{-1}$.  Velocity and entropy are dimensionless.  Newton's gravitational constant is $G = 6.70\times 10^{-45} \text{ MeV}^{-2}$.}
\begin{equation}
    \mathcal{L}_{\text{QCD}} = \sum_{f=1}^6 \bar{q}^f_i(i\slashed{\partial}-m_i)^{ij}q^f_j - g(\bar{q}^f_i\gamma^{\mu}T^a_{ij} q^f_j)A_{\mu}^{a} - \frac{1}{4}G_{\mu\nu}^a G^{a\,\mu\nu},
    \label{eq:QCD_L}
\end{equation}
where
$T^a = \lambda^a/2$ where $\lambda_a$ are the Gell-Mann matrices, that is, the $3\times 3$ traceless, Hermitian matrices that generate SU(3) (see Ref.~\cite{Halzen:1984mc} or \cite{Zee:2016fuk}).  The field tensor is  
\begin{equation}
    G_{\mu\nu}^a = \partial_{\mu}A_{\nu}^a-\partial_{\nu}A_{\mu}^a-gf_{abc}A_{\mu}^bA_{\nu}^c,
    \label{eq:gluon_field_tensor}
\end{equation}
where $f_{abc}$ are the structure constants \cite{Halzen:1984mc,Zee:2016fuk} of SU(3), defined through the equation $[T_a,T_b] = i f_{abc}T_c$.  

The Lagrangian can be expanded by using the definition of the gluon field tensor $G_{\mu\nu}^a$ [Eq.~(\ref{eq:gluon_field_tensor})] which becomes (schematically) $\mathcal{L} \sim ``\,\bar{q}q\," + ``\,G^2\," + g\,``\,\bar{q}qG\," + g\,``\,G^3\," + g^2\,``\,G^4\,"$.  Notice that there is just one coupling, $g$.  This is a result of the interaction being through a gauge field.  All of the allowed interactions in QCD can be read off from this schematic Lagrangian.  The first three interactions are much like QED - quarks and gluons have their kinetic terms and can couple to each other trilinearly with strength $g$.  However, the non-abelian gauge theory also allows gluons to couple to themselves, which makes QCD considerably more complex than QED \cite{Halzen:1984mc}.

\hiddensubsection{Asymptotic freedom}
The charge of a particle is modified by the polarization of the vacuum in which the charge sits.  Thinking pictorially for a moment, an electron placed in vacuum will polarize the vacuum, attracting virtual positrons which will surround the electron, reducing its ``visible'' charge.  When an observer tries to measure the charge, if they very weakly hit the screened electron with a probe, then the probe will see an electron with significantly reduced negative charge.  However if the observer throws a high-energy particle at the electron, it will penetrate the screening positrons, and will ``see'' an electron with higher negative charge.  In the theory of QED, this is expressed by the growth of the QED coupling with increased energy scales.

QCD acts in the opposite way - color charges are anti-screened.  Placing a color charge (say, a red quark) in vacuum will polarize the vacuum, but it will attract more virtual red charges, not anti-red\footnote{This phenomenon is traceable to the presence of gluon loops \cite{Griffiths:2008zz}.}.  Thus, when an observer weakly throws a particle at the red color charge, it sees a whole sea of red color charges adding up to a substantial net color charge.  However an observer throwing a very high energy particle at the red color charge will penetrate a large fraction of the sea of red color charges, and will only observe a small net color charge.  This counter-intuitive behavior is asymptotic freedom.  The QCD coupling decreases in magnitude with increased energy \cite{Halzen:1984mc}.

Asymptotic freedom can be formally derived by calculating the QCD beta function and noticing that it is negative.  This calculation was done for QCD by Politzer, Gross, and Wilczek \cite{Politzer:1973fx,Gross:1973id}.  Coleman and Gross later showed that only non-abelian gauge theories can be asymptotically free \cite{Coleman:1973sx}.  

Asymptotic freedom allows the use of perturbation theory to do high-energy QCD calculations, as the coupling is weak at high energy.  For example, QCD processes in colliders are calculated perturbatively in any particle physics or field theory textbook \cite{Halzen:1984mc,Thomson:2013zua,Greiner:2002ui,Schwartz:2013pla}.  At low energies, the QCD coupling becomes very large, and perturbation theory is inapplicable.  Here, the coupling is strong enough to confine quarks into hadrons, which are color-neutral.  As perturbation theory is impossible at low energies - relevant to nuclear physics - one must resort to models of QCD to make progress.

\section{The phase diagram of QCD}
\begin{figure}[h]
  \centering
  \includegraphics[scale=.7]{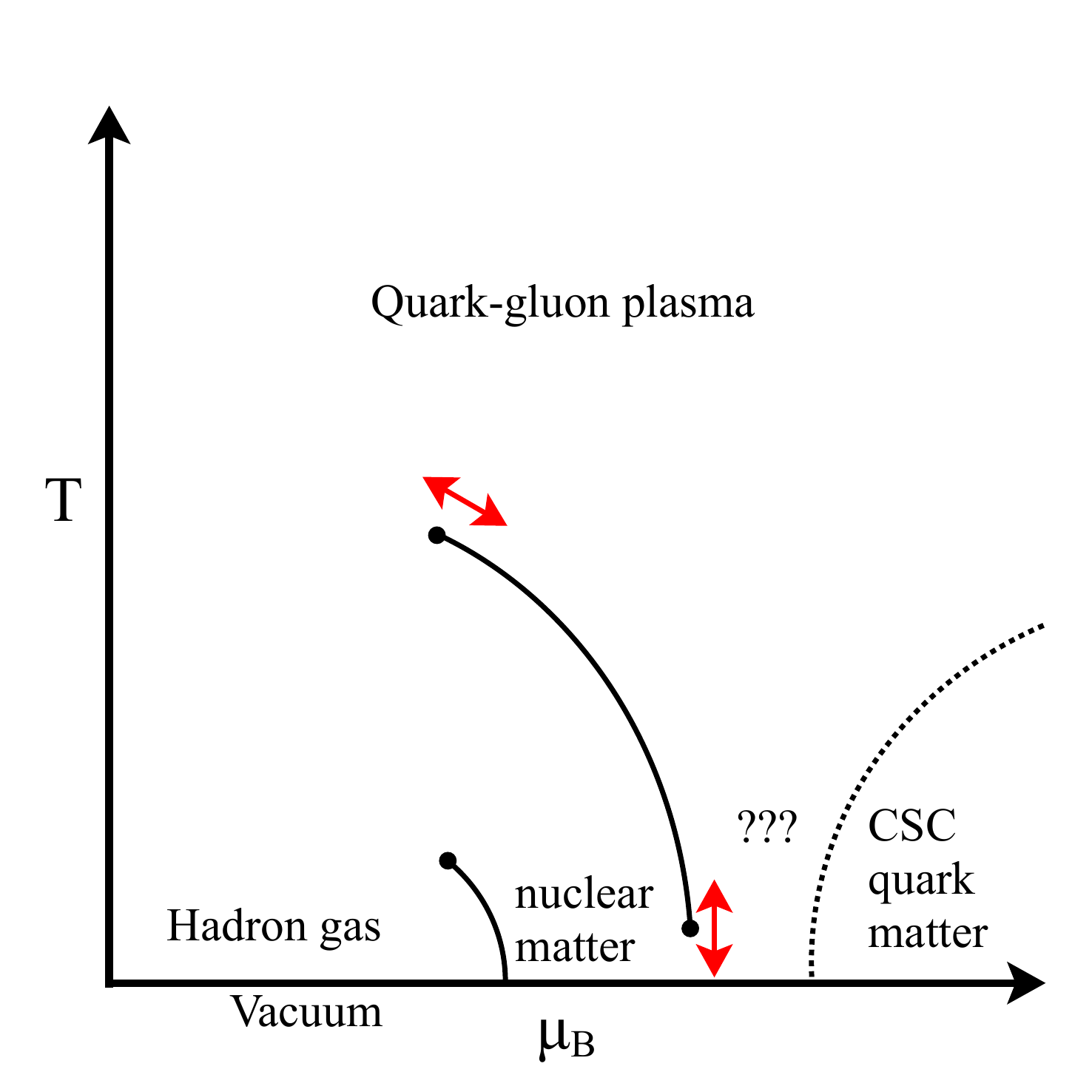}
  \caption{Schematic version of the QCD phase diagram at zero isospin chemical potential.  Matter described in this diagram is an equal mixture of up and down quarks (or neutrons and protons in the confined phase) and is governed only by QCD, not the weak interaction or electromagnetism.  The interior of the diagram is uncertain, for example, the first-order phase transition between the hadronic phase and the quark-gluon plasma has not yet been detected, and if it exists, it may have one or two critical points, whose locations are uncertain (this is indicated by the red arrows).  The dotted boundary between the color-superconducting (CSC) quark matter and quark-gluon plasma phases indicates that the type of transition between the two phases is uncertain.}
  \label{fig:qcd_phase_diagram}
\end{figure}

QCD can be studied coupled to external conditions, such as temperature, baryon chemical potential, and isospin chemical potentials.  These external ``knobs'' allow one to study the strong interaction as it would appear in a real-world system.  QCD, like many models, has different behavior at different values of these external parameters and so it is useful to construct a phase diagram to organize the behavior of QCD in different physical regimes.  This is analogous to how the behavior of water as a function of temperature and pressure is depicted in a phase diagram.  The typical presentation of the QCD phase diagram is with baryon chemical potential on the horizontal axis and temperature on the vertical axis, where the isospin chemical potential is set to zero, and contributions from the four heaviest quark flavors are neglected.  That is, the most common version of the QCD phase diagram is the phase diagram of an equal mixture of up and down quarks.  A schematic version of the QCD phase diagram is given in Fig.~\ref{fig:qcd_phase_diagram}.

In principle, studying 2-flavor QCD at finite baryon and isospin chemical potentials and finite temperature amounts to adding up and down quark chemical potentials to the QCD Lagrangian \cite{Alford:2007xm,Kogut:1999iv,Cohen:2015soa,Schmitt:2010pn}
\begin{align}
    \mathcal{L}_{\text{2-flavor QCD}} &= \bar{u}_i(i\slashed{\partial}+\mu_u\gamma^0-m_u)^{ij}u_j +  \bar{d}_i(i\slashed{\partial}+\mu_d\gamma^0-m_d)^{ij}d_j\label{eq:QCD_L}\\
    &- g(\bar{u}_i\gamma^{\mu}T^a_{ij} u_j + \bar{d}_i\gamma^{\mu}T^a_{ij} d_j)A_{\mu}^{a} - \frac{1}{4}G_{\mu\nu}^a G^{a\,\mu\nu},\nonumber
\end{align}
where the up and down quark chemical potentials $\mu_u$ and $\mu_d$ are determined by the baryon and isospin chemical potentials \cite{Wu:2020knd} via the relationships $\mu_u=(1/3)\mu_B+(1/2)\mu_I$ and $\mu_d = (1/3)\mu_B - (1/2)\mu_I$, which indicate that three quarks make one baryon, and each quark carries plus or minus one-half unit of isospin (this is just a convention).  At this point, we would like to calculate the QCD partition function, which is a functional integral over quark and gluon fields, and then we would calculate the pressure from the partition function (we use grand canonical ensemble, see Appendix \ref{sec:GCE_thermo}).  While the pressure is continuous as a function of temperature and chemical potentials, its derivatives (the baryon density and entropy density, see Appendix \ref{sec:GCE_thermo}) might not be.  Discontinuities in the thermodynamic derivatives indicate phase transitions, and so the phase diagram could be determined if we could calculate the partition function of QCD.  Of course, at present this is not possible, and so we have to rely on models and experiment to map out the QCD phase diagram.  

For the remainder of this chapter, we will survey the regions of the QCD phase diagram, Fig.~\ref{fig:qcd_phase_diagram}.  It should be noted upfront that there is consensus about the behavior of QCD matter only at $\mu_B=0$ and at asymptotically large $\mu_B$ at low temperature.  The rest of the QCD phase diagram is conjecture, inferred by models and some experiments.
\hiddensubsection{Hadron gas}
At low density and low temperature, QCD matter is in the hadron gas state.  In this state, quarks are confined and thus all states in this regime are color neutral.  If the temperature were zero, then no matter would be present at all - that is, the line on the phase diagram at zero temperature and $\mu_B \lesssim 1 \text{ GeV}$ represents the vacuum.  At finite temperature, particle states are populated.  With increasing temperature, heavier particle states have increasing probability of occupation.  

This region of the QCD phase diagram is often described with the hadron resonance gas model, which assumes that the QCD matter is the sum of independent Fermi or Bose gases corresponding to each hadron and hadron resonance in the particle data book below a certain mass \cite{Steinert:2018zni}.  Thus, this model is reliant on our knowledge of the spectrum of QCD \cite{Monnai:2019hkn,Karsch:2003vd,doi:10.1080/00107510110063843}.
\hiddensubsection{Quark-gluon plasma}
At sufficiently high temperatures, quarks are no longer confined into color-neutral hadrons, and instead QCD matter is a plasma of quarks and gluons.  At arbitrarily large temperatures, asymptotic freedom indicates that the coupling between the quark and gluon quasi-particles becomes weak, but the high temperature theory still contains non-perturbative physics \cite{Svetitsky:1982gs,DeGrand:1986uf} (see also the review \cite{Ogilvie:2012is}) and a weak-coupling treatment requires resummations of diagrams of all loop orders \cite{Ghiglieri:2020dpq}.  As the temperature decreases, the coupling increases and close to the deconfinement temperature, the coupling is strong enough that the quasi-particle mean free path is shorter than the de-Broglie wavelength of the quasi-particle, indicating that the quark-gluon plasma is so strongly coupled that it cannot be described in terms of quasi-particles.  In this regime, the quark-gluon plasma is a strongly coupled fluid, well described by hydrodynamics \cite{Busza:2018rrf,Romatschke:2017ejr}.
\hiddensubsection{Dense QCD matter}
While at high temperature and low density, lattice QCD (see Refs.~\cite{Zee:2003mt,DeGrand:2006zz,Gattringer:2010zz,Hjorth-Jensen:2017gss,Lin:2015dga} for review) is successful and serves as an anchor for models like the hadron resonance gas, at finite baryon density the sign problem prevents the use of lattice techniques.  The calculation of any observable in lattice QCD involves a calculation of the partition function, which is a functional integral over all possible configurations of the quark and gluon fields.  This extremely high dimensional integral is evaluated with Monte Carlo methods, which involve sampling the integrand according to a weighting function.  At finite baryon density, the weighting function gains an imaginary part, causing the integrand to be highly oscillatory and thus impossible to accurately evaluate with current computational resources \cite{Aarts:2015tyj,Stephanov:2007fk}.  Some progress has been made on mitigating the sign problem in certain situations \cite{Alford:1998sd,Fodor:2001au, deForcrand:2002hgr, Allton:2002zi,Alexandru:2015sua,Nishimura:2016yue,Medina:2017xbn}, and it is interesting to note that the sign problem does not appear in QCD at finite isospin chemical potential (with $\mu_B=0$) \cite{Son:2000xc}.  

Below we examine the QCD phase diagram at zero temperature.
\hiddensubsubsection{Nuclear matter onset}
When the baryon chemical potential is less than the mass of a nucleon ($m_N \approx 939 \text{ MeV}$), it is not possible to create neutrons and protons at zero temperature, and thus the state of QCD in the region is the vacuum.  However, when the baryon chemical potential reaches the mass of the nucleon\footnote{The chemical potential needed for the production of nuclear matter is slightly less than the nucleon mass, as the nuclear matter has binding energy of 16 MeV that lowers the threshold to around $939-16 = 923 \text{ MeV}$ \cite{Fukushima:2013rx}.  This binding energy occurs at a baryon density of around $n_0 \approx 0.16 \text{ fm}^{-3}$, which is termed nuclear saturation density \cite{dickhoff2008many,fetter2012quantum}.}, uniform nuclear matter becomes the equilibrium state of QCD.  

Nuclear matter is a liquid with neutron and proton quasiparticles.  As we are exploring the QCD phase diagram at zero isospin chemical potential, the nuclear matter is symmetric, meaning it has equal numbers of neutrons and protons.  In contrast to the hadron gas phase, the neutrons and protons are quasiparticles and interact with each other via the strong force.  There are many models of the nucleon-nucleon force, typically divided into nonrelativistic constructions of 2- and 3-body nucleon potentials or nucleons interacting via meson exchange as in a relativistic mean field theory \cite{Oertel:2016bki,Burgio:2018mcr}.  We will describe these models in Sec.~\ref{sec:nucl_matter_EoSs}.

At sufficiently low temperatures, nuclear matter is well described by a relativistic Fermi liquid theory with neutron and proton quasi-particles \cite{Friman:2019ncm,Baym:1975va}.  In addition, at sufficiently low temperatures (on the order of 1 MeV \cite{Chamel:2017wwp,Sedrakian:2018ydt}), pairing between nucleons is expected, analogous with BCS pairing of electrons \cite{Bardeen:1957mv}, as the nucleon-nucleon potential is attractive at long-range \cite{Sedrakian:2018ydt}.  Thus, nuclear matter is superfluid at nuclear densities and low temperatures.
\hiddensubsubsection{Dense quark matter}
\label{sec:dense_quark_matter}
At sufficiently high density, the nucleon quasiparticles in nuclear matter must touch, as they have a finite size.  If we assume a nucleon is a sphere of radius $R=0.8\text{ fm}$ \cite{basdevant2005fundamentals}, then two adjacent neutrons touching have a linear density of $0.625 \text{ fm}^{-1}$, which translates to a volume density of $n_{\text{touch}}=0.24\text{ fm}^{-3} = 1.5n_0$.  Taking into account more efficient packing arrangements \cite{callister2010materials} would push this touching density up to $n_{\text{touch}} = (4\sqrt{2}R^3)^{-1} \approx 2.2n_0$.  At densities much larger than this, it seems clear that quarks must be the relevant degree of freedom, not nucleons, and so quark matter becomes the ground state of QCD at high densities.

Though we know little about the transition from hadronic to quark matter, we expect at high enough densities, quark matter will be a Fermi liquid, with up and down quark quasiparticles.  As the density increases, the typical momentum transfer in an interaction between quarks near the Fermi surface becomes large\footnote{Small-angle scattering with low momentum transfer is cut off in the infrared by Landau damping of the gluons \cite{Son:1998uk}.} so asymptotic freedom dictates that at sufficiently high density quarks near their Fermi surface are weakly interacting \cite{Alford:2007xm}.  Additionally, the interaction between quarks on their Fermi surfaces is attractive, and so it is predicted that, below some critical temperature (which is several tens of MeV \cite{Alford:2007xm,Schmitt:2017efp,Sedrakian:2016ufm,Aguilera:2004ag}), quarks will form Cooper pairs, creating a superconducting state called a color superconductor.  One model for this type of matter is the 2SC (2-flavor, superconducting) phase \cite{Alford:2007xm}, where the up and down quarks can pair in any of the three combinations $uu,dd,ud$ \cite{shuryak2018quantum}.
\hiddensubsection{Phase transitions and critical points}
Phases in a phase diagram are separated by n\textsuperscript{th}-order phase transition lines, which in the case of the QCD phase diagram in Fig.~\ref{fig:qcd_phase_diagram}, are described by functions $\mu_B = \mu_B(T)$.  A first-order phase transition indicates a sharp change between the two phases, which shows up as a discontinuity in the first derivative of the pressure between the two phases, while n\textsuperscript{th}-order transitions indicate a discontinuity in the n\textsuperscript{th} derivative of the pressure.  Phase transitions of order greater than one are also called continuous phase transitions.  On a phase transition line, the phases on either side coexist.  Phase transition lines either terminate in a critical point or at the boundary of the phase diagram.  They can also terminate by intersecting another phase transition line, forming a triple point \cite{Goldenfeld:1992qy,nishimori2010elements}.

There could also be a crossover between two phases, where one can go from one phase to another by going around the critical endpoint, circumventing the n\textsuperscript{th}-order phase transition line.  A crossover is not really a phase transition, rather, it is a sign that two adjacent phases become indistinguishable in some region of the phase diagram (for example, liquid and gaseous water become indistinguishable at sufficiently large temperature.)
\hiddensubsubsection{Deconfinement phase transition at high temperature}
\label{sec:deconfinement}
The hadron gas state cannot exist at infinite temperature.  This was first pointed out by Hagedorn \cite{Hagedorn:1965st}, who noticed that the density of hadronic states in QCD increases exponentially in mass.  As the temperature of the hadron resonance gas increases, higher mass hadron states are increasingly populated and above the Hagedorn temperature $T_H$ (somewhere between 150-200 MeV), there would be so many hadronic states that any energy added to the system would excite further resonances, but would not increase the temperature of the system.  This shows up as a divergence in the partition function for $T>T_H$ \cite{Rafelski:2016hnq,Yagi:2005yb,Fukushima:2013rx,Fukushima:2010bq}.  Thus QCD matter must be in a different phase without hadronic degrees of freedom at temperatures above the Hagedorn temperature.

The deconfinement of quarks due to increasing temperature is analogous to deconfinement due to increasing density, explained earlier in this section.  In this case, high temperature leads to the increased overlap of the hadrons and hadron resonances, and at high enough temperature one can no longer connect quarks to specific hadrons, making quarks the proper degree of freedom.  This transition can be modeled as a percolation transition \cite{Yagi:2005yb,Satz:2008kb}, where one essentially asks the question ``how much overlap must hadrons have before they become quarks?''\footnote{As an example of percolation, Baym asks the question: if you make a block-pile out of blocks that are either copper or wood, what percentage of copper blocks is needed before the block-pile becomes a conductor? \cite{Baym:1979etb}.}

Lattice calculations show that the transition from a hadron gas to the quark-gluon plasma at $\mu_B=0$ is a smooth crossover\footnote{Some thermodynamic quantities do change quite rapidly as temperature exceeds 150 MeV.  For example, the energy density \cite{Satz:2008kb}.} \cite{Borsanyi:2013bia,Borsanyi:2010cj,Bazavov:2014pvz,Aoki:2006we}.  It is strongly suspected that the transition becomes first-order at some finite baryon chemical potential.  The current expectation is a first-order transition line running from high temperature and low density to low temperature and high density, shown in Fig. \ref{fig:qcd_phase_diagram}.  The first-order line is expected to terminate in a critical endpoint on the high temperature-low density end, and the low temperature-high density end of the line either continues down to zero temperature, or ends in a critical endpoint itself \cite{Busza:2018rrf,Stephanov:2004wx}.  These critical points are depicted in Fig.~\ref{fig:qcd_phase_diagram}.  However, at present there is no direct evidence for the presence of this critical line or its associated critical endpoint(s).  There is little theoretical agreement on the location of the high-temperature critical endpoint, though relativistic heavy ion collisions have ruled out a region of the phase diagram - see Sec.~\ref{sec:relativistic_HIC}.  Recent calculations predict it could lie at chemical potentials of more than 400 MeV and temperatures below 150 MeV - see Ref.~\cite{Stephanov:2007fk} for a summary of older predictions, and \cite{1790911,Critelli:2017oub,Ferreira:2018sun,Xu:2018kck} for more modern predictions.
\hiddensubsubsection{Liquid-gas transition}
There is a first-order phase transition at zero temperature between the vacuum and the onset of nuclear matter, where the baryon density (a first partial derivative of the pressure) switches sharply from 0 to nuclear density $n_0 = 0.16 \text{ fm}^{-3}$.  The first-order line extends to finite temperature, now separating a dense (liquid) hadron phase from a dilute hadron gas.  On the phase transition line, the liquid nuclear matter coexists with the hadron gas.  The first-order line ends in a critical point, and there is some evidence from low-energy heavy ion collisions that the critical point lies near $\mu_B \approx 924 \text{ MeV}$ and $T \approx 5-20 \text{ MeV}$ \cite{Fukushima:2013rx,Fukushima:2010bq,doi:10.1080/00107510110063843,Chomaz:2004nw,Csernai:1986qf}.
\hiddensubsubsection{High-density phase transitions}
As discussed earlier in this section, it is expected that as density increases, the quasiparticle nucleons in nuclear matter will become closer together until they eventually touch.  We estimated this would occur around $n_B \approx 2.2n_0$ if the (hard-sphere) nucleons form a face-centered cubic structure, which is the most efficient packing arrangement of uniform spheres.  Just like the deconfinement transition at high temperature, the quark-hadron transition has been modeled as a percolation transition \cite{Baym:1979etb,Satz:1998kg,Satz:2008kb,Celik:1980td}.  Additionally, some models of QCD, including the NJL model \cite{Klevansky:1992qe} and a random matrix model \cite{Halasz:1998qr}, predict a first-order phase transition between hadronic and quark matter, and thus a jump in baryon density between the two phases.

More recently, it has been proposed that the first-order line between hadronic matter and quark matter has two critical endpoints, one at high density and low temperature and the other at low density and high temperature \cite{Baym:2017whm,Hatsuda:2006ps,Zhang:2008ima}.  We depict this possibility in Fig.~\ref{fig:qcd_phase_diagram}.  This would mean that at very low temperature, below the critical endpoint, there is a smooth crossover between nuclear matter and quark matter.  This scenario is called quark-hadron continuity \cite{Baym:2017whm,Schafer:1998ef}.  It has also been proposed that in the crossover region, QCD matter goes through a series of spatially inhomogeneous phases as it transitions from nuclear to quark matter \cite{Baym:2017whm,Buballa:2014tba,Fukushima:2013rx}.
\hiddensubsection{Other two-dimensional planes of QCD phase diagram}
QCD has been studied at finite isospin chemical potential (with $\mu_B=0$), first by \cite{Son:2000xc,Son:2000by} and later by \cite{Cohen:2015soa}.  The most recent analysis \cite{Brandt:2017oyy} predicts four phases.  At low temperatures and densities there is a hadronic gas, which at high temperatures deconfines to a quark-gluon plasma.  Above a critical value of the isospin chemical potential, there is a second-order phase transition to a pion-condensed phase, which only exists up to a certain temperature where it smoothly transitions to a superconducting phase, which at high enough temperature turns into a quark-gluon plasma.  

Toublan and Kogut have studied QCD at finite baryon and isospin chemical potentials \cite{Toublan:2003tt}.  QCD at finite strangeness chemical potential has also been studied, as this regime is useful for heavy ion collisions \cite{Monnai:2019hkn,Toublan:2004ks,Fu:2018qsk,Noronha-Hostler:2019ayj,Mukherjee:2018yft}.
\section{Probes of the QCD phase diagram}
\hiddensubsection{Early universe}
A few microseconds after the big bang, the universe was very dense and very hot.  The hadronic content at this time was a quark-gluon plasma, with a nearly even ratio of quarks to antiquarks, meaning its baryon chemical potential was close to zero.  As the universe expanded, it cooled and eventually reached temperatures of 150-200 MeV, at which point it passed through the confinement crossover and the quarks became confined in hadrons.  The early universe provides evidence that the QCD phase transition at $\mu_B\approx 0$ is not strongly first-order, as that would disrupt big bang nucleosynthesis \cite{Thomas:1993md,Busza:2018rrf}.  
\hiddensubsection{Heavy ion collisions}
\hiddensubsubsection{Low energy}
Low-energy heavy ion collisions can be used to see the liquid-gas phase transition.  Unlike current heavy ion collisions at RHIC or the LHC, these collisions are not energetic enough to create a quark-gluon plasma.  There is some evidence that these collisions have detected signatures of the liquid-gas critical point, specifically, the observation \cite{Chomaz:2004nw} of a particular distribution of nuclear fragments \cite{bertulani2007nuclear}.  
\hiddensubsubsection{Relativistic}
\label{sec:relativistic_HIC}
\label{sec:RHIC}
\begin{figure}[h]
  \centering
  \includegraphics[scale=.35]{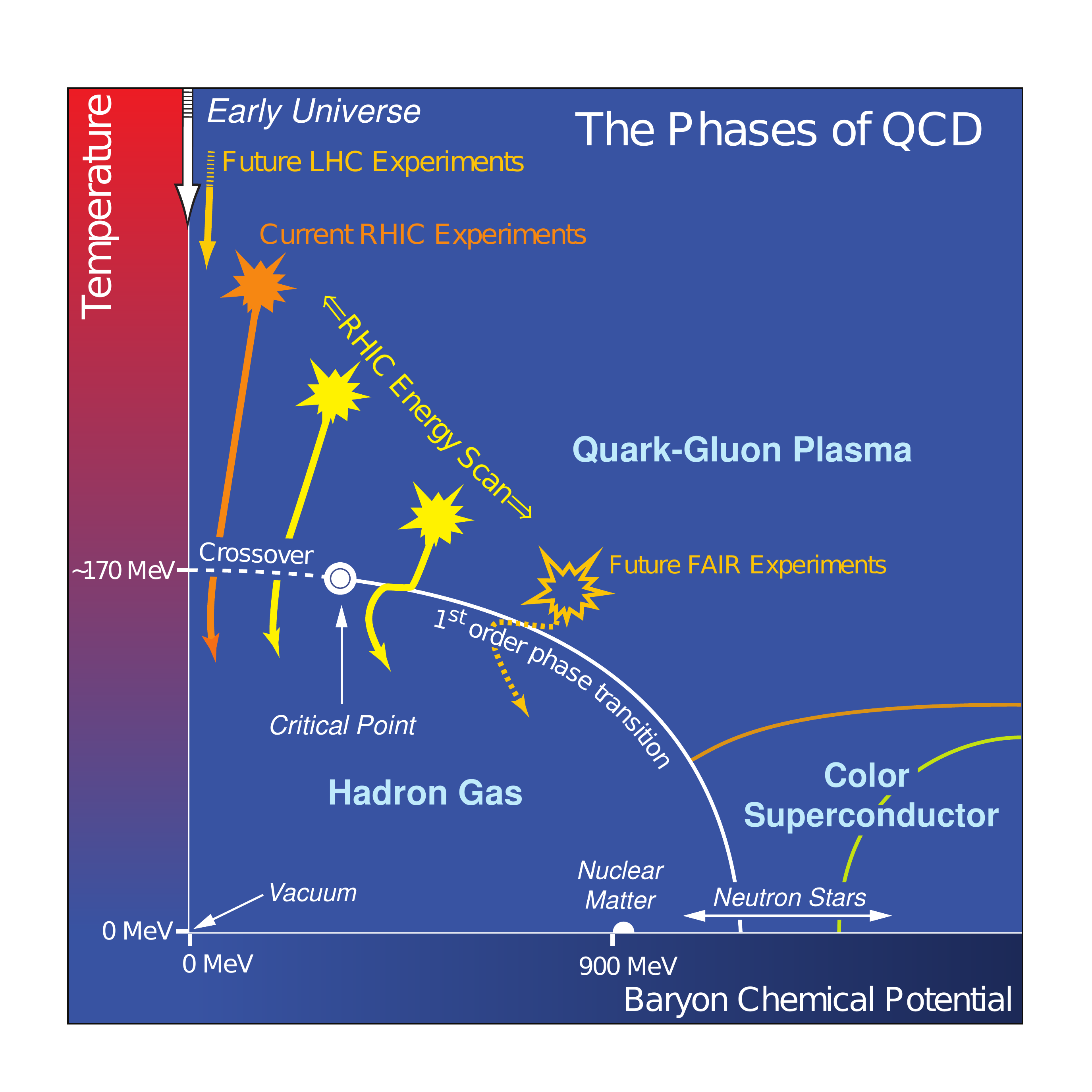}
  \caption{Beam energy scan coverage of the QCD phase diagram, indicating the desired path through the QCD phase diagram of the QCD matter in a heavy ion collision, from quark-gluon plasma to a hadron gas.  Figure reproduced from \cite{longrangeplan}.}
  \label{fig:beam_scans}
\end{figure}
Relativistic heavy ion collisions have center of mass energy\footnote{This is usually quoted \textit{per nucleon} \cite{Yagi:2005yb}.} of greater than the rest mass of the two nuclei\footnote{A proton has mass 940 MeV, a gold nucleus 183 GeV, and a lead nucleus 194 GeV.}, so that the nuclei move with speeds close to the speed of light and thus are significantly Lorentz contracted \cite{Romatschke:2017ejr}.  Because the overlapping nuclei have such high energy density, a quark-gluon plasma is produced and is then stretched longitudinally in the wake of the two receding nuclei.  Heavy ion collisions at RHIC provided the first evidence of the strongly-coupled nature of the quark-gluon plasma, including elliptic flow which showed that the QGP behaved like a fluid \cite{Heinz:2008tv}.  The energy of the collision determines the baryon chemical potential in the quark-gluon plasma.  High-energy collisions (like those at the LHC) produce QGP close to zero baryon chemical potential, while collisions at lower energy (like those at RHIC) have higher baryon content.  

\begin{figure}[h]
  \centering
  \includegraphics[scale=.4]{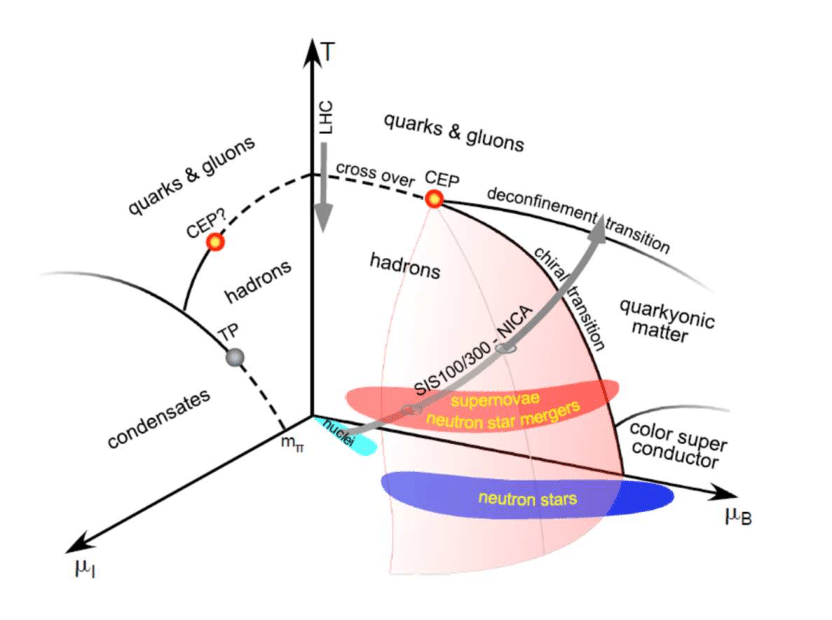}
  \caption{Schematic phase diagram of QCD with axes for temperature, baryon chemical potential and isospin chemical potential.  Also shown are regions of the phase diagram probed by different experiments and astrophysical situations.  Figure reproduced from \cite{NuPECC17-LRP}.}
  \label{fig:3d_phase_diagram}
\end{figure}

RHIC has the ability to adjust its beam energy, and so it is currently running ``beam energy scans'' in an effort to locate the QCD critical point (the one located at higher temperature).  The experiment is set up for nuclei to collide at a certain center of mass energy, creating QGP at a certain temperature and baryon density.  The QGP expands, cooling and decreasing its density as indicated in Fig.~\ref{fig:beam_scans}, crossing into the hadron gas region.  So far, heavy ion collisions have not detected the first-order phase transition line.  See Ref.~\cite{Keane:2017kdq} for a summary of results of the first beam energy scan at RHIC.  Relativistic heavy ion collisions probe the high temperature and low baryon density region of the QCD phase diagram, but they are aiming to reach higher densities in search of the critical point \cite{Brewer:2018abr}.
\hiddensubsection{Neutron stars and neutron star mergers}
\label{sec:ns_and_ns_mergers}
As we will discuss in detail in Ch.~\ref{sec:NS}, neutron stars probe the high density and low temperature portion of the phase diagram, as they are made of cold nuclear matter and perhaps cold quark matter.  In addition, they do not have an equal number of neutrons and protons, so they actually probe QCD at finite isospin chemical potential, which can be viewed as a third axis of the QCD phase diagram, shown in Fig.~\ref{fig:3d_phase_diagram}.  Neutron stars, like all astrophysical objects, are governed not only by the strong interaction but also by the weak and electromagnetic interactions.  The weak interaction sets the ratio of neutrons to protons in the nuclear matter.  Due to electromagnetism, neutron stars, like all long-lived astrophysical objects, are charge neutral\footnote{Glendenning calculates that the excess number of charged particles per baryon must be less than $10^{-36}$ or else the astrophysical object will eject material due to Coulomb repulsion \cite{glendenning2000compact}.}.  

When two neutron stars merge, their temperature rises rapidly and they probe a hotter, but still very dense region of the phase diagram, as indicated in Fig.~\ref{fig:3d_phase_diagram}.  Neutron star mergers will be discussed in detail in Ch.~\ref{sec:mergers}, and for the rest of this thesis.

\chapter{Compact stars and the nuclear equation of state}
\label{sec:NS}
\pagestyle{myheadings}
\section{Neutron star observations}
\label{sec:ns_observations}
We now know that neutron stars are the remnants of main-sequence stars that underwent core collapse supernovae.  Core collapse occurs once a main-sequence star has extracted all of the energy that it could from fusing elements in its core, and now, left with nothing but an iron core from which no energy can be gained, begins gravitational collapse.  As the stellar material collapses, its density increases and eventually it reaches nuclear density, where a combination of neutron degeneracy pressure and neutron-neutron hard-core repulsion make the material stiff enough to halt the gravitational collapse, ejecting much of the remainder of the infalling material.  This bounce is only possible if the original star was between about 8-30 solar masses.  If the star is lighter than that, the collapse never occurs, and if it is heavier, then not even the neutron repulsion can prevent the star from collapsing to a black hole.  The dense material left behind in the core collapse supernova is a neutron star \cite{glendenning2000compact,carroll2017introduction,Cerda-Duran:2018efz}.  

Soon after James Chadwick discovered the neutron in 1932, Baade and Zwicky predicted the existence of neutron stars, and even their birth in core collapse supernovae \cite{Baade254,Baade259}.  In 1939, Tolman \cite{Tolman:1939jz} and Oppenheimer and Volkoff \cite{Oppenheimer:1939ne} calculated, assuming a very simplistic model of nuclear matter, the structure of neutron stars (see Sec.~\ref{sec:tov}) and predicted neutron stars of 0.3-0.7 solar masses, with radii of 3-20 kilometers. The first neutron star observation occurred in 1967 when Hewish and Bell observed with their radio telescope a periodic signal occurring on intervals of just over one second \cite{Hewish:1968bj}.  They called the source of this signal a pulsar, and we now know that pulsars are rapidly rotating neutron stars with high magnetic fields.  The neutron star magnetic field causes electromagnetic beams to be emitted along an axis that is different from the rotation axis, and when that axis sweeps across earth as the star rotates, we see a periodic electromagnetic signal in the radio frequency range.  A review of pulsar physics and the history of neutron star and pulsar discoveries is given in \cite{glendenning2000compact,Shapiro:1983du}.

Neutron star observation is still of great interest at present.  In particular, it is very important to measure the mass and radii of neutron stars, which puts constraints on the nuclear equation of state (see Sec.~\ref{sec:EoS}).  While most neutron stars have masses of around 1.4 solar masses, three neutron stars with masses of around 2 solar masses have been found \cite{Demorest:2010bx,Antoniadis:2013pzd,Cromartie:2019kug} using Shapiro delay, a measurement that takes advantage of the warping of spacetime around a compact object \cite{thorne2017modern,zee2013einstein}.  These findings eliminated many models of the nuclear equation of state \cite{Haensel:2016yga,Lattimer:2010uk,Lattimer:2019eez}.

The NICER experiment, an x-ray telescope on the International Space Station, has recently measured the radius of one neutron star by modeling how light emitted from the back of the star makes it to the front of the star due to the extreme curvature of spacetime.  The radius was found to be likely between 11-14 km \cite{Riley_2019,Raaijmakers_2019,Miller_2019}.  

The gravitational waves and electromagnetic signals that we observe from neutron star mergers has significantly improved our understanding of nuclear matter, but we defer a discussion of mergers to Ch.~\ref{sec:mergers}.
\section{Nuclear equation of state}
\label{sec:EoS}
The equilibrium behavior of matter is a fundamental question in physics, and is only understood in the mild conditions we encounter here on Earth.  We will focus on matter that is charge-neutral and in equilibrium with respect to the weak interaction (called beta equilibrium), as this is the matter of astrophysical interest\footnote{For example, we discussed in Sec.~\ref{sec:ns_and_ns_mergers} the necessity for charge neutrality of neutron stars.  Further, isolated neutron stars must be close to beta equilibrium, because if their particle content was rapidly changing then it is unlikely they would stay intact for millions of years, as we expect that they do \cite{2010PhyU...53.1235Y,1995MNRAS.276L..21C}.}.  The information about the ground state of this matter as a function of baryon density and temperature is encapsulated in the equation of state of the matter.   The equation of state contains the complete thermodynamic description of the matter, including the pressure and its thermodynamic derivatives (see Appendix \ref{sec:GCE_thermo}) as a function of temperature and baryon density (or baryon chemical potential).  

Constructing an equation of state involves first choosing particle degrees of freedom and second, choosing how to model their interaction.  These steps are done implicitly when writing down a Hamiltonian or Lagrangian\footnote{An exception would be the QCD Lagrangian, which has quark and gluon degrees of freedom, however we know that at low temperatures, hadrons are the relevant degrees of freedom, not quarks.}.  The equation of state is the thermodynamic characterization of matter made of the chosen particles interacting in a specific way.  For example, the ideal gas law $P = nT$ is an equation of state, as it encapsulates the thermodynamic behavior of a non-interacting, one-component classical gas \cite{zee2013einstein}.  Often, the relationship between the energy density and pressure $\varepsilon = \varepsilon(P)$ is also called the equation of state.

Understanding the behavior of charge-neutral, beta-equilibrated nuclear matter is not exactly the same as understanding the QCD phase diagram, because the QCD phase diagram neglects the electromagnetic and weak interactions, which have important contributions to the nuclear matter in neutron stars.  With that said, uncertainty in the QCD phase diagram is by far the greatest obstacle to understanding the matter in neutron stars.

\begin{figure}[h]
  \centering
  \includegraphics[scale=1]{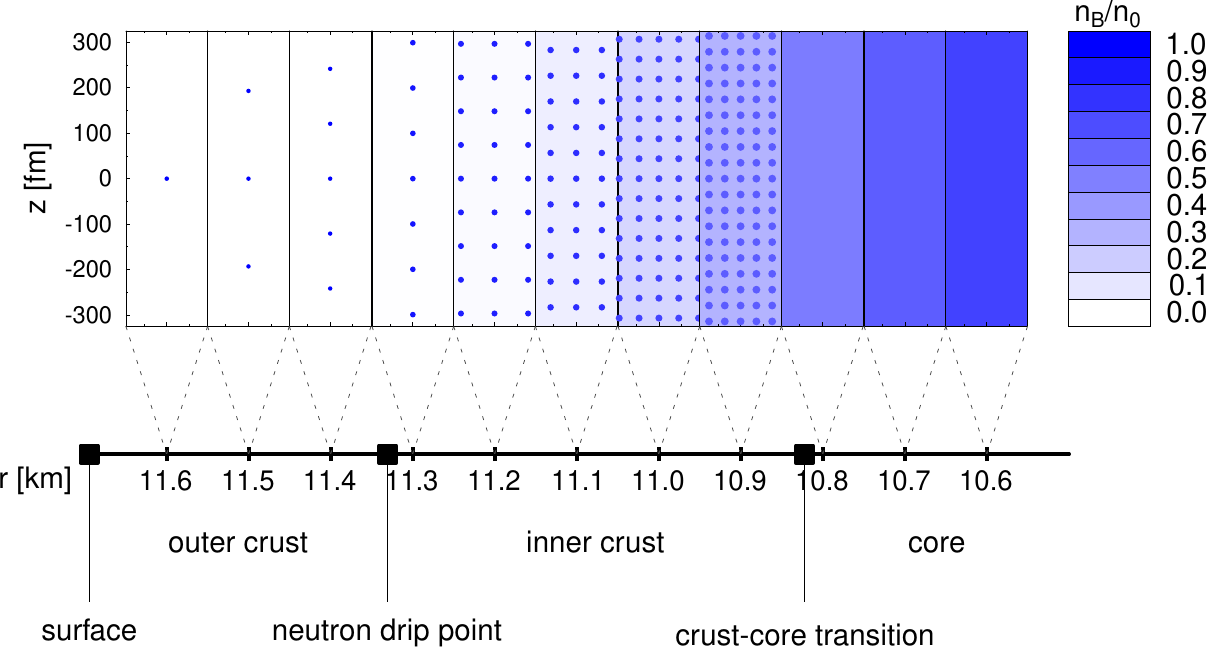}
  \caption{Depiction of the ground state of charge neutral, beta equilibrated nuclear matter as a function of density, showing the transition from a lattice of spherical nuclei to uniform nuclear matter as density increases.  Figure reproduced from \cite{Oertel:2016bki}.}
  \label{fig:oertel_nuclear_matter}
\end{figure}

Neutron stars provide the ideal physical situation to study the behavior of high-density nuclear matter.  As we will discuss in Sec.~\ref{sec:tov}, the baryon density at the edge of a neutron star (the crust) starts out small, and grows as the core is approached.  So, a neutron star should exhibit the entire range of equilibrium states of nuclear matter from the crust to the core.  A schematic illustration of the low density region of a neutron star is shown in Fig.~\ref{fig:oertel_nuclear_matter}.  At the edge of the star, there is a low density lattice of nuclei, but as the core is approached, the density increases (due to the weight of the stellar material above) and eventually the nuclei get close enough together to form uniform nuclear matter.

Below, we will discuss the charge-neutral, beta-equilibrated nuclear matter equation of state at zero temperature, which is relevant for neutron stars.  Then we will discuss extensions to finite temperature, which are particularly relevant for neutron star mergers.  Throughout the rest of this section, I will discuss equations of state that we use throughout the rest of this thesis.
\hiddensubsubsection{Nuclear matter at low density and the neutron star crust}
\label{sec:NS_crust}
The ground state of nuclear matter at low density is known to us.  At terrestrial densities, nuclear matter arranges itself into a lattice of nuclei with electrons localized around each nucleus, forming a shell structure.  The lattice takes whatever structure minimizes the Coulomb energy\footnote{The density is too low for the short-ranged nuclear force to have an impact, except for determining the mass of the nucleus.}.  On Earth, we have many solids, for example, gold, lead, iron, and silicon.  The solid with the lowest energy is the ground state of charge neutral, beta equilibrated nuclear matter, and all other states are only metastable (albeit with very long lifetimes).  We know from nuclear physics that \ce{^{56}Fe} is the nucleus with the lowest energy per nucleon \cite{Shapiro:1983du,Hartle:2003yu}, and thus the ground state of nuclear matter at low densities is a lattice of atoms of \ce{^{56}Fe}.

The ground state of nuclear matter at low densities was calculated by Feynman, Metropolis, and Teller in \cite{Feynman:1949zz}, where they used Thomas-Fermi theory to model the electronic structure to see how the electrons delocalized as the density increased.  This happens because an increase in density increases the momentum of the electrons, and eventually an electron has enough energy to escape the atom.  The Feynman, Metropolis, and Teller calculation was valid from terrestrial densities of several $\text{g/cm}^3$ up to $1000 \text{ g/cm}^3$, at which point the iron atoms are essentially completely deionized and the electrons form a Fermi sea.

A calculation of the equation of state of nuclear matter from $1000 \text{ g/cm}^3$ to $4.3\times 10^{11} \text{ g/cm}^3$ was done by Baym, Pethick, and Sutherland (called the BPS equation of state) \cite{Baym:1971pw}.  They assume the nuclei form a BCC lattice (which is the minimum energy configuration for a one-component crystal) and they determine which isotope $(A,Z)$ minimizes the energy.  The energy density of a one-component BCC lattice of nuclei with a Fermi sea of electrons is
\begin{equation}
    \varepsilon(A,Z,n_N) = n_N(W_N + W_L) + \varepsilon_e(n_e),
    \label{eq:coulomb_energy}
\end{equation}
where $A$ is the total number of nucleons in the nucleus, $Z$ is the number of protons in the nucleus, $n_N$ is the number density of nuclei, $W_N$ is the energy of an isolated nucleus (due to the strong interaction)\footnote{This quantity comes from experiment, as we are unable to calculate nuclear masses from QCD.  Even to this day, the masses of many neutron-rich nuclei are unknown, and upcoming experiments like those at FRIB \cite{Balantekin:2014opa} will try to make inroads on that section of the isotopic chart.}, and $W_L$ is the lattice energy per nucleus, which is $W_L\approx -1.44 Z^2e^2n_N^{1/3}$ for a BCC lattice.  The electron energy density at zero temperature is
\begin{equation}
    \varepsilon_e = 2\int_0^{k_{Fe}}\frac{\mathop{d^3k}}{(2\pi)^3}\sqrt{k^2+m_e^2} = \frac{1}{8\pi^2}\left[k_{Fe}\sqrt{k_{Fe}^2+m_e^2}\left(2k_{Fe}^2+m_e^2\right)-m_e^4\arcsinh{\left(\frac{k_{Fe}}{m_e}\right)}\right],
\end{equation}
where the electron Fermi momentum $k_{Fe} = (3\pi^2n_e)^{1/3}$.  Since $n_N = n_B/A$ and $n_e = n_B(Z/A)$ (due to charge neutrality), the expression for the system energy density [Eq.~(\ref{eq:coulomb_energy})] is a function $\varepsilon=\varepsilon(n_B,A,Z)$.  $W_N(A,Z)$ is taken from nuclear mass tables, for each possible isotope (A,Z).  The minimum energy state is determined by calculating the energy density $\varepsilon=\varepsilon(n_B,A,Z)$ for a given density and for each possible combination of $A$ and $Z$, and seeing which isotope has the lowest energy.  Through this calculation, Baym, Pethick, and Sutherland found that nuclear matter is a BCC lattice of \ce{^{56}Fe} at low density, and then as density increases the nuclei become increasingly neutron rich, going through the sequence \ce{^{56}Fe},\ce{^{62}Ni},\ce{^{64}Ni},\ce{^{84}Se},\ce{^{82}Ge},\ce{^{80}Zn},\ce{^{78}Ni},\ce{^{76}Fe},\ce{^{124}Mo},\ce{^{122}Zr},\ce{^{120}Sr}, and \ce{^{118}Kr}.  Throughout this sequence, the proton fraction $x_p = n_p/(n_p+n_n)$ goes from 0.46 down to 0.31.  Dense matter is increasingly neutron-rich, a trend that continues to densities well beyond the crust.

We will use the BPS equation of state to model the crust of a compact object in Sec.~\ref{sec:strange_dwarf}.  The BPS equation of state is used to model neutron star crusts even today, though improvements have been made, stemming from an increased understanding of exotic nuclei \cite{Oertel:2016bki}.

As the density increases beyond $4.1\times 10^{11} \text{ g/cm}^3$, the nuclei become so neutron-rich that the outer-shell neutrons\footnote{In the sense of the shell model of the nucleus \cite{dickhoff2008many}.} become unbound and now the ground state of matter is a lattice of neutron-rich nuclei with Fermi seas of electrons and neutrons.  This density is called the neutron drip density \cite{Baym:2017whm,Shapiro:1983du,Baym:1971pw,Hartle:2003yu}.  
\hiddensubsubsection{Nuclear pasta and the inner crust}
As the density increases past neutron drip, eventually the nuclei get sufficiently close together that adjacent nuclei are now within the interaction range of the strong force.  Now, the equilibrium structure of matter is determined by competition between the strong force and the Coulomb force.  This new competition between forces defines the inner crust of the neutron star.  

The competition between the two forces seems to result in a transition from spherical nuclei to uniform nuclear matter, with many new nuclear shapes in between, including cylinders and slabs, and hence this type of matter is called nuclear pasta \cite{Caplan:2016uvu,Oyamatsu:1993zz,Ravenhall:1983uh}.  These shapes appear for a range of densities around $0.2n_0$ to $0.7n_0$, depending on the model of the nuclear force used.  For some models of the strong interaction, pasta phases do not appear at all.  Both classical \cite{Schneider:2013dwa} and quantum \cite{Fattoyev:2017zhb,Nandi:2017aqq} simulations have been used to map out the array of possible pasta structures.
\hiddensubsubsection{Uniform nuclear matter and the neutron star mantle}
\label{sec:nucl_matter_EoSs}
At sufficiently high densities (around nuclear saturation density $n_0$), the nuclei (or pasta structures) are so close together that they become uniform nuclear matter.  In this regime, adjacent nucleons are well within the range of the strong force, and so force between nucleons is that of the strong interaction.  Electromagnetism and the weak interaction help to control the composition of the matter, which is a uniform fluid of neutrons, protons, and electrons.  The mantle and perhaps the core of a neutron star is made up of this matter.  As the mantle and core of a neutron star comprise most of the volume of the star, often neutron star matter as a whole is treated with the equation of state of uniform nuclear matter.  The equation of state of nuclear matter is typically calculated by constructing a non-relativistic model of the nuclear potential or by constructing a relativistic mean field theory.  We will go into detail about certain equations of state in the remainder of this section, as we will use those equations of state to model neutron star matter throughout the rest of this thesis.

\begin{figure}[h]
  \centering
  \includegraphics[scale=.6]{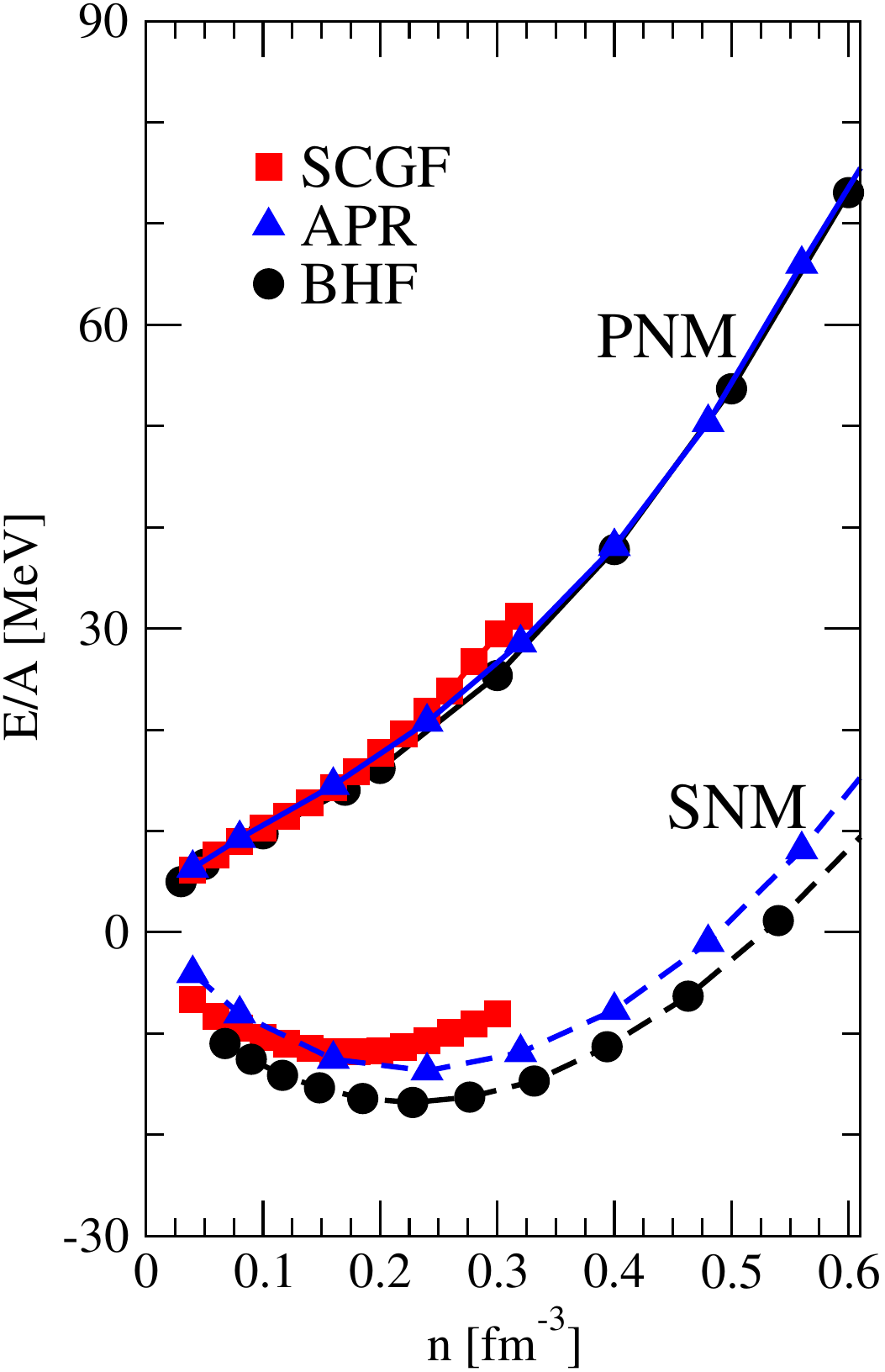}
  \caption{Energies of symmetric nuclear matter and neutron matter predicted by the APR equation of state as a function of baryon number density, shown in blue triangles.  The energy in beta equilibrium is found by interpolating between these two extremes, as explained in the text.  Figure reproduced from \cite{Burgio:2018mcr}.}
  \label{fig:APR_energies}
\end{figure}

\textit{Non-relativistic nuclear potentials}: One approach to constructing an equation of state for uniform nuclear matter is to start with a model of the interaction between nucleons, as is done in the APR equation of state \cite{Akmal:1998cf}.  The nucleon interaction is assumed to consist of interactions between two nucleons, along with interactions between three nucleons, along with interactions between four nucleons, and so on.  Chiral effective field theory \cite{Epelbaum:2008ga,Weinberg:1990rz} provides a way to systematically organize the nucleon n-body interactions, and it suggests that at densities up to $2n_0$, 4-body and higher interactions are small \cite{Baym:2017whm,Epelbaum:2008ga}.  Thus the nuclear Hamiltonian can be written as
\begin{equation}
    H = \sum_i \frac{p_i^2}{2m} + \sum_{i<j}v_{ij}+\sum_{i<j<k}V_{ijk},
    \label{eq:H_APR}
\end{equation}
where $v_{ij}$ is a nucleon two-body interaction and $V_{ijk}$ is a three-body interaction \cite{Wiringa:1988tp}.  The APR calculation used the A18 two-body potential \cite{Wiringa:1994wb} and the UIX three-body potential \cite{Pudliner:1995wk}.  These potentials are linear combinations of functions of the separation between the nucleons as well as their spin, isospin, and angular momentum, weighted by coefficients that are fit to nucleon scattering data and properties of the deuteron.  The energy of neutron matter and symmetric nuclear matter as a function of baryon density are calculated from this Hamiltonian using the variational principle \cite{Wiringa:1988tp}, and are plotted in Fig.~\ref{fig:APR_energies}.
The APR equation of state governing nuclear matter contains neutrons and protons, but also electrons and muons\footnote{At zero temperature, leptons of mass $m_i$ appear when $\mu_e > m_i$.  As we will see in the coming sections, $\mu_e$ lies between about 100-300 MeV in nuclear matter at neutron star densities, so the electron ($m_e = 0.511 \text{ MeV}$) and the muon ($m_{\mu} = 106 \text{ MeV}$) appear in neutron star matter.  The tau lepton ($m_{\tau}\approx 1800 \text{ MeV}$) has no chance of appearing in a neutron star.}.  The electrons and muons are treated as free Fermi gases (see Appendix \ref{sec:free_fermi_gas}).  The condition of charge neutrality ($n_p = n_e + n_{\mu}$) is enforced, and the energy of beta-equilibrated nuclear matter at a given baryon density is found by interpolating\footnote{The energy is assumed \cite{Akmal:1998cf} to depend on the proton fraction $x_p$ via the combination $1-2x_p$.} between the energies of $x_p=0$ and $x_p=0.5$ matter that are shown in Fig.~\ref{fig:APR_energies} until the beta equilibrium condition ($\mu_n = \mu_p + \mu_e$ and $\mu_e = \mu_{\mu}$) is satisfied\footnote{This expression for beta equilibrium will be explained in detail in Ch.~\ref{sec:beta_equilibrium}.}.  The APR equation of state is only intended to be valid at low temperature\footnote{Cold equations of state like APR can be extended to finite temperature, see Ref.~\cite{Raithel:2019gws}.}, so zero-temperature thermodynamic identities can be used.  The baryon chemical potentials in the beta equilibrium condition are found from the energies through the thermodynamic identity $\mu_i = \partial E/\partial N_i$.  The total energy is the sum of the energies of the neutrons and protons found through the interpolation plus the electron and muon energies.  The total pressure can be found from the energy through the identity $P = n_B^2\partial E(n_B)/\partial n_B$, and then the equation of state $\varepsilon = \varepsilon(P)$ can be calculated \cite{Akmal:1998cf,Wiringa:1988tp}.

\textit{Relativistic mean field theory}: Relativistic mean field (RMF) theory is a theory of neutrons and protons that interact via the strong interaction, which is modeled as meson-exchange.  To figure out which mesons are exchanged, we look for the lightest mesons in the particle data book which do not change strangeness (as we are only interested here in the force between nucleons, which have no strangeness).  The lightest such mesons are the pion, the sigma (or $f_0(500)$ resonance \cite{Pelaez:2015qba}), the rho, and the omega. In the mean field approximation in infinite nuclear matter, the ground state expectation value of the pion vanishes because the pion changes parity.  Thus, the pion does not affect the equation of state $\varepsilon = \varepsilon(P)$, so we will neglect it\footnote{However, when calculating the rate of certain particle processes, the pion-nucleon coupling should be considered in the Lagrangian.  For example, nucleon-nucleon scattering proceeds via light meson exchange, and the pion is the lightest meson so the one-pion-exchange diagrams will be the dominant contribution to the rate \cite{OPE,1979ApJ...232..541F,Hannestad:1997gc,Vidana:2018bdi,Ofengeim:2019fjy}.} \cite{glendenning2000compact}.

In the RMF formalism, we write down a Lagrangian with neutrons, protons, sigmas, omegas, and rhos, which can interact in various ways and with different coupling strengths.  Different choices of interactions and coupling strengths will give rise to different equations of state.  Of course, the Lagrangian will include free electron and muon fields, as well as neutrinos if studying neutrino-trapped nuclear matter (see Sec.~\ref{sec:nu_mfp}).  Dutra \textit{et al.} \cite{Dutra:2014qga} created a classification system for the different types of RMFs, and then discussed their compatibility with experiment \cite{Dutra:2014qga,Dutra:2015hxa,Lourenco:2018dvh}.  In this thesis, we will use the DD2 \cite{Typel:2009sy,Hempel:2009mc}, IUF \cite{Hempel:2009mc,Fattoyev:2010mx,RocaMaza:2008ja}, and NL$\rho$ \cite{Liu:2001iz} RMFs, which are type 5,4, and 2, respectively.  

The original RMF was the Walecka model, which included ``nucleons'' (there was no distinction between neutrons and protons) which interacted by exchanging sigma and omega mesons \cite{Walecka:1974qa}.  The rho meson was later added to allow for differentiation between neutrons and protons.  To force the RMF to better reproduce nuclear data, one of two additions was made.  Some RMFs added nonlinear couplings of the sigma, omega, and rho mesons, and adjusted the coupling constants to fit nuclear observables (this choice corresponds to models of type 2, 3, and 4 in Dutra \textit{et al.}'s classification).  Other RMFs forsook nonlinear meson couplings and turned the nucleon-meson coupling constants into functions of the baryon density (type 5 in Dutra \textit{et al.}'s classification)\footnote{This approach requires the addition of ``rearrangement'' terms in the Lagrangian \cite{Typel:2018cap,1995PhLB..345..355L}.}

I will focus here on the NL$\rho$ equation of state, and refer the reader to the discussion of Dutra \textit{et al.} \cite{Dutra:2014qga} for the full details of the other possible RMF theories.   The thermodynamic data from several of these RMF theories is tabulated on CompOSE, an online repository \cite{CompOSE}.

The NL$\rho$ Lagrangian is\footnote{We use parameter set 1 in \cite{Liu:2001iz}, which includes the sigma, omega, and rho mesons, but neglects the $\delta$ meson.  Application of this RMF to calculation of weak interaction rates is given in Ref.~\cite{Fu:2008zzg}.}
\begin{align}
    \mathcal{L} &= \bar{\psi}_n\left(i\gamma_{\mu}\partial^{\mu}-(m-g_{\sigma}\sigma)-g_{\omega}\gamma_{\mu}\omega^{\mu}-g_{\rho}\gamma^{\mu}\mathbf{t}\cdot\mathbf{\rho}_{\mu}\right)\psi_n\nonumber\\
    &+ \bar{\psi}_p\left(i\gamma_{\mu}\partial^{\mu}-(m-g_{\sigma}\sigma)-g_{\omega}\gamma_{\mu}\omega^{\mu}-g_{\rho}\gamma^{\mu}\mathbf{t}\cdot\mathbf{\rho}_{\mu}\right)\psi_p + \frac{1}{2}(\partial_{\mu}\sigma\partial^{\mu}\sigma-m_{\sigma}^2\sigma^2)\nonumber\\
    &-\frac{1}{3}bm(g_{\sigma}\sigma)^3-\frac{1}{4}c(g_{\sigma}\sigma)^4-\frac{1}{4}\omega_{\mu\nu}\omega^{\mu\nu}+\frac{1}{2}m_{\omega}^2\omega_{\mu}\omega^{\mu}-\frac{1}{4}\rho_{\mu\nu}\rho^{\mu\nu}+\frac{1}{2}m_{\rho}^2\mathbf{\rho}_{\mu}\cdot\mathbf{\rho}^{\mu}\nonumber\\
    &+\bar{\psi}_e(i\gamma_{\mu}\partial^{\mu}-m_e)\psi_e + \bar{\psi}_{\nu}(i\gamma_{\mu}\partial^{\mu})\psi_{\nu},
\end{align}
where $\psi_i$ is a 4-component Dirac spinor, $\omega_{\mu\nu}=\partial_{\mu}\omega_{\nu}-\partial_{\nu}\omega_{\mu}$ and $\rho_{\mu\nu}=\partial_{\mu}\mathbf{\rho}_{\nu}-\partial_{\nu}\mathbf{\rho}_{\mu}$.  We have assumed massless neutrinos and the neutron and proton have the same mass $m$.  From this Lagrangian it is apparent why the rho meson distinguishes between neutrons and protons - it is dotted into a vector of isospin operators $\mathbf{t}$.

This Lagrangian leads to equations of motion (see Ref.~\cite{glendenning2000compact} for a more complete discussion) for the neutron, proton, sigma, omega, and rho fields.  Since we are interested in the static, ground state of nuclear matter, the meson fields are assumed to be static and equal to their ground state expectation values (thus, they become mean fields in which the neutrons and protons interact), which are then determined through their (now simplified) equations of motion.  The neutron and proton equations of motion become modified version of the Dirac equation
\begin{align}
    &\left[i\slashed{\partial}-(m-g_{\sigma}\sigma)-g_{\omega}\gamma^0\omega_0+\frac{1}{2}g_{\rho}\gamma^0\rho_{03}\right]\psi_n=0\\
    &\left[i\slashed{\partial}-(m-g_{\sigma}\sigma)-g_{\omega}\gamma^0\omega_0-\frac{1}{2}g_{\rho}\gamma^0\rho_{03}\right]\psi_p=0.
\end{align}
We define the neutron and proton effective mass $m_* = m-g_{\sigma}\sigma$ and the nuclear mean fields
\begin{align}
    U_n &= g_{\omega}\omega_0 - \frac{1}{2}g_{\rho}\rho_{03}\label{eq:Un}\\
    U_p &= g_{\omega}\omega_0 + \frac{1}{2}g_{\rho}\rho_{03},\label{eq:Up}
\end{align}
which will become useful soon.  The neutron and proton have the same effective mass (inclusion of the $\delta$ meson splits the effective masses \cite{Liu:2001iz}), but do not experience the same nuclear mean field.  

Fourier transforming, we find the energy dispersion relations 
\begin{align}
    E_n &= \sqrt{p^2+m_*^2}+U_n\label{eq:En}\\
    E_p &= \sqrt{p^2+m_*^2}+U_p.\label{eq:Ep}
\end{align}
In the mean field approximation, neutrons and protons behave like free particles with effective (Dirac) masses $m_*$ and with effective chemical potentials $\mu_n^* = \mu_n - U_n$ and $\mu_p^* = \mu_p - U_p$.  The nucleon spinor $\psi_n$ or $\psi_p$ is exactly the free nucleon spinor, but with effective mass $m_*$ and effective energy $E_n^*$ or $E_p^*$, where $E_n^* = E_n-U_n$ and $E_p^* = E_p - U_p$. The nucleon four-vector to be used in matrix element calculations is $p_{\mu} = (E^*,\mathbf{p})$.  This formalism is discussed in \cite{Fu:2008zzg}, but see also \cite{Roberts:2016mwj,Leinson:2002bw,Leinson:2002bv}.

With the removal of the spacetime dependence of the meson fields, the meson equations of motion become
\begin{align}
    g_{\sigma}\sigma &= \left(\frac{g_{\sigma}}{m_{\sigma}}\right)^2\left[n_p^s+n_n^s-bm(g_{\sigma}\sigma)^2-c(g_{\sigma}\sigma)^3\right]\label{eq:rmf1}\\
    g_{\omega}\omega_0 &= \left(\frac{g_{\omega}}{m_{\omega}}\right)^2(n_p+n_n)\label{eq:rmf2}\\
    g_{\rho}\rho_{03} &= \left(\frac{g_{\rho}}{m_{\rho}}\right)^2\frac{1}{2}(n_p-n_n)\label{eq:rmf3},
\end{align}
where
\begin{align}
    n_N^s &= 2\int\frac{\mathop{d^3p}}{(2\pi)^3}\frac{m_*}{E_N^*}\left(1+e^{\beta(E_N-\mu_N)}\right)^{-1}\\
    n_N &= 2\int\frac{\mathop{d^3p}}{(2\pi)^3}\left(1+e^{\beta(E_N-\mu_N)}\right)^{-1}
\end{align}
are the scalar density\footnote{Called the scalar density because it is the source term of the scalar (sigma) relativistic mean field equation.} and the (familiar) number density for neutrons and protons.  Note that $E_N - \mu_N = E^*_N - \mu^*_N$.  The solution of these mean field equations turns out to only depend on the ratio $g_i/m_i$ for each of the three mesons, so $m_i$ can be eliminated from each of the mean field equations leaving the couplings $g_i$ as the unknowns.

In matter with just neutrons, protons, and electrons, the equation of state can be found by solving a system of 6 equations (numerically).  Three equations are the relativistic mean field equations (\ref{eq:rmf1},\ref{eq:rmf2},\ref{eq:rmf3}), and then $n_e=n_p$, $n_B = n_n+n_p$, and $\mu_n=\mu_p+\mu_e$.  Once you specify the temperature $T$ and the baryon density $n_B$, the quantities $\{g_{\sigma}\sigma, g_{\omega}\omega_0, g_{\rho}\rho_{03}, \mu_n, \mu_p, \mu_e\}$ are obtained from the solution of the system of equations.

For matter with neutrons, protons, electrons, and neutrinos, the equation of state can be found by solving 7 equations (numerically).  These equations are the same as for the no-neutrino case, except for the beta equilibrium condition is now $\mu_n+\mu_{\nu}=\mu_p+\mu_e$ (see Sec.\ref{sec:both_beta_eq_conditions}), and there's an extra equation $n_e+n_{\nu}=Y_L(n_n+n_p)$.  Now, the temperature, baryon density, and lepton fraction $Y_L$ must be specified to solve for $\{g_{\sigma}\sigma, g_{\omega}\omega_0, g_{\rho}\rho_{03}, \mu_n, \mu_p, \mu_e,\mu_{\nu}\}$.

\begin{figure}[h]
  \centering
  \includegraphics[scale=.6]{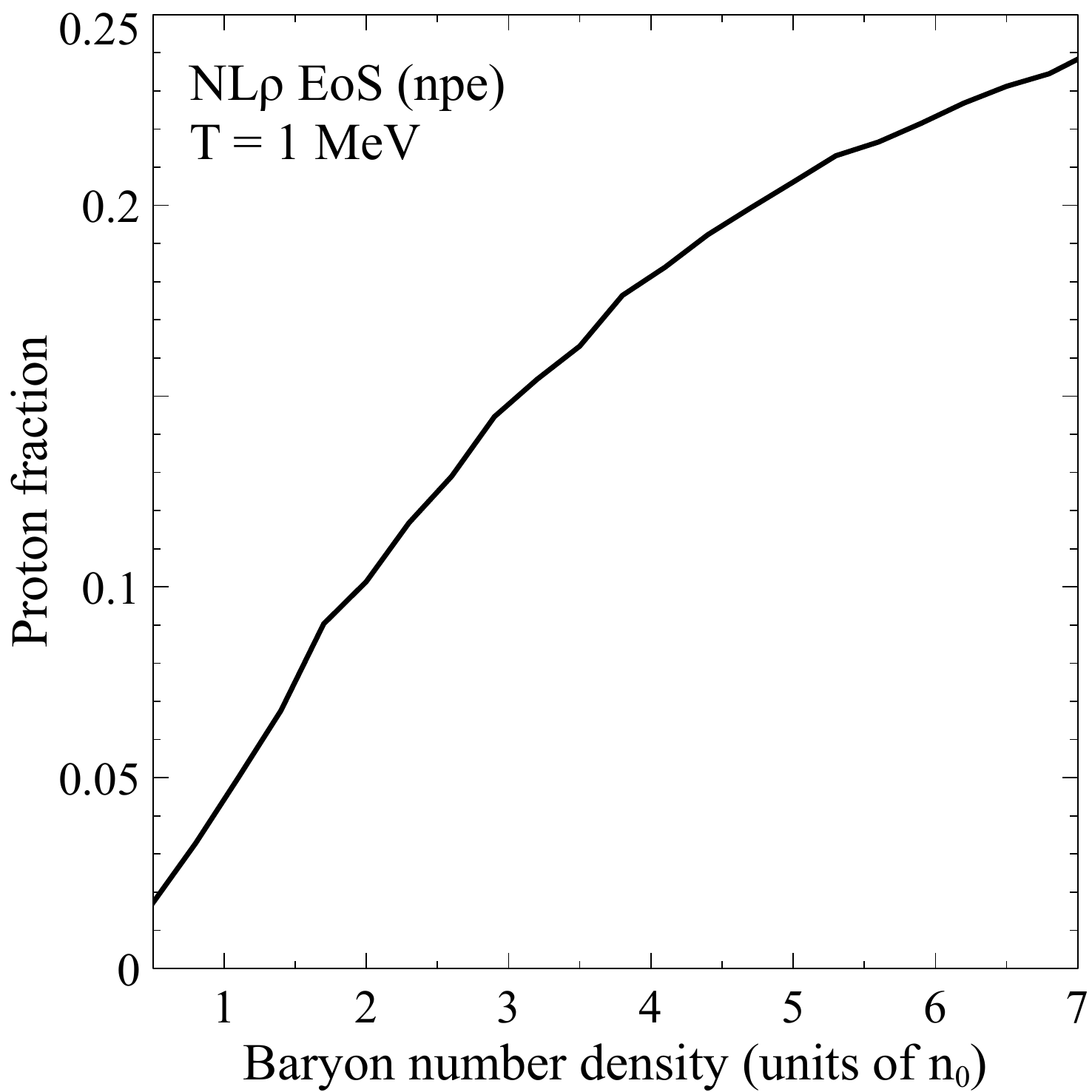}
  \caption{Proton fraction in npe matter with the NL$\rho$ equation of state, at T=1 MeV.}
  \label{fig:NLrho_xp}
\end{figure}
\begin{figure}[h]
  \centering
  \includegraphics[scale=.6]{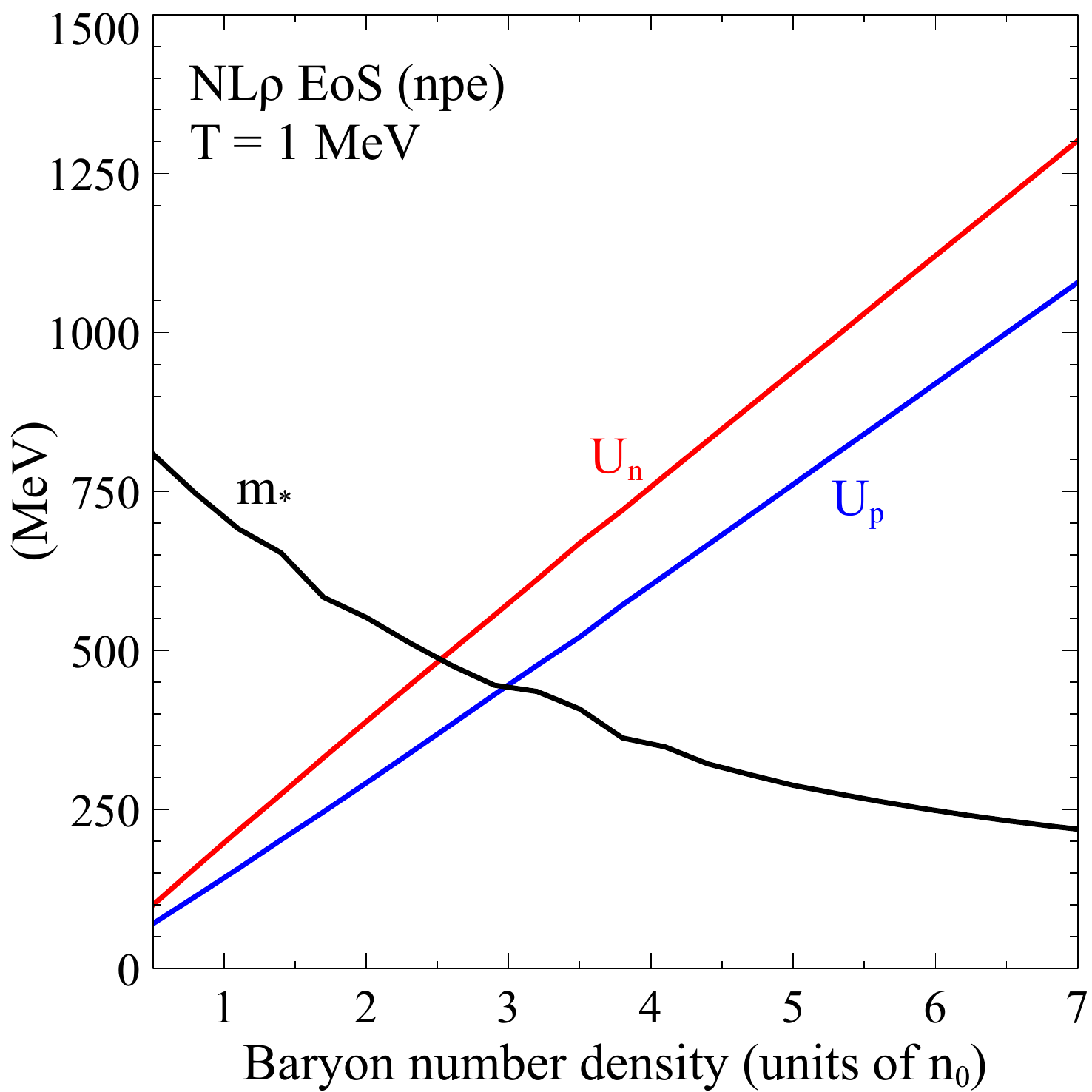}
  \caption{Nucleon effective mass $m_*$ and neutron and proton mean fields $U_n$ and $U_p$, for matter matter with the NL$\rho$ equation of state, at T=1 MeV.  As density increases, the difference in mean fields $U_n-U_p$ grows, which has kinematic consequences for weak interactions \cite{Roberts:2016mwj}.  The nucleon effective mass decreases at high densities, which means that at high densities, nucleons can no longer be treated non-relativistically \cite{Reddy:1997yr}.}
  \label{fig:NLrho_other}
\end{figure}

We have plotted the proton fraction predicted by the NL$\rho$ equation of state as a function of density in Fig.~\ref{fig:NLrho_xp}.  Also, the effective mass and nuclear mean fields are plotted in Fig.~\ref{fig:NLrho_other}.  We see that above nuclear saturation density, this equation of state predicts neutron-rich nuclear matter that has a low Dirac effective mass at high density.  Both nuclear mean fields grow with density, as does the difference between them, which has kinematic consequences for weak interactions \cite{Roberts:2016mwj}.  For more on the nuclear mean fields, see Ref.~\cite{Hempel:2014ssa}.
\hiddensubsubsection{Exotic phases and the neutron star core}
\label{sec:exotic_phases}
As the density increases past a few times nuclear saturation density, it is likely that other degrees of freedom start to appear in nuclear matter, including quarks and hyperons (baryons with net strangeness) \cite{Blaschke:2018mqw,Weber:2019xvv,Tanimoto:2019tsl,Alford:2004pf,Alford:2019oge}.  The nature of the quark-hadron transition at high density is unknown.  It could be a first-order phase transition, which occurs at a specific pressure.  Depending on the value of the quark-hadron surface tension, which is unknown, the transition could have a sharp jump in density between phases, or a very gradual change, in the case of a mixed phase.  Alternatively, the phase transition could be a crossover, which has phase coexistence over a range of densities.  Another option would be quarkyonic matter \cite{McLerran:2018hbz}.  Phase transitions in dense nuclear matter are discussed in \cite{glendenning2000compact,Han:2019bub,Baym:2017whm}.  
\section{TOV equation and Mass-Radius curves}
\label{sec:tov}
In this section, we derive equations governing the structure of spherical stars and discuss the impact of the equation of state on the stellar structure.  

In Newtonian mechanics\footnote{Newtonian mechanics applies quite well to our Sun and fairly well to white dwarfs.  It fails when describing neutron stars, and obviously it cannot account for black holes.}, a sphere of matter holds together because for any given spherical shell of the material, the forces that act on it balance.  First, there is the gravitational force inwards due to the mass enclosed by the shell.  Second, there are forces due to pressure from material above and below the shell (this should be considered one force, due to a pressure gradient on the shell) \cite{carroll2017introduction}.  If we consider a spherical shell of thickness $\mathop{dr}$ at distance $r$ from the center, the shell has mass $\mathop{dm}(r)=4\pi r^2\rho (r)\mathop{dr}$, where $\rho (r)$ is the rest mass density at position $r$.  The inward gravitational force on the shell due to the mass $m(r)$ that it encloses is $-Gm(r)\mathop{dm}/r^2 = -4\pi Gm(r)\rho (r)\mathop{dr}$.  The outward force due to the pressure gradient is $-A\mathop{dP} = -4\pi r^2\mathop{dP}$.  In hydrostatic equilibrium, these forces balance, yielding the (coupled) Newtonian equations for stellar structure
\begin{align}
    \frac{\mathop{dP}}{\mathop{dr}}&=-\frac{Gm(r)\rho(r)}{r^2},\\
     \frac{\mathop{dm}}{\mathop{dr}} &= 4\pi r^2\rho(r).
\end{align}
From these stellar structure equations, we can see that the enclosed mass grows from the core to the crust, which is intuitive, but also that the pressure decreases monotonically from core to crust.  The distance $r=R$ at which the pressure reaches zero defines the radius of the star.  The enclosed mass at that radius is the total mass of the star.  To solve these equations, the equation of state $\rho=\rho(P)$ of the stellar material must be introduced.  These coupled equations are integrated from core to crust, where the initial conditions at the core are $m(r=0)=0$ and $P(r=0) = P_c$, where $P_c$ is called the central pressure (and can be converted to a central density via the equation of state).  

The extension of these equations to General Relativity begins with a metric ansatz for a static and spherically symmetric spacetime 
\begin{equation}
    \mathop{ds^2} = g_{\mu\nu}\mathop{dx^{\mu}}\mathop{dx^{\nu}} = e^{2\nu(r)}\mathop{dt^2}-e^{2\lambda(r)}\mathop{dr^2}-r^2(\mathop{d\theta^2}+\sin^2{\theta}\mathop{d\phi^2}).\label{eq:metric_ansatz}
\end{equation}
The relativistic stellar matter is described as a perfect fluid, meaning that its stress-energy tensor only depends on the energy density $\varepsilon$ and pressure $P$ of the fluid, and so the metric ansatz and the perfect fluid stress-energy tensor $T^{\mu\nu}$ can be plugged into Einstein's equations 
\begin{equation}
    R^{\mu\nu} - \frac{1}{2}R g^{\mu\nu}=8\pi G T^{\mu\nu},
\end{equation}
yielding the general relativistic structure equations
\begin{align}
\frac{\mathop{dP}}{\mathop{dr}} &= -\frac{Gm(r)\varepsilon(r)}{r^2}\left(1+\frac{P(r)}{\varepsilon(r)}\right)\left(1+\frac{4\pi r^3 P(r)}{m(r)}\right)\left(1-\frac{2Gm(r)}{r}\right)^{-1}\label{eq:TOV}  \\
\frac{\mathop{dm}}{\mathop{dr}} &= 4\pi r^2\varepsilon(r).\label{eq:m(r)}
\end{align}
These are called the Tolman, Oppenheimer, Volkoff (TOV) equations \cite{Tolman:1939jz,Oppenheimer:1939ne}.  The details of the derivation are given in \cite{glendenning2000compact}.  Again, an equation of state $\varepsilon = \varepsilon(P)$ is needed to solve the equations, which is done the same way as in the Newtonian case.  Here, the enclosed mass is defined through
\begin{equation}
    e^{2\lambda(r)} = \left(1-\frac{2Gm(r)}{r}\right)^{-1}.
\end{equation}

Eqs.~(\ref{eq:TOV}) and (\ref{eq:m(r)}) with an equation of state $\varepsilon=\varepsilon(P)$ can be solved together.  Einstein's equations also produce a final equation 
\begin{equation}
\frac{\mathop{d\nu}}{\mathop{dr}} = \frac{G}{r}\left(\frac{m(r)+4\pi r^3P(r)}{r-2Gm(r)}\right)   \label{eq:dvdr} 
\end{equation}
for $\nu(r)$, which was part of the spacetime metric ansatz [Eq.~(\ref{eq:metric_ansatz})].  Once Eqs.~(\ref{eq:TOV}) and (\ref{eq:m(r)}) have been solved, Eq.~(\ref{eq:dvdr}) can be solved with the boundary condition $\nu(R)=(1/2)\ln{\left(1-2GM/R\right)}$, where the star has mass $M$ and radius $R$.  This boundary condition ensures that the spacetime inside the star connects to the Schwarzchild solution outside of the star \cite{glendenning2000compact}.

A given equation of state allows many stellar equilibrium configurations, each of which corresponds to a possible star that we could expect to find in the sky, if matter is indeed governed by that particular equation of state.  The set of allowed stellar configurations is indexed by one parameter, the central pressure (or density).  The set of all allowed stellar configurations can be displayed as a mass-radius curve, parametrized by central pressure.  Each point on the curve corresponds to a choice of central pressure, giving rise to a star of a particular mass and radius.  

\begin{figure}[h]
  \centering
  \includegraphics[scale=.5]{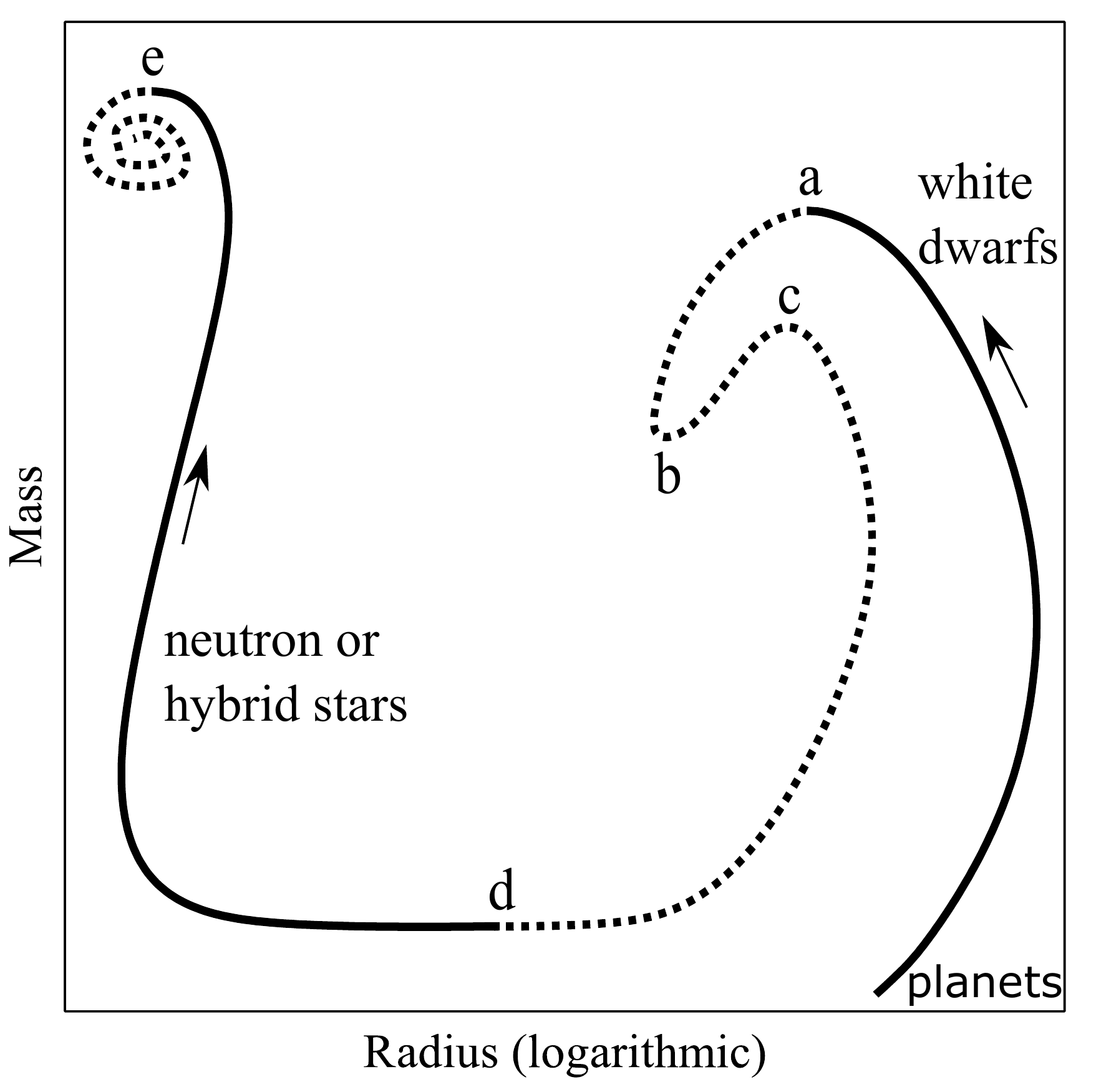}
  \caption{Schematic mass-radius diagram arising from a typical nuclear equation of state.  The central pressure (or density) is small at the bottom right of the plot, and increasing as the curve continues in the direction of the arrows.  Solid lines denote stable configurations, dotted lines indicate unstable configurations.  Letters (a) through (e) label the extrema of the curve, which will be referred to in Sec.~\ref{sec:radial_osc}.  Figure reproduced from \cite{Alford:2017vca}.}
  \label{fig:schematic_MR}
\end{figure}

A schematic view of the mass radius curve\footnote{Quantitative versions of this curve are given in \cite{Alford:2017vca,Shapiro:1983du,Hartle:2003yu,Misner:1974qy,1965gtgc.book.....H,thorne2017modern}.} for stars made of a typical nuclear equation of state is shown in Fig.~\ref{fig:schematic_MR}.  Each point on the curve corresponds to a star with a specific central pressure - low central pressure configurations start on the bottom right corner of the plot and as the curve winds from low mass, large radius configurations to high mass small radius configurations, the central pressure rises.  Even though all points on the curve are equilibrium configurations, only some of them (shown in solid lines) are stable to radial oscillations.  This will be discussed in Sec.~\ref{sec:radial_osc}.  

The configurations at low central density are planets, and as the central density increases they become white dwarfs.  Even the core of the most massive white dwarf (and thus the one with the highest central density) is orders of magnitude less dense than nuclear saturation density.  White dwarfs are smaller than stars like our sun, but still have radii of thousands of kilometers.  They are supported by electron degeneracy pressure.  However, they have a maximum mass (known to be about 1.44 solar masses \cite{carroll2017introduction}) called the Chandrasekhar mass, above which the electron degeneracy pressure is not enough to support the mass of the white dwarf \cite{glendenning2000compact,Shapiro:1983du,zeldovich1971stars}.  

For a wide range of central pressures above the Chandrasekhar limit, there are no stable configurations.  Eventually stability is regained, but at this point the stars are much smaller, with radii closer to 10 kilometers\footnote{We will see in Sec.~\ref{sec:strange_dwarf} that stable configurations can technically have radii of a couple hundred kilometers, however nobody has ever seen such an object and we know of no theoretical formation channels (like supernovae) that could produce such an object.}, and so they are generically called compact objects.  If the compact object is a neutron star, it supports itself through a combination of neutron degeneracy pressure and nucleon-nucleon repulsion at short distances.  However, it too has a maximum mass it can support before it collapses to a black hole.  From observations, we know that this maximum mass is at least two solar masses, as discussed in Sec.~\ref{sec:ns_observations}, but there is evidence to suggest the maximum mass is not too much larger than two solar masses \cite{Margalit:2017dij,Shibata:2019ctb}.  If there is a phase transition from nuclear to quark matter \cite{2000A&A...353L...9G,Schertler:2000xq,Alford:2013aca,Christian:2017jni,Montana:2018bkb}, or even two phase transitions \cite{Alford:2017qgh}, the mass-radius curve develops extra structure in the compact star branch.  

Knowing the mass-radius curve would significantly improve our understanding of nuclear matter, because it can (uniquely) be turned back into the equation of state $\varepsilon=\varepsilon(P)$ \cite{1992ApJ...398..569L}.
\section{Stability of compact objects}
\label{sec:radial_osc}
While all points on the mass-radius curve correspond to stars in hydrostatic equilibrium, only some of the stars are stable against radial oscillations and thus could be expected to be seen in nature.  A stellar configuration is stable only if all of its radial modes are stable.  There are two approaches to determine the stability of a star - one involves the static properties of the stellar sequence, namely the mass-radius curve, and the second involves dynamic properties of a particular star, namely, its radial eigenmodes \cite{1967hea3.conf..259T}.  We present both below.  
\hiddensubsection{Mass-radius stability criterion}
The mass-radius stability criterion was proposed in \cite{1965gtgc.book.....H} and further developed in \cite{1966ApJ...145..505B,1966ApJ...145..514M,1967hea3.conf..259T}.  Spherical stars of arbitrarily low central density are just terrestrial matter, and thus are stable to radial perturbations - they have zero unstable modes.  At each extremum in the mass-radius curve, one radial oscillation mode changes stability.  At each extremum where the mass-radius curve turns counterclockwise with increasing central pressure, one stable mode becomes unstable.  At each extremum where the mass-radius curve turns clockwise with increasing central pressure, one unstable mode becomes stable.  

We now apply this criterion to Fig.~\ref{fig:schematic_MR}.  At low central pressure, all modes are stable.  As the central pressure increases and the curve goes through point (a), the winding is counterclockwise and thus one mode becomes unstable, causing stars after point (a) to be unstable.  As the central pressure increases through point (b), the winding again is counterclockwise, so between (b) and (c) stars are unstable with two unstable oscillation modes.  At point (c), the winding is clockwise, so stars between (c) and (d) have one unstable mode.  The curve passes through (d) clockwise, so between (d) and (e) all modes are stable once more.  At point (e), the curve turns counterclockwise, and continues to do so up to arbitrarily high central density and so there are no stable stars above (e), as more and more modes turn unstable.  
\hiddensubsection{The Sturm-Liouville mode spectrum}
The most direct way of calculating stellar stability is by solving the Sturm-Liouville problem to find the radial eigenmodes.  The radial oscillations are described \cite{1964ApJ...140..417C,Misner:1974qy,glendenning2000compact} by
\begin{equation}
    \delta r_n(r,t) = \frac{e^{\nu(r)}}{r^2}u_n(r)e^{i\omega_n t},
\end{equation}
where $n$ indexed the radial eigenmode and $u_n(r)$ is a solution with eigenvalue $\omega_n^2$ to the Sturm-Liouville eigenvalue problem
\begin{equation}
\frac{\mathop{d}}{\mathop{dr}}\left(\Pi(r)\frac{\mathop{du_n}}{\mathop{dr}}\right)+(Q(r)+\omega_n^2W(r))u_n(r)=0,
\end{equation}
where
\begin{align}
    \Pi(r) &= \frac{e^{\lambda(r)+3\nu(r)}}{r^2}\Gamma(r)P(r),\nonumber\\
    Q(r) &= -4\frac{e^{\lambda(r)+3\nu(r)}}{r^3}\frac{\mathop{dP}}{\mathop{dr}}-8\pi\frac{e^{3\lambda(r)+3\nu(r)}}{r^2}P(r)(\varepsilon(r)+P(r))+\frac{e^{\lambda(r)+3\nu(r)}}{r^2(\varepsilon(r)+P(r))}\left(\frac{\mathop{dP}}{\mathop{dr}}\right)^2,\nonumber\\
    W(r) &= \frac{e^{3\lambda(r)+\nu(r)}}{r^2}(\varepsilon(r)+P(r)),\nonumber\\
    \Gamma(r) &= \frac{\varepsilon(r)+P(r)}{P(r)}\frac{\mathop{dP}}{\mathop{d\varepsilon}}.\nonumber
\end{align}
The boundary conditions for the eigenvalue problem are
\begin{align}
    u_n&\propto r^3 \quad \text{ at } \quad r=0,\\
    \frac{\mathop{du_n}}{\mathop{dr}}&=0 \quad \text{ at } \quad r=R.
\end{align}

The solutions to the Sturm-Liouville eigenvalue problem are a set of eigenfunctions $u_n(r)$ with eigenvalues $\omega_n^2$, which are the squared frequencies of the oscillation modes.  The eigenvalues are real due to the Sturm-Liouville nature of the problem, and form \cite{pryce1993numerical} a lower-bounded infinite sequence $\omega_0^2 < \omega_1^2 < \omega_2^2 < ...$.  If $\omega_n^2>0$, then the n\textsuperscript{th} mode has a real frequency and thus is a stable, oscillatory mode.  If, however, $\omega_n^2 < 0$, then the frequency of the n\textsuperscript{th} mode is imaginary and the mode is unstable.  
    
The overall stability of the star depends on just the lowest eigenvalue $\omega_0^2$.  If $\omega_0^2 > 0$, the lowest mode is stable and thus all other modes are stable because $\omega_n^2 > \omega_0^2$ for $n\geq 1$.  If $\omega_0^2 < 0$, then there is at least one mode that is unstable and thus the star is unstable.
\section{Strange dwarf stars and the strange matter hypothesis}
\begin{center}
{\textit{This section is based on my work with Mark Alford and Pratik Sachdeva, \cite{Alford:2017vca}.}}
\end{center}
\label{sec:strange_dwarf}
\hiddensubsection{Strange matter hypothesis}
In Sec.~\ref{sec:NS_crust} we discussed that at terrestrial densities, a bar of gold is a metastable state (with extraordinary long lifetime), since \ce{^{56}Fe} is the state with the lowest energy per nucleon.  In the seventies and eighties, Bodmer \cite{Bodmer:1971we} and Witten \cite{Witten:1984rs} proposed that the ground state of nuclear matter was not \ce{^{56}Fe}, but instead strange matter, a mixture of up, down, and strange quarks.  They proposed that \ce{^{56}Fe} was just a long-lived metastable state, not the true ground state.  
\hiddensubsection{Strange stars and strange dwarf stars}
If strange quark matter is the true ground state, then strange stars, stars made of strange quark matter, should exist.  This does not mean that neutron stars could not also exist, however \cite{glendenning2000compact}.  For this section, we will focus on strange stars with a nuclear matter crust.  Strange quark matter has a small net positive charge, because the strange quark has a much larger mass than the up and down quark, so its population is suppressed relative to the up and down quark populations.  As the strange quark has negative electric charge, the matter as a whole will have net positive charge and thus will need to include electrons to make it electrically neutral.  Alcock, Farhi, and Olinto noted that the electrons would extend slightly beyond the surface of the strange quark matter in a strange star, because they are not bound by the strong interaction.  Thus, strange stars have an electric dipole layer on their surface, which allows the star to suspend a nuclear matter crust several hundred femtometers above the strange matter \cite{1986ApJ...310..261A}.

\begin{figure}[h]
  \centering
  \includegraphics[scale=.6]{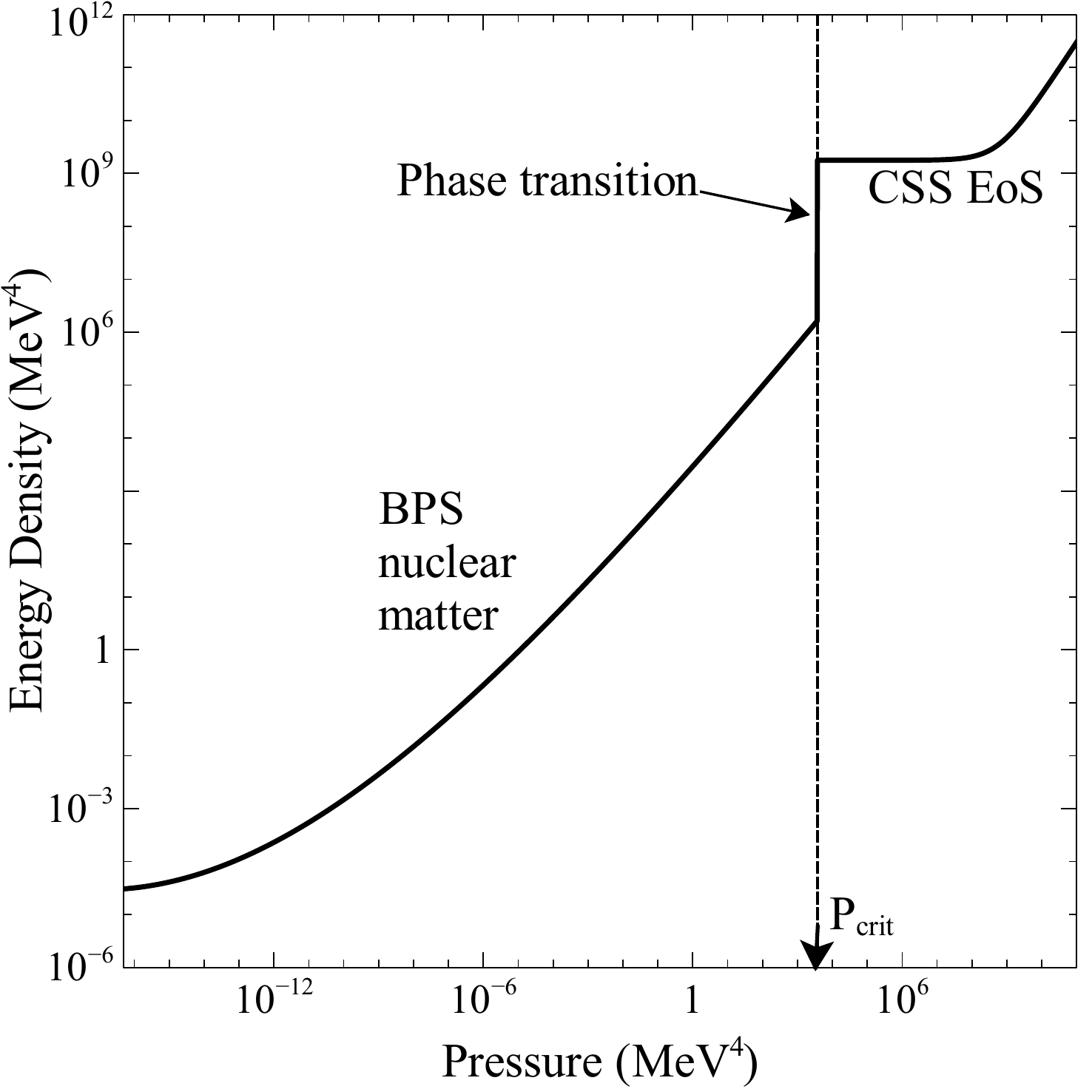}
  \caption{Equation of state Eq.~(\ref{eq:hybrid_eos}) with a sharp, first-order transition at the neutron drip pressure from BPS matter to a strange quark matter phase.}
  \label{fig:hybrid_eos}
\end{figure}

The strange star with a nuclear matter crust can be modeled by an effective equation of state \cite{1992ApJ...400..647G}
\begin{equation}
\varepsilon(P) =  \begin{cases} 
      \varepsilon_{\text{BPS}}(P) & P\leqslant P_{\text{crit}} \\
      kP+4B &  P> P_{\text{crit}},
   \end{cases}
\label{eq:hybrid_eos}
\end{equation}
where below a certain pressure $P_{\text{crit}}$, the matter is a nuclear lattice with a relativistic electron Fermi sea (described by the BPS equation of state, see Sec.~\ref{sec:NS_crust}), while above $P_{\text{crit}}$ the matter is strange quark matter described with the constant-speed-of-sound (CSS) parametrization \cite{Alford:2013aca,Chamel:2012ea,Zdunik:2012dj}.  In contrast to equations of state discussed in previous sections, this one is not supposed to represent the ground state of nuclear matter.  Instead, it just realizes the possibility of having a star made of strange matter with a metastable nuclear crust suspended on top of it.  Following Glendenning \textit{et al.} \cite{Glendenning:1994sp,Glendenning:1994zb}, we will take the transition pressure $P_{\text{crit}}=P_{\text{drip}} = 3724 \text{ MeV}$, which is the pressure above which neutrons begin to drip out of nuclei in the Coulomb lattice (see Sec.~\ref{sec:NS_crust})\footnote{The neutron drip pressure is the maximum allowed pressure the suspended nuclear crust can attain, because once free neutrons appear in the nuclear crust they will fall into the strange matter core \cite{glendenning2000compact}.}.  For the CSS phase, they choose $k=3$ and $B^{1/4} = 145 \text{ MeV}$.  This equation of state is plotted in Fig.~\ref{fig:hybrid_eos}.

In Refs.~\cite{Glendenning:1994sp,Glendenning:1994zb} (see also \cite{glendenning2000compact}), Glendenning \textit{et al.} constructed a sequence of stellar configurations using the equation of state Eq.~(\ref{eq:hybrid_eos}).  They constructed a mass-radius curve for this sequence and it looked (schematically) like Fig.~\ref{fig:schematic_MR}.  However, they claimed that in addition to the regions before points (a) and between points (d) and (e), the region of the mass-radius curve between (c) and (d) was also stable to radial oscillations.  They termed this new family of stars ``strange dwarfs'', as they are of a similar size to white dwarfs, but contain a small strange quark matter core.  The fraction of the star in the quark matter phase grows as the central pressure increases.  If these strange dwarf stars were stable, they would represent the first known contradiction between the mass-radius curve stability analysis (strange dwarfs are unstable according to the mass-radius criteria) and the Sturm-Liouville eigenmode analysis, which Glendenning \textit{et al.} claimed yield stable strange dwarfs because they find $\omega_0^2>0$.

\begin{figure}[h]
  \centering
  \includegraphics[scale=.6]{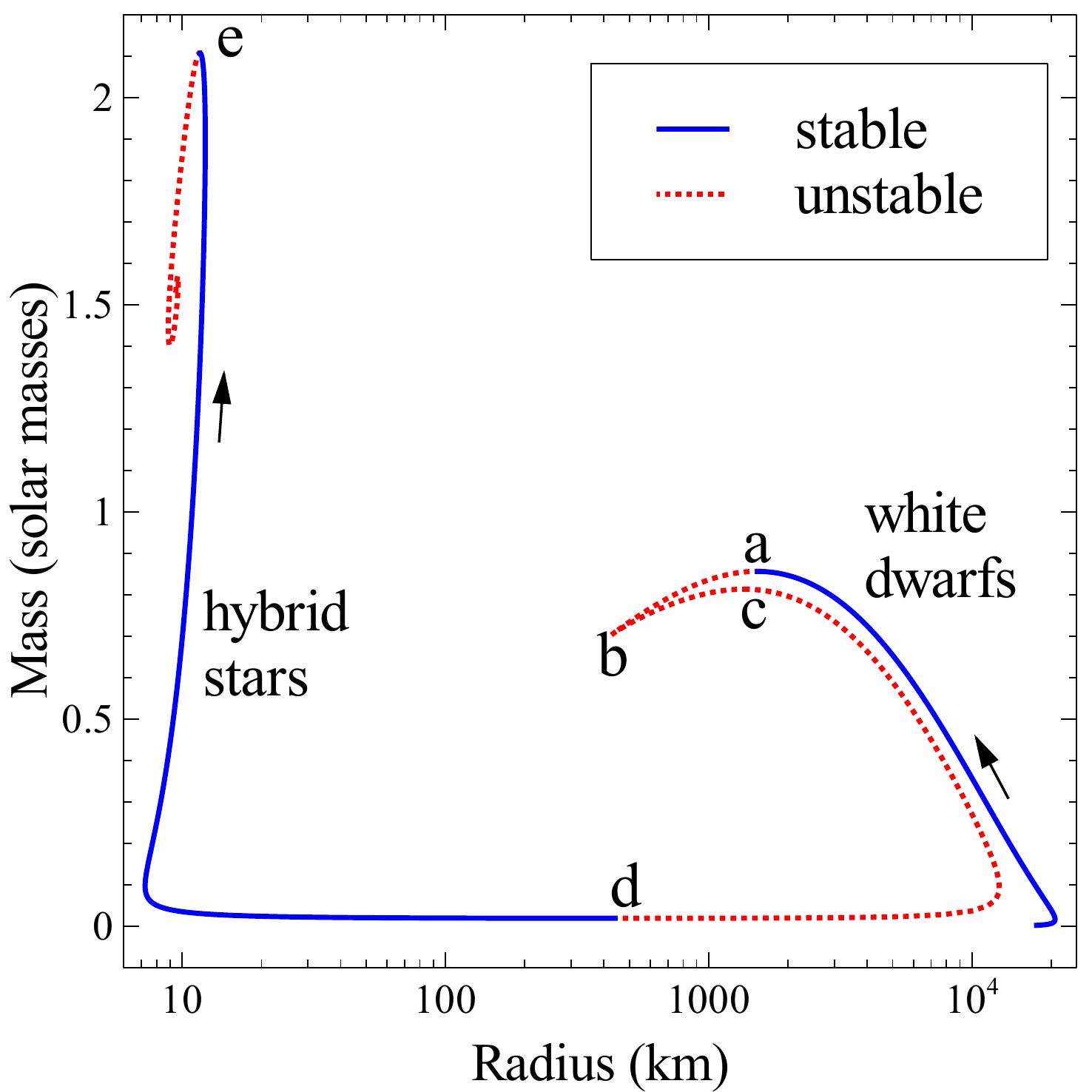}
  \caption{Mass-radius plot for hybrid stars with equation of state Eq.~(\ref{eq:reg_eos}).  Solid lines indicate stable configurations according to the mass-radius criterion and dashed lines indicate unstable configurations.  The arrows indicate the direction of increasing central pressure.}
  \label{fig:hybrid_MR_curve}
\end{figure}

In \cite{Alford:2017vca}, we use a version of the EoS (\ref{eq:hybrid_eos}) where the phase transition is smoothed out over a range of pressures $\delta P$
\begin{equation}
\varepsilon(P) = \frac{1}{2}\left( 1 - \tanh{\left(  \frac{P-P_{\text{crit}}}{\delta P} \right)}  \right)   \varepsilon_{\text{BPS}}(P)+ \frac{1}{2} \left( 1+\tanh{\left(  \frac{P-P_{\text{crit}}}{\delta P} \right)} \right) \left(kP+4B \right). 
\label{eq:reg_eos}
\end{equation}
Using this regulated phase transition allows us to use the traditional techniques and boundary conditions for solving the Sturm-Liouville problem for the radial modes\footnote{Pereira and Rueda \cite{Pereira:2015sua} have developed the matching conditions between the two regions of the star in the case of a true discontinuity in the equation of state.  See also \cite{Pereira:2017rmp}.  These matching conditions have also been formulated in the Newtonian limit in \cite{1989A&A...217..137H}.}.  The mass-radius curve for this equation of state (with $\delta P = 100 \text{ MeV}^4$) is shown in Fig.~\ref{fig:hybrid_MR_curve}.  The point on the mass radius curve (b) is where quark matter starts to appear.  Stars with central pressures less than at (b) are made solely of BPS nuclear matter.  As the central pressure increases beyond its value at (b), the stars have progressively larger quark matter cores.  Stars between (c) and (d) are Glendenning \textit{et al.}'s strange dwarfs, neutron stars with a small strange quark matter core.  Stars between (d) and (e) are small, strange quark matter stars with a thin nuclear crust.

If there is a sharp first-order transition in the equation of state
then point ($b$) is a cusp in the $M(R)$ relation. For finite but very
small transition width $\delta P\lesssim 1 \text{ MeV}^4$ the cusp becomes a
minimum at which, according to the mass-radius criteria,
the second-lowest mode goes from stable to unstable
as central pressure increases.
In our calculation we use values of $\delta P$ in the range $10$ to
$100 \text{ MeV}^4$, in which case the mass radius relation develops a
more complicated structure at $b$ which may have multiple extrema
as the curve spirals and then ``uncoils'' again. 
This structure occurs in a very small range of masses and radii near $b$,
and is invisible on the scales shown in Fig.~\ref{fig:hybrid_MR_curve}.
The details of this structure depend on the exact profile of the regulated 
transition, but, as we will see,
(i) the lowest eigenmode remains negative so all these configurations are unstable; (ii) as central pressure increases through $b$, the net outcome
is that the second-lowest mode goes from stable to unstable;
(iii) this behavior is not relevant to the stability of strange dwarfs,
which lie between $c$ and $d$ on the mass-radius curve.

\begin{figure}[h]
  \centering
  \includegraphics[scale=.6]{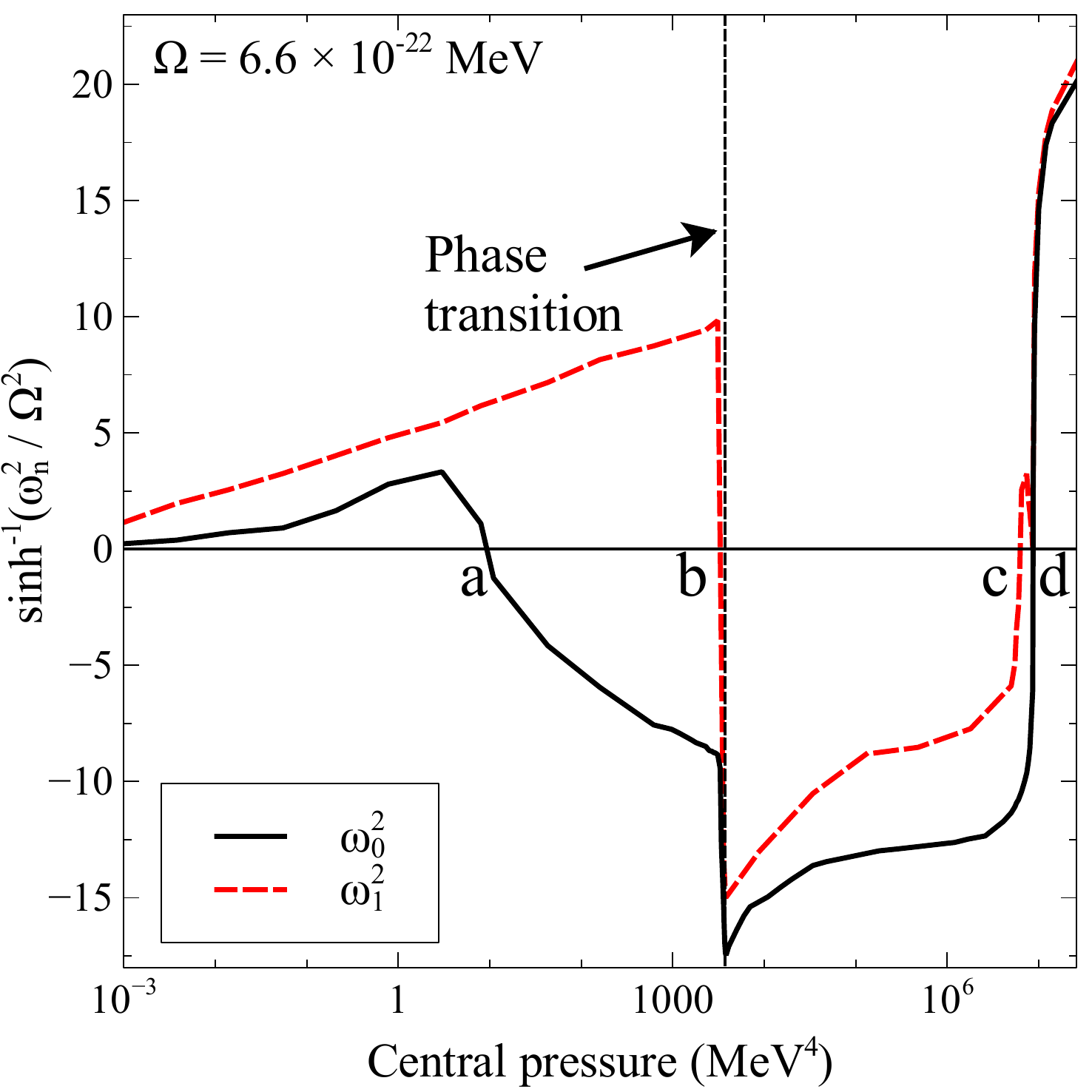}
  \caption{Squared frequencies for the two lowest radial eigenmodes for TOV solutions plotted in Fig.~\ref{fig:hybrid_MR_curve}.  We find that stellar configurations between (a) and (d) are unstable, because they have at least one unstable radial mode.}
  \label{fig:hybrid_star_SL}
\end{figure}

\begin{figure}[h]
  \centering
  \includegraphics[scale=.6]{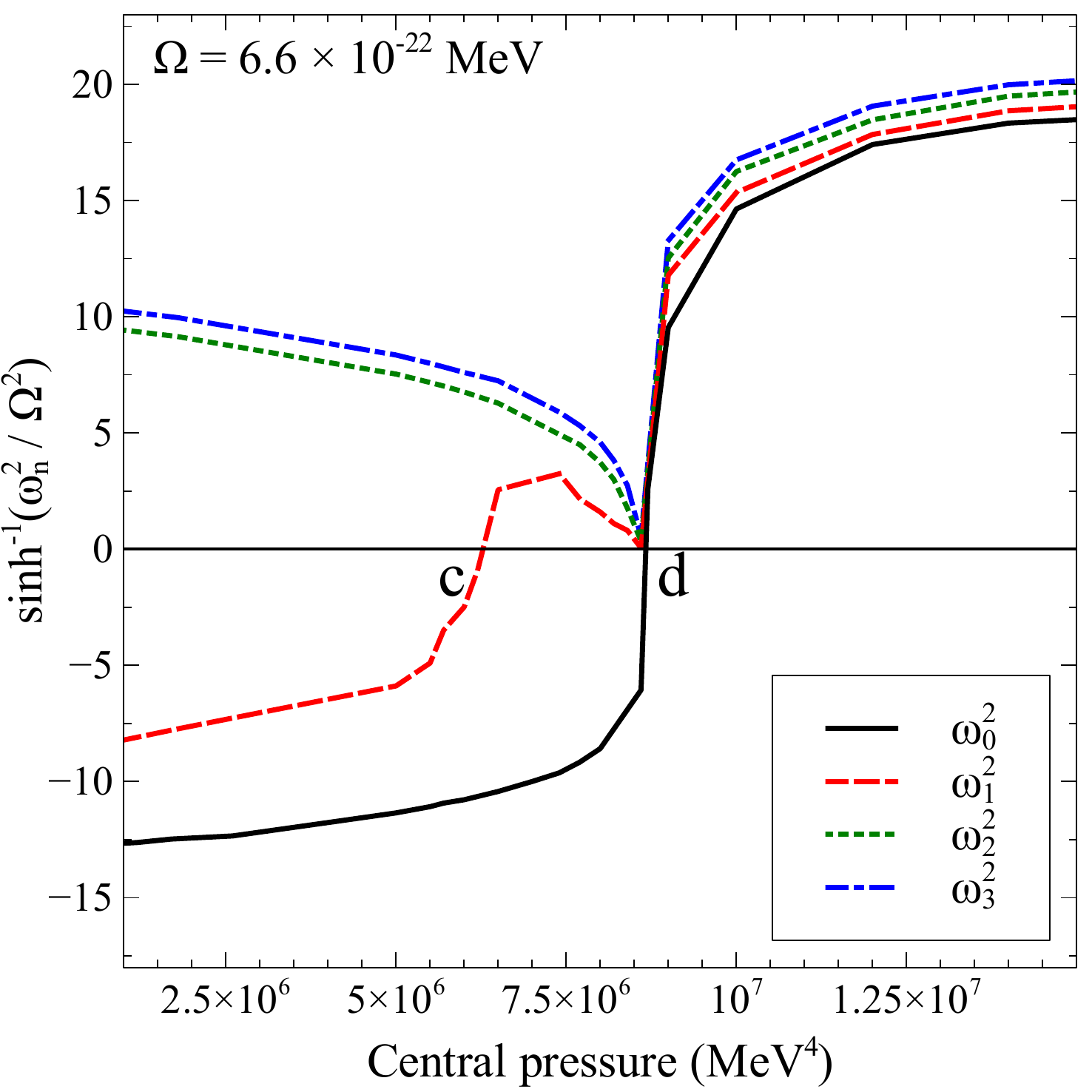}
  \caption{Squared frequencies for the four lowest eigenmodes for TOV solutions shown in Fig.~\ref{fig:hybrid_MR_curve}.  We have zoomed in on the region from (c) to (d) where the strange dwarfs were hypothesized to occur.  However, $\omega_0^2<0$ in this region, and thus strange dwarfs are unstable.}
  \label{fig:hybrid_star_SL_zoom}
\end{figure}

\begin{figure}[h]
  \centering
  \includegraphics[scale=.6]{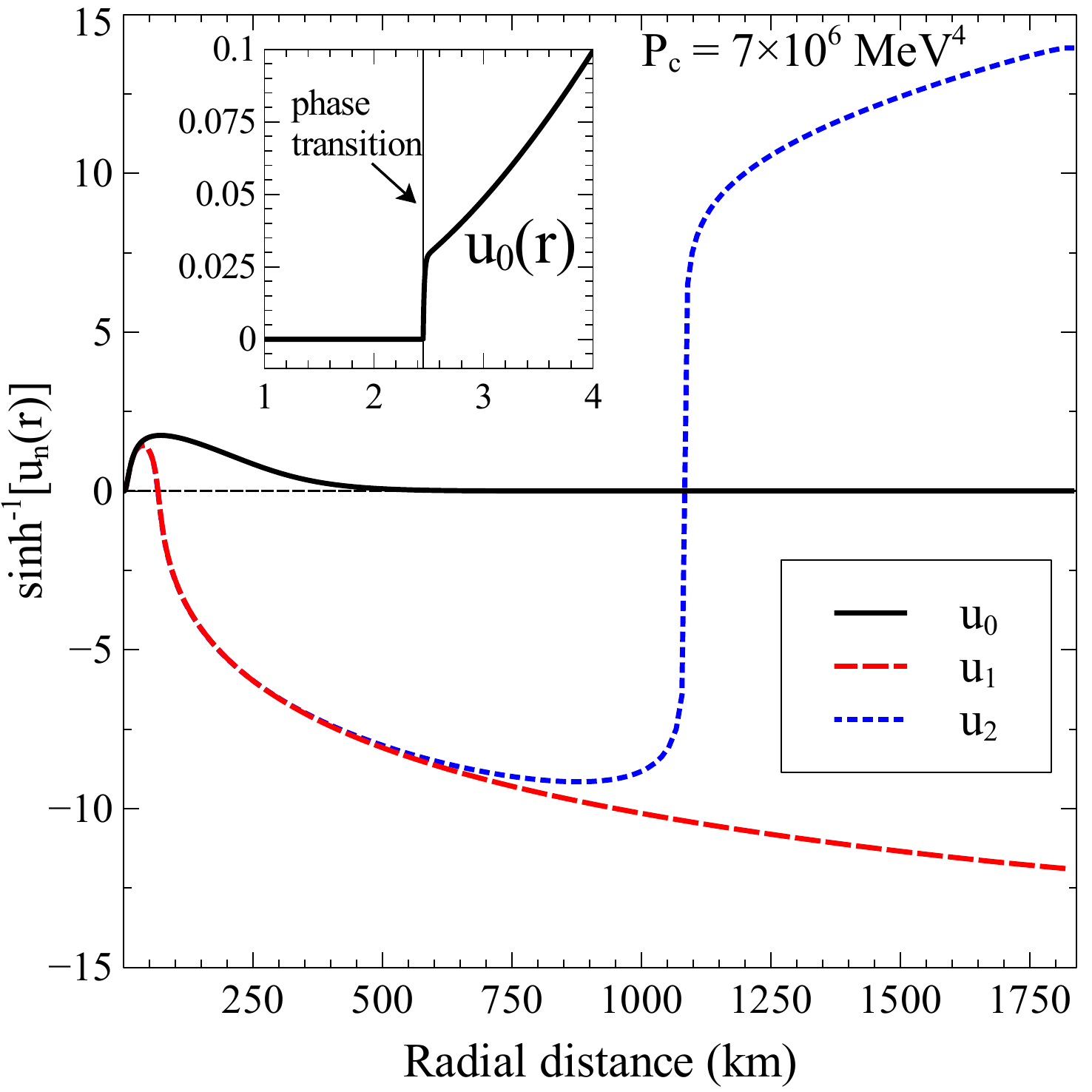}
  \caption{The first three eigenfunctions $u_n(r)$ for a TOV solution with a central pressure of $7\times 10^6 \text{ MeV}^4$, which lies between points (c) and (d) in Figs.~\ref{fig:hybrid_star_SL} and \ref{fig:hybrid_star_SL_zoom}.  For this TOV solution, the phase transition is located at $r = 2.4 \text{ km}$.  As expected for Sturm-Liouville eigenfunctions \cite{pryce1993numerical}, $u_n(r)$ has $n$ nodes.}
  \label{fig:hybrid_star_eigenfunctions}
\end{figure}

\begin{figure}[h]
  \centering
  \includegraphics[scale=.6]{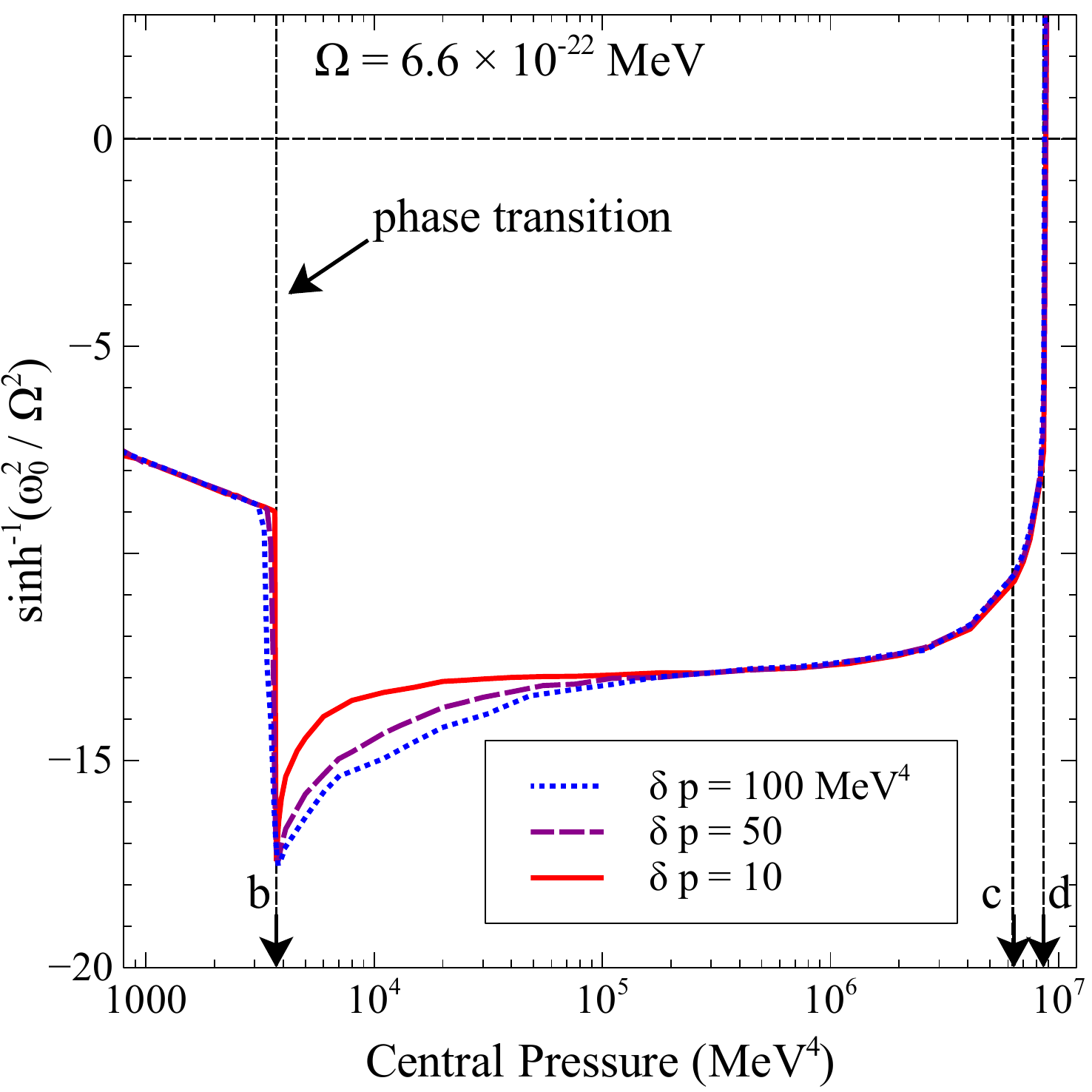}
  \caption{Dependence of the lowest eigenvalue on the regulator width $\delta P$.  As the regulator tends to zero width, and thus the phase transition becomes sharper, the lowest eigenmode remains unstable in the region between (c) and (d).}
  \label{fig:shrink_regulator}
\end{figure}

To study the stability of the strange dwarfs between (c) and (d), we solved the Sturm-Liouville problem for the radial eigenmodes using the equation of state Eq.~(\ref{eq:reg_eos}) with the regulated phase transition.  Our results are displayed in Figs.~\ref{fig:hybrid_star_SL}, \ref{fig:hybrid_star_SL_zoom}, \ref{fig:hybrid_star_eigenfunctions}, and \ref{fig:shrink_regulator}.

In Fig.~\ref{fig:hybrid_star_SL} we show the lowest two eigenvalues $\omega_0^2$ and $\omega_1^2$ as a function of central pressure for the entire stellar sequence depicted in Fig.~\ref{fig:hybrid_MR_curve}.  Our results are perfectly consistent with the mass-radius criteria.  Stars with central pressures less than at (a) or greater than at (d) are stable because all of their radial eigenmodes are stable.  

In Fig.~\ref{fig:hybrid_star_SL_zoom}, we focus on the range of central pressures where strange dwarfs were proposed to exist, between points (c) and (d) on the mass-radius curve.  In this range, the lowest eigenvalue is always negative, indicating that strange dwarfs are unstable.  Comparing with Fig.~2 in \cite{Glendenning:1994zb}, it seems likely that Glendenning \textit{et al.} mistook the second-lowest eigenvalue for the lowest one, giving them the impression that these configurations were stable.  

To check that we have found the lowest eigenmode, we show in Fig.~\ref{fig:hybrid_star_eigenfunctions} the first three eigenfunctions of a strange dwarf configuration, with $P_c = 7\times 10^6 \text{ MeV}^4$.  The n\textsuperscript{th} eigenfunction has $n$ nodes, as should be the case for a Sturm-Liouville problem \cite{pryce1993numerical}.

The eigenvalue spectra shown in Figs.~\ref{fig:hybrid_star_SL} and \ref{fig:hybrid_star_SL_zoom} were calculated with a regulator width of $\delta P = 100 \text{ MeV}^4$.  To show that the results carry over to the discontinuous limit, in Fig.~\ref{fig:shrink_regulator} we show the behavior of the lowest eigenvalue for different regulator widths.  The spectrum depends slightly on the regulator width around point (b) on the mass-radius curve, but between points (c) and (d) there is almost no dependence on the regulator and the lowest eigenvalue remains negative in the discontinuous limit. 

We conclude that the mass-radius criteria remains valid, as it agrees with the results of the Sturm-Liouville analysis of the eigenmodes, even in the case where the equation of state has a first-order phase transition.  Therefore, strange dwarf hybrid stars are unstable and will not be found in nature.  
\chapter{Transport and neutron star mergers}
\pagestyle{myheadings}
\label{sec:mergers}
For the remainder of this thesis, the topic will be neutron stars mergers and the nuclear matter in those mergers.  Each of the next three chapters is a separate project we have worked on related to these mergers.  First, we set the stage by giving a general introduction to binary neutron star mergers.
\section{Neutron star merger observations}
\begin{figure}[h]
  \centering
  \includegraphics[scale=.7]{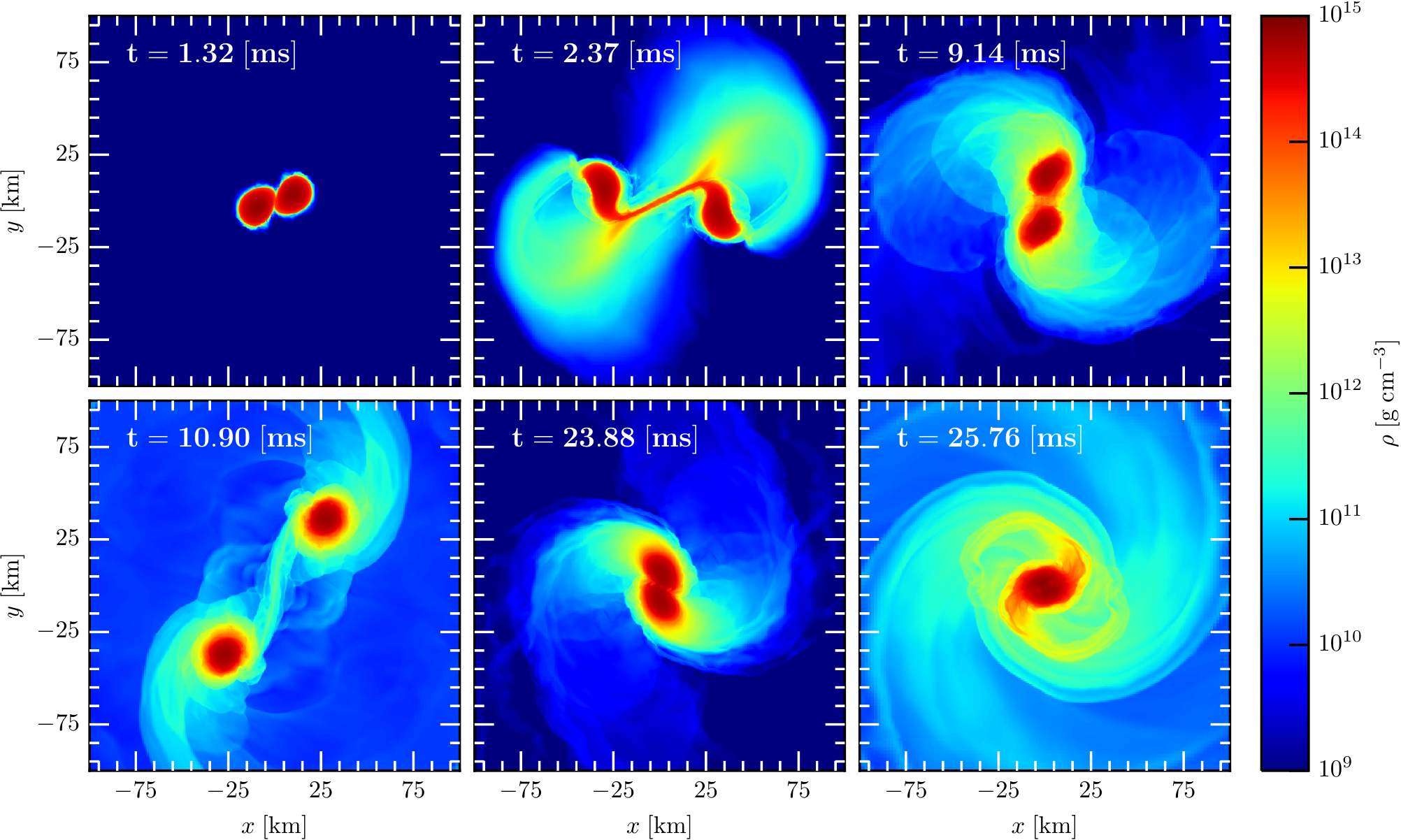}
  \caption{The results of a neutron star merger simulation.  The time from first-touch to combination into one object is about 25 milliseconds in this simulation.  Figure reproduced from \cite{Radice:2016dwd}.}
  \label{fig:radice_merger_panels}
\end{figure}

On August 17, 2017, the gravitational and electromagnetic signals from a neutron star merger, GW170817, were detected on Earth for the first time \cite{TheLIGOScientific:2017qsa}.  The gravitational wave signal was detected by LIGO, and two seconds after the merger a gamma ray burst was detected, installing GW170817 as the second member of a class of multimessenger astrophysical events, the first being supernova 1987a\footnote{SN1987A had a clear electromagnetic signal, but also a few neutrinos emitted from the supernova were detected \cite{Raffelt:1996wa}.}.  As a binary neutron star system evolves in time, the orbiting stars emit gravitational waves, removing energy from the orbit and bringing the stars closer together\footnote{The energy lost due to gravitational radiation for a circular binary system goes as $\mathop{dE}/\mathop{dt} \sim -\mu^2M^3/(a^5)$, where $\mu$ is the reduced mass, $M$ is the total mass, and $a$ is the separation between the two stars \cite{Misner:1974qy}.}.  As the stars move closer together, their orbital frequency increases.  The frequency of the gravitational waves emitted by the binary system is twice the orbital frequency \cite{thorne2017modern}, and so the frequency of emitted gravitational waves continues to increase as the orbit loses energy and the stars get closer together until the stars touch, merging into one object, either a neutron star or a black hole.  The part of this process before the two stars touch is called the inspiral.  In Fig.~\ref{fig:radice_merger_panels} we show results from a simulation depicting the various stages of a merger.  We will discuss simulations of mergers in Sec.~\ref{sec:simulations}.

\begin{figure}[h]
  \centering
  \includegraphics[scale=.5]{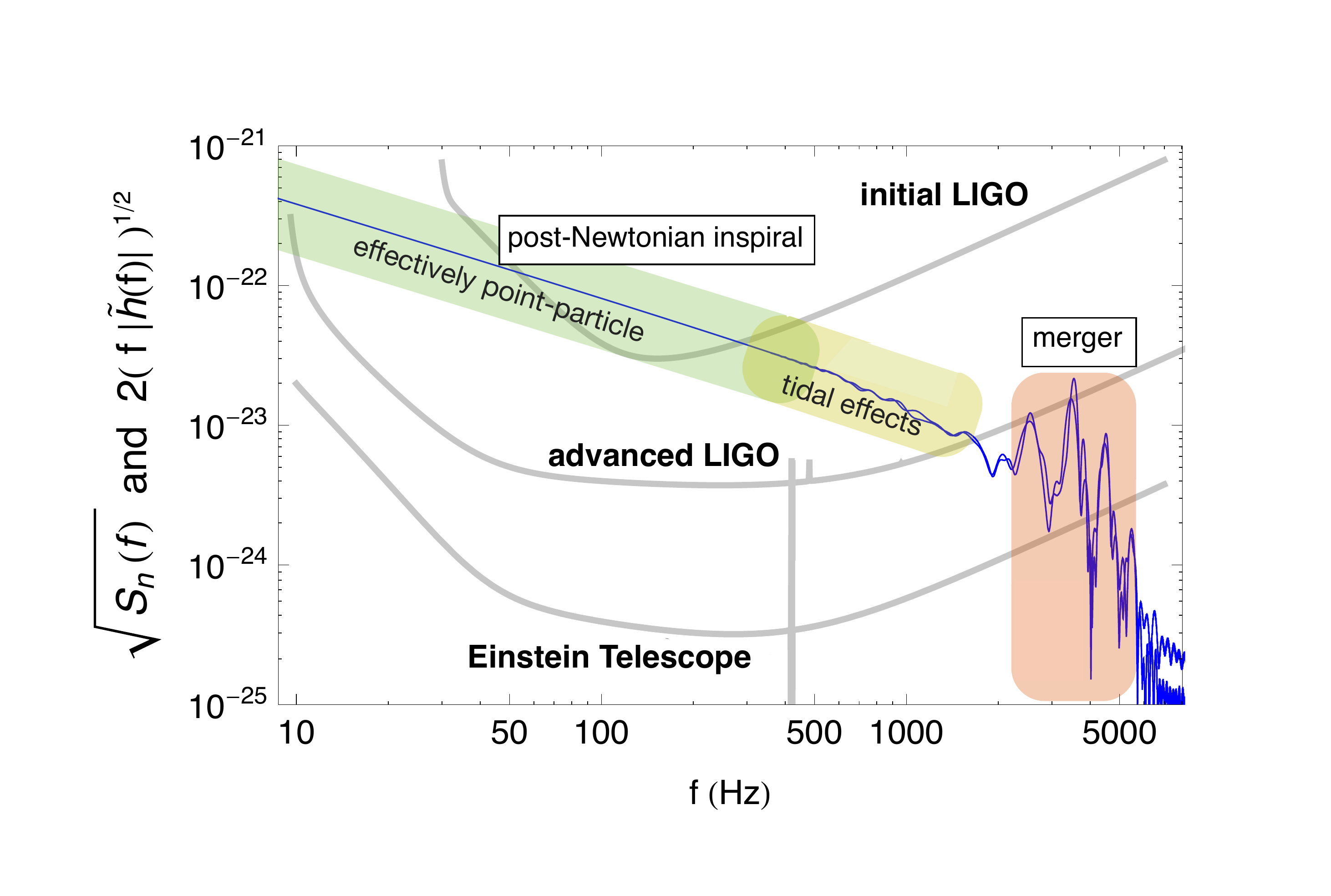}
  \caption{Sensitivity of LIGO over a wide range of gravitational wave frequencies - LIGO can detect gravitational waves above the given grey curve.  Superimposed is the gravitational wave signal as a function of frequency (which is a proxy for time, as discussed in the text).  Advanced LIGO can detect the inspiral, but not the post-merger signal.  Figure courtesy of J.~Read.  See also \cite{Read:2013zra,andersson2020gravitational}.}
  \label{fig:grav_wave_sim_signal}
\end{figure}

As the gravitational wave frequency increases throughout the inspiral phase, eventually it is high enough to be measured by LIGO.  Fig.~\ref{fig:grav_wave_sim_signal} shows the sensitivity of LIGO, Advanced LIGO, and the Einstein Telescope across a wide range of gravitational wave frequencies.  For a given detector, the grey line indicates the noise threshold - signals above the grey line at a particular frequency could be measured by the corresponding detector.  Superimposed is the gravitational wave signal as a function of gravitational wave frequency, which is a proxy for time because the frequency of the emitted gravitational waves increases as the inspiral progresses.  This figure shows us that Advanced LIGO, which detected GW170817, can only measure the inspiral, not the merger itself.  

As the two compact objects get close together, they begin to tidally deform one another due to the strong gravitational fields produced by both objects.  The amount of tidal deformation depends on the equation of state; a star that is stiff will not deform much, but one that is softer will experience greater deformation.  The degree of tidal deformation can be inferred from the gravitational wave signal.  Tidal deformation causes a phase shift in the signal because orbital energy goes into deforming the stars (as well as gravitational radiation).  Using the tidal deformation to study the equation of state is discussed extensively in \cite{Baiotti:2019sew}.

After the two neutron stars merge, unless there is prompt collapse to a black hole, a remnant neutron star will be created.  If the remnant has a mass larger than the TOV limiting mass, it survives for a period of time due to differential rotation, but it typically collapses within a few tens of milliseconds.  If the remnant has a mass less than the TOV limit, then it can survive indefinitely as a massive neutron star \cite{Baiotti:2016qnr,Lucca:2019ohp,Bernuzzi:2020tgt}.  The neutron star remnant experiences oscillations of fundamental modes with typical frequencies of a few kHz.  These oscillations emit gravitational waves, but unfortunately they are at frequencies at which Advanced LIGO is not sensitive.  However, in principle the equation of state can be deduced from the frequency of these gravitational waves, because each equation of state has a unique spectrum of oscillation modes \cite{Baiotti:2019sew}.  Similarly, the gravitational wave signal postmerger can determine if a phase transition from hadronic to quark matter was induced by the merger \cite{Bauswein:2018bma,Most:2018eaw}.

\section{Neutron star merger simulations}
\label{sec:simulations}
Starting two decades before GW170817, various groups have conducted numerical simulations of neutron star mergers.  Simulations are needed because of the complexity of the theories required to describe nuclear matter at high densities in dynamical, curved spacetime\footnote{This is true for the last several revolutions of the inspiral, where the neutron stars can no longer be treated like point particles (see Fig.~\ref{fig:grav_wave_sim_signal}).}.  Additionally, simulations are a playground in which the impact of adding or subtracting different types of physics can be explored.  Simulations predict observables like the gravitational wave signal or the optical signal, which can then be compared to observation to evaluate the simulation and its physical assumptions.

The backbone of a merger simulation is general relativity.  Einstein's equations are evolved from an initial configuration of two Schwarzschild solutions some distance apart.  A detailed account of the numerical solution of Einstein's equations is given in \cite{rezzolla2013relativistic}.  General relativity is also important to determine the gravitational radiation that escapes the system and is measured in our gravitational wave detectors.  

Just as in static neutron stars (see Sec.~\ref{sec:tov}), the matter in the merging neutron stars is modeled as a perfect fluid with some equation of state $\varepsilon = \varepsilon(P)$.  Viscosity can also be included in the stress-energy tensor\footnote{See, for example, Israel-Stewart theory \cite{rezzolla2013relativistic,Romatschke:2017ejr}.}.  Unlike isolated neutron stars, the fluid in a merger is not static, but obeys the equation of motion $\partial_{\mu}T^{\mu\nu}=0$, which is coupled with Einstein's equations because both contain the fluid stress-energy tensor.  Many simulations also couple these equations to Maxwell's equations in the form of magnetohydrodynamics (MHD) \cite{Baiotti:2016qnr}, as neutron stars have very strong magnetic fields and magnetic fields are expected to impact mergers.

As mergers represent violent, often far-from-equilibrium situations, a realistic nuclear equation of state must be specified both in and out of beta equilibrium.  Several such equations of state exist, many of them are tabulated on CompOSE \cite{CompOSE}.  The equation of state must be constructed for a wide range of temperatures and densities (see Sec.~\ref{sec:thermo_conditions}).   These equations of state are reviewed in \cite{Oertel:2016bki}.

Nuclear matter in neutron star mergers is in thermal equilibrium, because the mean free path of neutrons, protons, and electrons are certainly much less than the dynamical scales which are certainly larger than a meter (we neglect the possibility of turbulence here).  Thus the neutrons, protons, and electrons collide very frequently compared to merger timescales and thus stay in thermal equilibrium.  However, the neutrino mean free path varies widely throughout the merger (as we will see in Sec.~\ref{sec:nu_mfp}), and often the neutrinos are not in thermal equilibrium.  Thus, neutrinos are treated separately from the rest of the nuclear matter, and are either treated as their own fluid or in a kinetic theory formalism \cite{Ardevol-Pulpillo:2018btx, Perego:2014qda,Galeazzi:2013mia,Sekiguchi:2012uc,Rosswog:2003rv,1999JCoAM.109..281M}.
\section{Transport in mergers}
Neutron star merger simulations already contain an immense amount of physics, but it is natural to ask if there is something dramatic that they are missing.  Transport processes and viscosity are both possibilities.  This question was considered by Alford \textit{et al.}, who did back-of-the-envelope calculations to determine the importance of thermal conductivity, shear viscosity, and bulk viscosity \cite{Alford:2017rxf}.  Electrical conductivity was examined in \cite{Harutyunyan:2018mpe}.  To determine the importance of a particular physical effect, we determine the timescale on which it acts.  If the timescale is comparable to the merger timescale of tens of milliseconds, then the process might be relevant in mergers.  Of course, if one is interested in a long-lived neutron star remnant, then the timescale for relevance of a physical process must be adjusted accordingly.  Gravitational radiation damps large-scale motion in the post-merger system on a timescale of tens of milliseconds \cite{Alford:2017rxf}, so a transport process involving large-scale motion is only relevant if it acts on a timescale less than that of gravitational radiation.

Thermal conductivity $\kappa$, shear viscosity $\eta$, and bulk viscosity $\zeta$ enter the hydrodynamic equations via entropy production.  Entropy is conserved in a perfect fluid \cite{zee2013einstein}, but in a non-perfect fluid, entropy is generated by terms proportional to $\kappa$, $\eta$, and $\zeta$ (see Refs.~\cite{Schmitt:2017efp,weinberg1972gravitation} for the full details).  Thermal conductivity produces entropy when there are thermal gradients in the system, shear viscosity produces entropy when there is shear flow, and bulk viscosity produces entropy when the velocity field has a nonzero divergence, meaning the fluid is getting compressed or expanded.  We will discuss bulk viscosity in Ch.~\ref{sec:bulk_viscosity} and thermal conductivity in Ch.~\ref{sec:axions}.
\section{Thermodynamic conditions in mergers}
\label{sec:thermo_conditions}
Neutron star merger simulations, which typically do not take viscosity or transport processes into account,\footnote{However, simulations have an inherent viscosity that comes from numerical error leading to energy non-conservation \cite{Radice:2015qva}.  Also, the simulations track gravitational radiation, which damps large-scale motion of matter in the merger.} give us an idea of the thermodynamic conditions encountered by the nuclear matter, which is of vital importance for studying transport because many transport properties are strongly temperature-dependent.  It is well established that neutron stars reach densities of several times nuclear saturation density \cite{Abbott:2018exr,Landry:2018prl}, but before the merger they have temperatures below 1 MeV.  After the two stars touch, the temperature rises dramatically to tens of MeV, and the density rises by perhaps a factor of two.

\begin{figure}[h]
  \centering
  \includegraphics[scale=.6]{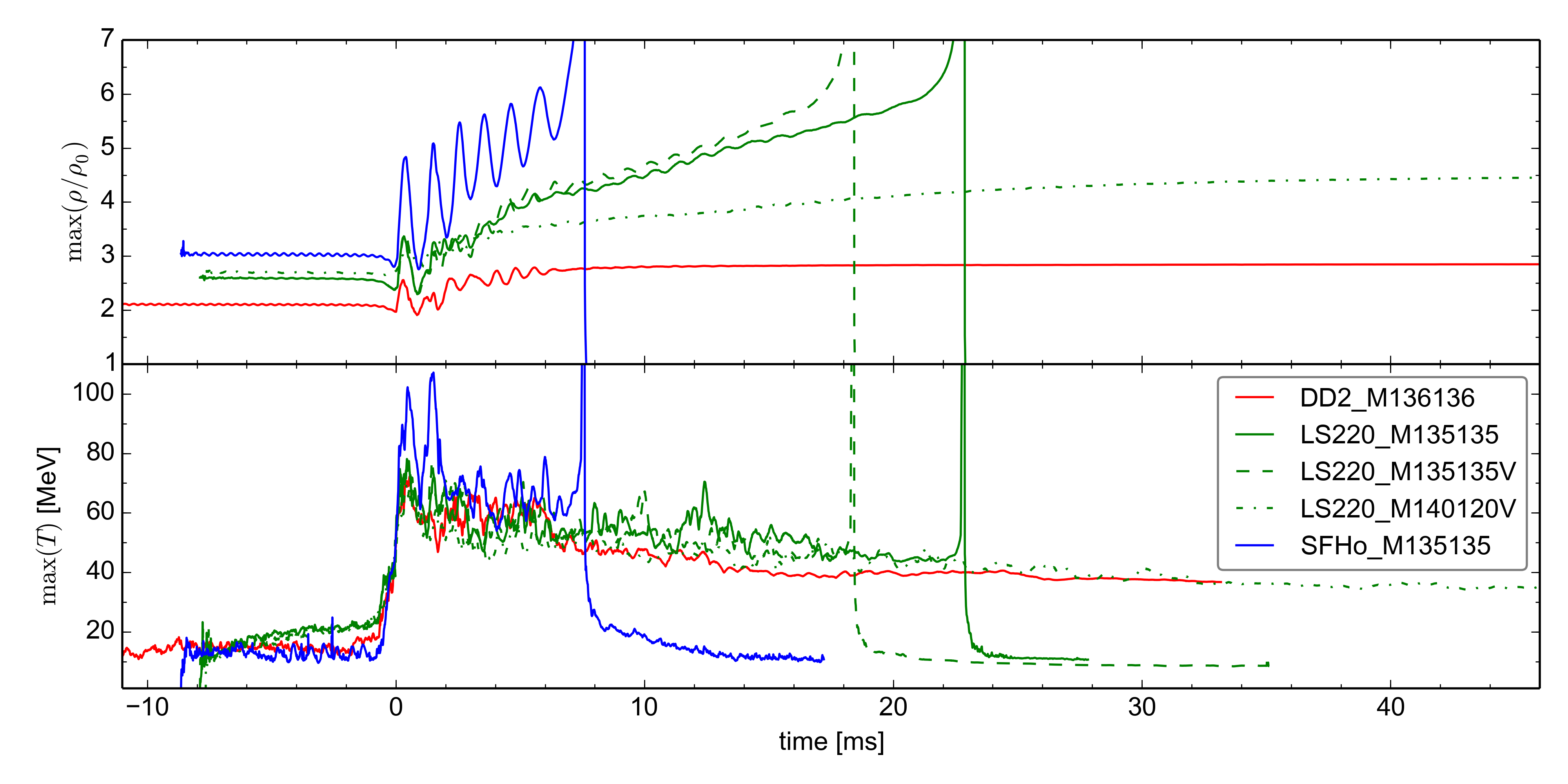}
  \caption{Maximum baryon density and temperature present in the merger as a function of time, where $t=0$ is when the two stars touch.  Each color corresponds to a different equation of state.  Figure reproduced from \cite{Perego:2019adq}.}
  \label{fig:max_nb_T}
\end{figure}

For example, Fig.~\ref{fig:max_nb_T} shows a plot from Perego, Bernuzzi, and Radice \cite{Perego:2019adq} which tracks the maximum temperature and density throughout a merger simulation.  Each different color line corresponds to a simulation run with a different equation of state.  We see that before the merger ($t<0$), the maximum density is just the density of the core of the more massive of the two inspiraling neutron stars.  When the stars touch at $t=0$, the maximum density changes wildly for several milliseconds, indicating that the nuclear matter is undergoing severe mechanical oscillations.  For some equations of state, the maximum density diverges as a black hole is formed.  For others, where a hypermassive remnant is formed, the density slowly increases as the hypermassive remnant ceases to differentially rotate and realizes its instability.  After about 20 milliseconds, a black hole is formed.  For still other equations of state, a stable neutron star is formed, which indeed has a larger core density than the two original stars.  As for the temperature, the inspiraling neutron stars start out with some temperature, which is artificially large in Fig.~\ref{fig:max_nb_T} for numerical reasons.  When the stars touch, the temperature spikes to 60-100 MeV depending on the equation of state, and then decreases as time progresses - an exception being when a black hole is formed, which causes the temperature to diverge.  

\begin{figure*}[t!]
\begin{minipage}[t]{0.5\linewidth}
\includegraphics[width=.95\linewidth]{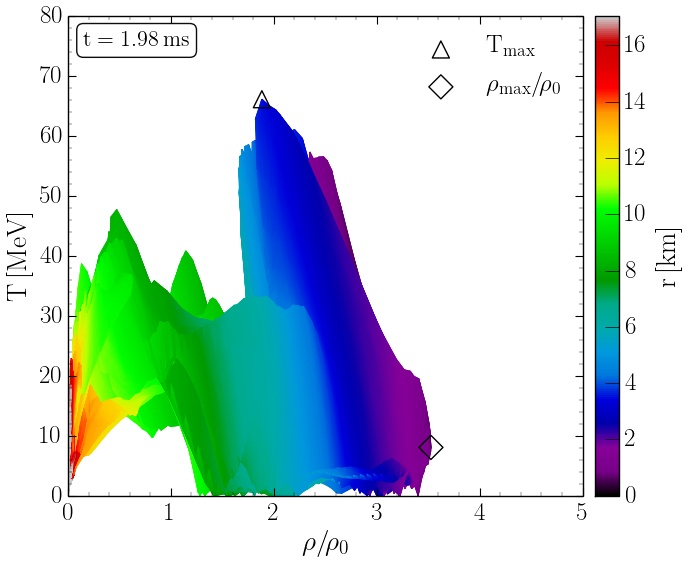}
\end{minipage}\hfill%
\begin{minipage}[t]{0.5\linewidth}
\includegraphics[width=.95\linewidth]{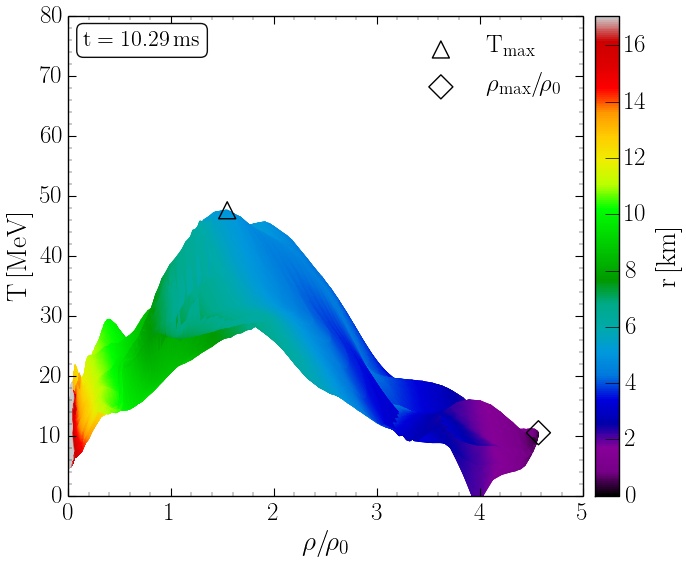}
\end{minipage}%
\caption{Temperature and densities attained by nuclear matter throughout a neutron star merger.  After ten milliseconds, this simulation indicates that the matter has settled into a configuration with cold, low density matter on the outskirts of the merger, hot medium-density matter in the outer core, and cold, high density matter in the core.  Figure courtesy of M.~Hanauske and the Rezzolla group, see also \cite{Hanauske:2019qgs}.}
\label{fig:T_nb_plane}
\end{figure*}
\begin{figure}[H]
  \centering
  \includegraphics[scale=1]{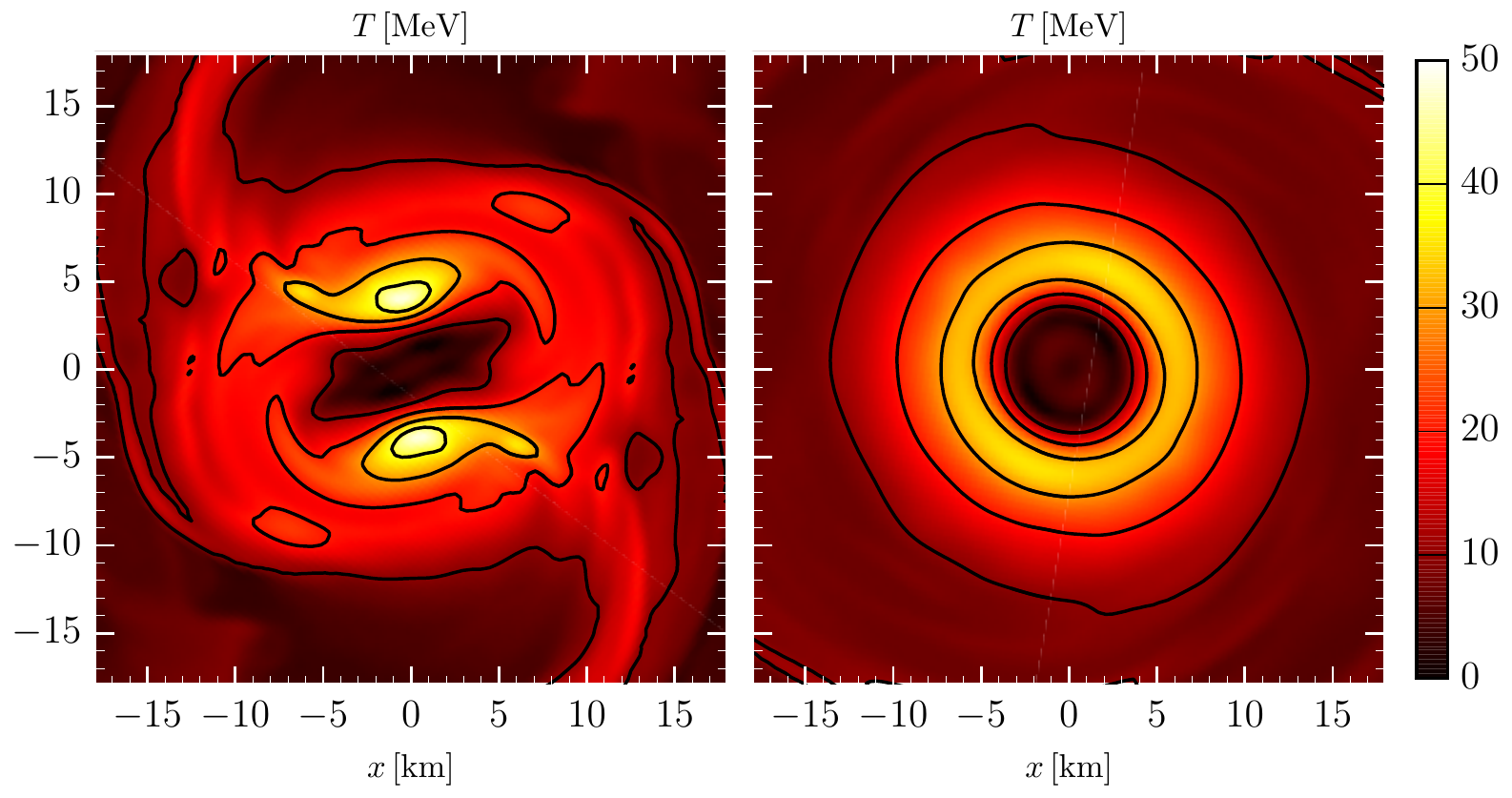}
  \caption{Temperature distribution in neutron star merger 6.71 ms postmerger (left panel) and 23.83 ms postmerger (right panel).  The right panel indicates that the region of the merger than experiences extremely large temperatures (above, say, 20 MeV) is a spherical shell which is 1-2 km wide.  Figure reproduced from \cite{Hanauske:2016gia}.}
  \label{fig:T_distribution}
\end{figure}

It is also informative to look at the temperature and density distributions throughout the merging neutron stars.  In Fig.~\ref{fig:T_nb_plane} we show the results of simulations by the Frankfurt group, where they have recorded the temperature and density of fluid elements in the merger at a few snapshots in time, in particular 1.98 ms postmerger (left panel) and 10.29 ms postmerger (right panel).  Right at the time of merger, the temperature increases dramatically, up to 70 MeV according to the left panel of Fig.~\ref{fig:T_nb_plane}.  After several milliseconds of chaos, the merger has settled down into an arch-like structure in the $\{n_B,T\}$ plane, where the outskirts of the merger are relatively cold and low density, the outer core of the merger is hot and moderate density, and the core is cold and high density.  This is illustrated in Fig.~\ref{fig:T_distribution}, which shows that the highest temperatures are reached not in the core of the merger remnant, but in a spherical shell 1-2 km thick that is in the outer-core region \cite{Hanauske:2016gia}, which has density of $1-2n_0$.  See similar figures in \cite{Perego:2019adq}. 

\chapter{Beta equilibrium in neutrino-transparent nuclear matter}
\pagestyle{myheadings}
\label{sec:beta_equilibrium}
\begin{center}
{\textit{This section is based on my work with Mark Alford, \cite{Alford:2018lhf}}. \\\copyright 2018 American Physical Society}.
\end{center}
\section{Neutrinos in nuclear matter}
\label{sec:nu_mfp}
The weak interactions that beta equilibrate nuclear matter\footnote{These are called Urca processes $n\rightarrow p + e^- + \bar{\nu}$ and $e^- + p\rightarrow n + \nu$, which we discuss in the next section.} produce electron neutrinos and antineutrinos.  Neutrinos and antineutrinos can participate in both charged current and neutral current processes, the rate of which depends on the density, temperature, and composition of matter with which the neutrinos are interacting.  If the neutrinos interact often, they will have a short mean free path and if they interact rarely they will have a long mean free path.  If the neutrino mean free path is very long compared to the size of a neutron star, then neutrinos that are created in the neutron star free-stream from the star and do not build up a population.  If the neutrino mean free path is short compared to the system size, then the neutrinos build up a population with some chemical potential $\mu_{\nu}$ according to the Fermi-Dirac distribution.  The value of $\mu_{\nu}$ is determined by the beta equilibrium condition, which is explained in Sec.~\ref{sec:both_beta_eq_conditions}.

\begin{figure}[h]
  \centering
  \includegraphics[scale=.6]{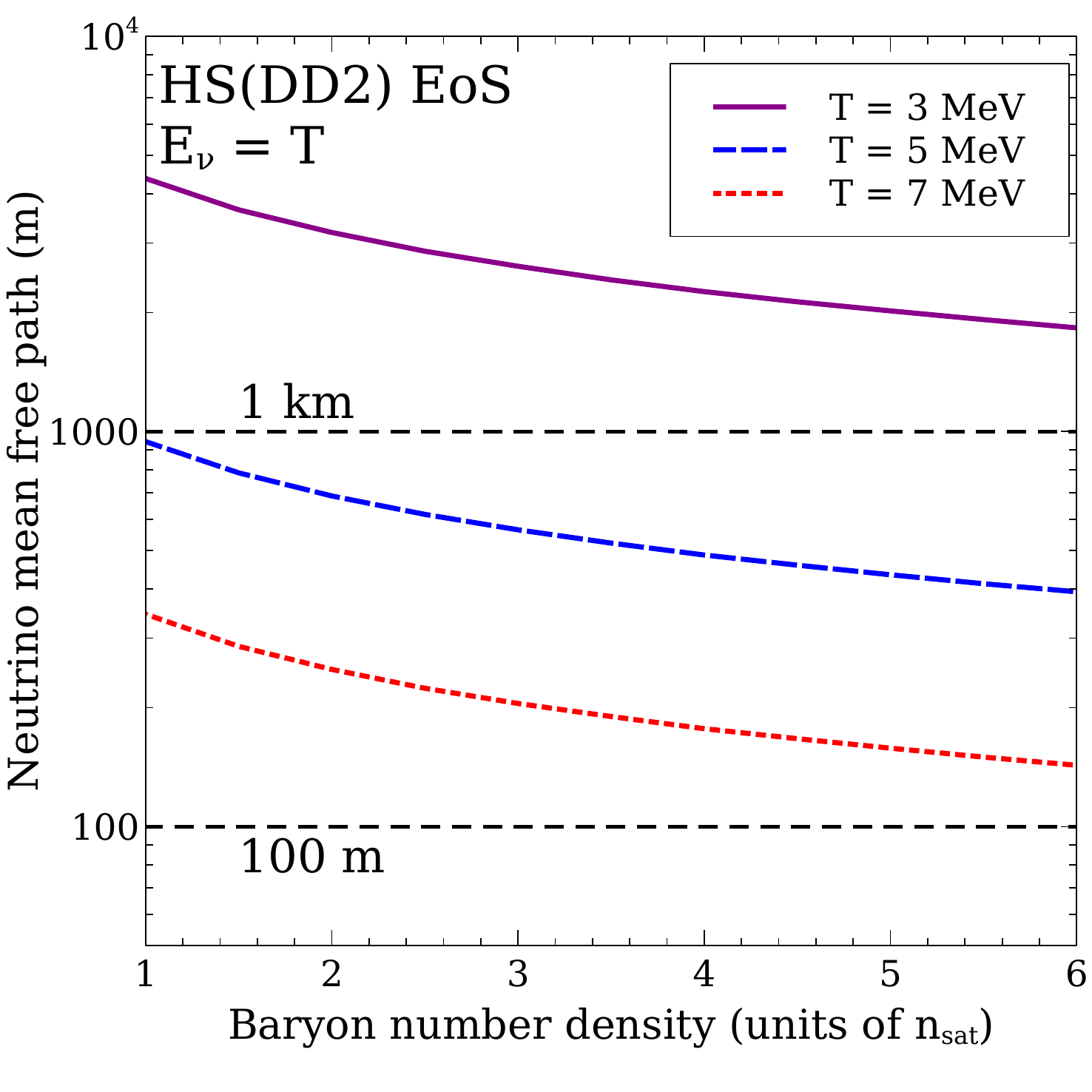}
  \caption{Mean free path of neutrino with energy $E_{\nu}=T$ in nuclear matter described by the DD2 equation of state.  Neutrinos become trapped in the merger as the temperature increases above 5 MeV.}
  \label{fig:nu_mfp}
\end{figure}

The mean free path of neutrinos and anti-neutrinos has been calculated for several decades with increasing levels of sophistication \cite{Sawyer:1975js,1979ApJ...230..859S,1985ApJS...58..771B,Haensel:1987zz,Reddy:1997yr,Burrows:2004vq}.  Most recently, kinematic effects relating the neutron and proton mean fields have been taken into account \cite{Roberts:2016mwj,Roberts:2012um} and a software package for calculating neutrino mean free paths has been developed \cite{nuopac}.  We used this software to calculate the neutrino mean free path using the nuclear mean fields from the DD2 equation of state (see Sec.~\ref{sec:nucl_matter_EoSs}), as a function of temperature and density.  The results are plotted in Fig.~\ref{fig:nu_mfp}.  The neutrino mean free path also depends on the energy of the neutrino, which we have set to the temperature, signifying thermal neutrinos.  At a temperature of 3 MeV, neutrinos have a mean free path of a few kilometers, so they would easily escape from the merger.  As the temperature rises above 5 MeV, the mean free path drops below one kilometer, where we consider neutrinos to be trapped.  At these densities, the process trapping neutrinos is neutral current scattering $n+\nu \rightarrow n+\nu$.

\section{Beta equilibrium and flavor equilibration}
\label{sec:both_beta_eq_conditions}
As discussed in Sec.~\ref{sec:EoS}, a long-lived astrophysical object must be close to beta equilibrium (also called chemical equilibrium), meaning its particle content does not change over time.  The particle content in nuclear matter is changed by weak interactions, called Urca processes in this context.  Let us first consider neutrino-trapped nuclear matter, where there are Fermi seas of neutrons, protons, electrons, and either neutrinos or antineutrinos (this depends on the lepton fraction $Y_L = (n_e + n_{\nu})/n_B$, which depends on the history of the system).  In this system, there are two direct Urca (dUrca) processes
\begin{align}
    n&\leftrightarrow p+e^-+\bar{\nu}\qquad \text{dUrca neutron decay}\\
    e^-+p &\leftrightarrow n + \nu \qquad \text{dUrca electron capture},
\end{align}
but also four modified Urca (mUrca) processes
\begin{align}
    n+N&\leftrightarrow p+e^-+\bar{\nu}+N\qquad \text{mUrca neutron decay}\\
    e^-+p +N&\leftrightarrow n + \nu +N\qquad \text{mUrca electron capture},
\end{align}
where $N$ is a spectator neutron or proton.  The modified Urca processes are important when direct Urca is kinematically forbidden (see Sec.~\ref{sec:durca_FS}).  If there is a large neutrino Fermi sea, then the anti-neutrino processes are suppressed and vice-versa.  In beta-equilibrated nuclear matter, the forward and backward rates for each of these processes balance - this is called detailed balance.  A closed system in chemical equilibrium has a constraint on the chemical potentials of its constituent particles.  For example, if you have system of particle species $A$, $B$, $C$, and $D$ in chemical equilibrium, where a reaction $aA+bB\leftrightarrow cC+dD$ takes place ($a$, $b$, $c$, and $d$ represent stoichiometric coefficients), then $a\mu_A+n\mu_B=c\mu_C+d\mu_D$ \cite{glendenning2000compact,thorne2017modern}.  We apply this first to the pair production process, for example, neutron bremsstrahlung $n+n\leftrightarrow n+n+\nu+\bar{\nu}$.  If the system of neutrons, protons, electrons, and neutrinos is in equilibrium, then $2\mu_n = 2\mu_n + \mu_{\nu}+\mu_{\bar{\nu}}$ and thus neutrinos and antineutrinos have chemical potentials of the same magnitude but opposite sign.  This is true for any particle and its antiparticle.  Now if we apply the chemical equilibrium condition to any of the Urca processes, we find that the beta equilibrium condition in neutrino-trapped nuclear matter is\footnote{Recall that we use relativistic definitions for the chemical potentials, so they include the rest mass.}
\begin{equation}
    \mu_n + \mu_{\nu} = \mu_p + \mu_e.
    \label{eq:beq_nu_trapped}
\end{equation}

Neutrino-transparent nuclear matter contains Fermi seas of neutrons, protons, and electrons, but not neutrinos.  In addition, neutrinos and anti-neutrinos do not participate in the initial state of any process, because their mean free path is larger than the system size.  Thus, only the direct Urca processes 
\begin{align}
    n&\rightarrow p+e^-+\bar{\nu}\qquad \text{dUrca neutron decay}\label{eq:dU_1}\\
    e^-+p &\rightarrow n + \nu \qquad \text{dUrca electron capture},
\end{align}
and the modified Urca processes
\begin{align}
    n+N&\rightarrow p+e^-+\bar{\nu}+N\qquad \text{mUrca neutron decay}\\
    e^-+p +N&\rightarrow n + \nu +N\qquad \text{mUrca electron capture},\label{eq:mU_2}
\end{align}
which have neutrinos or antinuetrinos in the final state can proceed.  In this case, detailed balance no longer applies because there are no processes that are inverses of each other in neutrino-transparent matter\footnote{For a similar discussion in the context of hot plasmas, see Refs.~\cite{Yuan:2005tj,liu2011steady}.}.  It is typically assumed that the neutrino or antineutrino is kinematically negligible, (the conventional wisdom is that its energy is typically a few times the temperature \cite{1992A&A...262..131H}), and so it is common to think of the direct Urca processes as approximately $n  \leftrightarrow p + e^-$, in which case neutron decay and electron capture \textit{are} inverse reactions and detailed balance dictates that their rates are equal, yielding the beta equilibrium condition in neutrino-transparent nuclear matter,
\begin{equation}
    \mu_n = \mu_p + \mu_e.
    \label{eq:beq_nu_trans}
\end{equation}
Because of the reasons above, we will call this the low-temperature beta equilibrium condition.

Matter remains transparent to neutrinos at temperatures of up to about 5 MeV, but by this point the neutrino is no longer kinematically negligible.  We will see in Fig. \ref{fig:FS_deviation} that the neutrino needs to have an energy of 10-15 MeV for the direct Urca electron capture process to proceed below the direct Urca threshold.  In the next sections, we study the validity of the low-temperature beta equilibrium condition [Eq.~(\ref{eq:beq_nu_trans})] in neutrino-transparent nuclear matter.  To do this, we will calculate the rate (number of reactions per time per volume) of the Urca processes and see if they balance when the beta equilibrium condition [Eq.~(\ref{eq:beq_nu_trans})] is imposed.
\section{Urca processes and the Fermi surface approximation}
\label{sec:urca_and_FS}
We consider only the six Urca processes [Eq.~(\ref{eq:dU_1})-(\ref{eq:mU_2})].  Weak interactions involving positrons are negligible, since the electron chemical potential is always above $100 \text{ MeV}$ for the densities that we will consider, and so the positron occupation is suppressed by a factor of more than $\exp(-100 \text{ MeV}/T)$.  Additionally, for simplicity we neglect Urca processes involving muons, because even though those processes are not negligible, they do not qualitatively change the conclusions that we present here.

We now obtain the standard expressions for the rate of the direct and modified Urca processes in matter with the APR equation of state.  We will assume ultra-relativistic electrons and neutrinos, but nucleons that are non-relativistic, with dispersion relation
\begin{equation}
\label{eq:dispersion}
E_i = m_{\text{eff},i} + \dfrac{p_i^2}{2m_i}
\end{equation}
where, following Roberts \textit{et al.} \cite{Roberts:2012um}, at each density $m_{\text{eff},i}$ is chosen such that the Fermi energy $E_{F,i} \equiv E_i(p_{F_i})$ matches the chemical potential $\mu_i$ from the APR equation of state, which is a simple way of taking into account the nuclear mean field.  For the kinetic mass $m_i$ we use the rest mass in vacuum.
\hiddensubsection{Direct Urca}
\label{sec:durca_FS}
The rates of the two direct Urca processes are given by the phase space integrals \cite{Yakovlev:2000jp,Villain:2005ns}
\begin{align}
\Gamma_{dU,nd} &= \int \dfrac{\mathop{d^3p_n}}{\left(2\pi\right)^3}\dfrac{\mathop{d^3p_p}}{\left(2\pi\right)^3}\dfrac{\mathop{d^3p_e}}{\left(2\pi\right)^3}\dfrac{\mathop{d^3p_{\nu}}}{\left(2\pi\right)^3}\frac{\sum_{\text{spins}}\vert\mathcal{M}\vert^2}{2^4E_nE_pE_eE_{\nu}}\left(2\pi\right)^4\delta^4(p_n-p_p-p_e-p_{\nu}) f_n\left(1-f_p\right)\left(1-f_e\right)\label{eq:ndecay} \\
\Gamma_{dU,ec} &= \int \dfrac{\mathop{d^3p_n}}{\left(2\pi\right)^3}\dfrac{\mathop{d^3p_p}}{\left(2\pi\right)^3}\dfrac{\mathop{d^3p_e}}{\left(2\pi\right)^3}\dfrac{\mathop{d^3p_{\nu}}}{\left(2\pi\right)^3}\frac{\sum_{\text{spins}}\vert\mathcal{M}\vert^2}{2^4E_nE_pE_eE_{\nu}} \left(2\pi\right)^4\delta^4(p_n-p_p-p_e+p_{\nu}) \left(1-f_n\right)f_pf_e,\label{eq:ecapture}
\end{align}
where $f_i$ are the Fermi-Dirac distributions for $n,p,\text{ or }e$, and the matrix element in the approximation of non-relativistic nucleons is 
\begin{equation}
\frac{\sum_{\text{spins}}\vert\mathcal{M}\vert^2}{2^4E_nE_pE_eE_{\nu}} =  2G^2\left(1+3g_A^2+\left(1-g_A^2\right)\dfrac{\mathbf{p}_e\cdot\mathbf{p}_{\nu}}{E_eE_{\nu}}\right),
\end{equation}
where $G^2 = G_F^2\cos^2{\theta_c} = 1.1\times10^{-22} \text{ MeV}^{-4}$, where $G_F$ is the Fermi coupling constant and $\theta_C$ is the Cabibbo angle, and the axial vector coupling constant $g_A = 1.26$.  See Appendix \ref{sec:durca_M_derivation} for a derivation of this matrix element.

The direct Urca rate is commonly evaluated in strongly degenerate systems, for example, cold neutron stars, in which case the rate can be calculated analytically using the Fermi surface (FS) approximation.  In strongly degenerate systems, only particles near their respective Fermi surfaces can participate in processes, so in the Fermi surface approximation, all momentum magnitudes in the phase space integral are set equal to the appropriate Fermi momentum.  As the neutrino has no Fermi surface, it has negligible momentum and it is set to zero.  The phase space integral is converted to spherical coordinates, and then split into an angular part and a momentum magnitude (or energy) part\footnote{This is called phase space decomposition \cite{Shapiro:1983du}.}.  While the momenta are set to their respective Fermi momenta, the particle energies are integrated over, consistent with the thermal blurring of the Fermi surface (although inconsistent with the momentum magnitudes having fixed values, because energy and momentum are related through the particle dispersion relation $E=E(p)$.).  The energy integral is evaluated in \cite{Shapiro:1983du,Yakovlev:2000jp} and the angular integral in \cite{Kaminker:2016ayg}.  When we impose the low-temperature neutrino-transparent beta equilibrium condition [Eq.~(\ref{eq:beq_nu_trans})], we find that the direct Urca neutron decay and electron capture rates balance and are given by \cite{Yakovlev:2000jp,2000A&A...357.1157H,Kaminker:2016ayg}
\begin{align}
\label{eq:dUrca_FS}
\Gamma_{dU,nd} &= \Gamma_{dU,ec} =A_{dU} G^2\left(1+3g_A^2\right) m_nm_pp_{Fe}\vartheta_{dU}T^5 \\
\vartheta_{dU} &\equiv \left\{
 \begin{array}{ll}
  0 & \text{if}\ p_{Fn}>p_{Fp}+p_{Fe}\\
 1 & \text{if}\  p_{Fn}<p_{Fp}+p_{Fe},
 \end{array} \right. \nonumber \\[2ex]
A_{dU} &\equiv 3\left(\pi^2\zeta(3)+15\zeta(5)\right)/(16\pi^5)\approx 0.0170.\nonumber
\end{align}

\begin{figure}[h]
  \centering
  \includegraphics[scale=.6]{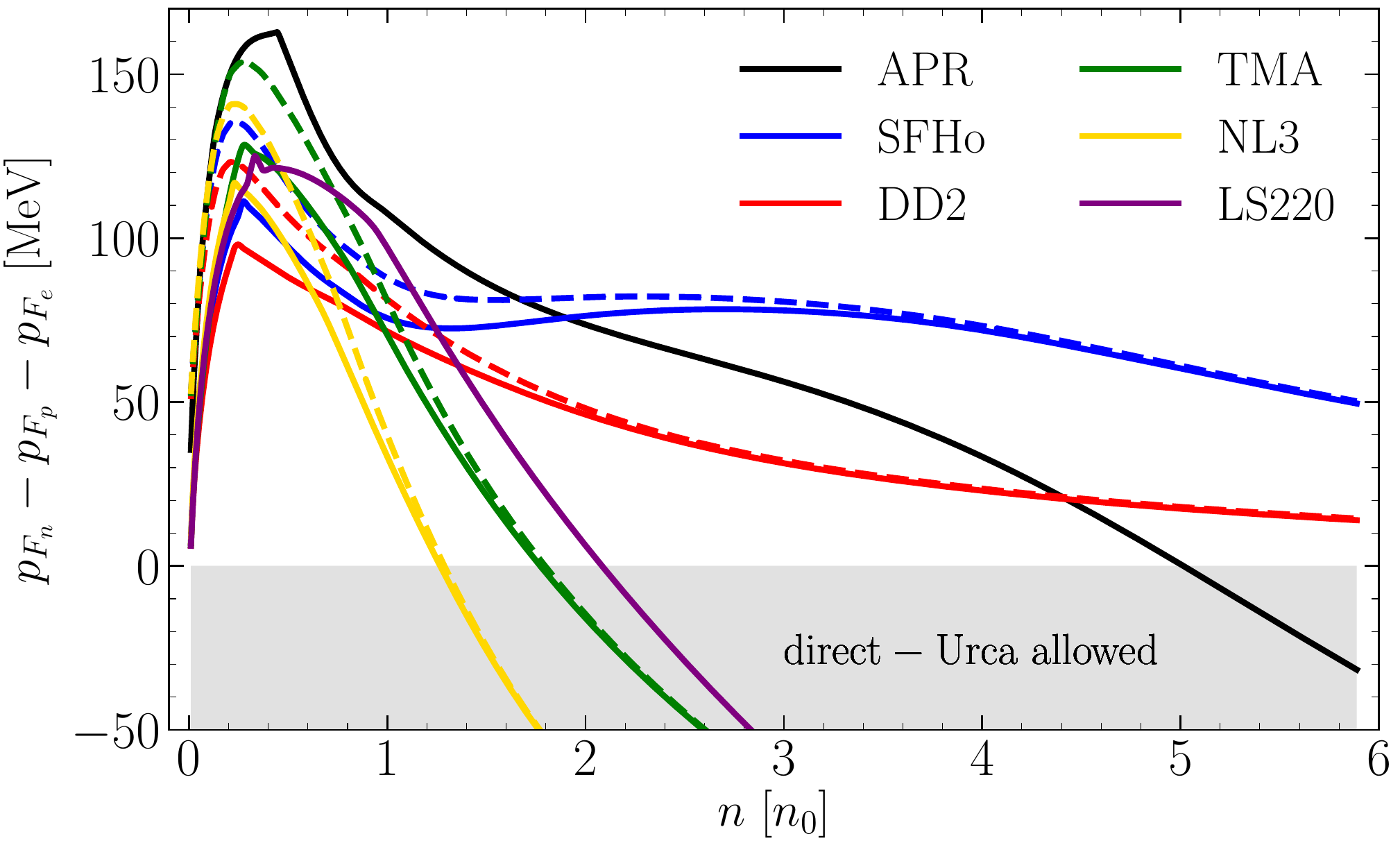}
  \caption{Momentum mismatch $p_{Fn}-p_{Fp}-p_{Fe}$ for different equations of state.  When $p_{Fn}-p_{Fp}-p_{Fe}<0$, direct Urca is kinematically allowed.  Figure reproduced from \cite{Alford:2017rxf}.}
  \label{fig:dUrca_threshold}
\end{figure}

We see from this expression that in the Fermi surface approximation, direct Urca has a threshold - it only operates at densities where $p_{Fn}<p_{Fp}+p_{Fe}$.  In nuclear matter, direct Urca proceeds for densities above the threshold density, as the proton fraction increases at densities above $n_0$.  For densities below the threshold density, the electron and proton Fermi momenta are not large enough to add up to the neutron Fermi momentum, and so direct Urca is kinematically forbidden.  As density increases, the proton and electron Fermi momenta grow more quickly than the neutron Fermi momentum and when the threshold is reached, they, when coaligned, add up to exactly the neutron Fermi momentum.  Above the direct Urca threshold, the proton and electron Fermi momenta can add up to the neutron Fermi momentum even when they are not co-aligned \cite{Yakovlev:2000jp}.  Different equations of state have different direct Urca thresholds.  Fig.~\ref{fig:dUrca_threshold} shows the momentum mismatch $p_{Fn}-p_{Fp}-p_{Fe}$ for several different equations of state.  Where the momentum mismatch is negative, the direct Urca process is allowed.  Some equations of state never allow direct Urca to proceed, at least, as long as the Fermi surface approximation is applicable.

\hiddensubsection{Modified Urca}
When direct Urca is kinematically forbidden, modified Urca is the dominant weak interaction.  When direct Urca is allowed, modified Urca is subdominant, as we will see.  The four modified Urca rates are given by phase space integrals analogous to Eq.~(\ref{eq:ndecay}) and (\ref{eq:ecapture}).  For the modified Urca processes, which involve strong interactions between the nucleons, we use the matrix elements given by Yakovlev \textit{et al.} \cite{Yakovlev:2000jp} and Friman and Maxwell \cite{1979ApJ...232..541F}, which involve a long-range one-pion exchange interaction \cite{OPE}.  The matrix element for neutron decay and electron capture with a neutron spectator (n-spectator modified Urca) is given by
\begin{equation}
S\frac{\sum_{\text{spins}}\vert\mathcal{M}_n\vert}{2^6E_nE_pE_eE_{\nu}E_{N1}E_{N2}} = 42G^2\dfrac{f^4}{m_{\pi}^4}\dfrac{g_A^2}{E_e^2}\dfrac{p_{Fn}^4}{\left(p_{Fn}^2+m_{\pi}^2\right)^2},
\end{equation}
and the matrix element for neutron decay and electron capture with a proton spectator (p-spectator modified Urca) is given by
\begin{equation}
S\frac{\sum_{\text{spins}}\vert\mathcal{M}_p\vert}{2^6E_nE_pE_eE_{\nu}E_{N1}E_{N2}} = 48G^2\dfrac{f^4}{m_{\pi}^4}\dfrac{g_A^2}{E_e^2}\dfrac{(p_{Fn}-p_{Fp})^4}{\bigl((p_{Fn}-p_{Fp})^2+m_{\pi}^2\bigr)^2},
\end{equation}
with the pion-nucleon coupling constant $f \approx 1$.  In both cases, $S=1/2$ because there is one set of identical particles in each process.

When the low-temperature beta equilibrium condition (\ref{eq:beq_nu_trans}) is used, the n-spectator modified Urca neutron decay and electron capture rates are equal and given by  \cite{Yakovlev:2000jp,Kaminker:2016ayg,Shapiro:1983du,Haensel:2001mw}
\begin{align}
\label{eq:mUrca_n}
& \Gamma_{\text{mU,n}}
 = A_{mU}G^2f^4g_A^2\dfrac{m_n^3m_p}{m_{\pi}^4} \dfrac{p_{Fn}^4p_{Fp}}{\left(p_{Fn}^2+m_{\pi}^2\right)^2} \, \vartheta_n  \, T^7 \ , \\[2ex]
\vartheta_n &\equiv \left\{
 \begin{array}{ll}
  1 & \text{if}\ p_{Fn}>p_{Fp}+p_{Fe}\\
 1-\dfrac{3}{8}\dfrac{(p_{Fp}+p_{Fe}-p_{Fn})^2}{p_{Fp}p_{Fe}} & \text{if}\  p_{Fn}<p_{Fp}+p_{Fe}.
 \end{array} \right. \nonumber
\end{align}
See Sec.~6 of \cite{Kaminker:2016ayg} for a comprehensive discussion of the integrals involved in the Fermi surface approximation of the modified Urca rates, including clarification of errors and omissions in the literature.

The p-spectator modified Urca neutron decay and electron capture rates are equal to each other when (\ref{eq:beq_nu_trans}) holds, and are given by \cite{Yakovlev:2000jp,Kaminker:2016ayg,Shapiro:1983du,Haensel:2001mw}
\begin{equation}
\label{eq:mUrca_p}
\Gamma_{\text{mU,p}} = \dfrac{A_{mU}}{7}G^2f^4g_A^2 \dfrac{m_nm_p^3}{m_{\pi}^4}
\dfrac{p_{Fn} (p_{Fn}\!-\!p_{Fp})^4}{\bigl((p_{Fn}\!-\!p_{Fp})^2+m_{\pi}^2\bigr)^2}\, \vartheta_p  T^7
\end{equation}
\begin{align}
\vartheta_p &\equiv \left\{
\begin{array}{ll}
\qquad 0 &\text{if} \ p_{Fn}>3p_{Fp}+p_{Fe}\\[3ex]
 \dfrac{(3p_{Fp}+p_{Fe}-p_{Fn})^2}{p_{Fn}p_{Fe}} & \text{if}\ \begin{array}{l} p_{Fn}>3p_{Fp}-p_{Fe}\\p_{Fn}<3p_{Fp}+p_{Fe}\end{array} \\[3ex]
 4\dfrac{3p_{Fp}-p_{Fn}}{p_{Fn}} 
   & \text{if}\ \begin{array}{l} 3 p_{Fp}-p_{Fe}>p_{Fn} \\ p_{Fn} >p_{Fp}+p_{Fe}\end{array} \\[3ex]
 2+3\dfrac{2p_{Fp}-p_{Fn}}{p_{Fe}}
  -\,3\dfrac{(p_{Fp}-p_{Fe})^2}{p_{Fn}p_{Fe}} & \text{if}\ p_{Fn}<p_{Fp}+p_{Fe} \ .
\end{array}\right. \nonumber
\end{align}
where $A_{mU} \approx 7\times 2300 / (64\pi^9)\approx .0084. $  We see that the p-spectator modified Urca process does have a threshold density, in this case the density where $p_{F_n}=3p_{F_p}+p_{F_e}$, which occurs at a proton fraction $x_p = 1/65$.  Thus, the p-spectator modified Urca process is only prohibited at extremely low densities \cite{Yakovlev:2000jp}, well below nuclear saturation density, which is the minimum density that we consider here.  
In Appendix \ref{sec:out_of_equil_mU}, we give the Fermi-surface approximation for the modified Urca rates when Eq.~(\ref{eq:beq_nu_trans}) is violated by an amount $\xi = (\mu_n-\mu_p-\mu_e)/T$.

In the $T\to 0$ limit, where the Fermi surface approximation is valid,
the standard low-temperature beta equilibrium condition holds: when
Eq.~(\ref{eq:beq_nu_trans}) is obeyed, the neutron decay and electron capture rates balance for both direct and modified Urca processes.

In the upper panels of Fig \ref{fig:urca_rates}, we have plotted, among other curves that we explain in Sec.~\ref{sec:exact}, the Fermi-surface approximation of the two direct Urca (in dotted, green) and four modified Urca (labeled ``mU'', in blue) rates in APR matter for $T = 500 \text{ keV}$ and $5 \text{ MeV}$ respectively.  For the APR equation of state the direct Urca threshold density is around $5n_0$.  Above threshold, the direct Urca neutron decay and electron capture rates are identical and dominate over the modified Urca processes which have no threshold in the density range we consider.  Below threshold, neither direct Urca process is allowed and so the four modified Urca processes dominate.  The two n-spectator modified Urca processes are slightly more important than the two p-spectator modified Urca processes.  As long as the Fermi surface approximation is used, and Eq.~(\ref{eq:beq_nu_trans}) is imposed, the proton-producing Urca processes balance the neutron-producing Urca processes exactly at all densities and temperatures.

\section{Beyond the Fermi surface approximation}
\label{sec:beyond_FS}
Nuclear matter is transparent to neutrinos for temperatures up to about 5 MeV, but at such temperatures, the Fermi surface approximation might no longer be valid.  The energy distribution of fermions has width $T$ near the Fermi energy \cite{swendsen2020introduction}, which translates - for nonrelativistic particles like the neutron and proton - into a momentum-space blurring of the Fermi surface $\Delta p = (m/p_F)\Delta E = mT/p_F$.  At neutron star densities with $T=5\text{ MeV}$, $\Delta p$ is largest for the protons, and could be up to 30 MeV, significantly easing the kinematic barriers to direct Urca.   

For densities below the direct Urca threshold, we can estimate the range of validity of the Fermi surface approximation by noting that it will become invalid when the exponential suppression of direct Urca processes involving particles away from their Fermi surface is not so severe as to make those processes negligible relative to modified Urca. In direct Urca processes the proton is expected to play a crucial role, since it is the most non-relativistic fermion which means that the energy $E_p$
of a proton rises very slowly as the momentum $p_p$ of the proton
deviates from its Fermi surface: $E_p-E_{F_p}\sim (p_p-p_{F_p})p_{F_p}/m$.  For particles on their Fermi surfaces, the momentum
mismatch for direct Urca at densities around $3n_0$ in nuclear matter
described by the APR equation of state (Sec.~\ref{sec:nucl_matter_EoSs}) is
$p_{\text{miss}}=p_{Fn}-p_{Fp}-p_{Fe}\approx 50 \text{ MeV}$ (see Fig.~\ref{fig:dUrca_threshold}), and the proton Fermi momentum is about 220 MeV.  The energy cost of finding a proton that is  $p_{\text{miss}}$ from its Fermi surface
is $p_{\text{miss}} p_{Fp}/m \approx 12 \text{ MeV}$, so we might expect that direct Urca electron capture, where the probability of finding a proton from above its Fermi surface includes a Boltzmann factor, becomes unsuppressed at temperatures of order 10 MeV, and that it starts to compete with modified Urca at even lower temperatures.  

In fact, the modified Urca rate is approximately a factor of $\left(m_n T/(3 m_{\pi}^2) \right)^2$ smaller than the above-threshold direct Urca rate (see Eq.~(\ref{eq:dUrca_FS}) and (\ref{eq:mUrca_n})).  Thus, the below-threshold direct Urca electron capture rate would begin to compete with modified Urca when
\begin{equation}
e^{-(E_p-E_{F_p})/T} \approx \left( \dfrac{m_n T}{3m_{\pi}^2} \right)^{\!2},
\end{equation}
which, for a proton with $E_p-E_{F_p} = 12 \text{ MeV}$, is when the temperature is between 1 and 2 MeV.

As we will show in a calculation of the full phase space integral for direct Urca, this is a fair estimate.  The Fermi surface approximation starts to become invalid at temperatures $T\gtrsim 1 \text{ MeV}$, which is still in the neutrino-transparent regime.  As proofs of the neutrino-transparent beta equilibrium condition rely on either the Fermi surface approximation (Sec.~\ref{sec:durca_FS}) or the ability to kinematically neglect the neutrino \cite{Yuan:2005tj}, we study the validity of the neutrino-transparent beta equilibrium condition at temperatures of around several MeV, where we expect possible corrections to appear.  This is relevant to neutron star mergers, which likely contain dense matter at temperatures of a few MeV (see Fig.~\ref{fig:T_nb_plane}).
\hiddensubsection{Particles away from their Fermi surface}
To discuss the rates it is useful to introduce the concept of the single particle free energy, defined as $\gamma_i(p) \equiv E_i(p) - \mu_i = E_i(p) - E_{F\!,i}$ (see Eq.~(\ref{eq:dispersion}) and subsequent discussion).  The single particle free energy tells us how far in energy a given state is from its Fermi surface.

At densities below the threshold density, the direct Urca process becomes Boltzmann suppressed because after imposing energy and momentum conservation the phase space integral is dominated by processes whose initial state includes particles above their Fermi surface or whose final state requires holes below their Fermi surface. In both cases the Fermi-Dirac factors in the rate expression provide a suppression factor of $\exp(-|\gamma_i|/T)$.

\begin{figure}[h]
\centering
\includegraphics[scale=.6]{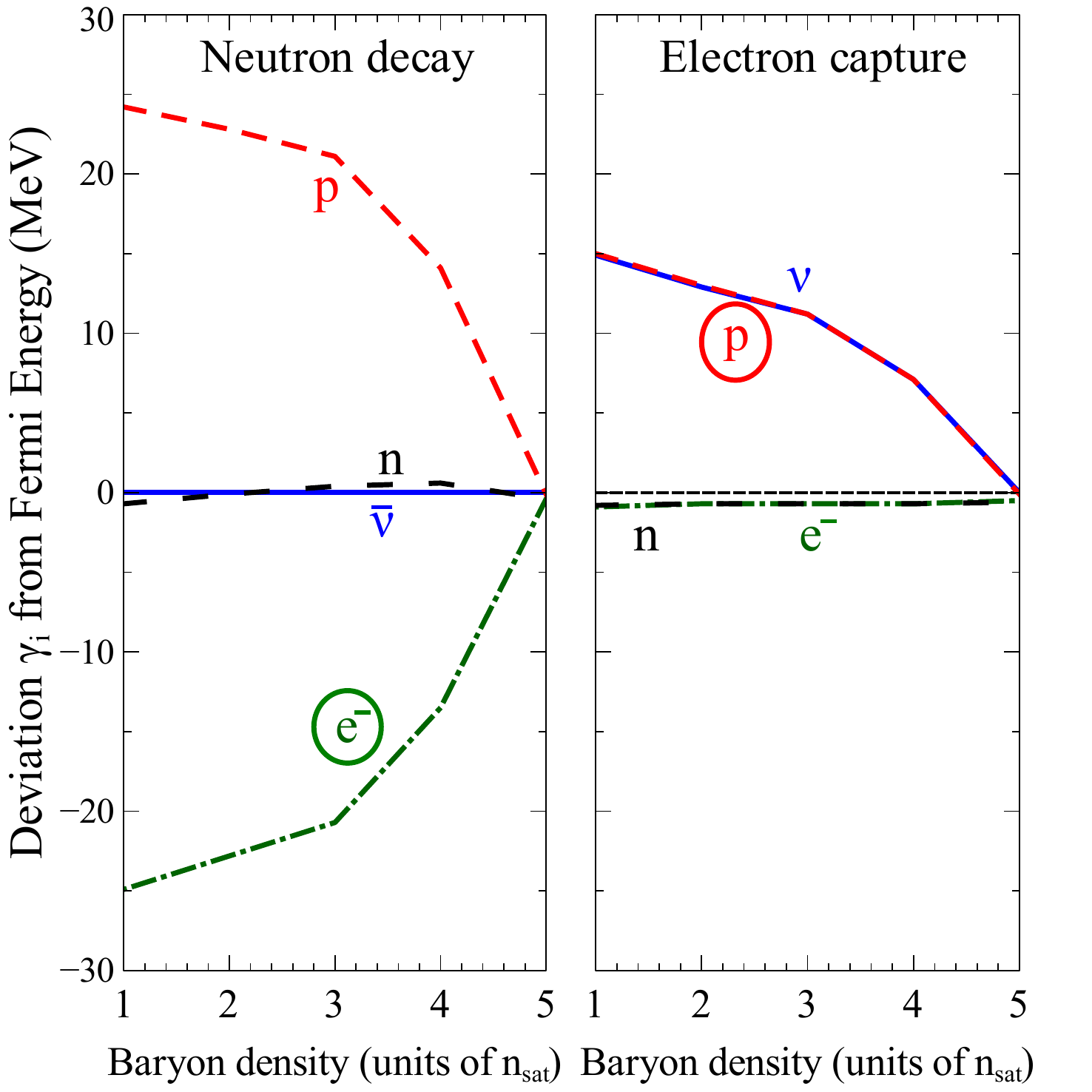}
\caption{Energy relative to their Fermi energy (defined as ``$\gamma$'')
for particles participating in direct Urca reactions in APR nuclear matter obeying the standard low-temperature condition Eq.~(\ref{eq:beq_nu_trans}) for beta equilibrium.  At each density we choose the momenta and energies of participating particles (consistent with energy and momentum conservation) that maximizes the product of their Fermi-Dirac factors.  Above threshold, all particles can have $\gamma=0$. Below threshold, the plot shows the least Boltzmann-suppressed processes.  The circles indicate the particles that cause the Boltzmann suppression.}
\label{fig:FS_deviation}
\end{figure}

To see how strong the resultant Boltzmann suppression will be, we show in Fig.~\ref{fig:FS_deviation} the typical single particle free energy $\gamma_i$ for the particles participating in neutron decay (left panel) and electron capture (right panel) at various densities of nuclear matter described by the APR equation of state, and obeying the low-temperature criterion for beta equilibration Eq.~(\ref{eq:beq_nu_trans}).  To obtain the typical momenta and energies at a given density we impose energy and momentum conservation to reduce the momentum space integral to the lowest possible dimension and find the point at which the product of Fermi-Dirac factors attains its maximum value.  We emphasize that these typical momenta and energies are independent of temperature.  Temperature merely influences the strength of the Boltzmann suppression due to particles with finite single particle free energies $\gamma_i$ participating in Urca reactions. 

Above the direct Urca threshold density (about $5n_0$ for the APR equation of state) we
find, as expected, that particles on their Fermi surface (i.e.~with $\gamma=0$) can participate in direct Urca processes while conserving energy and momentum. Below the direct Urca threshold density, however, this is no longer true.

\hiddensubsection{Below-threshold direct Urca neutron decay}

For direct Urca neutron decay, the kinematic obstacle is that although a neutron on its Fermi surface has the same free energy as a proton on its Fermi surface and an electron on its Fermi surface (they all have $\gamma=0$), the neutron's momentum is larger than the co-linear sum of the proton and electron momenta.  We see in Fig.~\ref{fig:FS_deviation} (left panel) that the best available option below threshold
is for a neutron on its Fermi surface to decay into a proton that is above its Fermi surface by an amount $\gamma_p$ and an electron that is below its Fermi surface by the same amount, $\gamma_e=-\gamma_p$.  The energies of the proton and electron still add up to the energy of the neutron, but a co-linear proton and electron now have more momentum 
then when they were both on their Fermi surfaces because the proton's momentum rises rapidly as $\gamma_p$ becomes more positive (because the proton is non-relativistic with a low Fermi velocity) whereas the electron's momentum drops more slowly as $\gamma_e$ becomes more negative, because the electron is relativistic.  This ``best available option'' has a Boltzmann suppression factor of $\exp(-|\gamma_e|/T)$ because the final state electron is trying to occupy a state in the already mostly occupied electron Fermi sea.  From Fig.~\ref{fig:FS_deviation} we see that for the APR equation of state the value of $|\gamma_e|$ for this process is around 20 to 25 MeV at lower
densities and then drops quickly to zero as we approach the direct Urca threshold.
\hiddensubsection{Below-threshold direct Urca electron capture}
For direct Urca electron capture, the kinematic obstacle is that a proton on its Fermi surface combined with an electron on its Fermi surface does not have enough momentum to produce a neutron on its Fermi surface.  We see in Fig.~\ref{fig:FS_deviation} (right panel) that the best available option below threshold is for a proton above its Fermi surface to combine with an electron at its Fermi surface. Because the proton is nonrelativistic this combination has enough momentum to create a neutron on its Fermi surface, and the excess energy $\gamma_p$ (and the remaining momentum) goes in to 
the final state neutrino.  This process has a Boltzmann suppression factor of
$\exp(-|\gamma_p|/T)$ because we are unlikely to find initial state protons far above the proton Fermi surface.  From Fig.~\ref{fig:FS_deviation} we see that for the APR equation of state the value of $\vert \gamma_p\vert$ for this process is around 10 to 15 MeV at lower densities and then drops to zero as we approach the direct Urca threshold.
The suppression is less than for neutron decay because the neutrino momentum can be directed opposite to the neutron momentum, so it helps to reduce the amount by which the proton needs to be above its Fermi surface.

\begin{figure*}[t!]
\begin{minipage}[t]{.5\linewidth}
\includegraphics[width=.95\linewidth]{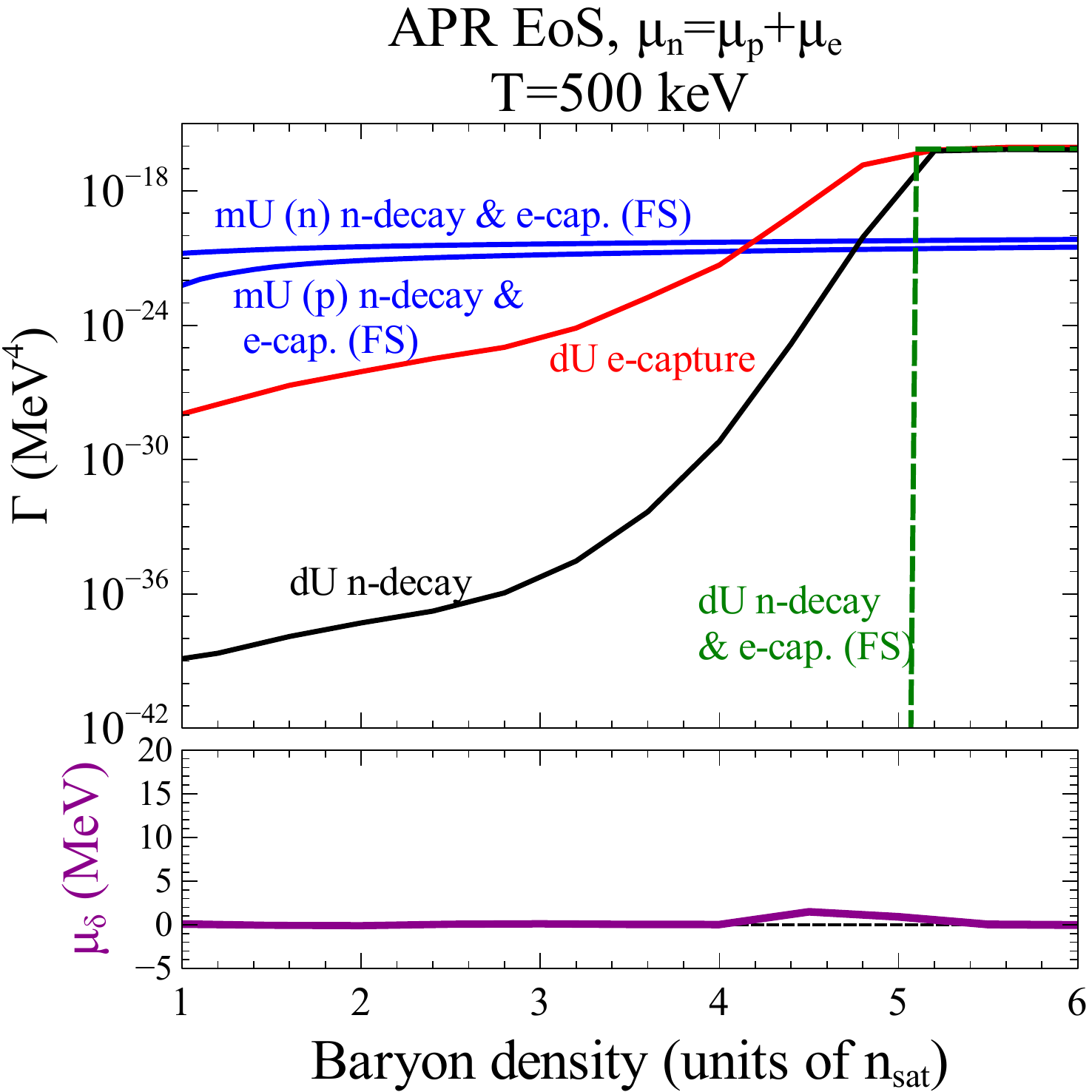}
\end{minipage}\hfill%
\begin{minipage}[t]{0.5\linewidth}
\includegraphics[width=.95\linewidth]{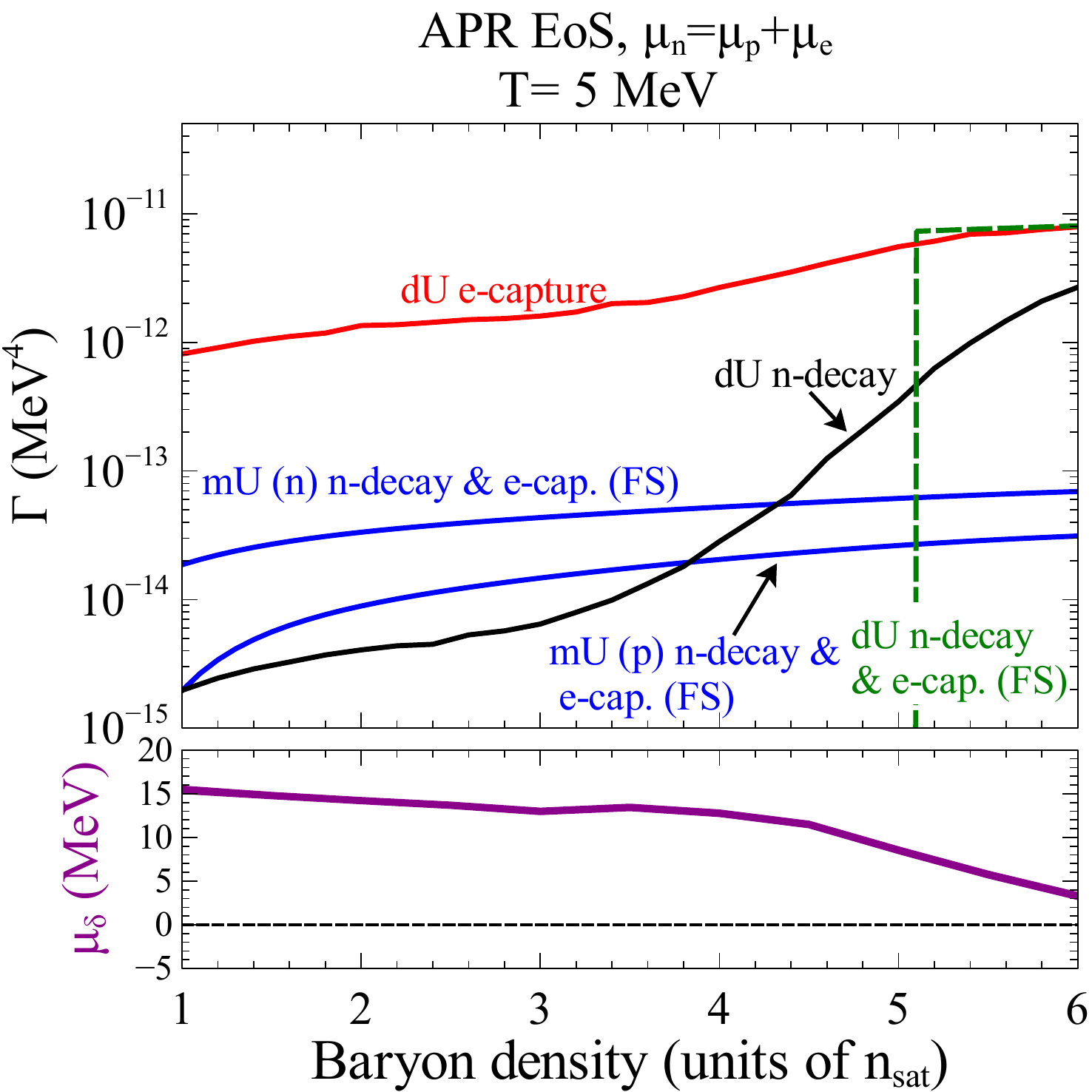}
\end{minipage}%
\caption{Exact direct Urca and Fermi-surface approximate modified Urca rates at $T = 500 \text{ keV}$ (left panel) and $T = 5 \text{ MeV}$ (right panel), obeying the neutrino-transparent beta equilibrium condition [Eq.~(\ref{eq:beq_nu_trans})].  Far above the direct Urca threshold, the two direct Urca rates balance each other, and they also match the Fermi surface approximate direct Urca rate.  At $T=500 \text{ keV}$, the direct Urca rates fall off exponentially below the direct Urca threshold, and modified Urca dominates.  At $T = 5 \text{ MeV}$, the direct Urca electron capture rate does not exponentially fall off below the direct Urca threshold, and instead dominates at all densities.  The bottom panels of both plots show the deviation $\mu_{\delta}$ from low-temperature beta equilibrium needed to achieve true beta equilibrium.}
\label{fig:urca_rates}
\end{figure*} 
\hiddensubsection{Relevance of below-threshold direct Urca}

We learn from the calculation presented in Fig.~\ref{fig:FS_deviation} that,
below the threshold density, direct Urca processes are Boltzmann suppressed by
a factor $\exp(-\gamma/T)$ where $\gamma$ is in the 10 to 20 MeV range at lower
densities, dropping to zero as the threshold is approached. For typical
neutron star temperatures $T\lesssim 100 \text{ keV}$ the Boltzmann suppression is
overwhelming, and direct Urca processes can be safely neglected compared with
modified Urca. However, as we will show in the next section, at the temperatures 
characteristic of neutron star mergers this is not the case.

We note that a similar analysis of the Fermi-Dirac factors can be done with
the modified Urca process, but it simply reproduces the expected finding that
at any density the dominant contribution comes from particles close to their
Fermi surfaces, so the Fermi surface approximation is always valid for
modified Urca and there is never any Boltzmann suppression of the rate.  This is due to the presence of the spectator nucleon.

In Sec.~\ref{sec:beyond_FS}, we estimated that 
at densities below the direct Urca threshold
the Boltzmann-suppressed direct Urca electron capture rate would match the modified Urca rate once the temperature rose to around $1$ or $2 \text{ MeV}$.   As we will see in the next section, a full calculation
confirms this estimate, showing that at $T\gtrsim 1 \text{ MeV}$ the
contribution of below-threshold direct Urca processes leads to corrections
to the low-temperature criterion for beta equilibrium.
Since the dominant contribution to the below-threshold direct Urca rates comes
from particles that are far from their Fermi surfaces, we now calculate the
direct Urca rates exactly, performing the entire momentum space integral.
\section{Exact direct Urca calculation} 
\label{sec:exact}
Instead of assuming that all particles lie on their Fermi surfaces, we numerically evaluate the direct Urca rate integrals (\ref{eq:ndecay}) and (\ref{eq:ecapture}) with non-relativistic nucleons, but without any further approximation, allowing the particles to have any set of momenta that is consistent with energy-momentum conservation.  The details of the calculation, which reduces to a three dimensional numerical integral, are given 
in Appendix~\ref{sec:rate-integral}.
\hiddensubsection{Urca rates}
Tne upper panels of Fig.~\ref{fig:urca_rates} show the rates of various Urca processes in APR nuclear matter that obeys the low-temperature beta equilibrium criterion Eq.~(\ref{eq:beq_nu_trans}).  As we will see, at $T\gtrsim 1 \text{ MeV}$ the Fermi surface approximation starts breaking down and Eq.~(\ref{eq:beq_nu_trans}) is no longer the
correct criterion for beta equilibrium: an additional chemical potential is needed to achieve beta equilibrium, and its magnitude $\mu_{\delta}$ is shown in the lower panel.

In Fig.~\ref{fig:urca_rates}, we see that at a temperature of $500 \text{ keV}$, the Fermi surface approximation is reasonably accurate.  Well above the direct Urca threshold density, the neutron decay and electron capture rates are almost identical and agree well with the Fermi surface approximation, so that when the low-temperature beta equilibrium criterion Eq.~(\ref{eq:beq_nu_trans}) is obeyed the net rate of neutron or proton creation is zero. Consequently, to the accuracy of our calculation ($\mu_{\delta}$ is accurate to about $\pm 150 \text{ keV}$, described in Sec.~\ref{sec:full}) there is no need for any additional chemical potential to enforce beta equilibrium.  As the density drops below the threshold value, the direct Urca rates drop below the modified Urca rate and become negligible, and the modified Urca rates for neutron decay and electron capture are identical so again there is no net creation of neutrons or protons, and no noticeable 
modification to the low-temperature beta equilibrium criterion.  However, it is interesting to note that below (and even slightly above) threshold the direct Urca rates for neutron decay and electron capture are not the same.  The deviation increases as the density goes further below threshold.  The size of the discrepancy agrees with our analysis in Fig.~\ref{fig:FS_deviation}, where we determine the exponential suppression of each direct Urca rate, due to the Fermi-Dirac factors.  Only right below threshold is there a region where the two direct Urca rates are different, but both are larger than the modified Urca rates, requiring a finite but small $\mu_{\delta}$ to establish true beta equilibrium.  As temperature decreases further, this effect will vanish, and the low-temperature beta equilibrium criterion (\ref{eq:beq_nu_trans}) will be increasingly valid.  

In Fig.~\ref{fig:urca_rates} we see that when we increase the temperature to $T=5 \text{ MeV}$ the Fermi surface approximation becomes unreliable.  The direct Urca electron capture rate agrees with the Fermi surface result above threshold, but it also shows no suppression below threshold, dominating over modified Urca and direct Urca neutron decay at all densities for which APR is well defined. This means that when $\mu_n=\mu_p+\mu_e$ [Eq.~(\ref{eq:beq_nu_trans})], there is a nonzero net rate of proton to neutron conversion, implying that the system is not actually in beta equilibrium. 
\hiddensubsection{Full criterion for beta equilibrium}
\label{sec:full}
The fact that electron capture is much less suppressed than neutron decay at $T\gtrsim 1 \text{ MeV}$ means that the system will be driven away from the state that obeys the standard low-temperature beta equilibrium criterion. The predominance of electron capture drives the neutron Fermi energy up and the proton Fermi energy down, effectively introducing an additional chemical potential that couples to the third component of isospin.  The general criterion for beta equilibrium is
\begin{equation}
\mu_n = \mu_p + \mu_e + \mu_{\delta}.
\label{eq:beta-general}
\end{equation}
As the proton density drops, the rates of electron capture and neutron decay move towards each other and eventually balance when $\mu_{\delta}$ reaches its equilibrium value, which depends on the density and the temperature. This value of $\mu_{\delta}$ at $T = 5 \text{ MeV}$ is shown in the bottom right panel of Fig.~\ref{fig:urca_rates}.  We see that below the direct Urca threshold $\mu_{\delta}$ is about 15 MeV and it decreases but remains non-negligible above threshold as well.

\begin{figure}[h]
\centering
\includegraphics[scale=0.6]{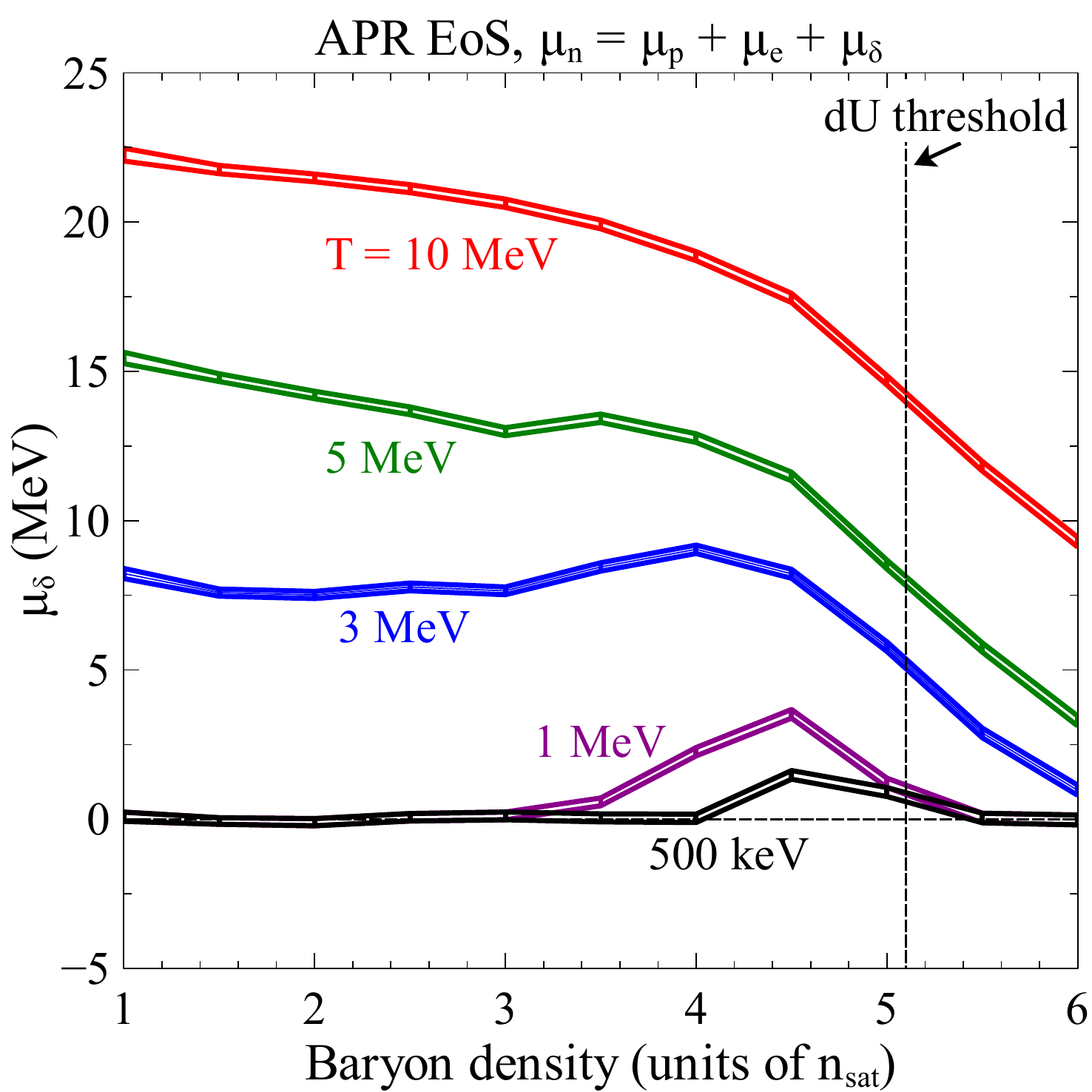}
\caption{
The chemical potential $\mu_{\delta}$ needed to achieve true beta equilibrium in APR matter at various temperatures.  For a given temperature, the upper and lower curves indicate the range of values of $\mu_{\delta}$ that are consistent with our estimates of the theoretical uncertainty.  Further details on the uncertainty are given in the text.}
\label{fig:dmu_plot}
\end{figure} 

In Fig.~\ref{fig:dmu_plot}, we show, for several temperatures, the magnitude of the additional chemical potential $\mu_{\delta}$ needed to achieve true beta equilibrium.  At low temperatures, below 1 MeV, the standard criterion [Eq.~(\ref{eq:beq_nu_trans})] is correct to within about 1 MeV the only noticeable correction, $\mu_{\delta} \approx 1.5 \text{ MeV}$, occurring right below threshold where direct Urca electron capture begins to dominate over modified Urca. After that, however, the correction term rises quickly with temperature: at $T=5 \text{ MeV}$ we need $\mu_{\delta}\sim 15 \text{ MeV}$ and at $T=10 \text{ MeV}$ we need $\mu_{\delta}\sim 23 \text{ MeV}$.  Although $\mu_{\delta}$ drops with density once we reach the direct Urca threshold, that decrease becomes quite slow at these higher temperatures.  If the neutrino trapping temperature is indeed around 5 MeV (see Sec.~\ref{sec:nu_mfp}), then our calculations of $\mu_{\delta}$ are only physically relevant at temperatures below 5 MeV.  Above the neutrino trapping temperature we expect that neutrinos are in statistical equilibrium with a chemical potential $\mu_{\nu}$ obeying the detailed balance relation $\mu_n + \mu_{\nu} = \mu_p + \mu_e$, as discussed in Sec.~\ref{sec:both_beta_eq_conditions}.

As discussed in Sec.~\ref{sec:nucl_matter_EoSs}, the APR equation of state is based on variational calculations of the energy (as a function of density) of pure neutron matter (PNM, $x_p = 0$) and symmetric nuclear matter (SNM, $x_p = 0.5$).  An interpolation scheme was used to get to the intermediate proton fraction, for a given density, that satisfies the low-temperature beta equilibrium condition (\ref{eq:beq_nu_trans}).  Thus, all thermodynamic quantities calculated within the APR framework are functions of both baryon density and proton fraction.  To find the value of $\mu_{\delta} = \mu_n-\mu_p-\mu_e$ necessary to achieve true beta equilibrium at a given density and temperature, we varied the proton fraction in discrete steps until we found a proton fraction that at which there was net neutron production, and an adjacent-step proton fraction at which there was net neutron destruction, giving us an upper and lower bound on $\mu_{\delta}$. These upper and lower bounds provide the theoretical error on $\mu_{\delta}$.  The uncertainty in the direct Urca rate calculations, discussed at the end of Appendix \ref{sec:rate-integral}, is smaller than the binning of $\mu_{\delta}$, which is $\approx 300 \text{ keV}$.  

\begin{figure}[h]
\centering
\includegraphics[scale=0.6]{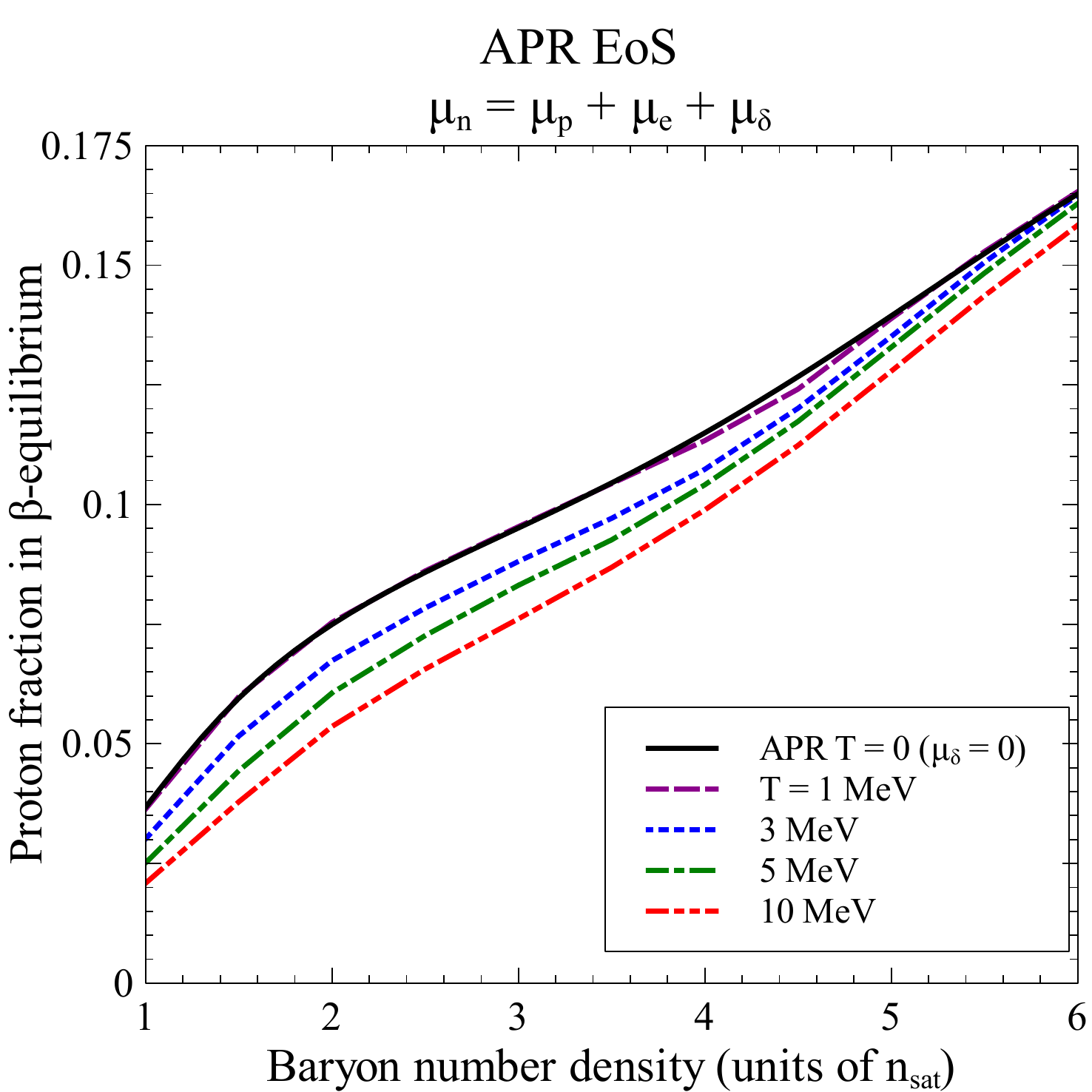}
\caption{Proton fraction in neutrino-transparent nuclear matter in true beta equilibrium, for several different temperatures.  As temperature increases, the beta-equilibrated nuclear matter becomes more neutron-rich than predicted by the low-temperature beta equilibrium condition.}
\label{fig:new_xp}
\end{figure}

The temperature-dependent correction $\mu_{\delta}$ to the beta equilibrium condition arises from the temperature dependence of the proton fraction in true beta equilibrium. In Fig.~\ref{fig:new_xp} we plot the proton fraction in true beta equilibrium as a function of density, for various temperatures.  As temperature rises above 1 MeV, the proton fraction drops. This reflects the predominance of electron capture over neutron decay seen in Fig.~\ref{fig:urca_rates}.

\section{Conclusions}
We have shown that the standard low-temperature criterion Eq.~(\ref{eq:beq_nu_trans}) for beta equilibrium breaks down in neutrino-transparent nuclear matter at densities above nuclear saturation density and temperatures above about 1 MeV. An additional chemical potential Eq.~(\ref{eq:beta-general}) is required to obtain true equilibrium under the weak interactions.  The ultimate reason for this is that neutrinos are not in thermal equilibrium, so reactions that drive the system to equilibrium are not exact inverses of each other (neutrinos can only occur in final states) so the principle of detailed balance does not apply.  Our calculations for nuclear matter
obeying the APR equation of state show (Fig.~\ref{fig:dmu_plot}) that the chemical potential $\mu_{\delta} = \mu_n-\mu_p-\mu_e$ required to obtain beta equilibrium
becomes greater than about 5 MeV as the temperature rises above 1 MeV, and reaches a maximum value around 23 MeV at temperatures of 10 MeV.

\begin{figure}[h]
\centering
\includegraphics[scale=0.6]{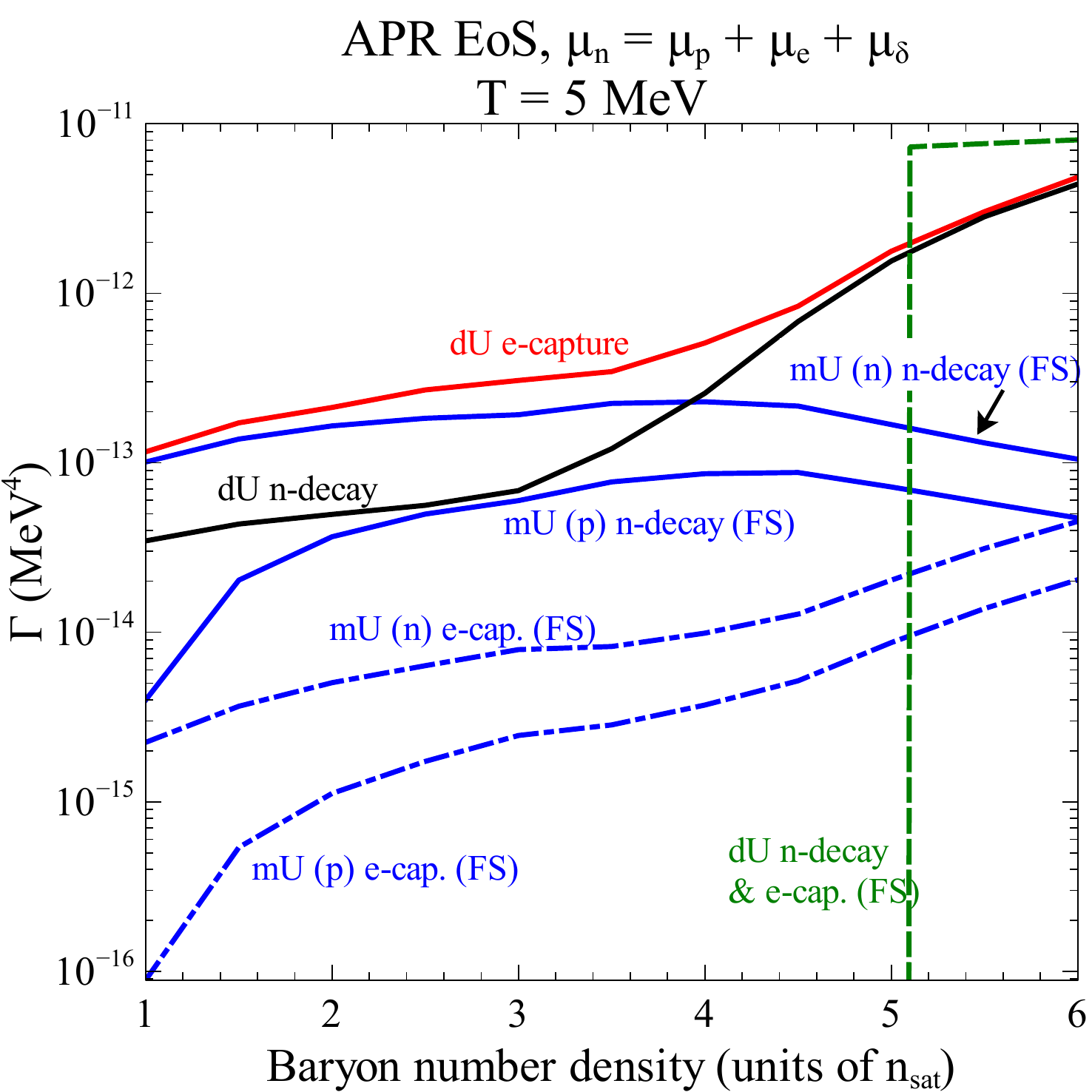}
\caption{
Urca rates in true beta equilibrium, $\mu_n = \mu_p + \mu_e + \mu_{\delta}$.  Above threshold, the two direct Urca processes dominate.  Below threshold, direct Urca electron capture balances against modified Urca neutron decay (n-spectator) and, to a lesser extent, direct Urca neutron decay.
}
\label{fig:beta-true}
\end{figure}

We have recalculated the Urca rates for APR nuclear matter at
$T=5\text{ MeV}$ in true beta equilibrium.  The results are given in Fig.~\ref{fig:beta-true}, which shows that there are three processes that play the central role in beta equilibration. Above threshold, direct Urca neutron decay balances with direct Urca electron capture.  Far below threshold, neutron-spectator modified Urca neutron decay competes with direct Urca electron capture, but as threshold is approached from below, direct Urca neutron decay becomes increasingly important, eventually becoming more important than neutron-spectator modified Urca neutron decay.

\begin{figure}[h]
\centering
\includegraphics[scale=0.6]{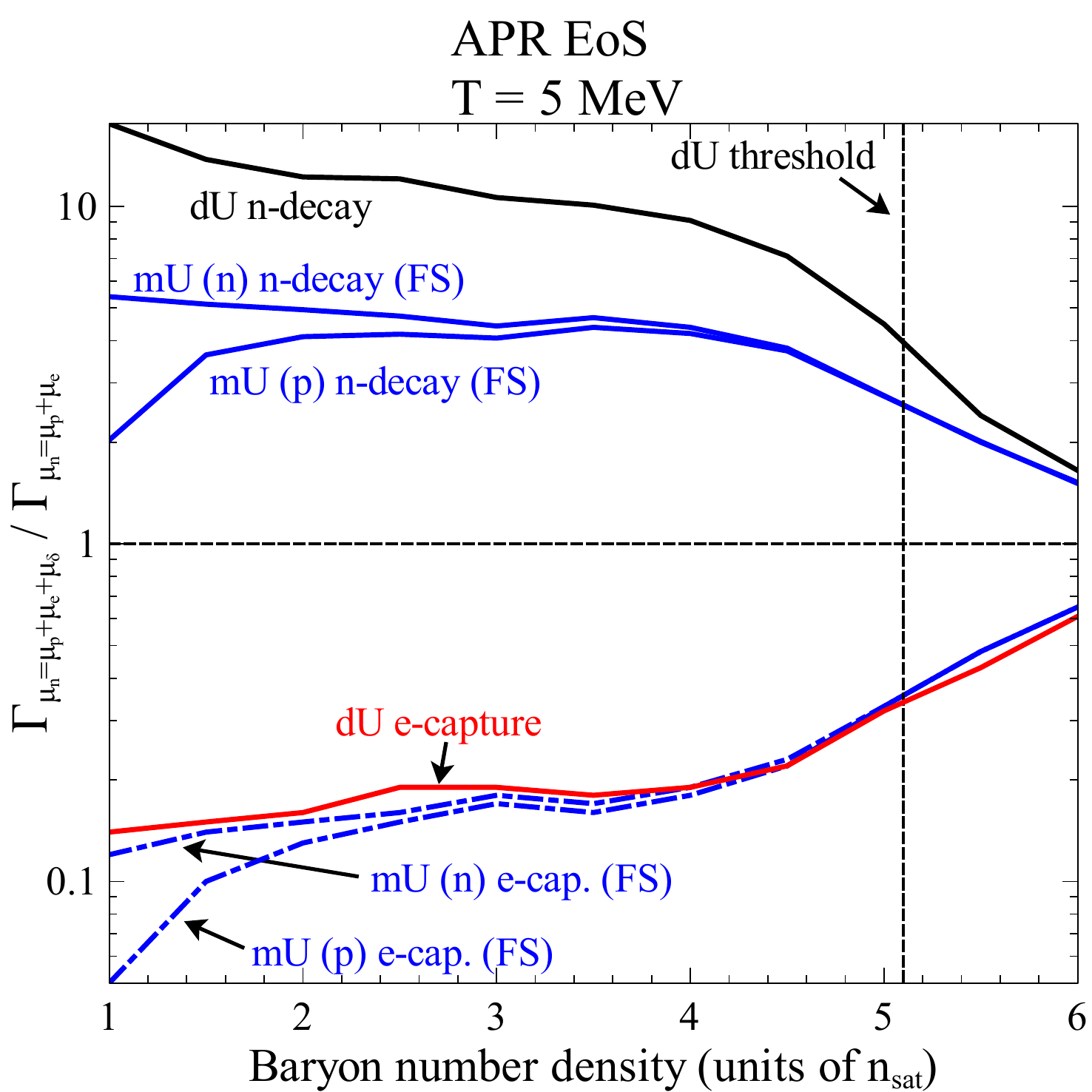}
\caption{Fractional change in the six Urca rates at $T = 5 \text{ MeV}$ when we change $\mu_{\delta}$ from zero to the value for true beta equilibrium.  Below threshold, the change is the most prominent, and as density increases to far above the threshold density, the true beta equilibrium condition approaches the behavior of the low-temperature beta equilibrium condition (\ref{eq:beq_nu_trans}).  
}
\label{fig:frac_change}
\end{figure}

In Fig.~\ref{fig:frac_change}, we show the fractional change in the six direct Urca rates when the correct beta equilibrium condition is imposed, compared with the low-temperature criterion for beta equilibrium (\ref{eq:beq_nu_trans}).  Far below threshold, the Urca rates increase or decrease by a factor of 10, and far above threshold, the Urca rates approach their low-temperature beta equilibrium values. 

In this work we did not consider particle processes involving muons, although the APR equation of state includes contributions from muons.  Muons can participate in Urca processes and also in leptonic processes \cite{Alford:2010jf} such as $\mu^- \rightarrow e^-\ +\ \bar{\nu}_e\ +\ \nu_{\mu}$.  A complete treatment of the Urca processes in neutrino-transparent nuclear matter would introduce another chemical potential $\mu_{f}$ (which couples to lepton flavor, differentiating electrons from muons) whose value at a given temperature and density is determined by balancing the six muon Urca process rates.

In Ch.~\ref{sec:bulk_viscosity}, we will discuss how this modification to the beta equilibrium condition affects the bulk viscosity in neutrino-transparent nuclear matter in conditions encountered in neutron star mergers.
\chapter{Bulk viscosity from weak interactions in neutron star mergers}
\begin{figure}[h]
\centering
\includegraphics[scale=0.6]{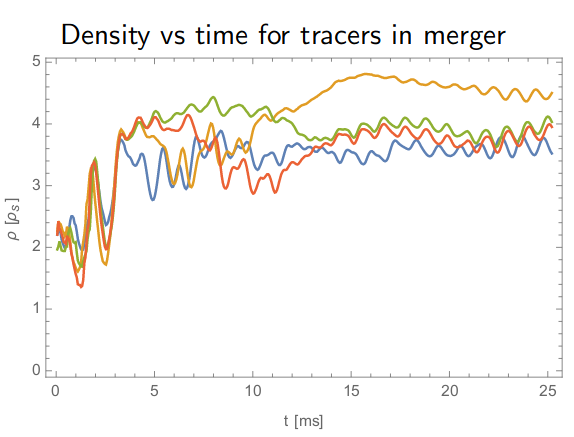}
\caption{Density of several comoving fluid elements during a neutron star merger.  Fluid elements initially experience large density oscillations right when the two neutron stars touch ($t=0$), but after 5 milliseconds the oscillation amplitude decreases.  Figure courtesy of M.~Hanauske and the Rezzolla group.}
\label{fig:density_oscillations}
\end{figure}

\pagestyle{myheadings}
\label{sec:bulk_viscosity}
\begin{center}
{\textit{This section is based on my work with Mark Alford, \cite{Alford:2019qtm}}.
\\\copyright 2019 American Physical Society}
\end{center}
When two neutron stars merge, simulations indicate that the nuclear fluid that makes up the neutron stars experiences wild oscillations in density, changing density by up to 50\%.  The oscillations occur on a millisecond timescale\footnote{This timescale can be estimated by considering the time it takes sound to cross a neutron star.  The speed of sound in a neutron star is some fraction of the speed of light, say $c/3$ on average, and so the time for sound to cross a neutron star 25 km in diameter is $t=x/v = 0.25 \text{ ms}$, which translates into a frequency of 4 kHz.}.  One simulation tracked fluid elements throughout the merger and plotted their density as a function of time, seen in Fig.~\ref{fig:density_oscillations}.  The presence of fluid elements undergoing density oscillations makes it possible that bulk viscosity will play a role in neutron star mergers\footnote{Bulk viscosity is already considered an important process in isolated neutron stars undergoing oscillations \cite{Ofengeim:2019fjy,1987ApJ...314..234C,2000A&A...357.1157H,Alford:2010gw}.}.

In this chapter, we calculate the bulk viscosity of neutrino-transparent nuclear matter\footnote{The bulk viscosity in neutrino-trapped nuclear matter is calculated in \cite{Alford:2019kdw}.} undergoing uniform, small-amplitude density oscillations in thermodynamic conditions relevant to neutron star mergers.  Bulk viscosity is a resonance phenomenon, and we map out the thermodynamic conditions where bulk viscosity is near its resonant maximum.  We estimate the timescale in which bulk viscosity dissipates energy from density oscillations and examine its relevance for neutron star mergers.  We take advantage of our results for the Urca rate in neutrino-transparent nuclear matter obtained in Ch.~\ref{sec:beta_equilibrium}.
\section{Bulk viscosity}
\label{sec:bv}
We will consider a fluid element of nuclear matter undergoing a small-amplitude oscillation in baryon density
\begin{equation}
    n_B(t) = \bar{n}_B + \delta n_B \cos{(\omega t)},
\end{equation}
which corresponds to an oscillation in volume
\begin{equation}
    V(t) = \bar{V} + \delta V \cos{(\omega t)},
\end{equation}
since the total baryon number $N_B=n_BV$ is conserved.  We consider adiabatic oscillations (meaning entropy per baryon $\sigma$ is constant) in this chapter, because there is negligible heat flow between adjacent fluid elements during the merger. This is valid as long as the thermal equilibration time in the absence of neutrinos is much longer than about 10\,ms. From Eq.~(1) of Ref.~\cite{Alford:2017rxf} this will be true as long as density oscillations (and the resultant thermal gradients) have wavelengths longer than about a meter. This criterion is obeyed in current simulations, whose spatial resolution is tens of meters at best.

By definition, the energy dissipated due to bulk viscosity $\zeta$ is
\begin{equation}
    \frac{\mathop{d\varepsilon}}{\mathop{dt}} = -\zeta (\nabla\cdot \mathbf{v})^2.
\end{equation}
Using the continuity equation (In the Lagrangian formalism that tracks fluid elements \cite{rezzolla2013relativistic}),
\begin{equation}
    \frac{\partial n_B}{\partial t}+n_B\nabla\cdot \mathbf{v}=0,
\end{equation}
we find that
\begin{equation}
    \frac{\mathop{d\varepsilon}}{\mathop{dt}}=-\zeta \left(\frac{\partial n_B/\partial t}{n_B}\right)^2.
\end{equation}
Therefore, the average energy dissipated in one oscillation is
\begin{equation}
\bigg\langle \frac{\mathop{d\varepsilon}}{\mathop{dt}} \bigg\rangle = -\frac{1}{2}\omega^2\left(\frac{\delta n_B}{n_B}\right)^2\zeta.
\end{equation}

Energy is dissipated from the oscillation because of $p\mathop{dV}$ work.  When the fluid element is compressed, the beta equilibrium proton fraction changes, causing the Urca processes to change the proton fraction to the new equilibrium value.  The weak interactions take a finite amount of time, and thus the baryon density (or volume) and pressure (which is determined by the particle content) are out of phase.  The average energy dissipation over a cycle can be calculated in this approach.

The pressure, assumed to be out of phase with the baryon density, can be written as
\begin{equation}
    P(t) = \bar{P} + \Re(\delta P)\cos{(\omega t)} - \Im(\delta P)\sin{(\omega t)},
    \label{eq:pressure_1}
\end{equation}
where $\delta P$ is a complex number (if it were real, then pressure and baryon density would be in phase and there would be no energy dissipation).  For small oscillations, the energy dissipated is
\begin{equation}
    \mathop{d\varepsilon} = \frac{P}{V}\mathop{dV} = -\frac{P}{n_B}\mathop{dn_B},
\end{equation}
and the average energy dissipated per cycle is
\begin{equation}
    \bigg\langle \frac{\mathop{d\varepsilon}}{\mathop{dt}} \bigg\rangle = -\frac{\omega}{2\pi}\int_0^{2\pi/\omega}\mathop{dt} \frac{P(t)}{n_B(t)}\frac{\mathop{dn_B}}{\mathop{dt}}.
\end{equation}
We again approximate small oscillations ($n_{B}(t)\approx \bar{n}_B$), and we notice that only the term proportional to $\sin{(\omega t)}$ in $P(t)$ survives the integral.  So,
\begin{equation}
    \bigg\langle \frac{\mathop{d\varepsilon}}{\mathop{dt}} \bigg\rangle = -\frac{\omega^2}{2\pi}\frac{\delta n_B}{n_B}\Im(\delta P)\int_0^{2\pi/\omega}\mathop{dt} \sin^2{(\omega t)} = -\frac{\omega}{2}\frac{\delta n_B}{n_B}\Im(\delta P).
\end{equation}
Setting these two expressions for the energy dissipation equal, we find
\begin{equation}
    \zeta =  \frac{\Im(\delta P)}{\omega} \frac{n_B}{\delta n_B}.
    \label{eq:preliminary_zeta}
\end{equation}
With an eye towards incorporating the Urca process rates, we want to write $\delta P$ in terms of $\delta x_p$.  We are about to do a number of Taylor expansions around beta equilibrium, so it is important that we understand that there are three independent variables in this section - the baryon density $n_B$, the proton fraction $x_p$, and the entropy per baryon $\sigma = S/N_B = s/n_B$.  The entropy per baryon is constant throughout the oscillation, and thus does not appear in Taylor expansions.  

The pressure expanded around equilibrium is
\begin{equation}
    P \approx \bar{P} + \frac{\partial P}{\partial x_p}\bigg\vert_{n_B,\sigma}\Re(\delta x_p e^{i\omega t})+\frac{\partial P}{\partial n_B}\bigg\vert_{x_p,\sigma} \Re(\delta n_B e^{i\omega t}).
\end{equation}
Equating this with Eq.~(\ref{eq:pressure_1}) and matching the terms proportional to $\sin{(\omega t)}$, we find 
\begin{equation}
    \Im(\delta P) = \frac{\partial P}{\partial x_p}\bigg\vert_{n_B,\sigma}\Im(\delta x_p).
    \label{eq:dpdxp}
\end{equation}
Now we want to turn $\delta x_p$ into quantities related to the weak interactions that restore beta equilibrium.  We have
\begin{equation}
    n_B \frac{\mathop{dx_p}}{\mathop{dt}} = \Gamma_{n \rightarrow p}-\Gamma_{p \rightarrow n} \approx \lambda \mu_{\delta}.
    \label{eq:lambda_xp}
\end{equation}
The Urca rates $\Gamma$ that we calculated in Ch.~\ref{sec:beta_equilibrium} (divided by baryon density) change the proton fraction.  Since we are assuming small oscillations, the nuclear matter is never pushed that far out of beta equilibrium\footnote{Oscillations where $\mu_{\Delta} \ll T$ are called subthermal, and are by far the most frequently studied case of bulk viscosity.  Suprathermal oscillations $\mu_{\Delta}\gg T$ as well as oscillations where $\mu_{\Delta} \sim T$ are studied in \cite{Alford:2010gw}.} and thus we can use the linear approximation of the Urca rates near beta equilibrium, where we define 
\begin{equation}
    \lambda = \frac{\partial (\Gamma_{n\rightarrow p}-\Gamma_{p\rightarrow n})}{\partial \mu_{\Delta}}\bigg\rvert_{\mu_{\Delta}=0}.
\end{equation}
Now we write 
\begin{equation}
    x_p(t) = \bar{x}_p + \Re(\delta x_p e^{i\omega t}) = \bar{x}_p + \Re(\delta x_p)\cos{(\omega t)} - \Im(\delta x_p)\sin{(\omega t)}
    \label{eq:xp(t)}
\end{equation}
and expand $\mu_{\Delta}$ around beta equilibrium
\begin{equation}
\mu_{\Delta} \approx \frac{\partial \mu_{\Delta}}{\partial n_B}\bigg\vert_{x_p,\sigma} \delta n_B \cos{(\omega t)} + \frac{\partial \mu_{\Delta}}{\partial x_p}\bigg\vert_{n_B,\sigma} (\Re(\delta x_p)\cos{(\omega t)}-\Im(\delta x_p)\sin{(\omega t)}),\label{eq:mu_delta(t)}
\end{equation}
where we've used the expression for $\Re(\delta x_p e^{i\omega t})$ from Eq.~(\ref{eq:xp(t)}).  Plugging Eq.~(\ref{eq:xp(t)}) and (\ref{eq:mu_delta(t)}) into Eq.~(\ref{eq:lambda_xp}), and then matching the $\sin$ terms and $\cos$ terms, then solving for $\Im(\delta x_p)$, we find
\begin{equation}
    \Im(\delta x_p) = - \frac{\lambda \frac{\partial \mu_{\Delta}}{\partial n_B}\big\vert_{x_p,\sigma}\delta n_B}{n_B\omega + \frac{\lambda^2}{n_B\omega}\left(\frac{\partial \mu_{\Delta}}{\partial x_p}\big\vert_{n_B,\sigma}\right)^2}.
    \label{eq:imxp}
\end{equation}
Combining Eqs.~(\ref{eq:preliminary_zeta}), (\ref{eq:dpdxp}), and (\ref{eq:imxp}), we find
\begin{equation}
\zeta = \frac{\lambda \frac{\partial P}{\partial x_p}\big\vert_{n_B,\sigma}\frac{\partial \mu_{\Delta}}{\partial n_B}\big\vert_{x_p,\sigma}}{\omega^2 + \frac{\lambda^2}{n_B^2}\left(\frac{\partial \mu_{\Delta}}{\partial x_p}\big\vert_{n_B,\sigma}\right)^2}.
\end{equation}
Using a Maxwell relation (see Appendix \ref{sec:maxwell}), we can show that 
\begin{equation}
\frac{\partial P}{\partial x_p} \bigg\vert_{n_B,\sigma} = -n_B^2\frac{\partial\mu_{\Delta}}{\partial n_B}\bigg\vert_{x_p,\sigma}.\label{eq:maxwell}
\end{equation}
We also define the adiabatic susceptibilities\footnote{Instructions for turning isothermal susceptibilities into adiabatic susceptibilities are given in Appendix \ref{app:adiabatic}.  $B$ and $C$ are actually \textit{inverse} susceptibilities, as they are inverses of second derivatives of the pressure - see Appendix \ref{sec:GCE_thermo}.}
\begin{align}
    B &= -\frac{1}{n_B}\frac{\partial \mu_{\Delta}}{\partial x_p}\bigg\vert_{n_B,\sigma},\\
    C &= n_B \frac{\partial \mu_{\Delta}}{\partial n_B}\bigg\vert_{x_p,\sigma}.
\end{align}
Finally, we define the rate of beta equilibration $\gamma = B\lambda$, and now the bulk viscosity can be written in the form
\begin{equation}
    \zeta = \frac{C^2}{B}\frac{\gamma}{\omega^2+\gamma^2}. 
    \label{eq:bv}
\end{equation}

\section{Reequilibration rates}
As we can see from Eq.~(\ref{eq:bv}), bulk viscosity is a resonance phenomenon.  For a fixed oscillation frequency $\omega$, the bulk viscosity peaks when the reequilibration rate $\gamma = \omega$.  This is why we assumed that particle content was the relevant quantity that equilibrated in response to density oscillations.  The density oscillations happen on a millisecond timescale, as can weak interactions.  Strong and electromagnetic interactions happen orders of magnitude too quickly\footnote{Strong interactions are the relevant equilibration process in heavy ion collisions, however \cite{Czajka:2017wdo,Schafer:2009dj}.}.  

The rate of beta equilibration $\gamma = B\lambda$ can take advantage of the Urca rates we calculated in Ch.~\ref{sec:beta_equilibrium}.  The total rate of beta equilibration is the sum of the individual rates
\begin{equation}
    \lambda = \lambda_{dU}+\lambda_{mU\!,n}+\lambda_{mU\!,p},
    \label{eq:lambda}
\end{equation}
which are given by, in the FS approximation \cite{Yakovlev:2000jp,1992PhRvD..45.4708H,1995ApJ...442..749R,1979ApJ...232..541F,PhysRevD.39.3804}
\begin{align}
    \lambda_{dU} &= \frac{17}{240\pi}G^2(1+3g_A^2)m_nm_pp_{Fe}T^4,\\
    \lambda_{mU\!,n} &= \frac{367}{1152\pi^3}G^2g_A^2f^4 \frac{m_n^3 m_p}{m_{\pi}^4}\frac{p_{Fn}^4 p_{Fp}}{(p_{Fn}^2+m_{\pi}^2)^2}\vartheta_n T^6 \ ,\\
    \lambda_{mU\!,p}&= \frac{367}{8064\pi^3}G^2 g_A^2 f^4 \frac{m_n m_p^3}{m_{\pi}^4}\frac{p_{Fn}(p_{Fn}-p_{Fp})^4}{((p_{Fn}-p_{Fp})^2+m_{\pi}^2)^2}\vartheta_p T^6.
\end{align}

We will use the full phase integration for the direct Urca (see Sec.~\ref{sec:exact}) and the Fermi surface approximation for the modified Urca rates.  We will use the true beta equilibrium condition for neutrino transparent nuclear matter, Eq.~(\ref{eq:beta-general}), so $\mu_{\Delta}$ is now defined as
\begin{equation}
    \mu_{\Delta} \equiv \mu_n - \mu_p - \mu_e - \mu_{\delta}.
\end{equation}

\begin{figure}[h]
\centering
\includegraphics[width=.6\textwidth]{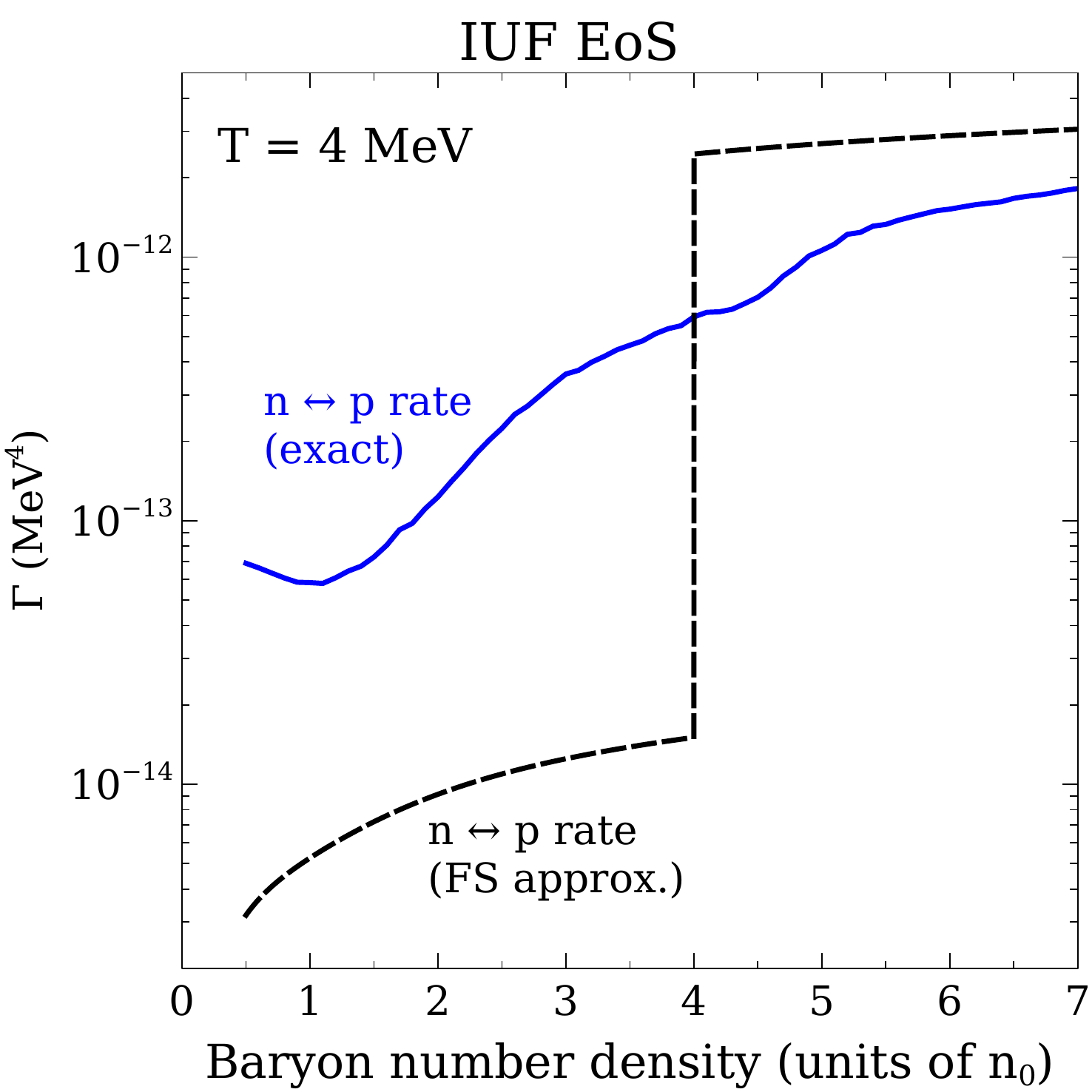}
\caption{Total Urca rates (direct plus modified) in beta equilibrium for the IUF EoS at $T=4 \text{ MeV}$.  The dashed (black) curve is the Fermi Surface approximation to the Urca rates, using the low-temperature beta equilibrium criterion Eq.~(\ref{eq:beq_nu_trans}). The solid (blue) curve is the total Urca rate with the full phase space integral, and using the general beta equilibrium condition Eq.~(\ref{eq:beta-general})}.
\label{fig:urca_rates_IUF}
\end{figure}

We use two relativistic mean field theories to describe nuclear matter in this chapter: DD2 and IUF.  See Sec.~\ref{sec:nucl_matter_EoSs} for an explanation of RMFs.  We use the same dispersion relations as in Sec.~\ref{sec:urca_and_FS}.  The rate of the Urca process at $T = 4 \text{ MeV}$ (a temperature high enough that the FS approximation is not valid) with the IUF equation of state is plotted in Fig.~\ref{fig:urca_rates_IUF}.

\section{Bulk viscosity from 1 kHz density oscillations}
\label{sec:bv_results}
\begin{figure*}[t!]
    \centering
    \begin{subfigure}[t]{0.5\textwidth}
        \centering
        \includegraphics[width=.95\textwidth]{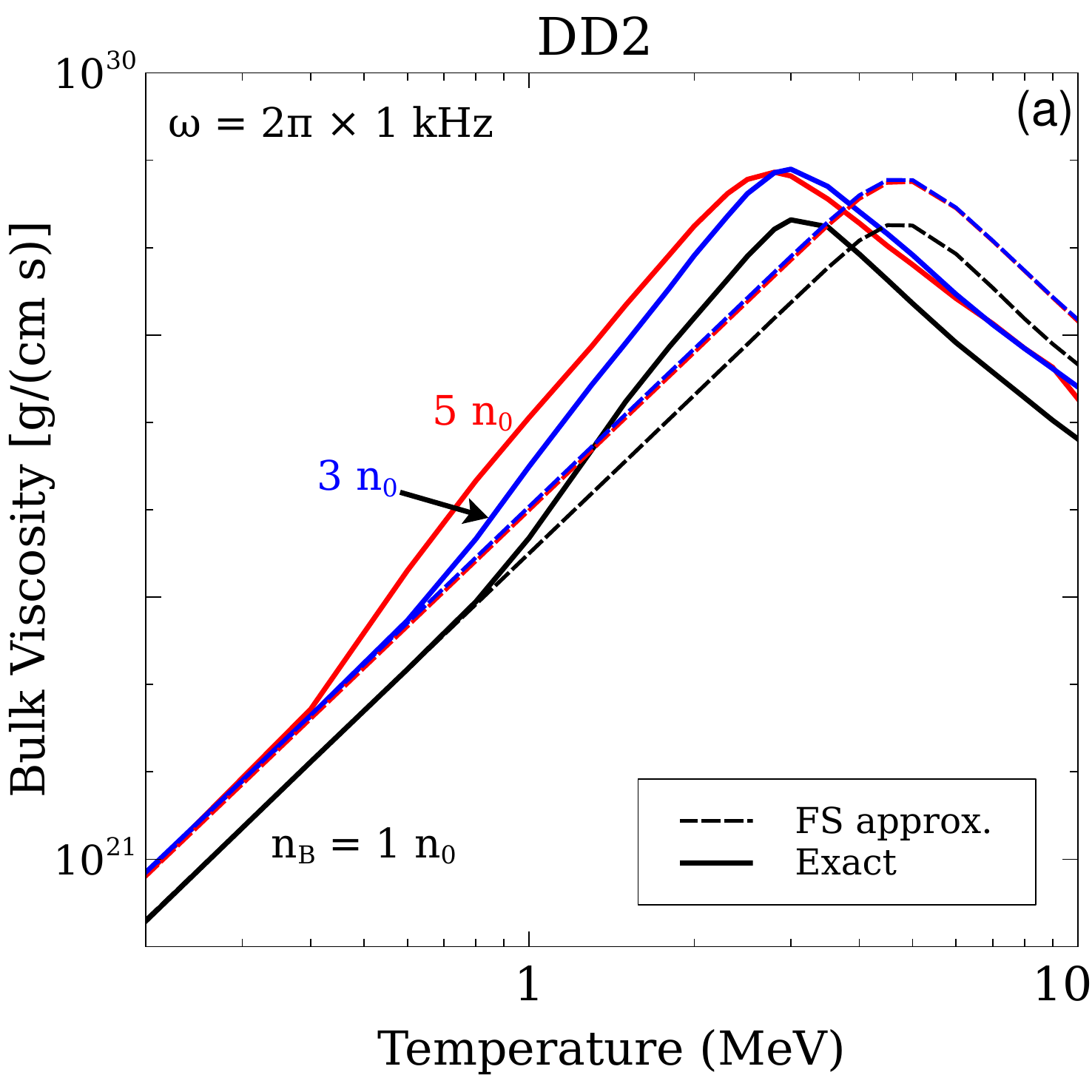}
    \end{subfigure}%
    \begin{subfigure}[t]{0.5\textwidth}
        \centering
        \includegraphics[width=.95\textwidth]{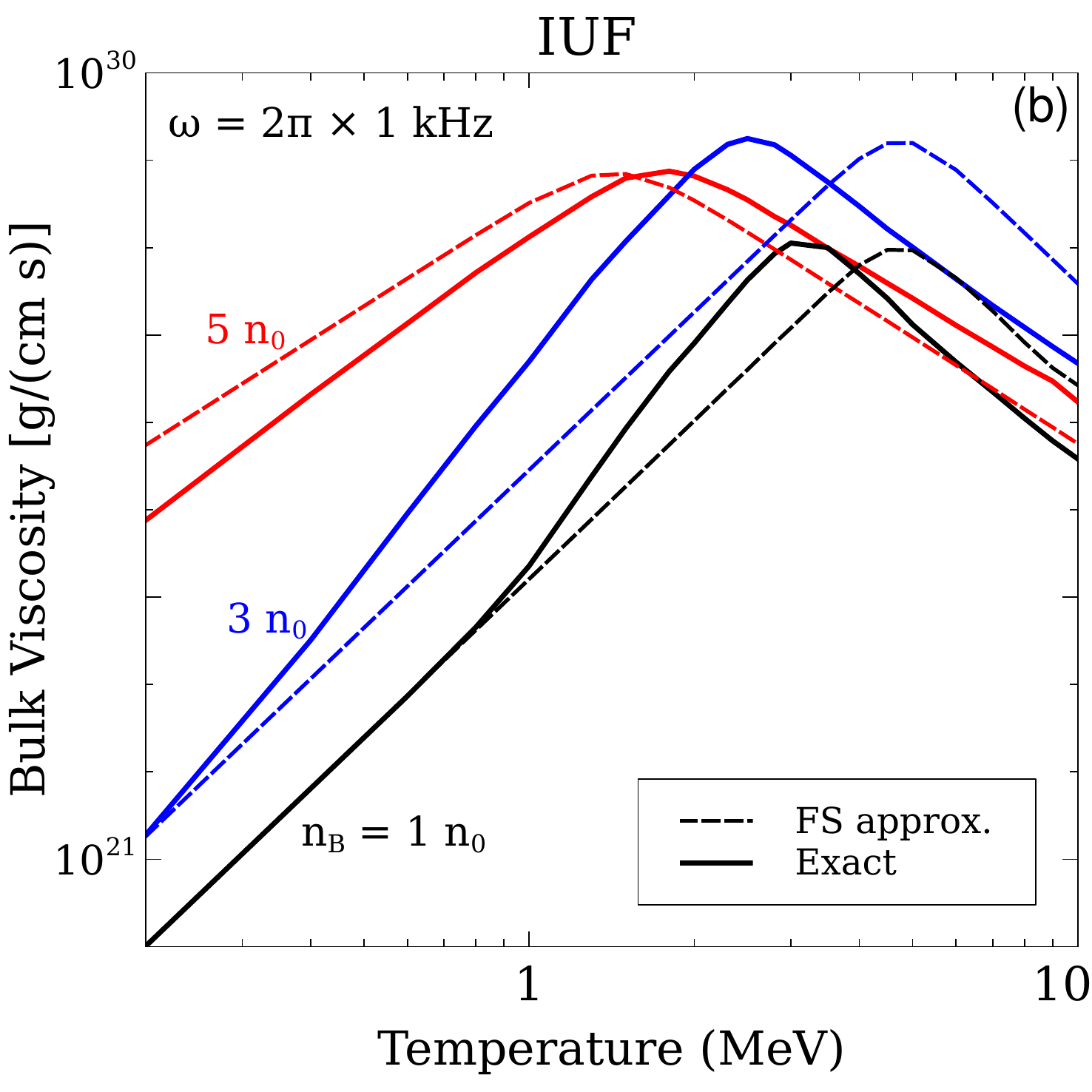}
    \end{subfigure}
    \caption{Bulk viscosity of nuclear matter as a function of temperature, for densities of $n_0$, $3n_0$, $5n_0$ when undergoing a density oscillation at 1 kHz. The equation of state is DD2 (a) or IUF (b).  Thin, dotted lines are the Fermi Surface approximation.  Thick, solid lines use the exact Urca rates.}
    \label{fig:bv_2d}
\end{figure*}

In Fig.~\ref{fig:bv_2d}, we show the bulk viscosity of nuclear matter with the DD2 and IUF EoS, when subjected to a 1 kHz density oscillation, which is a typical frequency for neutron star mergers (\cite{Alford:2017rxf} and Fig.~\ref{fig:density_oscillations}).  The dashed lines are the bulk viscosity with Urca rates calculated in the Fermi Surface approximation while the solid lines use the exact Urca rates.  

In Fig.~\ref{fig:bv_2d}(a), corresponding to the DD2 EoS, the exact bulk viscosity peaks at a  temperature that is 1-2 MeV lower than would be predicted by the Fermi Surface approximation.  This is because DD2 never allows direct Urca (the threshold is at infinite density), and we know (see Fig.~\ref{fig:urca_rates_IUF}) that the Fermi Surface approximation underestimates the below-threshold Urca rate.  This means that in the Fermi Surface approximation the temperature must be pushed up to a higher value in order for the equilibration rate to match the oscillation frequency, which is where the resonant peak occurs.

In Fig.~\ref{fig:bv_2d}(b), corresponding to the IUF EoS, which has a direct Urca threshold near $4n_0$.  Here we see two distinct behaviors.  For densities $n_0$ and $3 n_0$, which are below threshold, the behavior is similar to that seen for DD2: the Fermi Surface approximation only includes modified Urca processes, but the exact calculation includes below-threshold direct Urca processes which increase the total rate, moving the resonant peak to lower temperatures.  Above the threshold density, the Fermi Surface approximation for direct Urca overestimates the total Urca rate, since the exact phase space integration leads to only a gradual opening of the phase space around the direct Urca threshold, hence the resonant peak moves to higher temperatures than predicted by the Fermi Surface approximation.

As can be seen from Eq.~(\ref{eq:bv}), the maximum value of bulk viscosity at a frequency $\omega$ is 
\begin{equation}
    \zeta_{\text{max}} = \frac{C^2}{2B\omega}.
    \label{eq:bv-max}
\end{equation}
\begin{figure*}[t!]
    \centering
    \begin{subfigure}[t]{0.5\textwidth}
        \centering
        \includegraphics[width=.95\textwidth]{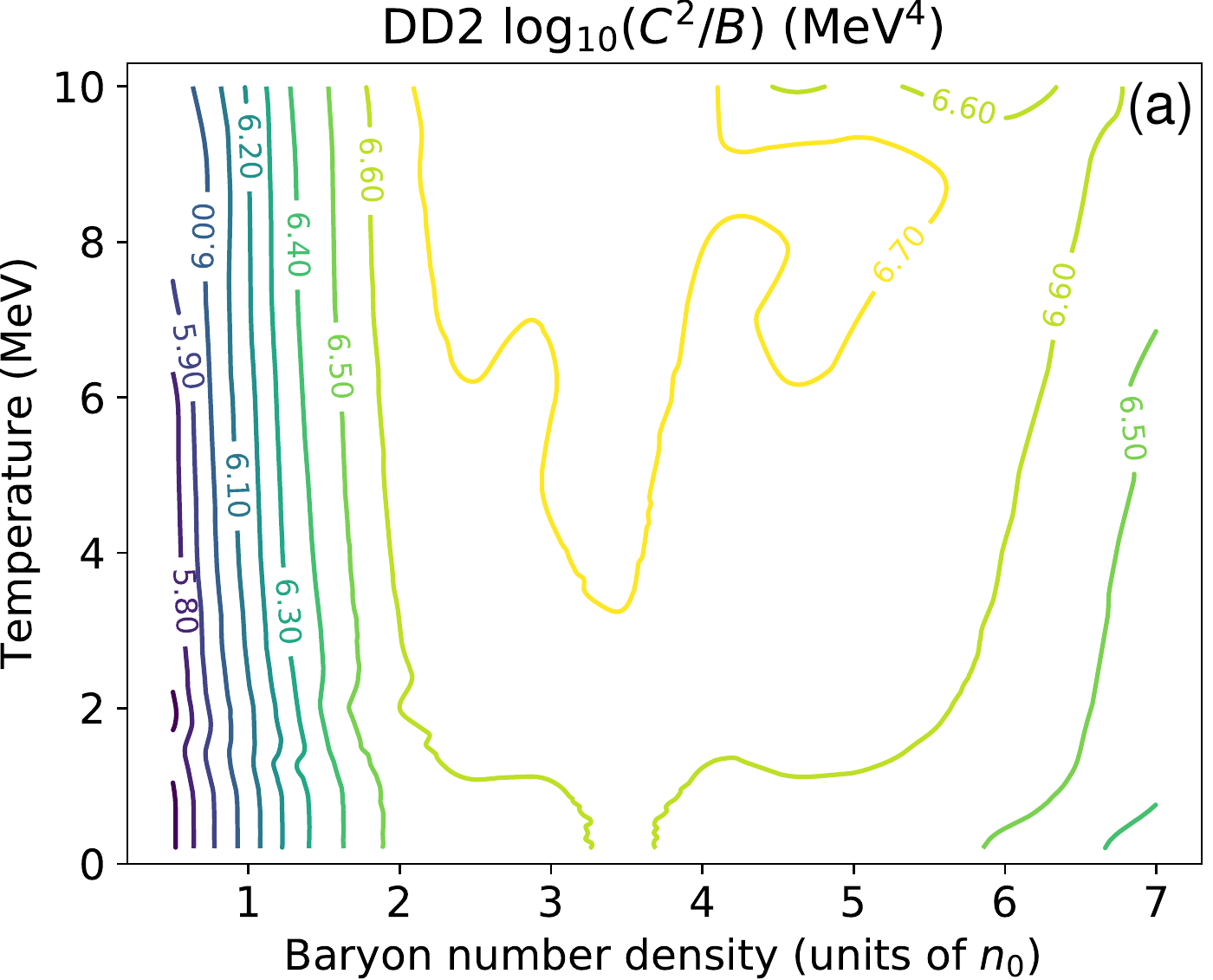}
    \end{subfigure}%
    \begin{subfigure}[t]{0.5\textwidth}
        \centering
        \includegraphics[width=.95\textwidth]{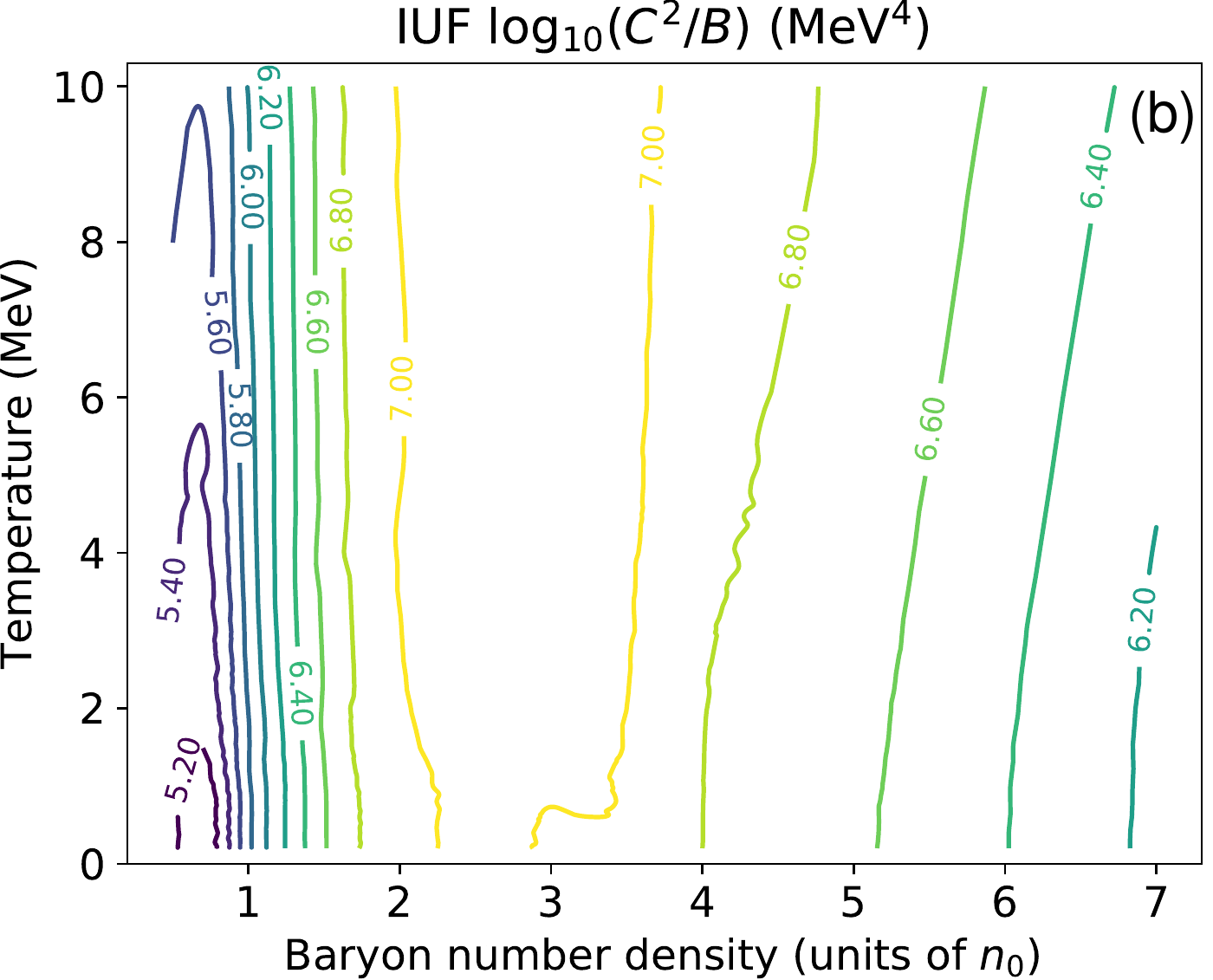}
    \end{subfigure}
    \caption{Logarithmic plot of the ratio of susceptibilities $C^2/B = 2\omega \zeta_{\text{max}}$ that determines the maximum bulk viscosity at a given oscillation frequency. We show results for the DD2 (a) and IUF (b) equations of state, calculated in beta equilibrium [Eq.~(\ref{eq:beta-general})]. }
    \label{fig:c2b}
\end{figure*}

In Fig.~\ref{fig:c2b} we plot $C^2/B = 2\omega\zeta_{\text{max}}$ for a representative range of densities and temperatures for which nuclear matter is likely neutrino-transparent.  We see that for a given frequency, the maximum value of bulk viscosity varies by 1-2 orders of magnitude, and depends more strongly on density than on temperature.  Most notably, we can see that $C^2/B$ rises rapidly at low densities, then levels off at $n\sim 2 n_{\rm sat}$ to a value about an order of magnitude larger then its value at $n=n_{\rm sat}$.  This could already be seen in Fig.~\ref{fig:bv_2d}.

\begin{figure*}[t!]
    \centering
    \begin{subfigure}[t]{0.5\textwidth}
        \centering
        \includegraphics[width=.95\textwidth]{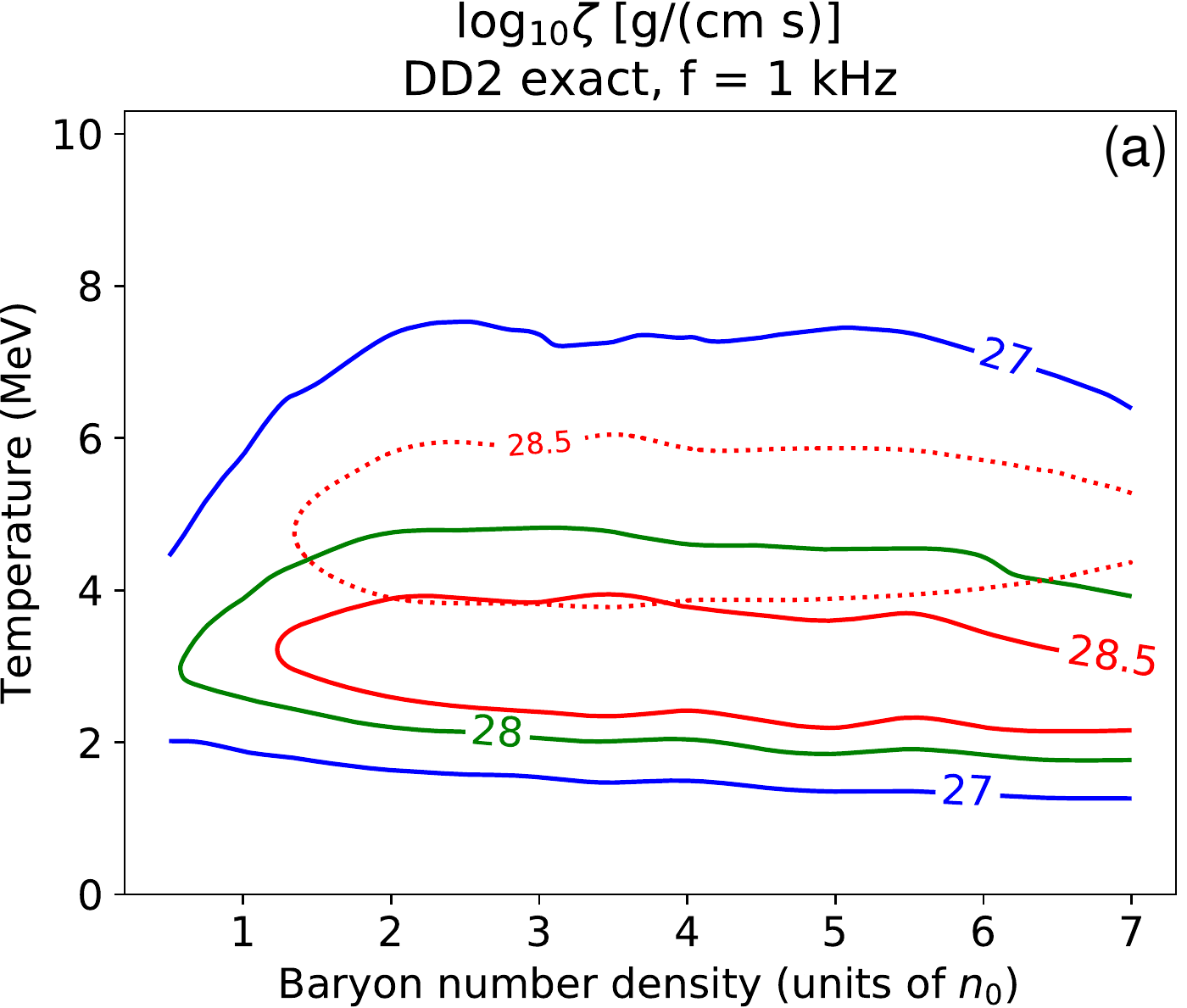}
    \end{subfigure}%
    \begin{subfigure}[t]{0.5\textwidth}
        \centering
        \includegraphics[width=.95\textwidth]{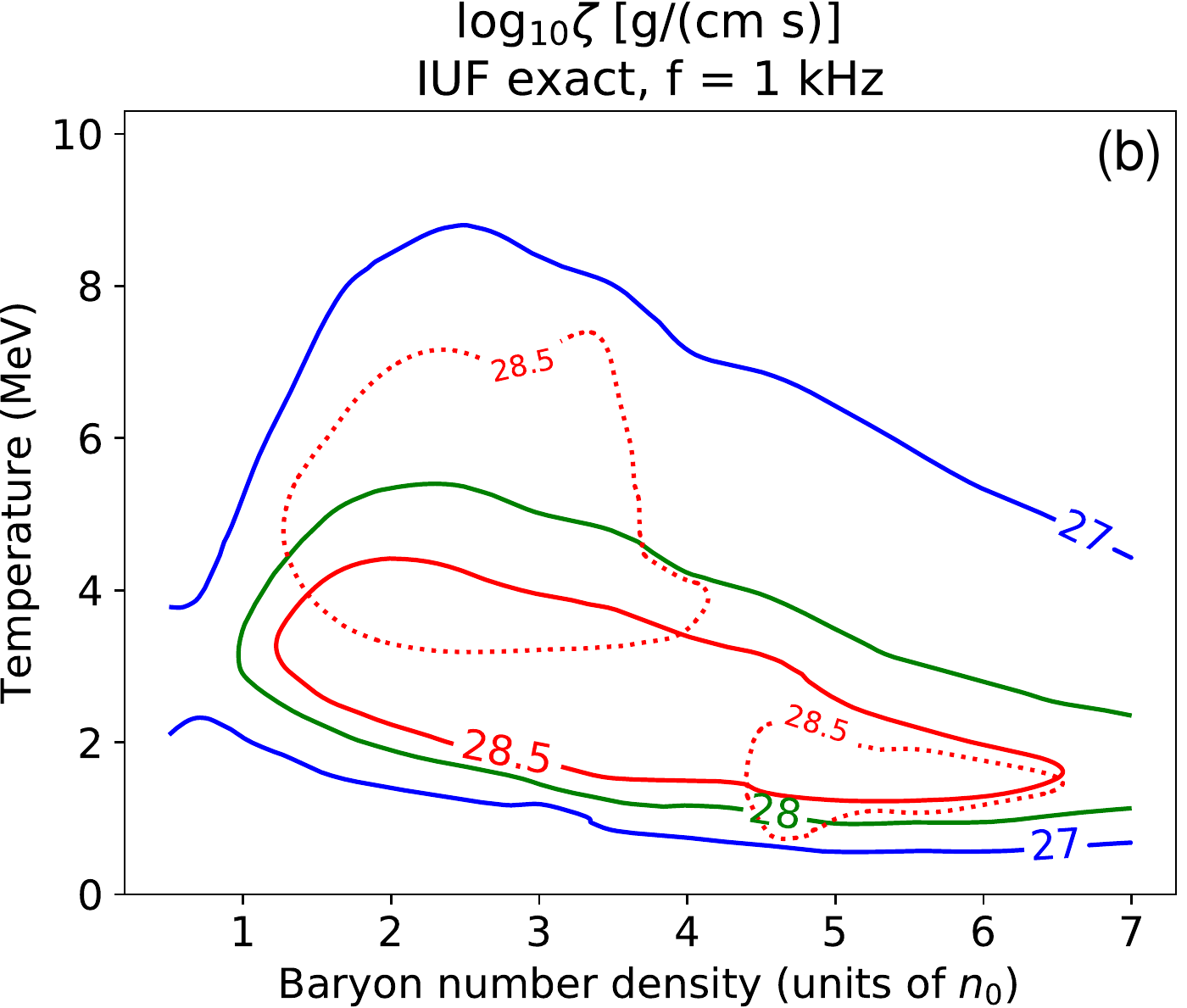}
    \end{subfigure}
    \caption{Bulk viscosity as a function of density and temperature, for the DD2 (a) and IUF (b) EoSs.  The full phase space integral for the direct Urca rate is used in the solid line contours, while the 28.5 dashed contour uses the FS approximation.}
    \label{fig:bv3d}
\end{figure*}

In Fig.~\ref{fig:bv3d}, we plot the bulk viscosity as a function of density and temperature (the curves in Fig.~\ref{fig:bv_2d} are cross-sections through Fig.~\ref{fig:bv3d}).  For a fixed density, as the temperature rises, the beta equilibration rate $\gamma$ rises rapidly because of the increase in available phase space.  At temperatures of a few MeV, the reequilibration rate closely matches the oscillation frequency of 1 kHz, then bulk viscosity reaches a maximum.  At higher temperatures, the reequilbration is too fast and the bulk viscosity drops.

We see that for the DD2 EoS, the bulk viscosity peak is at a temperature of about 3 MeV for all densities, which is a lower temperature than predicted by the Fermi surface approximation.  For IUF, the FS approximation would suggest two different peaks in bulk viscosity: one below the direct Urca threshold corresponding to the near-equality of the modified Urca rate and the density oscillation frequency, and one above the threshold, corresponding to the near-equality of the direct Urca rate and the density oscillation frequency.  However, the gradual opening of the direct Urca threshold coming from the exact direct Urca calculation melds these two peaks into one broad peak.  At low density, the peak is at 3-4 MeV, but as density increases it moves down to 2 MeV.
\hiddensubsection{Energy dissipation time}
The most direct indicator of the importance of bulk viscous damping is the dissipation time $\tau_{\rm diss}$ for density oscillations. Since the merging stars settle down into a massive remnant in tens of milliseconds, bulk viscous damping will be important if $\tau_{\rm diss}$ is tens of milliseconds or less.  To calculate the dissipation time, we need the energy of an oscillation and the rate at which that energy  is dissipated by bulk viscosity.  The energy density of an adiabatic baryon density oscillation $n_B(t)=n_B + (\delta n_B)\sin{(\omega t)}$ is \cite{Alford:2017rxf}
\begin{equation}
    \varepsilon = \frac{1}{2} (\delta n_B)^2  \frac{\partial^2\varepsilon}{\partial n_B^2}\bigg\vert_{x_p,s/n_{B}} 
    = \frac{\kappa_S^{-1}}{2}\left(\frac{\delta n_B}{n_B}\right)^2,
    \label{eq:osc-energy}
\end{equation}
where $\kappa_S$ is the adiabatic compressibility \cite{schroederintroduction,CompOSE}
\begin{equation}
    \kappa_S^{-1} = n_B \frac{\partial P}{\partial n_B}\bigg\vert_{x_p,s/n_B}.
\end{equation}

\begin{figure}[h]
\centering
\includegraphics[scale=0.6]{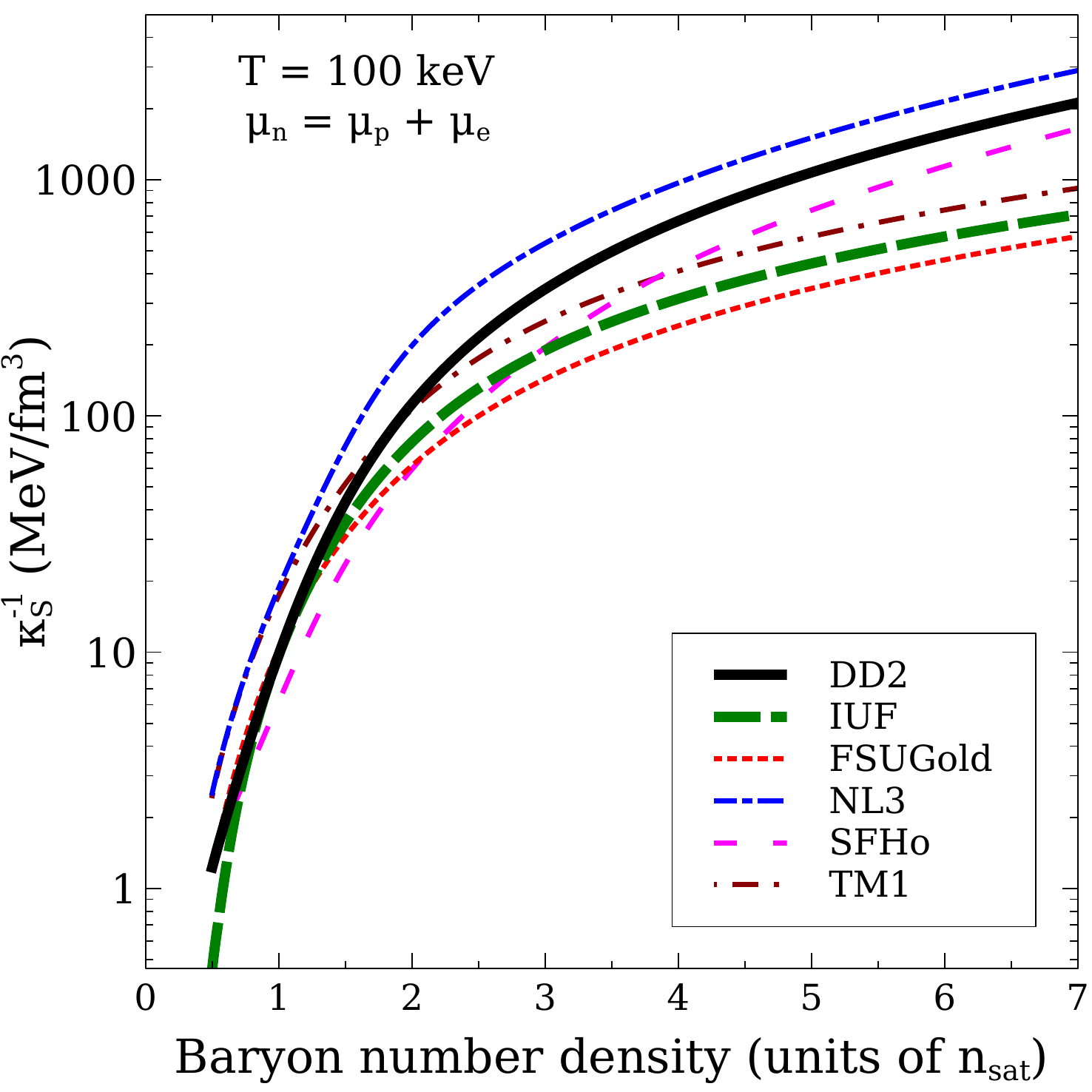}
\caption{ Adiabatic inverse compressibility $(\kappa_S)^{-1}$ at low temperature, versus density, for several EoSs derived from relativistic mean-field theories.  Each is tabulated on CompOSE.}
\label{fig:K}
\end{figure}

We note that the adiabatic compressibility depends on the EoS.  However, it is a common feature of all nucleonic EoSs that  nuclear matter becomes more incompressible at high densities, so the inverse compressibility  $1/\kappa_S$ rises with density, as shown in Fig.~\ref{fig:K} for a range of EoSs including those used in this work. This means that at higher density, oscillations in the density store more energy.  

To facilitate comparison with previous work (for example, \cite{Alford:2017rxf}), we mention that the ``stiffness'' of nuclear matter is often described via the nuclear incompressibility $K$ \cite{Shapiro:1983du,glendenning2000compact,Schmitt:2010pn}.  $K$ is conventionally defined at saturation density, zero temperature, and for symmetric nuclear matter, and is approximately 250 MeV \cite{Oertel:2016bki}.  Some works have extended the definition of the nuclear incompressiblity to densities above nuclear saturation \cite{Dexheimer:2007mt}.  At zero temperature, $n_0$, and for symmetric nuclear matter, the adiabatic $\kappa_S$ can be related to the nuclear incompressibility $K$ by $K = 9/(\kappa_S n_0)$~\cite{CompOSE,Schmitt:2010pn}.

The rate of energy density dissipation is given by \cite{Alford:2010gw,Sawyer:1980wp} 
\begin{equation}
\frac{\mathop{d\varepsilon}}{\mathop{dt}} = \frac{\omega^2}{2}\left(\!\frac{\delta n_B}{n_B}\!\right)^{\!2} \zeta. 
\label{eq:osc-dissipation}
\end{equation}
Using Eq.~(\ref{eq:osc-energy}), the energy dissipation time is 
\begin{equation}
    \tau_{\text{diss}} \equiv \frac{\varepsilon}{d\varepsilon/dt} = \frac{(\kappa_S)^{-1}}{\omega^2 \zeta}.
    \label{eq:t-diss}
\end{equation}
Note that one can also define \cite{1990ApJ...363..603C} a decay time for the amplitude, which would be longer by a factor of two since the energy of an oscillation goes as the square of the amplitude.  
\begin{figure*}[t!]
    \centering
    \begin{subfigure}[t]{0.5\textwidth}
        \centering
        \includegraphics[width=.95\textwidth]{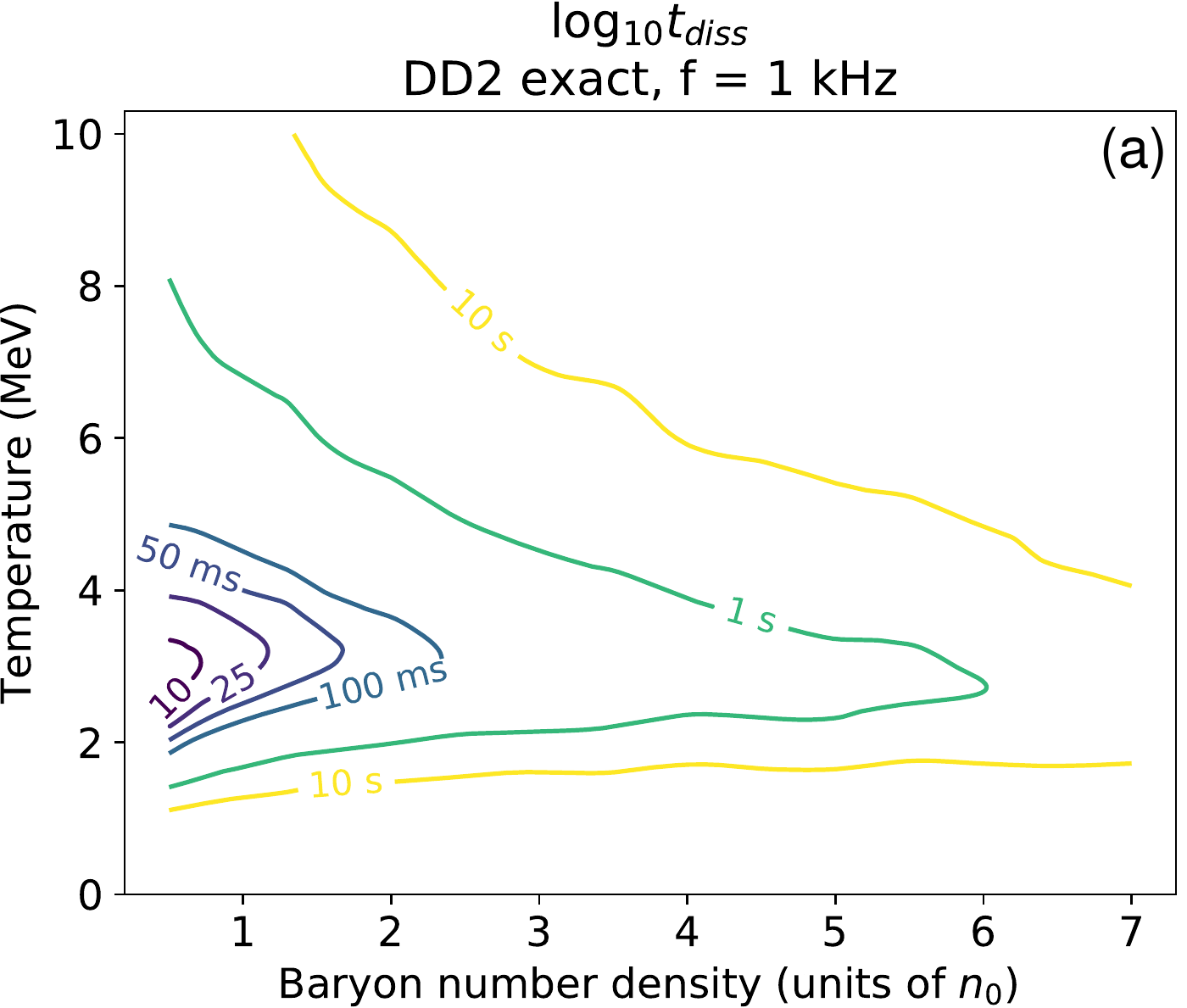}
    \end{subfigure}%
    \begin{subfigure}[t]{0.5\textwidth}
        \centering
        \includegraphics[width=.95\textwidth]{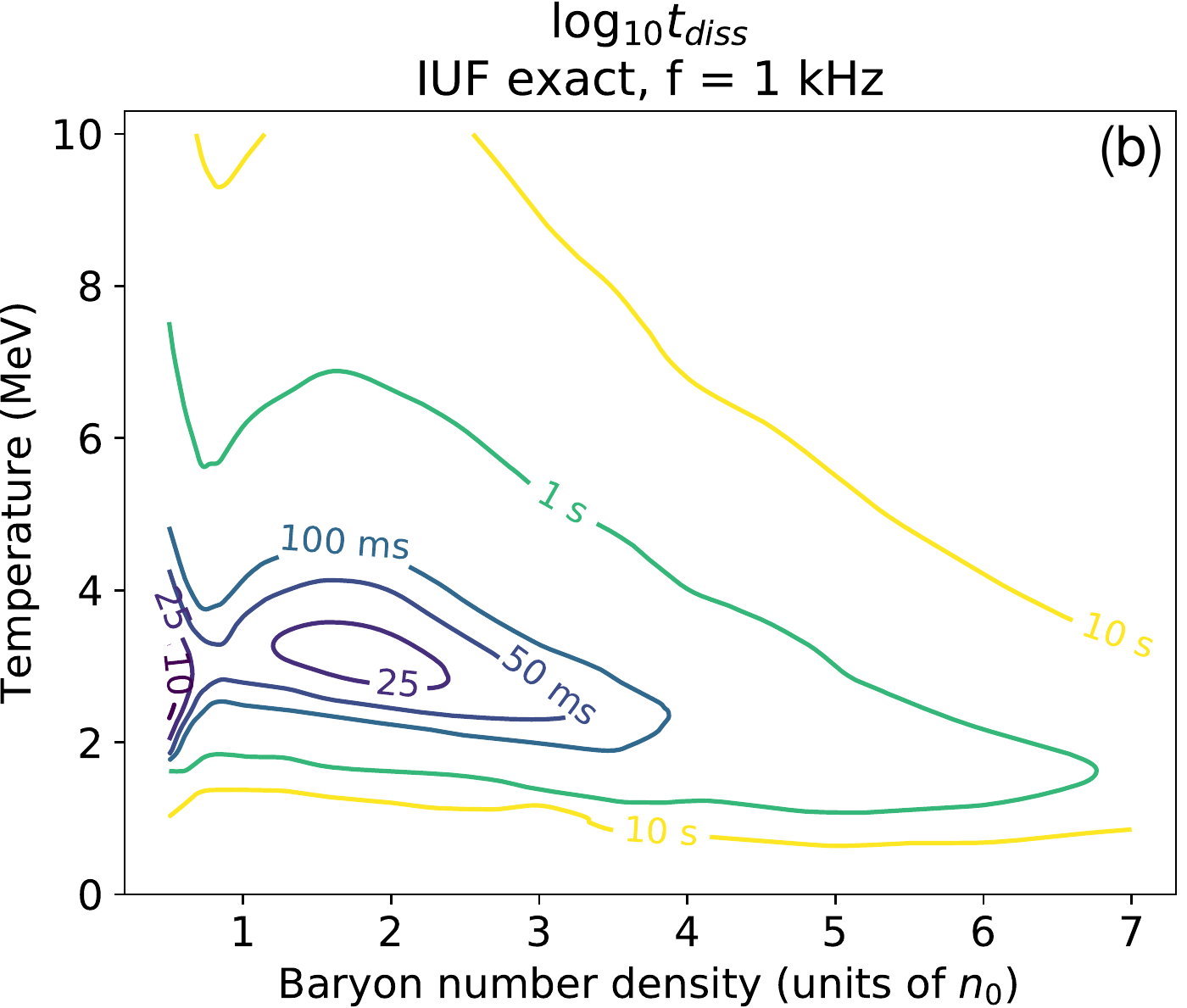}
    \end{subfigure}
    \caption{Dissipation time $\tau_{\rm diss}$ of a 1 kHz density oscillation, using the DD2 EoS (a) and IUF EoS (b), with the exact Urca rates.}
    \label{fig:tdiss_1kHz_exact}
\end{figure*}
\begin{figure*}[t!]
    \centering
    \begin{subfigure}[t]{0.5\textwidth}
        \centering
        \includegraphics[width=.95\textwidth]{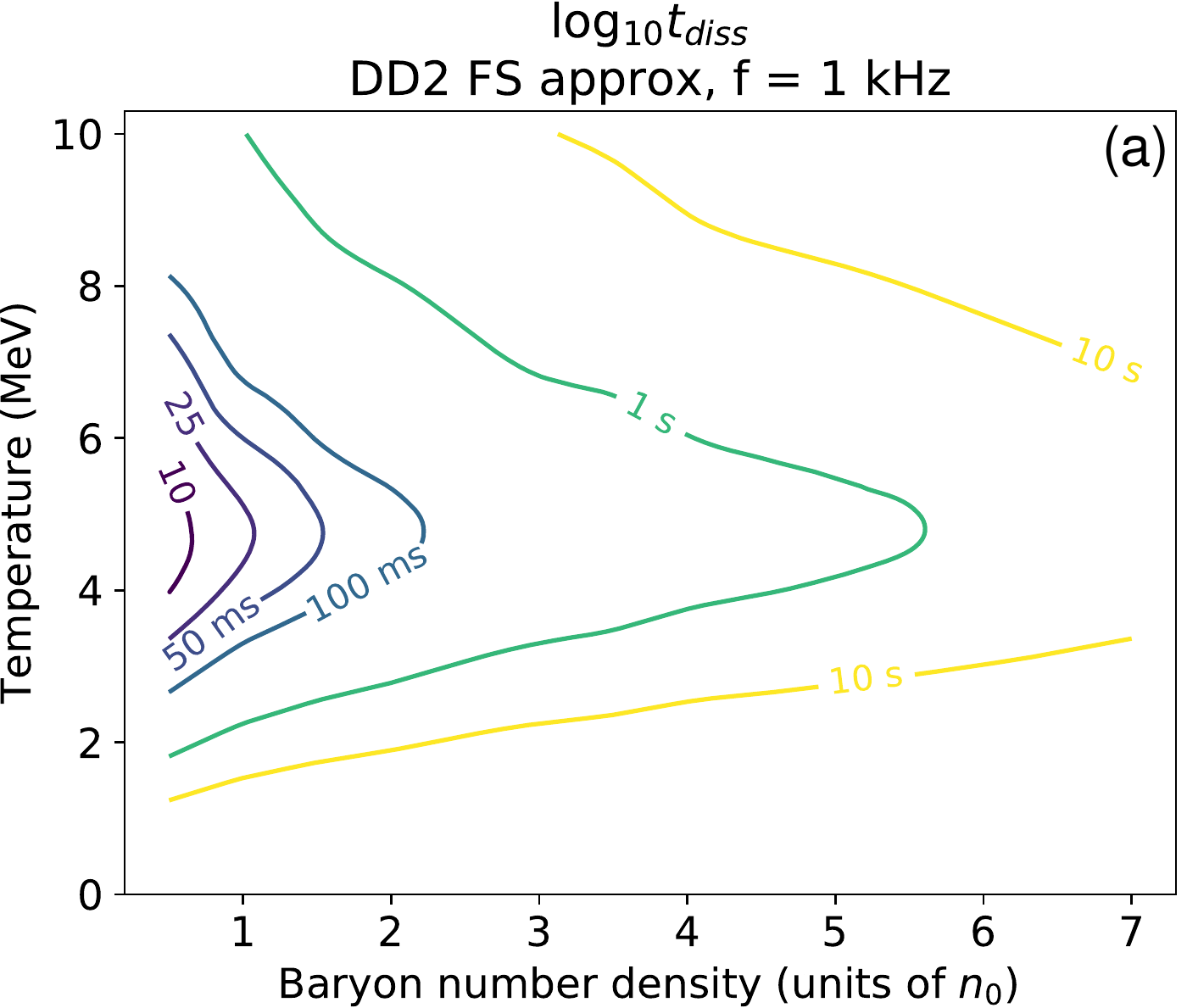}
    \end{subfigure}%
    \begin{subfigure}[t]{0.5\textwidth}
        \centering
        \includegraphics[width=.95\textwidth]{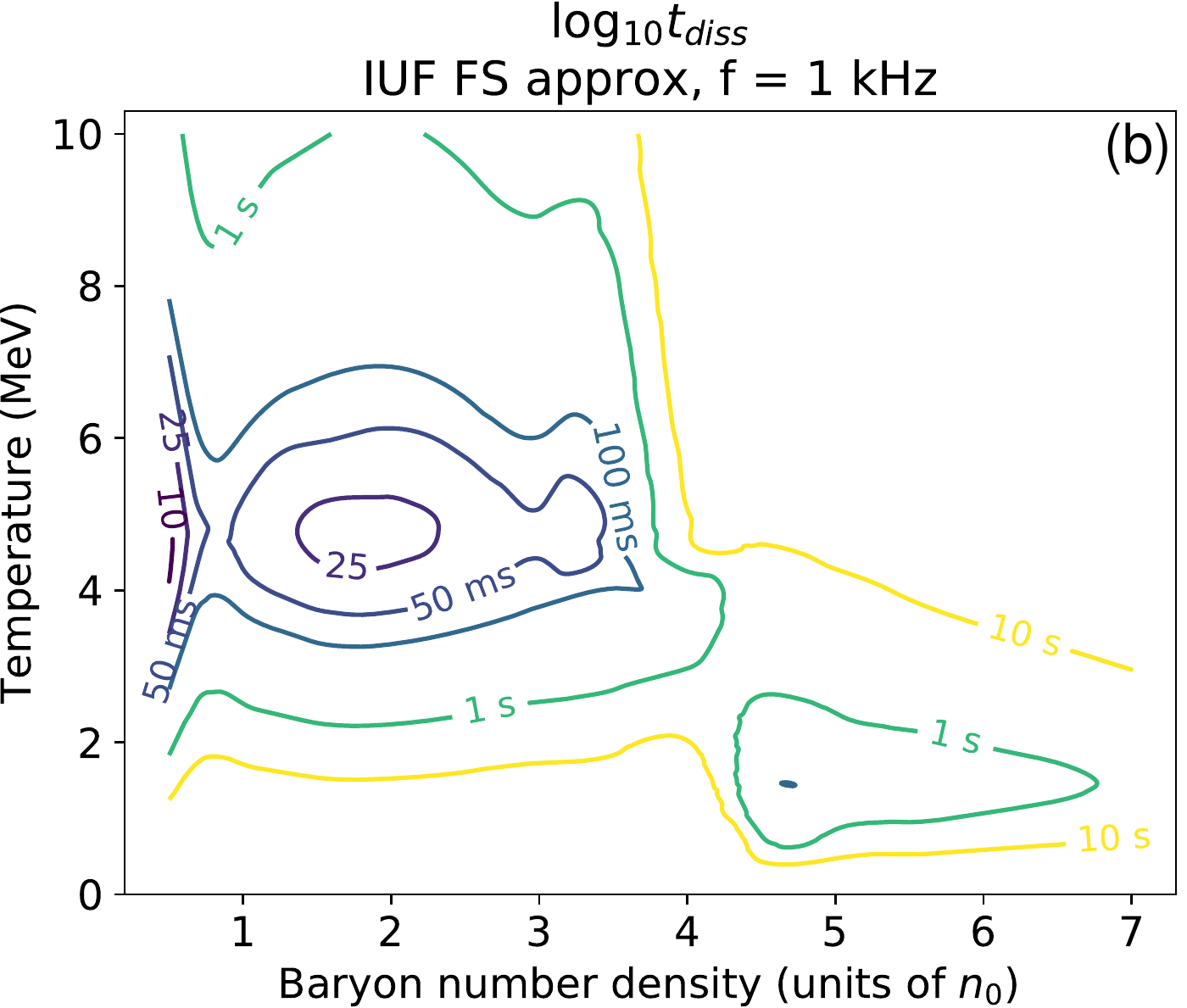}
    \end{subfigure}
    \caption{Dissipation time $\tau_{\rm diss}$ of a 1 kHz density oscillation, using the DD2 EoS (a) and the IUF EoS (b), calculated in the Fermi Surface approximation.}
    \label{fig:tdiss_1kHz_FS}
\end{figure*}

In Fig.~\ref{fig:tdiss_1kHz_exact}, we plot the dissipation time of a 1\,kHz density oscillation as a function of density and temperature for two different EoSs, using the exact Urca rates.  We first discuss the physical content and implications of the exact results (Fig.~\ref{fig:tdiss_1kHz_exact}), then compare them to the Fermi Surface approximation, shown in Fig.~\ref{fig:tdiss_1kHz_FS}.

\noindent {\em Temperature dependence}. The adiabatic compressibility is relatively independent of temperature, so the bulk viscosity dominates the temperature dependence of the dissipation time.  As discussed in Sec.~\ref{sec:bv_results}, for a given density, the bulk viscosity increases, reaches a resonant maximum when the beta reequilibration rate $\gamma$ matches the oscillation frequency $\omega$, and then decreases as temperature increases.  This leads to minimum dissipation time at approximately the temperature at which the bulk viscosity reaches its maximum, for a given density.

\noindent {\em Density dependence}. 
The adiabatic inverse compressibility strongly increases as a function of density, as seen in Fig.~\ref{fig:K}.  While the bulk viscosity was weakly dependent on density, the dissipation time at high density is strongly increased due to the several order-of-magnitude rise of the adiabatic inverse compressibility.  Physically, oscillations in high density nuclear matter have a lot of energy due to the high incompressiblity of dense nuclear matter [see Eq.~(\ref{eq:osc-energy})].  Thus, it takes correspondingly longer time for those high-energy oscillations to damp.  As a result of the behavior of the compressibility of nuclear matter, the minimum of dissipation time is likely to be located at a low density.

It is worth noting that the bulk viscosity varies non-monotonically with density.  It rises as density increases from $0.5n_0$, reaches a peak at several times $n_0$, and then falls off at high density.  This can be seen by noting that the maximum bulk viscosity is $\zeta_{\rm max} = (1/2\omega) C^2/B$ [Eq.~(\ref{eq:bv-max})], which is plotted in Fig.~\ref{fig:c2b}.  It is clear that the particular features of the rise and fall in bulk viscosity as a function of density depend on the EoS.  Throughout the range of densities that we consider, the bulk viscosity prefactor $C^2/B$ varies by 1-2 orders of magnitude.  However, the inverse compressibility rises by three orders of magnitude over that density range, so it has a more substantial effect on the density dependence of the dissipation time.

For the DD2 EoS, as seen in Fig.~\ref{fig:tdiss_1kHz_exact}(a), the minimum dissipation times lie around temperatures of 3 MeV for all densities, indicating that the reequilibration rate doesn't change strongly with density, which is expected since only modified Urca and below-threshold direct Urca are acting.  As a function of density, the dissipation times get longer as density increases.  This behavior comes from the dramatic monotonic rise of the inverse compressibility as a function of density.  The bulk viscosity prefactor $C^2/B$ rises by one order of magnitude from $0.5n_0$ to 3 or 4 $n_0$, and then slightly decreases at higher densities, but it doesn't vary rapidly enough to compete with the rise of the inverse compressibility, and thus the dissipation time rises monotonically with density.  DD2 has a minimum dissipation time of about 6 ms, which occurs only at low density ($0.5n_0$) at temperatures of just under 3 MeV.  Only fluid elements with densities under twice saturation density would dissipate energy on timescales relevant for mergers.  

As seen in Fig.~\ref{fig:tdiss_1kHz_exact}(b), the behavior of the dissipation time scale for the IUF EoS is more complicated.  The lowest dissipation times do occur at temperatures of around 3 MeV, since the resonant peak of bulk viscosity is around that temperature.  However, the nonmonotonic behavior of $C^2/B$ as a function of density is more dramatic for the IUF EoS than for DD2, so it competes with the rapidly rising inverse compressibility as density increases, leading to two minima in the dissipation time.  The first is at low density, where the nuclear inverse compressibility is decreasing rapidly as the density decreases to the lowest value for which we trust our equation of state, $n=0.5n_0$. There, energy dissipation can occur in as little as 5 ms.  There is also a local minimum around $n=2n_0$, where the bulk viscosity prefactor $C^2/B$ has a local maximum (see Fig.~\ref{fig:c2b}(b)) and dissipation times reach down to 19 ms.   For the IUF EoS, dissipation occurs on merger timescales in fluid elements up to four times saturation density, in contrast to the behavior of DD2.

It is interesting to compare the Fermi Surface approximate results (Fig.~\ref{fig:tdiss_1kHz_FS}) and the exact results (Fig.~\ref{fig:tdiss_1kHz_exact}) for each EoS.  For DD2, the use of the exact Urca rates just increases the total Urca rate and thus the bulk viscosity is maximized at a lower temperature than would be predicted by the Fermi Surface approximation.  For IUF, the Fermi Surface approximate result would predict a sharp change in the behavior of the bulk viscosity at the direct Urca threshold,  $n=4n_0$ (for a generic example of this behavior, see Fig.~1 in \cite{Haensel:2001mw}).  However, at the temperatures of interest to us the exact Urca rates show a gradual increase with density and thus the bulk viscosity does not change suddenly at the threshold density.    
\section{Higher frequency oscillations}
\begin{figure*}[t!]
    \centering
    \begin{subfigure}[t]{0.5\textwidth}
        \centering
        \includegraphics[width=.95\textwidth]{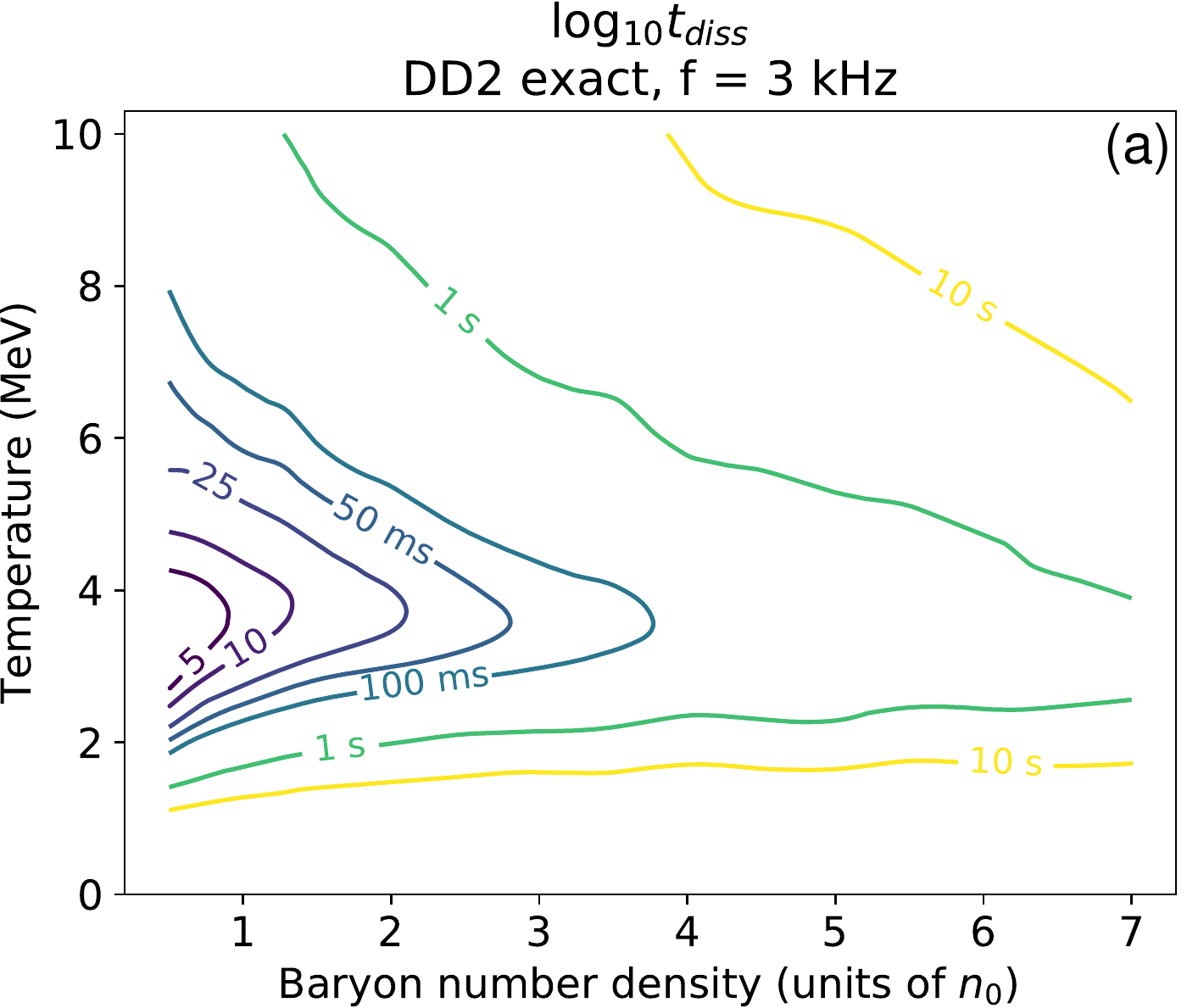}
    \end{subfigure}%
    \begin{subfigure}[t]{0.5\textwidth}
        \centering
        \includegraphics[width=.95\textwidth]{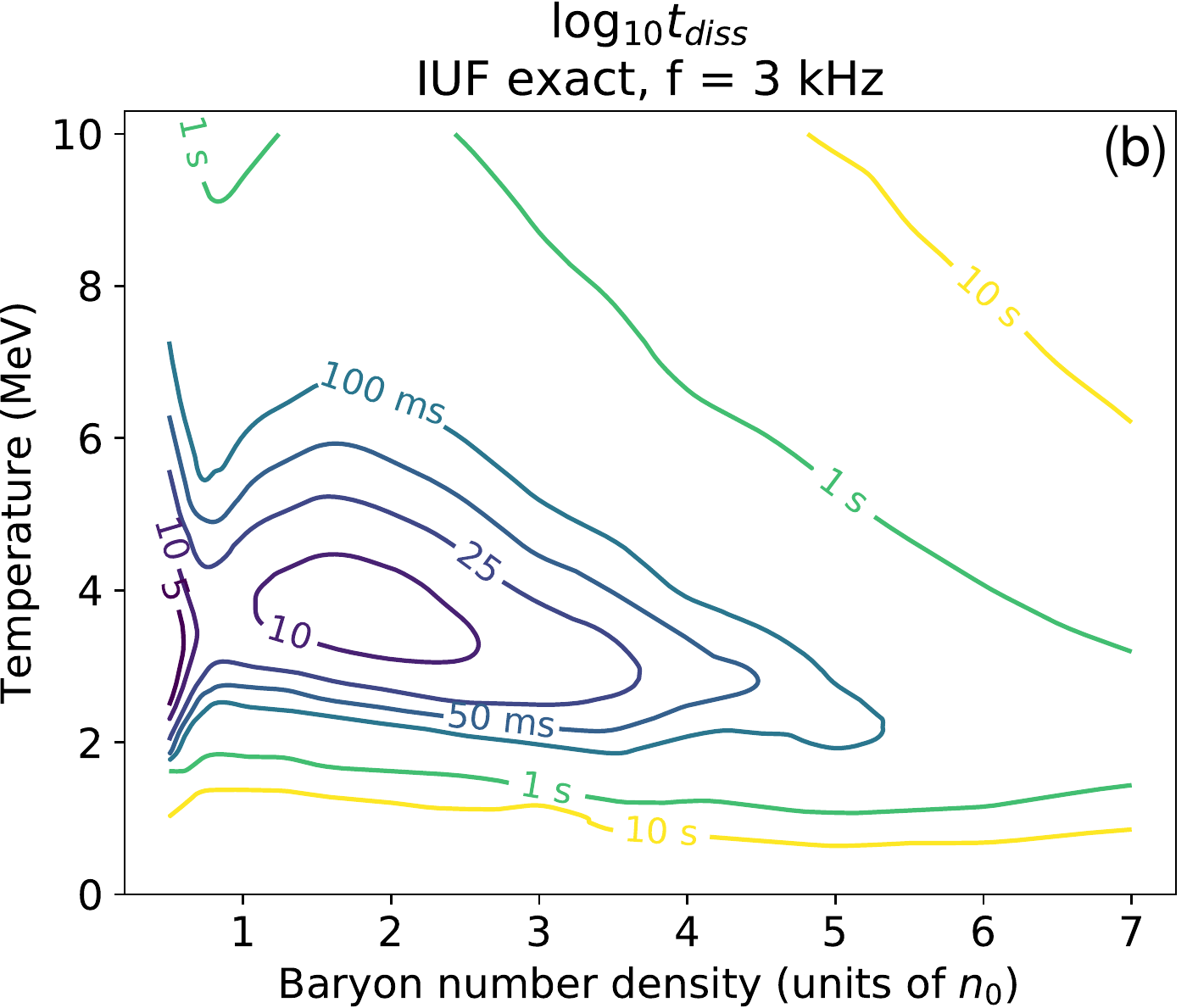}
    \end{subfigure}
    \caption{Dissipation time scale for 3 kHz oscillations in nuclear matter with the DD2 EoS (a) or IUF EoS (b).  The exact Urca rates are used.}
    \label{fig:3kHz}
\end{figure*}
\begin{figure*}[t!]
    \centering
    \begin{subfigure}[t]{0.5\textwidth}
        \centering
        \includegraphics[width=.95\textwidth]{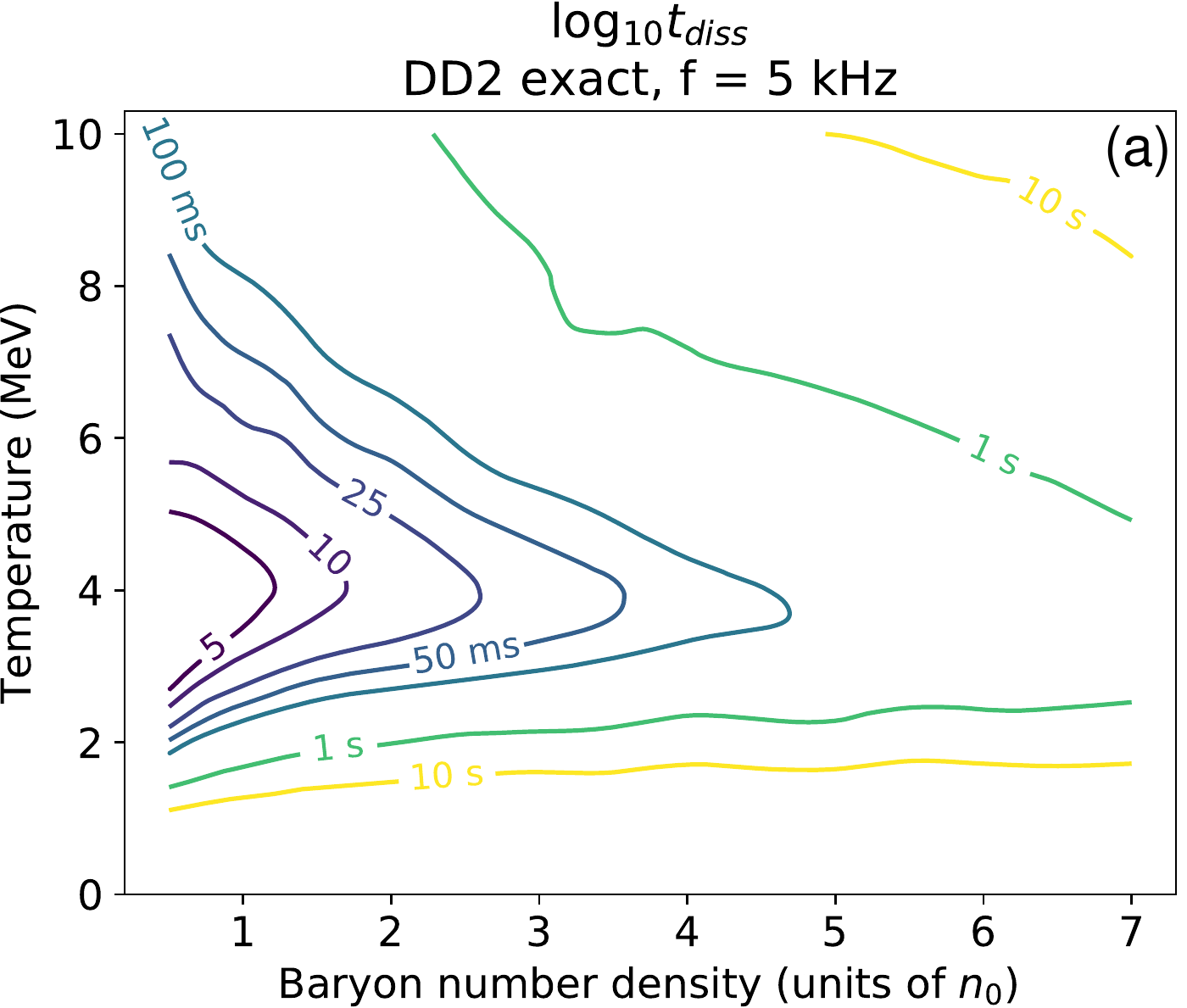}
    \end{subfigure}%
    \begin{subfigure}[t]{0.5\textwidth}
        \centering
        \includegraphics[width=.95\textwidth]{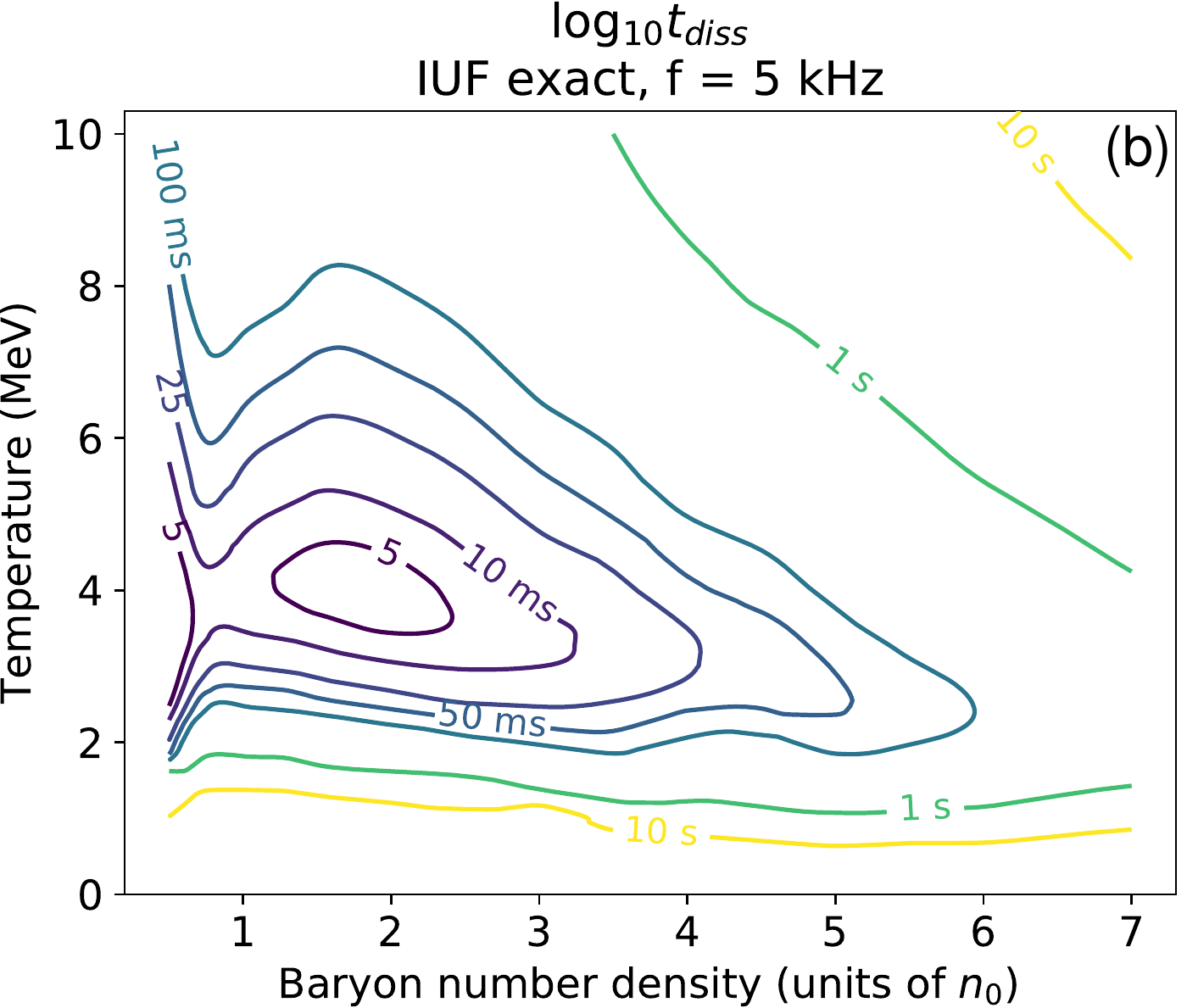}
    \end{subfigure}
    \caption{Dissipation time scale for 5 kHz oscillations in nuclear matter with the DD2 EoS (a) or IUF EoS (b).  The exact Urca rates are used.}
    \label{fig:5kHz}
\end{figure*}
There is evidence from simulations \cite{Hotokezaka:2013iia,Chaurasia:2018zhg,Clark:2015zxa,Rezzolla:2016nxn,Dietrich:2016hky,Maione:2017aux,Bauswein:2018bma} (see also the review \cite{Baiotti:2019sew}) that eccentric binary neutron star mergers excite oscillations at frequencies above 1 kHz.  We plot the dissipation times for 3 kHz and 5 kHz oscillations in Figs \ref{fig:3kHz} and \ref{fig:5kHz}.  We see that at these higher frequencies, bulk viscosity plays a bigger role, and density oscillations can be damped in as little as 1 ms, and for a broad range of temperatures and densities, oscillations can be damped in under 25 ms.

We note that, at a given density, a higher temperature is required to make the reequilibration rate $\gamma$ match a density oscillation which has a frequency above 1 kHz, and thus the region of maximum bulk viscosity is moved to higher temperatures.  For example, a 5 kHz density oscillation has maximum bulk viscosity (and thus minimum damping time) at about $T = 4 \text{ MeV}$ (see Fig.~\ref{fig:5kHz}), while a 1 kHz density oscillation has maximum bulk viscosity at around $T = 3 \text{ MeV}$ (see Fig.~\ref{fig:tdiss_1kHz_exact}).
\section{Conclusions}

We have calculated the bulk-viscous dissipation time in nuclear matter at temperatures and densities relevant to neutron star mergers. 
We assumed the material was transparent to neutrinos, which should be valid for temperatures up to about 5 MeV, and we studied the damping of oscillations with frequencies in the 1 kHz range, which are seen in simulations of mergers.
The main uncertainty in our result is the form of the
nuclear matter equation of state at supranuclear densities,
so we performed calculations for two different equations of state, one stiffer, DD2, and one softer, IUF.  Our main results are displayed in Fig \ref{fig:tdiss_1kHz_exact}.

Bulk viscous damping will play a significant role at densities and temperatures where the dissipation time is comparable to or less than the typical timescale of the merger, which is in the range of tens of milliseconds. Both equations of state show a similar overall pattern: bulk viscosity damps oscillations on timescales comparable to a merger for nuclear matter at temperatures of 2-4 MeV and for densities between $0.5n_0$ to $2n_0$, with IUF also exhibiting fast damping for densities up to $4n_0$.  Both EoSs have minimum dissipation times of about 5 ms, occurring at $0.5n_0$, while IUF has another local minimum of dissipation time, about 20 ms, occurring at $2n_0$.  The occurrence of dissipation times in the 10 ms range leads us to conclude that bulk viscous damping in neutrino-transparent nuclear matter should be seriously considered for inclusion in future simulations.  Bulk viscosity in neutrino-trapped nuclear matter is too small to impact mergers, because the Urca rates are too fast at trapping temperatures ($T \gtrsim \text{5-10 MeV}$) to match a 1 kHz density oscillation \cite{Alford:2019kdw}.  Between these extremes lies the regime where the spectrum of neutrinos includes a low-energy population that escapes, a high-energy tail that is trapped, and an intermediate energy range where the mean free path is comparable to the distance scale of the fluid flows, requiring explicit inclusion of neutrinos in the dynamics of the nuclear fluid \cite{Ardevol-Pulpillo:2018btx,Perego:2014qda,Galeazzi:2013mia,Sekiguchi:2012uc,Rosswog:2003rv,1999JCoAM.109..281M}.

Another limitation of our calculation is the assumption of low-amplitude density oscillations. We calculated the ``subthermal'' bulk viscosity, but simulations show high amplitude density oscillations \cite{Alford:2017rxf} for which the suprathermal bulk viscosity \cite{Alford:2010gw} is relevant. This could extend the region of large bulk viscosity down to lower temperatures, since suprathermal effects allow high-amplitude oscillations to experience the maximum bulk viscosity $\zeta_{\rm max}$ [Eq.~(\ref{eq:bv-max})] at lower temperatures \cite{Alford:2010gw}.

Our discovery of short bulk-viscous dissipation times at densities below nuclear saturation density, primarily due to the low inverse compressibility, underscores the need for a detailed understanding of the structure of nuclear matter below saturation density.  The DD2 and IUF EoSs predict uniform nuclear matter down to densities of 0.25 to 0.4 $n_0$ respectively, which is why we restricted our calculations to densities  above $0.5n_0$.  However,  a sequence of mixed ``pasta'' phases has been predicted at densities between $0.2$ and $0.7$ $n_0$ \cite{Grill:2014aea,Fattoyev:2017zhb,Pais:2015lma,Pais:2014hoa,Oyamatsu:1993zz}.  It has been noted \cite{Yakovlev:2018jia, Gusakov:2004mj} that the appearance of free protons in certain pasta phases would open up the direct Urca process, albeit with such a reduced rate that it would take temperatures of tens of MeV---which is well above the pasta melting temperature of a few MeV \cite{Roggero:2017pag}---to reach the resonant peak of bulk viscosity.  Thus, it is important to know how and at what densities and temperatures nuclear matter transforms from a uniform phase to a mixed phase.  Based on our findings above, we expect subthermal bulk viscosity to be large for these low densities, down to the density at which uniform nuclear matter transitions to a pasta or spherical nuclei phase. 

We did not consider Urca processes involving muons, and did not include muons in the EoSs.  The presence of muon Urca processes would increase the equilibration rate $\gamma$ for densities at which muons are present.  In addition, muon-electron conversion would give rise to a separate contribution to the bulk viscosity \cite{Alford:2010jf}.  The calculation of bulk viscous damping time in Ref.~\cite{Alford:2017rxf} uses EoSs that contain muons. Above the onset density for muons the nuclear matter susceptibilities are larger, which would lead to larger bulk viscosity and thus shorter dissipation times compared to the muonless EoSs considered in this work. We are therefore planning to perform a full study of bulk viscous dissipation in EoSs that include muons. 

There is evidence that properly including in-medium effects in the nucleon propagator can lead to a large increase in the modified Urca rate just below the direct Urca threshold \cite{Shternin:2018dcn}.  We have not included this in our analysis, but it could potentially lead to a shift of the resonant peak of bulk viscosity to lower temperatures for a range of densities near the direct Urca threshold.
\chapter{Thermal transport in neutron star mergers: axions as a candidate}
\pagestyle{myheadings}
\label{sec:axions}
\begin{center}
{\textit{This section is based on my work with Mark Alford, Jeff Fortin, and Kuver Sinha \cite{Harris:2020qim}}}.
\end{center}
\section{Introduction}
Thermal transport is due to particles with long mean free paths.  If a particle's mean free path is extremely short compared to the system size, say 1 fm compared to the size of a neutron star, the particle cannot transport anything, because it takes forever to travel any significant distance.  If a particle can travel a reasonable fraction of the system size, say, 10-100 meters in a neutron star, then it can help the system thermally equilibrate, that is, it allows fluid elements to exchange energy and reach a common temperature.  If a particle has an extremely long mean free path compared to the system, say greater than 10 kilometers compared to the size of a neutron star, then the particle acts as a way to take energy out of the system, cooling it.  

In neutron stars and neutron star mergers, neutrinos facilitate energy transport.  As we will see, neutrino cooling is only efficient at temperatures where neutrinos have a long mean free path, perhaps for $T \lesssim 5 \text{ MeV}$ (see Sec.~\ref{sec:nu_mfp}).  As we will discuss in Sec.~\ref{sec:axion_trapped}, trapped neutrinos are not very efficient at thermally equilibrating the matter in a merger over the relevant tens of millisecond timescales.  We analyze the possibility that axions \cite{Peccei:1977hh,Wilczek:1977pj,Weinberg:1977ma,Kolb:1990vq}, bosonic particles which are an interesting candidate for physics beyond the standard model, will play a role in cooling or thermally equilibrating neutron star mergers.  

Supernovae and cold, isolated neutron stars have long been used as laboratories to study beyond the standard model particles like the axion \cite{Chang:2018rso,Raffelt:1996wa,Sedrakian:2015krq}.  In this chapter, we discuss the possible impact of axions on neutron star mergers, a hotter and denser environment than individual neutron stars, and a denser and more neutron-rich environment than supernovae.  An early effort to include axions in simulations of neutron star mergers has been made by Dietrich and Clough \cite{Dietrich:2019shr}.  They model axion cooling of a merger by using standard axion emissivity expressions from Brinkmann and Turner \cite{Brinkmann:1988vi} in nuclear matter in both the non-degenerate and degenerate regimes.

To determine the role of axions in neutron star mergers, we consider a couple of possibilities.  If axions have a long mean free path (MFP) compared to the size of the merger (20-30 kilometers in diameter - see Figs.~\ref{fig:radice_merger_panels} and \ref{fig:T_distribution}, as well as \cite{Baiotti:2016qnr}), then they free-stream through the nuclear matter, taking energy away from the merger which results in cooling.  On the other hand, if axions have a relatively short mean free path, they would contribute to transport inside the merger.  

To calculate the mean free path we first discuss the production (or absorption) of axions (with field operator $a$) via bremsstrahlung from neutrons
(with field operator $\psi_n$). The relevant coupling term in the
Lagrangian is $\mathcal{L} = G_{an}(\partial_{\mu}a)\,\bar{\psi}_n \gamma^{\mu}\gamma_5\psi_n$.  The standard calculations of axion mean free paths and emissivities rely on the Fermi surface (FS) approximation: we propose an improvement to the Fermi surface approximation which extends its validity to semi-degenerate nuclear matter.  Our main result is a calculation of the axion emissivity and mean free path, where the only approximation is the assumption of a momentum-independent matrix element for the neutron bremsstrahlung process.  We keep the relativistic energy dispersion of the neutrons and we keep the axion momentum in the energy-momentum conserving delta function.  The full phase space integration is valid for degenerate neutrons as well as non-degenerate neutrons.  We then discuss the axion-transparent regime (where the axion mean free path is comparable to or larger than the system size) and axion-trapped regime of the merger (where the axion mean free path is much less than the system size). For the former, the temperature of fluid elements radiating axions as a function of time is computed and the characteristic cooling times are obtained in Figs.~\ref{fig:radiative_cooling_dens_temp} and \ref{fig:rad_cool_time_vs_G}. In the trapped regime, the timescale of thermal equilibration for a fluid element to transfer heat to its neighboring fluid elements is computed. The results are depicted in Fig.~\ref{fig:conductivecooling}. 

Throughout our work, we show the constraints coming from SN1987A on the axion-neutron coupling constant $G_{an}$ \cite{Graham:2015ouw,Raffelt:1996wa,PhysRevD.42.3297,PhysRevD.39.1020,Mayle:1987as,Mayle:1989yx,PhysRevLett.60.1793,Turner:1987by} and discuss the interplay of our results with those coming from supernova physics.  Generally, SN1987A bounds on $G_{an}$ prefer the axion-transparent regime. Since supernova bounds can vary considerably depending on the details of the core-collapse simulations (we refer to \cite{Bar:2019ifz} for a recent critical assessment), our approach is to treat the SN1987A constraint loosely, and thus we examine a range of axion-neutron couplings that extends somewhat above the upper bound, down to significantly below the upper bound.
\section{Axion production in nuclear matter}
\label{sec:axions_in_nuclear_matter}
Axions are proposed to couple to neutrons with the interaction term $\mathcal{L} = G_{an}\partial_{\mu}a\bar{\psi}_n\gamma^{\mu}\gamma_5\psi_n$ \cite{Brinkmann:1988vi}.  A neutron by itself cannot emit an axion because of energy-momentum conservation, so a spectator nucleon is required to donate energy/momentum to the processes to allow it to proceed.  The strong interaction between the spectator nucleon and the nucleon emitting the axion (throughout this paper, we assume that both nucleons are neutrons) is modeled by one-pion exchange (OPE) \cite{OPE} with Lagrangian  $\mathcal{L}_{n\pi} = i (2m_n/m_{\pi}) f \gamma^5 \pi_0 \bar{\psi}_n\psi_n$, where $f\approx 1$.  The neutron and pion masses here are their respective masses in vacuum.  Thus, axions can be created and absorbed by the neutron bremsstrahlung process $n + n \leftrightarrow n + n + a$.  This process is described at tree level by eight Feynman diagrams (see Fig.~4 of Ref.~\cite{Brinkmann:1988vi}), giving rise to the matrix element (derived in the appendix of \cite{Brinkmann:1988vi})
\begin{equation}
    S\sum_{\text{spins}} \vert \mathcal{M}\vert^2 = \frac{256}{3}\frac{f^4m_n^4G_{an}^2}{m_{\pi}^4}\label{eq:matrix_element}\left[ \frac{\mathbf{k}^4}{\left(\mathbf{k}^2+m_{\pi}^2\right)^2}+\frac{\mathbf{l}^4}{\left(\mathbf{l}^2+m_{\pi}^2\right)^2}+\frac{\mathbf{k}^2\mathbf{l}^2-3\left(\mathbf{k}\cdot\mathbf{l}\right)^2}{\left(\mathbf{k}^2+m_{\pi}^2\right)\left(\mathbf{l}^2+m_{\pi}^2\right)}    \right],
\end{equation}
where $\mathbf{k}$ and $\mathbf{l}$ are three-momentum transfers $\mathbf{k}=\mathbf{p_2}-\mathbf{p_4}$ and $\mathbf{l}=\mathbf{p_2}-\mathbf{p_3}$.  The symmetry factor for these diagrams is $S=1/4$ .  As above, the prefactors of the neutron mass $m_n$ and pion mass $m_{\pi}$ correspond to respective masses in vacuum, since they arise from the definitions of the couplings in the pion-neutron Lagrangian shown above.  The dot product term in the matrix element is often written as $\beta \equiv 3\langle (\mathbf{\hat{k}}\cdot\mathbf{\hat{l}})^2\rangle$, where the brackets denote an average over phase space, which is a common technique to simplify the matrix element \cite{Brinkmann:1988vi}.

For a QCD axion, the axion mass $m_a$ is related to the axion-neutron coupling strength $G_{an}$ through\footnote{This expression comes from Eq.~(3.2) in \cite{Raffelt:2006cw}, where $f_a$ is found in terms of $G_{an}$ by matching the coefficients of the axion-neutron interaction term in the Lagrangian (given in the beginning of this section and in Eq.~(3.6) in Ref.~\cite{Raffelt:2006cw}, taking $C_j \approx 1$.)  }
\begin{equation}
   m_a = 1.2\times 10^7 \text{ eV } \left(\frac{G_{an}}{\text{GeV}^{-1}}\right)
\end{equation}
Given current constraints on the axion-neutron coupling, the mass of the QCD axion must be well below 1 eV, which is much less than the typical momentum scales of order 100 MeV in neutron stars, thus we treat all ALPs as ultrarelativistic particles in our calculations.

We model the nuclear matter inside a neutron star with the $NL\rho$ EoS, which we discussed in Sec.~\ref{sec:nucl_matter_EoSs}.  The formalism for calculating the rate of a particle process in such a relativistic mean field theory is detailed in \cite{Fu:2008zzg}, which uses parameter set I of the model in \cite{Liu:2001iz}.  In the mean free path and emissivity calculations, $E^* \equiv \sqrt{p^2+m_*^2}$ should be used for the energies in the matrix element and in the energy factors in the denominator, while $E = E^* + U_n$ should be used in the energy delta function and the Fermi-Dirac factors \cite{Leinson:2002bw,Roberts:2016mwj,Fu:2008zzg}
\begin{equation}
    f_i = (1+e^{(E_i-\mu_n)/T})^{-1}.\label{eq:FD}
\end{equation}
Note that $E-\mu_n = E^*-\mu_n^*$.

\begin{figure}[h]
\centering
\includegraphics[scale=0.6]{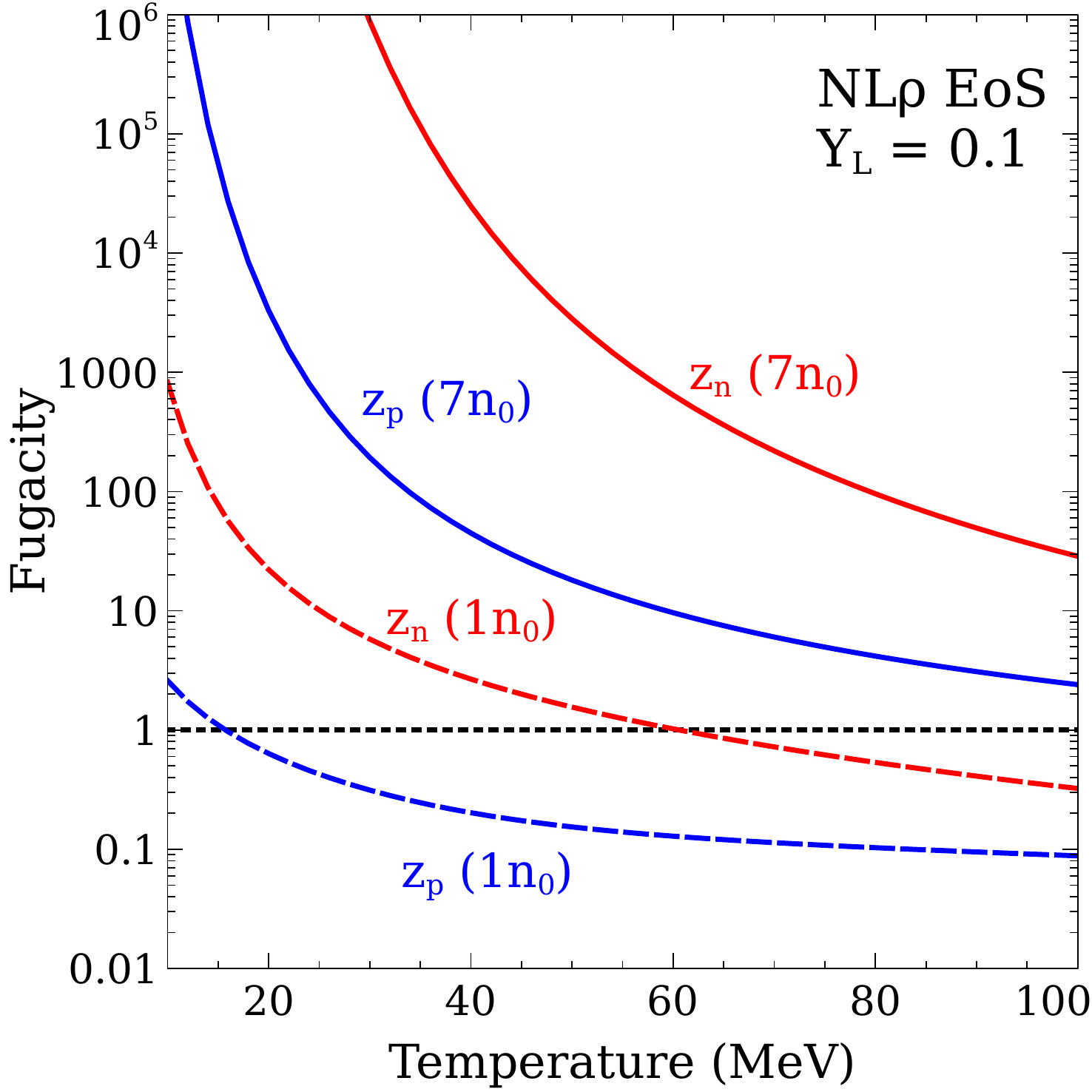}
\caption{Fugacities $z_i = e^{(\mu_i^*-m_{*,i})/T}$ of neutrons and protons in the $NL\rho$ EoS, with $Y_L = 0.1$.  A fugacity much larger than one indicates strongly degenerate particles, while a fugacity much smaller than one indicates non-degenerate particles.}
\label{fig:fugacity}
\end{figure}

In our calculations, we consider a lepton fraction of $Y_l = (n_{\nu}+n_e)/n_B = 0.1$, as this is a typical value for the neutrino-trapped region of a neutron star merger \cite{Alford:2019kdw} (though even this might be an overestimate \cite{Perego:2019adq,Most:2019onn}).  In Fig.~\ref{fig:fugacity}, we show the fugacities $z_i  = e^{(\mu_i^*-m_{*,i})/T}$ of neutrons and protons in this EoS at $Y_L = 0.1$.  A fugacity much larger than one indicates a very degenerate Fermi gas, while a fugacity much smaller than one indicates a highly non-degenerate Fermi gas \cite{Horowitz:2005zv,Fore:2019wib}.  Fig.~\ref{fig:fugacity} indicates that at
nuclear saturation density $n_0$, the protons are nondegenerate for nearly all considered temperatures, while the neutrons transition from degenerate to nondegenerate as the temperature goes above 50-60 MeV.  At $7n_0$, both types of nucleons are degenerate for all considered temperatures.

\section{Axion mean free path}
\label{sec:axion_mfp_intro}
The mean free path of an axion through nuclear matter depends on the temperature and density of the nuclear matter, but also the axion energy.  Often, we will consider axions with energy $\omega \approx 3T$, called ``thermal axions'', because this is the average energy of axions emitted via $n+n\rightarrow n+n+a$ from a fluid element of temperature $T$ \cite{Ishizuka:1989ts}.  In this section, we will compute the mean free path of axions, specializing to the case $\omega=3T$, and categorize thermodynamic conditions as either trapping axions or as allowing axions to free-stream.  As we are interested in neutron-star sized systems, we compare the mean free path to the system size, which is 20-30 kilometers in diameter.  If the mean free path of axions is less than 100 meters, we will consider them trapped, and if the mean free path is longer than 1 kilometer, then we will consider them free streaming.  These choices are somewhat arbitrary, and deserve further study.  Also, the intermediate region (mean free paths from 100\,m to 1\,km) is difficult to treat, as axions are neither trapped, forming a Bose sea, nor do they escape cleanly from the nuclear matter.

The mean free path $\lambda$ of an axion with energy $\omega$, due to absorption via $n+n+a\rightarrow n + n$, is given by \cite{PhysRevD.42.3297}
\begin{equation}
\lambda^{-1} = \int \frac{\mathop{d^3p}_1}{\left(2\pi\right)^3}\frac{\mathop{d^3p}_2}{\left(2\pi\right)^3}\frac{\mathop{d^3p}_3}{\left(2\pi\right)^3}\frac{\mathop{d^3p}_4}{\left(2\pi\right)^3}\frac{S\sum \vert \mathcal{M}\vert^2}{2^5 E_1^*E_2^*E_3^*E_4^*\omega}\left(2\pi\right)^4\delta^4(p_1+p_2-p_3-p_4+\omega)  f_1f_2(1-f_3)(1-f_4).\label{eq:MFP_integral}
\end{equation}
In the rest of this section, we will describe the results of calculating this MFP in various approximations, leaving the details to the appendix.  
\hiddensubsection{Relativistic, arbitrary degeneracy}
\label{sec:axion_mfp_rel}
While at nuclear saturation density the nucleon effective mass is about 3/4 of its vacuum value \cite{glendenning2000compact}, at high baryon densities the nucleon effective mass decreases to only a few hundred MeV, which is comparable to or even lower than the typical momentum values of the nucleons participating in bremsstrahlung.  For example, the NL$\rho$ EoS predicts a neutron Dirac effective mass of 228\,MeV at a density of $7n_0$, where the neutron Fermi momentum is 604\,MeV.  Thus, it is important to use the full dispersion relation [Eq.~(\ref{eq:En})] for the neutrons in the MFP calculation at high densities.  We are able to do the phase space integration while using the relativistic neutron dispersion relation, provided we assume the matrix element is momentum-independent (a common, though not always necessary, approximation assumed in the literature by \cite{Brinkmann:1988vi,PhysRevD.42.3297,Paul:2018msp,Lee:2018lcj}).  The matrix element becomes independent of momentum if we assume that the momentum transfer magnitudes $\mathbf{k}^2$ and $\mathbf{l}^2$ have some typical value $k_{\text{typ}}$.  Then the matrix element can be written as 
\begin{equation}
    S\sum_{\text{spins}}\vert\mathcal{M}\vert^2 \approx 256\frac{f^4m_n^4G_{an}^2}{m_{\pi}^4}\left(1-\frac{\beta}{3}\right)\left(1+\frac{m_{\pi}^2}{k_{\text{typ}}^2}\right)^{-2}.\label{eq:cst_matrix}
\end{equation}
The typical values of momentum transfer are $k_{\text{typ}} \sim \sqrt{3m_*T}$ in the non-degenerate regime and $k_{\text{typ}}\sim p_{Fn}$ in the degenerate regime.  For temperatures and densities where the $NL\rho$ EoS predicts degenerate neutrons, $k_{\text{typ}}\sim p_{Fn}$ takes values between 320-600 MeV, while where the EoS predicts non-degenerate neutrons, $k_{\text{typ}}\sim\sqrt{3m_*T}$ takes values between 375-470 MeV.  Thus the factor of $\left(1+m_{\pi}^2/k_{\text{typ}}^2\right)^{-2}$ ranges from 0.78 to 0.91.  In the degenerate regime, $\beta = 0$, while in the non-degenerate regime, $\beta \approx 1.0845$ \cite{Brinkmann:1988vi}.

The momentum-independent matrix element can be pulled out of the phase space integral, and now the integral can be reduced to a 6-dimensional integral to be done numerically.  In addition to using the relativistic dispersion relation for neutrons, our calculation is also novel in that it keeps the axion three-momentum in the momentum-conserving delta function.  Finally, we emphasize that this approach to the mean free path integral is valid for arbitrary neutron degeneracy, as we make no simplifications to the Fermi-Dirac factors.  Our final expression for the axion mean free path in the constant-matrix-element approximation is given in Eq.~(\ref{eq:MFPexactanswer}), and the details of the calculation are given in Appendix \ref{sec:rel_PS_MFP}.

\begin{figure}[h]
\centering
\includegraphics[scale=0.6]{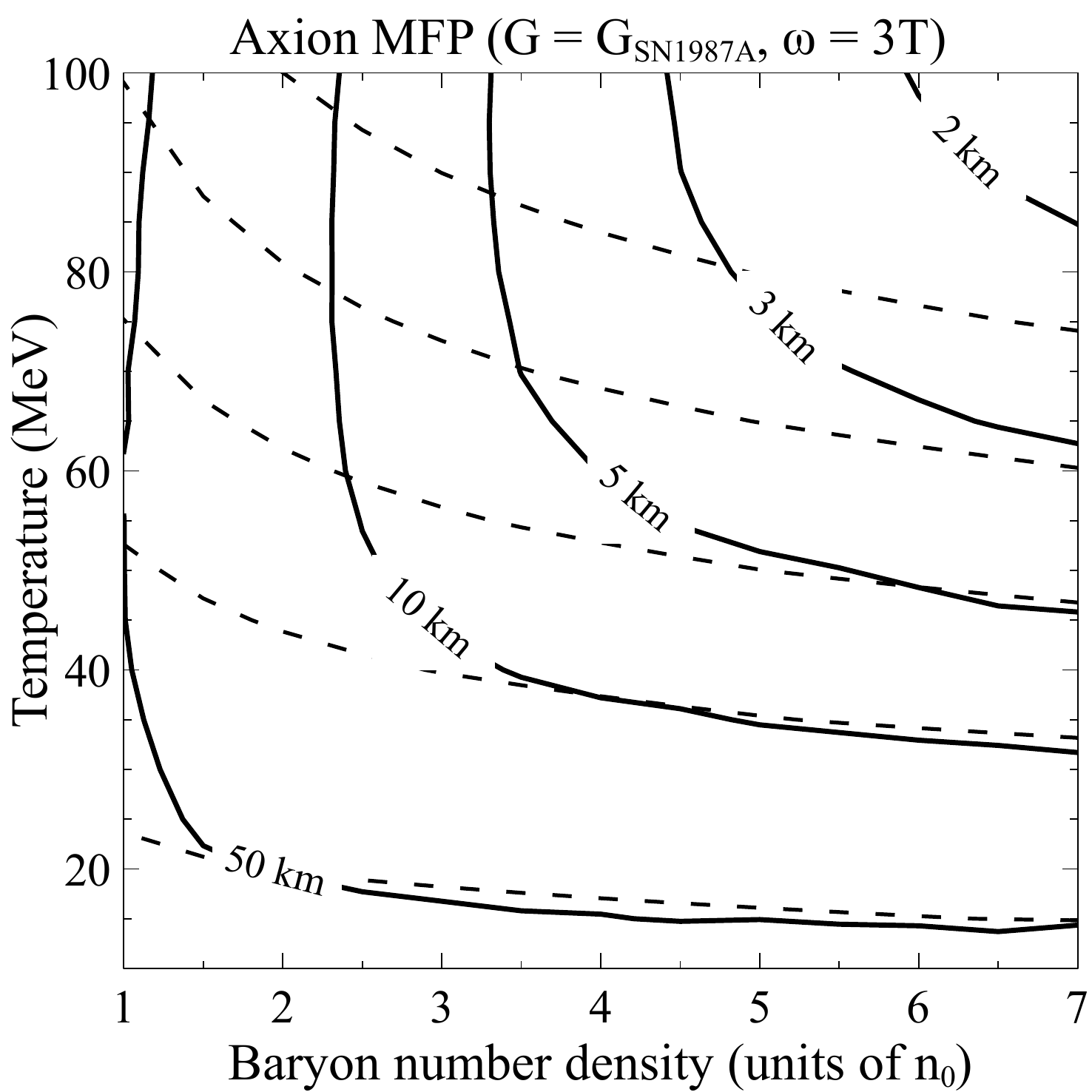}
\caption{Axion mean free path (for axions with energy $\omega = 3T$) due to absorption via $n+n+a\rightarrow n+n$ with an assumed axion-neutron coupling constant equal to the upper bound set by SN1987A.  The dashed contours correspond to the axion MFP calculated in the Fermi surface approximation, while the solid contours use the constant matrix element approximation of the axion mean free path.}
\label{fig:axionMFP}
\end{figure}

In Fig.~\ref{fig:axionMFP} we show a contour plot of the axion mean free path [Eq.~(\ref{eq:MFPexactanswer})].  We have assumed the axion energy $\omega = 3T$, and so Fig.~\ref{fig:axionMFP} does not depict the MFP of a fixed-energy axion, but of axions with progressively higher energies as the temperature increases.  The solid contours are the result of our constant-matrix-element approximation of the axion mean free path, where we have chosen $\beta=0$ for convenience.  The MFP is inversely proportional to the square of the unknown axion-neutron coupling constant, so we have chosen that coupling to be equal to the upper bound set by SN1987A.  Thus Fig.~\ref{fig:axionMFP} represents the smallest MFP allowed by SN1987A.  First, we see that the axion MFP is longer than 1 km for all thermodynamic conditions, indicating that axions will free-stream from the neutron star merger in which they are created.  Second, we see that the axion mean free path shrinks as matter becomes hotter and denser.  We see that at large neutron degeneracy (high density, low temperature), the mean free path of a thermal axion ($\omega=3T$) becomes relatively independent of density.

In Appendix \ref{sec:MFP_nonrel_PS}, we also present the phase space integral of the MFP while assuming non-relativistic neutrons.  In this case, the momentum dependence of the matrix element can be retained, and it is left inside the integral.  This calculation has been done before in the literature, but we present a version of the calculation well adapted to a relativistic mean field theory.
\hiddensubsection{Fermi surface approximation (degenerate neutrons)}
\label{sec:axion_mfp_FS}
The most common approximation of the full mean free path integral Eq.~(\ref{eq:MFP_integral}) is to assume the neutrons are strongly degenerate.  As can be seen in Fig.~\ref{fig:fugacity}, this assumption is valid at high densities like $7n_0$ for all temperatures encountered in neutron star mergers, but also even at lower densities like $n_0$, provided the temperature is below about 50 MeV.  We call this approximation the ``Fermi surface approximation'' because in degenerate nuclear matter only the particles near the Fermi surface can participate in any reactions.  The concept of the Fermi surface approximation is discussed in detail in \cite{Alford:2018lhf} and in Ch.~\ref{sec:beta_equilibrium} in the context of the Urca process.

The mean free path of an axion with energy $\omega$ due to (inverse) axion bremsstrahlung has been calculated in the Fermi surface approximation in \cite{Ishizuka:1989ts,Iwamoto:1992jp}, where they find 
\begin{equation}
\lambda_{FS}^{-1} = \frac{1}{18\pi^5}\frac{f^4G_{an}^2m_n^4}{m_{\pi}^4}p_{Fn}F(y)\frac{\omega^2+4\pi^2T^2}{1-e^{-\omega/T}},\label{eq:lambdaFS}
\end{equation}
where
\begin{equation}
F(y) = 4 - \frac{1}{1+y^2} + \frac{2y^2}{\sqrt{1+2y^2}}\arctan{\left(\frac{1}{\sqrt{1+2y^2}}\right)}-5y\arcsin{\left(\frac{1}{\sqrt{1+y^2}}\right)},
\label{eq:Fofy}
\end{equation}
with $y=m_{\pi}/(2p_{Fn})$.  The derivation of this formula is sketched in Appendix \ref{sec:mfp_FS_calculation}.  The axion MFP in the Fermi surface approximation is plotted in dotted lines in Fig.~\ref{fig:axionMFP}.  We see from Fig.~\ref{fig:axionMFP} that in conditions that are not strongly degenerate, the Fermi surface approximation significantly underestimates the mean free path compared to the constant-matrix-element approximation.

The virtue of the Fermi surface approximation is that it allows the 15 dimensional phase space integral to be done analytically.  However, during the course of the calculation, the lower endpoint of the integration over neutron energy (which comes from converting the phase space integral into spherical coordinates and then turning the momentum magnitude integral to an integral over energy - see Appendix \ref{sec:mfp_FS_calculation}) is extended to minus infinity.  Extending the integration bounds this way is valid in degenerate nuclear matter, because it adds only an exponentially small term \cite{Shapiro:1983du}.  However, as temperature increases, the extension of the integral gives rise to a sizeable error.  We propose here an improved Fermi surface approximation calculation which keeps the energy bounded by $U_n + m_* < E_n < \infty$.  From there it is possible to write the axion mean free path as a one dimensional integral
\begin{equation}
    \lambda^{-1}=\frac{4}{3\pi^5}\frac{f^4G_{an}^2m_n^4}{\omega m_{\pi}^4}p_{Fn}F(y)T^3K_1(\hat{y},\omega/T),\label{eq:lambda_new}
\end{equation}
where 
\begin{align}
    &K_1(\hat{y},\omega/T) \equiv \int_{-2\hat{y}}^{\infty}\mathop{du} \frac{1}{(1-e^u)(1-e^{-u-\omega/T})}\label{eq:K1}\\
    &\times\ln{\left\{\frac{\cosh{\left[(u+\hat{y}+\omega/T)/2\right]}}{\cosh{(\hat{y}/2)}}\right\}}\ln{\left\{\frac{\cosh{(\hat{y}/2)}}{\cosh{\left[(u+\hat{y})/2\right]}}\right\}},\nonumber
\end{align} and
$\hat{y}=(\mu_n^*-m_*)/T$.  The degeneracy parameter $\hat{y}$ is just the logarithm of the neutron fugacity $\hat{y}=\ln{z_n}$.  This expression for the mean free path follows the Fermi surface approximation, but better treats the lower endpoint of integration over the neutron energies, where the traditional treatment \cite{Shapiro:1983du} becomes increasingly poor.  The details are further explained in Appendix \ref{sec:mfp_FS_calculation}.  Our expression Eq.~(\ref{eq:lambda_new}) of course matches the Fermi surface approximation (\ref{eq:lambdaFS}) in the degenerate limit $\hat{y}\rightarrow \infty$.  

We emphasize that this proposed expression improves the behavior of the FS approximation in semi-degenerate conditions, but is definitely not valid for non-degenerate conditions.  After all, this approximation still assumes that only neutrons on their Fermi surface participate in the process.
 
\hiddensubsection{MFP dependence on the axion-neutron coupling}
\label{sec:axion_MFP_vs_G}
\begin{figure*}[t!]
\begin{minipage}[t]{0.5\linewidth}
\includegraphics[width=.95\linewidth]{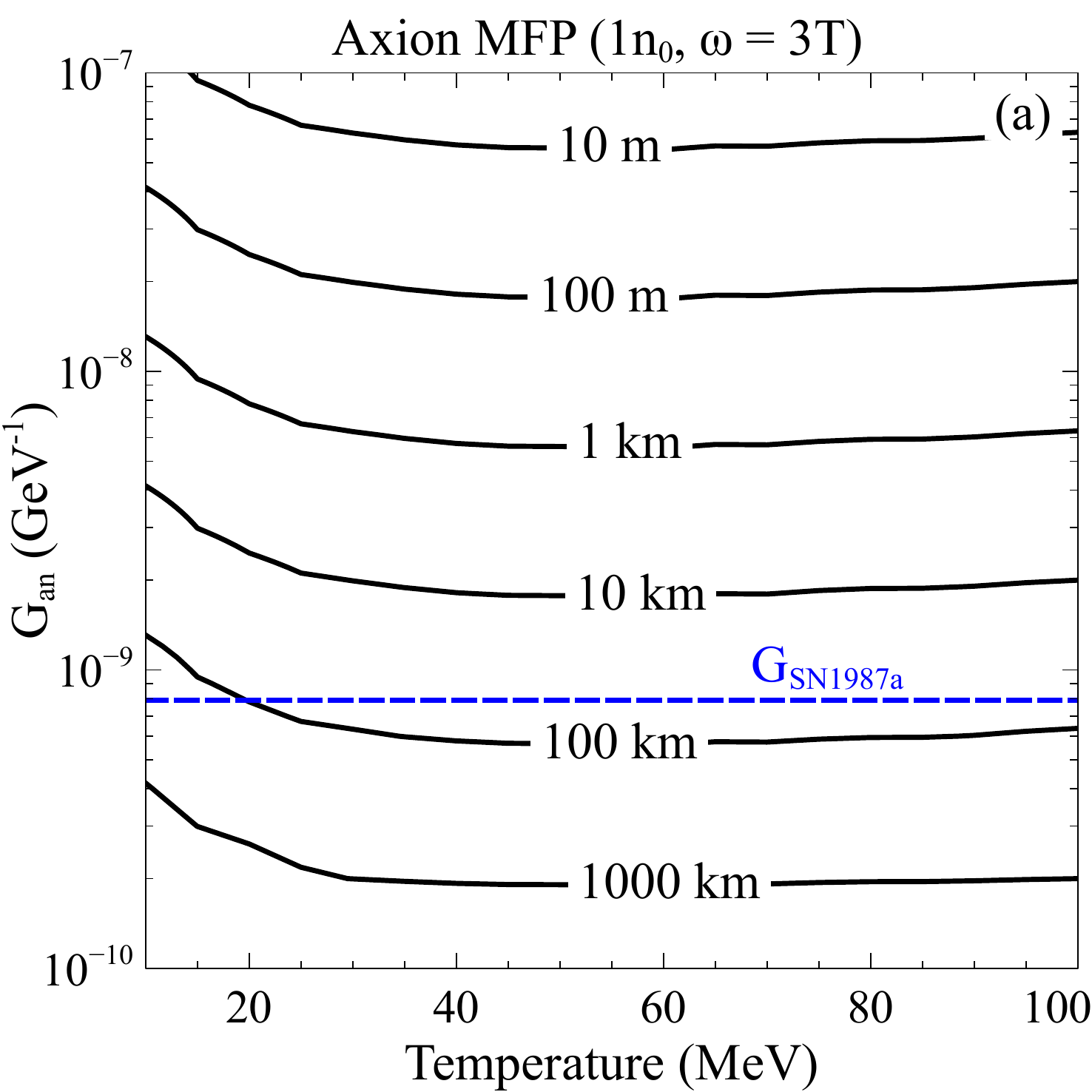}
\end{minipage}\hfill%
\begin{minipage}[t]{0.5\linewidth}
\includegraphics[width=.95\linewidth]{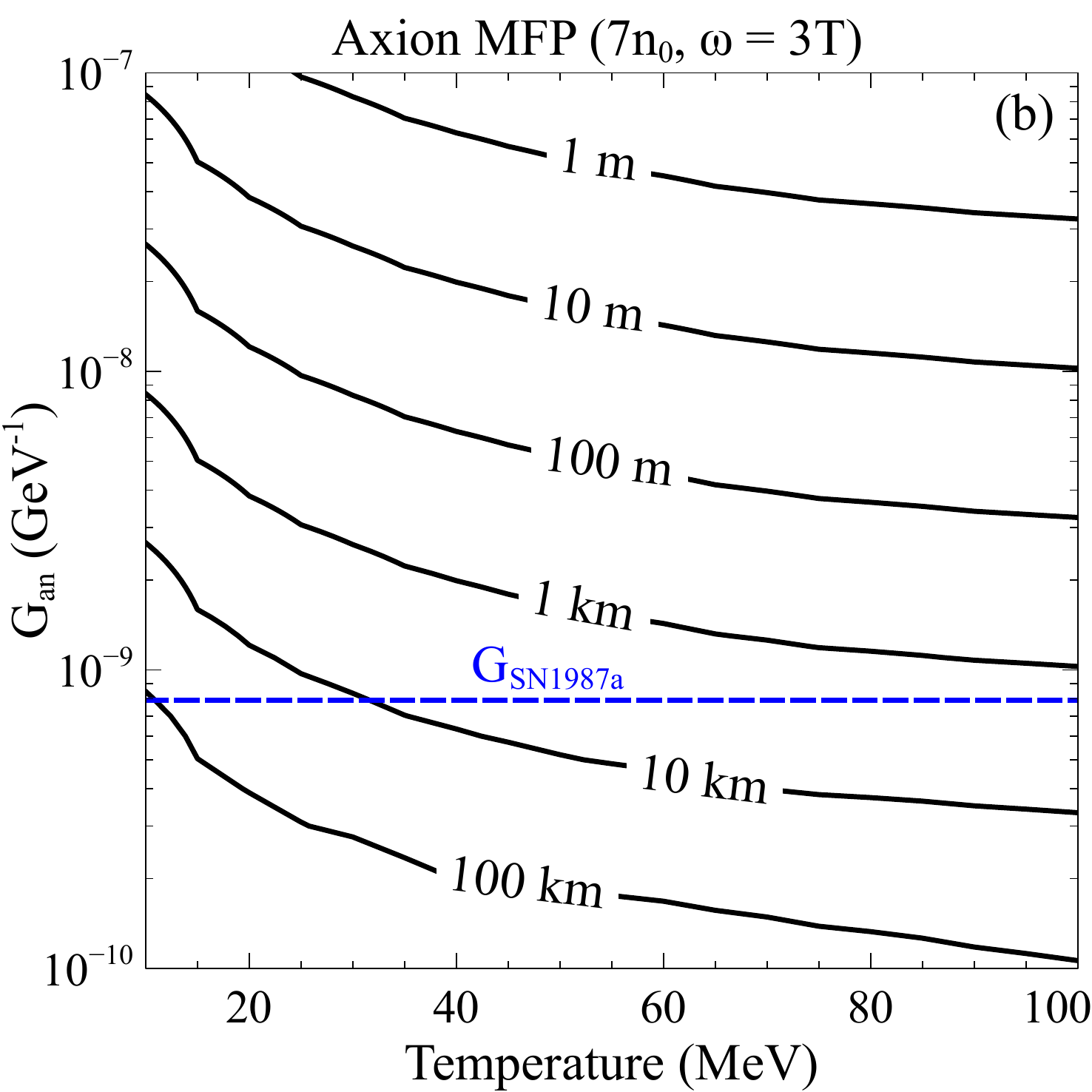}
\end{minipage}%
\caption{Axion mean free path (for axions with energy $\omega = 3T$) via $n+n+a\rightarrow n+n$ for densities of $1n_0$ (a) and $7n_0$ (b).  All couplings larger than the dotted blue line may be disallowed by the observation of SN1987A.  Thus, axions likely free-stream through neutron star mergers as their mean free path is above several kilometers regardless of the density.}
\label{fig:axionMFP_vs_T}
\end{figure*}

It is clear that if the axion-neutron coupling is less than or equal to the maximum value SN1987A will allow, all thermodynamic conditions encountered in mergers will fail to trap axions.  However, if the axion-neutron coupling was larger, the hotter, denser regions of the parameter space depicted in Fig.~\ref{fig:axionMFP} would begin to trap axions.  We depict this observation in Fig.~\ref{fig:axionMFP_vs_T}, using the constant-matrix-element approximation for the axion mean free path (and setting $\beta=0$ in the constant matrix element, for simplicity.)  We see that if the SN1987A bound is robust, for allowed values of the coupling, the axion mean free path is always comparable to or much bigger than the size of a neutron star and thus axions would free-stream from the star or merger.  However, if the SN1987A bounds are flexible due to uncertainties in modeling supernovae as well as uncertainties in the nuclear EoS (as is suggested by \cite{Bar:2019ifz,Fischer:2016cyd}, see also \cite{Giannotti:2017hny} for useful commentary), in the highest density and temperature conditions possibly reached in mergers, axions would be trapped for couplings just 4 times the SN1987 bound.  However, we note that there are many other independent proposed upper bounds on the axion-neutron coupling, coming from neutron star observations \cite{Berenji:2016jji,Sedrakian:2015krq,Paul:2018msp,Beznogov:2018fda,Lloyd:2019rxg}, all within about an order of magnitude in either direction of the SN1987A bound.
\section{Axion-transparent matter}
\label{sec:axion_transp_matter}
In the axion-transparent regime where the axion mean free path is comparable to or larger than the system size, an axion created inside the merger by neutron bremsstrahlung would escape, cooling down the merger.  This is analogous to neutrino cooling \cite{Yakovlev:2004iq,Potekhin:2015qsa}, which occurs in nuclear matter at temperatures below 5 or 10 MeV, which is the regime where it is transparent to neutrinos \cite{Alford:2018lhf,Roberts:2016mwj,Haensel:1987zz,1979ApJ...230..859S,Sawyer:1975js}.  We study regions of the merger above temperatures of 10 MeV, where neutrinos are trapped (and thus only cool via the relatively slow diffusion process \cite{1982ApJ...253..816G}) but axion emission could serve as an unexpected cooling mechanism.

We calculate the temperature of a fluid element radiating axions as a function of time, and in particular, find the characteristic cooling time at which the temperature has halved.  We can write a differential equation for the temperature $T$ as a function of time
\begin{equation}
    \frac{\mathop{dT}}{\mathop{dt}} = -\frac{Q}{c_V},\label{eq:dTdt}
\end{equation}
where $Q=\mathop{d\varepsilon}/\mathop{dt}$ is the axion emissivity (the energy emitted in axions per volume per time) and $c_V = \mathop{d\varepsilon}/\mathop{dT}$ is the specific heat of the nuclear matter per unit volume.  The specific heat is dominated by neutrons, as they have the largest particle fraction in the neutron star and thus the most possible low-energy excitations \cite{Alford:2017rxf}.  The specific heat\footnote{Technically, we should not use the specific heat for a degenerate Fermi gas as we do here, but instead for a Fermi gas at arbitrary degeneracy.  However, for the thermodynamic conditions encountered here, these differ by at most 14\%, the greatest difference occurring at large temperature.  Extending the traditional calculation \cite{1994ARep...38..247L} to the case of neutrons described by a relativistic mean field theory, the specific heat of a neutron gas of arbitrary degeneracy is $c_V = 2\int\frac{\mathop{d^3p}}{(2\pi)^3}(E^*-\mu_n^*)\frac{\mathop{d(f(E^*-\mu_n^*))}}{\mathop{dT}} = \frac{T^3}{4\pi^2}\int_{-\hat{y}_n}^{\infty}\mathop{dx}\frac{x^2(x+(\mu_n^*/T))\sqrt{(x+(\mu_n^*/T))^2-(m_*/T)^2}}{\cosh{(x/2)}^2}$, where $\hat{y}_n\equiv (\mu_n^*-m_*)/T$ and we have assumed that $m_*$ and $\mu_n^*$ do not depend on temperature.} is given by \cite{1994ARep...38..247L}
\begin{equation}
    c_V \approx \frac{1}{3}m_L p_{Fn}T,\label{eq:cV_fermion}
\end{equation}
where $m_L$ is the Landau effective mass of the neutron, which is related to its density of states at the Fermi surface \cite{Maslov:2015wba,Li:2018lpy}
\begin{equation}
    m_L = \frac{p_{Fn}}{(\mathop{dE}/\mathop{dp})\vert_{p_{Fn}}}=\sqrt{p_{Fn}^2+m_*^2}.\label{eq:Landau}
\end{equation}
The axion emissivity is given by the phase space integral \cite{Brinkmann:1988vi}
\begin{align}
&Q = \int \frac{\mathop{d^3p}_1}{\left(2\pi\right)^3}\frac{\mathop{d^3p}_2}{\left(2\pi\right)^3}\frac{\mathop{d^3p}_3}{\left(2\pi\right)^3}\frac{\mathop{d^3p}_4}{\left(2\pi\right)^3}\frac{\mathop{d^3\omega}}{\left(2\pi\right)^3}\frac{S\sum \vert \mathcal{M}\vert^2}{2^5 E_1^*E_2^*E_3^*E_4^*\omega}\omega\label{eq:emissivity_integral}\\
&\times\left(2\pi\right)^4\delta^4(p_1+p_2-p_3-p_4-\omega)f_1f_2\left(1-f_3\right)\left(1-f_4\right).\nonumber
\end{align}
In the rest of this section, we will discuss approximations of this axion emissivity phase space integral (just as we did for the axion MFP) and then we use our results to calculate the cooling time due to axion emission.
\hiddensubsection{Relativistic, arbitrary degeneracy}
\label{sec:axion_Q_rel}
Like the calculation of the axion MFP, the axion emissivity involves an integration over phase space and we will make the same set of approximations we made for the MFP in Sec.~\ref{sec:axion_mfp_rel}.  Thus, assuming a momentum-independent matrix element Eq.~(\ref{eq:cst_matrix}) and using relativistic dispersion relation Eq.~(\ref{eq:En}) for the neutrons, we do the phase space integral [Eq.~(\ref{eq:emissivity_integral})] and find that the axion emissivity can be reduced to a six-dimensional integral to be done numerically.  The expression [Eq.~(\ref{eq:exact_emissivity})] and its derivation are given in Appendix \ref{sec:Q_rel_PS}.  That expression is valid for neutrons of arbitrary degeneracy.

In Appendix \ref{sec:Q_nonrel_PS}, we also present the phase space integral of the axion emissivity while assuming non-relativistic neutrons.  In this case, the momentum-dependence of the matrix element can be retained, and it is left inside the integral.  This calculation has been done before in the literature, but, as with the axion MFP integration, we present a version of the calculation well-adapted to a relativistic mean field theory.
\hiddensubsection{Fermi surface approximation (degenerate neutrons)}
\label{sec:axion_Q_FS}
As in Sec.~\ref{sec:axion_mfp_FS}, the Fermi surface approximation can be applied to the axion emissivity if the neutrons are strongly degenerate, as in that case only neutrons near the Fermi surface will participate in the bremsstrahlung process.  The calculation of the axion emissivity in this regime was done first by Iwamoto \cite{PhysRevLett.53.1198}, and extended by \cite{Iwamoto:1992jp, Stoica:2009zh}.  The Fermi surface approximation for the axion emissivity is
\begin{equation}
    Q_{FS} = \frac{31}{2835\pi}\frac{f^4 G_{an}^2m_n^4}{m_{\pi}^4}p_{Fn}F(y)T^6,\label{eq:Q_FS}
\end{equation}
where $F(y)$ is given in Eq.~(\ref{eq:Fofy}).  The derivation of this formula is sketched in Appendix \ref{sec:Q_FS_derivation}.  

As with the mean free path, discussed in Sec.~\ref{sec:axion_mfp_FS}, the Fermi surface approximation of the emissivity extends the lower endpoint of integration of neutron energy down to $-\infty$.  We propose an improvement to the FS approximation which keeps the neutron energy bounded by $m_*+U_n<E_n<\infty$, at the cost of having an emissivity expression in terms of a two-dimensional integral instead of an analytic expression like Eq.~(\ref{eq:Q_FS}).  The axion emissivity in the improved FS approximation is
\begin{equation}
    Q = \frac{2}{3\pi^7}\frac{f^4G_{an}^2m_n^4}{m_{\pi}^4}p_{Fn}F(y)T^6K_2(\hat{y})\label{eq:Q_new}
\end{equation}
where
\begin{align}
    K_2(\hat{y}) &= \int_{-2\hat{y}}^{\infty}\mathop{du}\frac{1}{1-e^u}\ln{\left\{\frac{\cosh{(\hat{y}/2)}}{\cosh{\left[(u+\hat{y})/2\right]}}\right\}}\label{K2}\\
    &\times \int_0^{u+2\hat{y}}\mathop{dw}\frac{w^2}{1-e^{w-u}}\ln{\left\{\frac{\cosh{\left[(u+\hat{y}-w)/2\right]}}{\cosh{(\hat{y}/2)}}\right\}}.\nonumber
\end{align}
\hiddensubsection{Radiative cooling time dependence on axion-neutron coupling}
\label{sec:axion_cooling}

In the Fermi surface approximation, the differential equation (\ref{eq:dTdt}) for $T(t)$ can be solved exactly (assuming the neutron Landau effective mass does not depend on temperature, which is a reasonable approximation) and we find a fluid element that starts at temperature $T_0$ cools according to
\begin{equation}
    T(t)^{-4}=T_0^{-4}+\frac{124}{945\pi} \frac{f^4G_{an}^2m_n^4F(y)}{m_Lm_{\pi}^4}t,
\end{equation}
and thus has cooling time (to reach half of its initial temperature)
\begin{equation}
    \tau_{FS,1/2} \approx 12 \text{ s }\times \frac{\left(\dfrac{m_L}{0.8m_n}\right)}{\left(\dfrac{G_{an}}{G_{SN1987A}}\right)^{\!2}F(y)\left(\dfrac{T_0}{10\text{ MeV }}\right)^{\!4}} .\label{eq:t_cool_FS}
\end{equation}

\begin{figure}[h]
\centering
\includegraphics[scale=0.6]{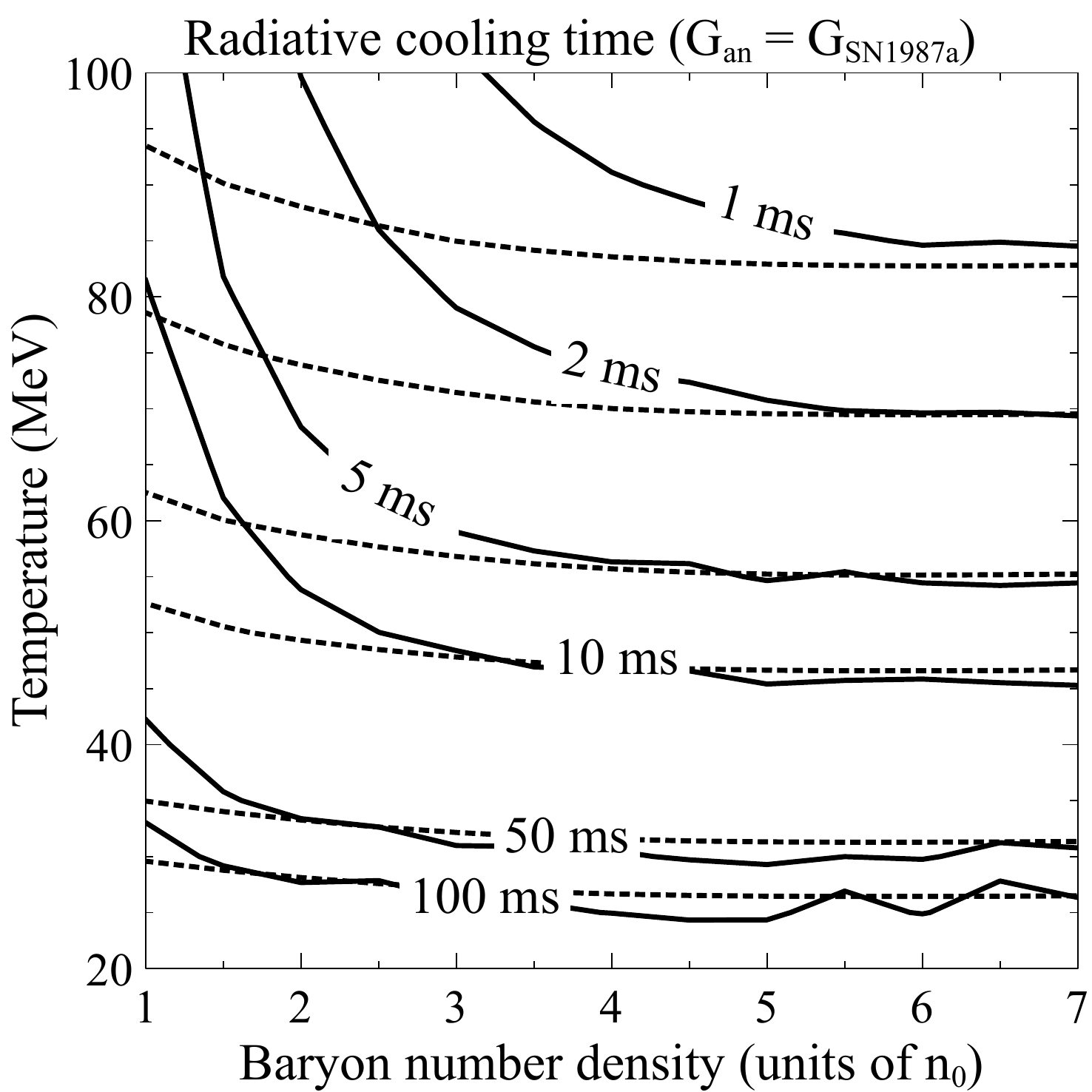}
\caption{Cooling time of nuclear matter due to radiating axions.  Solid lines use the constant-matrix-element approximation of the emissivity while dotted lines use the FS approximation.  The axion-neutron coupling is chosen to be the bound set by SN1987A.}
\label{fig:radiative_cooling_dens_temp}
\end{figure}

While there is no analytic result for the characteristic cooling time with our new expression for the emissivity Eq.~(\ref{eq:exact_emissivity}), we can solve the differential equation numerically and plot the characteristic cooling time as a function of density and temperature for a particular choice of axion-neutron coupling constant.  In Fig.~\ref{fig:radiative_cooling_dens_temp}, we plot the radiative cooling time due to axion emission, choosing the coupling constant to be the maximum value allowed by SN1987A.  The solid contours use the constant-matrix-element phase space integral for the emissivity (with $\beta=0$), while the dotted contours use the FS approximation.  We see that hotter and denser regions cool faster, because they emit axions at a higher rate.  The solid and dashed contours agree where the nuclear matter is strongly degenerate, which occurs at high density and low temperature.  This plot indicates that within the constraints set by SN1987A, axions can cool fluid elements in timescales relevant to neutron star mergers.  The treatment of the axion emissivity via the full phase space integration limits (compared to the FS approximation) the range of densities for which fast cooling can occur.  In particular, hot nuclear matter near saturation density has a significantly longer cooling time than predicted by the FS approximation.

\begin{figure*}[t!]
\begin{minipage}[t]{0.5\linewidth}
\includegraphics[width=.95\linewidth]{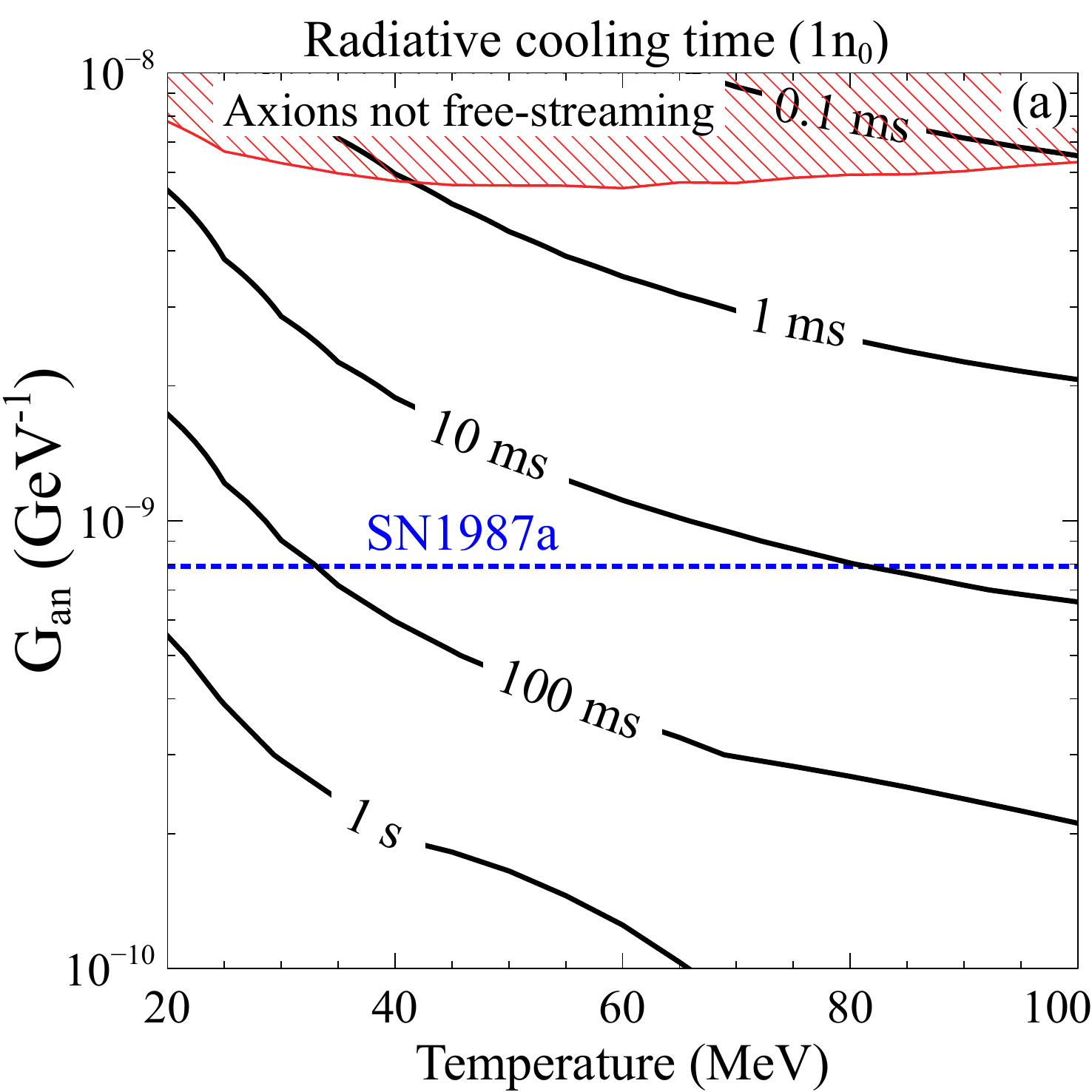}
\end{minipage}\hfill%
\begin{minipage}[t]{0.5\linewidth}
\includegraphics[width=.95\linewidth]{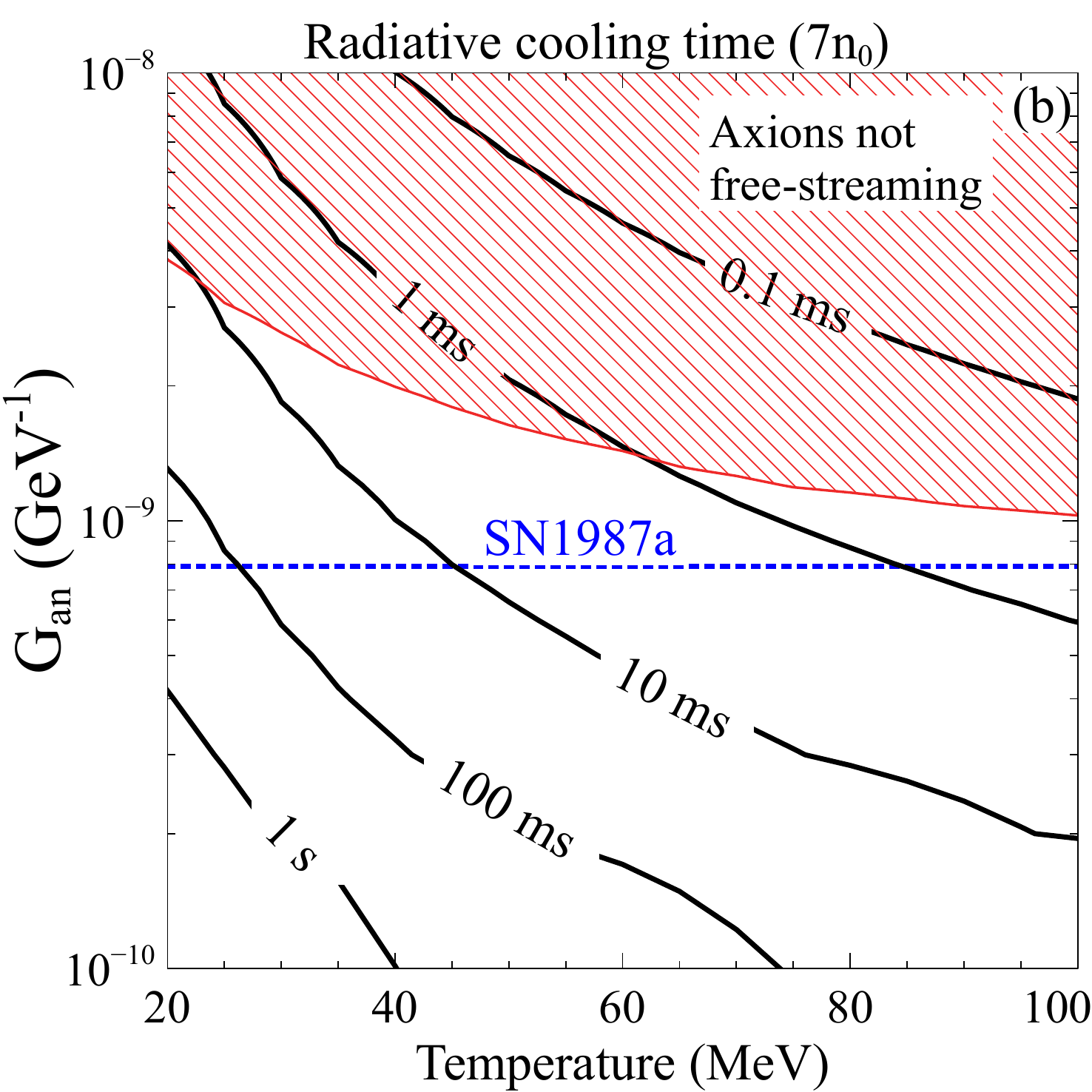}
\end{minipage}%
\caption{Radiative cooling time due to axion emission at densities of $1n_0$ (a) and $7n_0$ (b).  All couplings stronger than the dotted blue line are disallowed by the observation of SN1987A.  Considering couplings compatible with SN1987A, lower density regions cool somewhat slowly compared to merger timescales, but higher density regions could cool on timescales relevant for mergers.}
\label{fig:rad_cool_time_vs_G}
\end{figure*}
In Fig.~\ref{fig:rad_cool_time_vs_G}, we plot the radiative cooling time at two different densities, as a function of temperature and axion-neutron coupling. Radiative cooling is only relevant for thermodynamic conditions where the axions have a mean free path longer than a few kilometers.  At $1n_0$, it is difficult to get substantial cooling on merger timescales, but at high densities ($7n_0$), cooling times can be under 10 ms, and definitely under 100 ms for a wide range of temperatures.  
\section{Axion-trapped matter}
\label{sec:axion_trapped}
Based on our results in Fig.~\ref{fig:axionMFP}, we do not expect axions to be trapped in any part of a neutron star merger.  However, we present the following analysis of trapped axions for completeness, but also as an example of the contribution to thermal equilibration of the interior of a neutron star due to a boson that interacts with neutrons. 

If the mean free path of axions is much less than the system size, then the axions form a Bose gas inside the neutron star merger.  In this situation, the axions could transport energy around the star, smoothing out temperature gradients, much like neutrinos do when they are trapped.  We calculate the timescale of thermal equilibration for a fluid element to transfer heat to its neighboring fluid elements.  
From \cite{Alford:2017rxf}, a hot spot of volume $z^3$ in the merger has extra thermal energy (compared to its neighbors) $E_{th}\approx(\pi/6)c_V z^3 \Delta T$, and it conducts that energy away through its boundaries at rate $W_{th} = \pi \kappa \Delta T z$.  Thus, the timescale for heat conduction is
\begin{equation}
    \tau_{\kappa} = E_{th}/W_{th} = \frac{c_V z^2}{6\kappa}.\label{eq:t_conduct_general}
\end{equation}
As discussed in Sec.~\ref{sec:axion_transp_matter}, the specific heat $c_V$ is dominated by the neutrons, which have specific heat $c_V = (1/3)m_L p_{Fn} T$.  The thermal conductivity is the sum of the contributions $\kappa_i = (1/3)c_{V_i}v_i\lambda_i$ from each particle species.  The particles with both high density and long mean free path (but still less than the system size) will dominate the thermal conductivity.  We consider here only stars with temperatures above 10 MeV, which mean that neutrinos will be trapped and would traditionally dominate the thermal conductivity \cite{Alford:2017rxf}.  However, if the star also traps axions, then axions could take over the role of energy transportation.

The neutrinos have \cite{1982ApJ...253..816G} thermal conductivity $\kappa_{\nu} \approx n_{\nu}^{2/3}/(3G_F^2m_L^2n_e^{1/3}T)$, which implies that neutrinos re-establish thermal equilibrium between nearby fluid elements in time \cite{Alford:2017rxf}
\begin{equation}
    \tau_{\nu} = 700 \text{ ms}\left(\frac{0.1}{x_p}\right)^{1/3}\left(\frac{m_L}{0.8m_n}\right)^3\label{eq:conduct_nu}\left(\frac{\mu_e}{2\mu_{\nu}}\right)^2\left(\frac{z}{1 \text{ km}}\right)^2\left(\frac{T}{10 \text{ MeV}}\right)^2.\nonumber
\end{equation}
When two species contribute to thermal equilibration, their individual timescales add according to $\tau_{\kappa}^{-1} = \tau_{\kappa_a}^{-1}+\tau_{\kappa_{\nu}}^{-1}$.  As we will see, the equilibration timescale due to axions is much shorter (for the range of couplings we consider) than the timescale due to neutrinos, and so for the rest of this discussion we assume axions are the only species contribution to thermal equilibration and thus
\begin{equation}
    \tau_{\kappa} \approx \frac{c_{V_n} z^2}{6\kappa_a}.
\end{equation}

As the axions, when trapped, are a free Bose gas with zero chemical potential (they are equilibrated by the reaction $n+n+a\leftrightarrow n+n$), they have energy density $\varepsilon = (\pi^2/30)T^4$ and thus specific heat per unit volume $c_V = (2\pi^2/15)T^3$, and so their thermal conductivity is $\kappa_a = (2\pi^2/45)T^3\lambda_a$.  The axion conduction timescale is
\begin{equation}
    \tau_{a} = \frac{5}{4\pi^2}\frac{m_Lp_{Fn}z^2}{T^2\lambda},
    \label{eq:conductive_time_general}
\end{equation}
which in the Fermi surface approximation for the MFP, reduces to
\begin{equation}
    \tau_{\kappa,a,FS} \approx 2.757 f^4G_{an}^2m_Lp_{Fn}^2F(y)z^2.
    \label{eq:t_conduct_FS}
\end{equation}
Eq.~(\ref{eq:conductive_time_general}) indicates that the longer the axion mean free path, the shorter the timescale for thermal equilibration.  However, if the mean free path is too long, then there is no heat conduction due to axions.

To get an estimate for the thermal equilibration timescale, we consider the situation where a neutron star merger has only gradual temperature gradients, occurring on at least the 1 km scale.  For example, the hot spherical shell (Fig.~\ref{fig:T_distribution}) observed in many simulations \cite{Hanauske:2016gia, Perego:2019adq,Hanauske:2019qgs,Hanauske:2019czl,Hanauske:2019vsz} is 1-2 km thick.  If this is the case, then neutrino conduction has no effect on a neutron star merger, as heat conduction via neutrinos occurs on timescales greater than one second.

\begin{figure*}[t!]
\begin{minipage}[t]{0.5\linewidth}
\includegraphics[width=.95\linewidth]{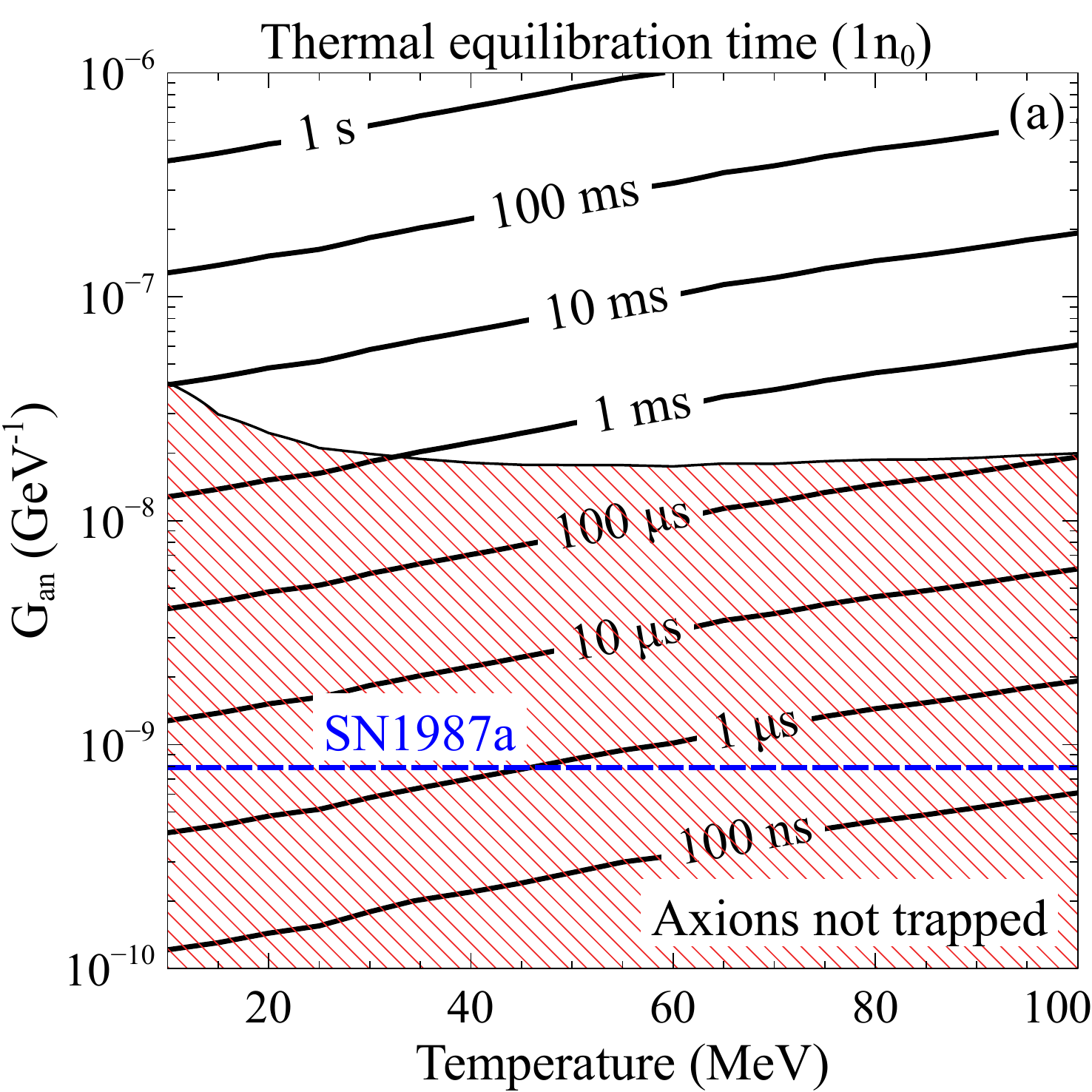}
\end{minipage}\hfill%
\begin{minipage}[t]{0.5\linewidth}
\includegraphics[width=.95\linewidth]{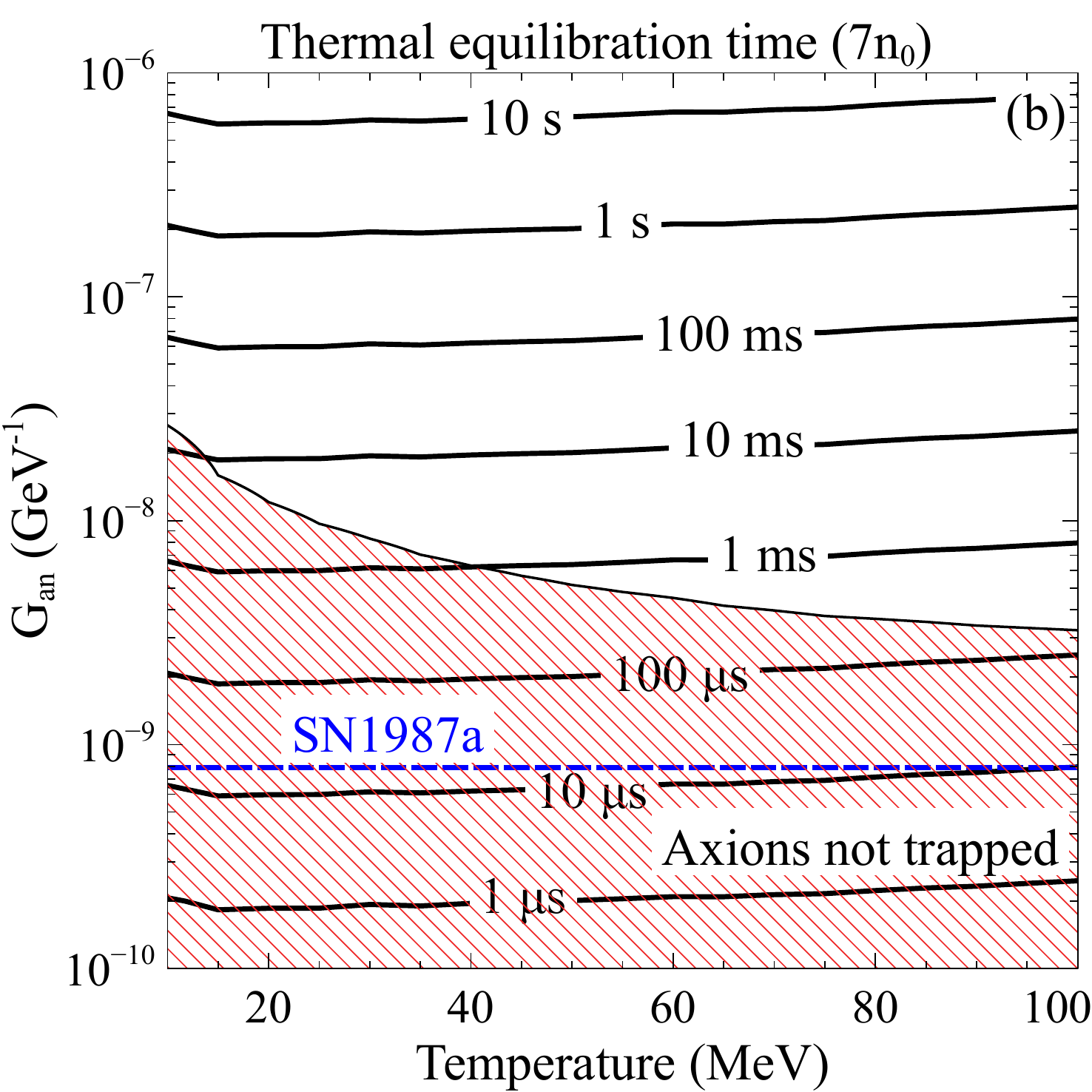}
\end{minipage}%
\caption{Thermal equilibration time due to axion conduction at a density of $1n_0$ (a) and $7n_0$ (b).  All couplings stronger than the dotted blue line are disallowed by the observation of SN1987A.  The red shaded region corresponds to couplings at which axions are not trapped, and so there is no axion conduction between neighboring fluid elements.}
\label{fig:conductivecooling}
\end{figure*}

According to Fig.~\ref{fig:conductivecooling}, at low axion-neutron couplings, the thermal equilibration due to axion conduction would occur very quickly because of the long axion mean free path.  However, the mean free path is too long - bigger than the size of a neutron star - and so neighboring fluid elements are unable to transfer heat to each other via axions.  As the coupling gets strong enough to trap axions, the thermal equilibration time is still fast enough to be relevant.  However, this only occurs at axion coupling constants more than an order of magnitude higher than the limit set by SN1987A, which is why thermal equilibration of neutron star mergers due to axions is unlikely.  
\section{Conclusion}
\label{sec:conclusion}
We have analyzed the impact of axions on neutron stars mergers in the case that they are trapped and in the case that they are free-streaming.  As part of this effort, we calculated the axion mean free path and emissivity due to the neutron bremsstrahlung process $n+n\leftrightarrow n+n+a$.  In contrast to previous calculations, we integrated over the entire phase space while using a relativistic treatment of the neutrons (although assuming the matrix element was momentum-independent).  In particular, we used a relativistic mean field theory to describe the nucleons, which means that we took into account the precipitous decrease in the Dirac effective mass of the nucleons as density increases above nuclear saturation density.

With our calculation of the axion mean free path, we were able to, for a given value of the axion-neutron coupling, divide the thermodynamic parameter space $\{n_B,T\}$ into regions where axions are trapped and where they are free-streaming (and of course, a difficult-to-treat region in between where they are neither).  We find that for axion-neutron couplings allowed by SN1987A, axions have a long mean free path in all thermodynamic conditions encountered in mergers (see Fig.~\ref{fig:axionMFP_vs_T}), and thus we expect them to free-stream from a neutron star merger.

We examine the time it would take for a fluid element of nuclear matter in a merger to cool to one half of its current temperature by radiating axions.  The result is depicted in Figs.~\ref{fig:radiative_cooling_dens_temp} and \ref{fig:rad_cool_time_vs_G}.  For allowed values of the axion-neutron coupling, we find that the hottest fluid elements in a merger could cool in timescales less than 10 ms, which would have an impact on the dynamics of a neutron star merger, in particular, reducing the thermal pressure of the possible remnant \cite{Dietrich:2019shr} and changing the values of temperature-dependent transport properties like bulk viscosity \cite{Alford:2017rxf,Alford:2019kdw, Alford:2019qtm}.  However, the radiative cooling time increases dramatically as the axion-neutron coupling decreases.

We also consider the possibility of axion trapping, where axion  diffusion might serve to thermally equilibrate the interior of a neutron star or merger.  The timescales of thermal equilibration are given in Fig.~\ref{fig:conductivecooling}.  The SN1987A bound rules out axion trapping, but we present the analysis of thermal equilibration as it is potentially useful for the study of thermal transport due to future proposed bosons in neutron stars.  

Finally, in Appendix \ref{sec:H} and Appendix \ref{sec:I} we present in detail our calculations of the axion mean free path and emissivity.  We first present the phase space integrals assuming relativistic neutrons of arbitrary degeneracy but a momentum-independent matrix element.  Then we present the phase space integrals assuming non-relativistic neutrons of arbitrary degeneracy and a momentum-dependent matrix element.  Finally, we present the Fermi surface approximation, which is valid for strongly degenerate neutrons, and we propose a better treatment of the energy integration in that approximation which extends the usefulness of the Fermi surface approximation to the semi-degenerate regime.  

In our calculations, we have neglected the interaction of protons and axions, which should further reduce the mean free path and increase the emissivity \cite{Brinkmann:1988vi,Paul:2018msp,Iwamoto:1992jp,Keil:1996ju,PhysRevD.52.1780}.  We have also used a very simplistic neutron-neutron interaction (one pion exchange), improvements to which have been discussed in \cite{Hanhart:2000ae,PhysRevD.52.1780,Keil:1996ju,Beznogov:2018fda,Bartl:2016iok}, which indicate that more sophisticated nuclear interactions decrease the emissivity and increase the axion mean free path at high densities.

Future inclusion of axions in merger simulations should be done consistently.  The axion opacity should be calculated using the same nuclear equation of state used in the merger simulation itself \cite{Horowitz:2003yx,Fischer:2018kdt}, and the nucleon effective mass should be taken into account as discussed in Sec.~\ref{sec:axion_mfp_rel}.  This parallels the improvements of the neutrino transport in supernovae and merger simulations over the past couple of decades \cite{Ardevol-Pulpillo:2018btx,Perego:2014qda,Galeazzi:2013mia,Sekiguchi:2012uc,Rosswog:2003rv,1999JCoAM.109..281M}.
\chapter{Conclusions}
\pagestyle{myheadings}
Mapping out the QCD phase diagram is a major goal in nuclear and particle physics, and neutron star mergers provide an environment to study matter that is both dense and hot.  The problem of understanding the physics of neutron star mergers is attacked from two directions.  LIGO and Virgo can measure the gravitational waves during the inspiral - future generations of gravitational wave detectors should be able to measure the post-merger signal \cite{Clark:2015zxa,Tsang:2019esi} - and telescopes can measure the electromagnetic signal that comes from the merger \cite{GBM:2017lvd}.  Neutron star mergers are sufficiently complicated that a second approach, numerical simulation, is needed to compliment observations.

Numerical simulations of mergers evolve Einstein's equations determining the spacetime metric throughout the inspiral and merger.  The neutron star matter is modeled as a fluid, evolving according to the equations of relativistic hydrodynamics.  The fluid has an equation of state that is derived from nuclear physics, which is one of the great sources of uncertainty in the modeling of mergers.  The simulations also include neutrino transport.  

In this thesis we investigated additional physical processes in merger conditions and determined their relevance to neutron star mergers.  We investigated the nature of beta equilibrium in the neutrino-transparent region of the merger in Ch.~\ref{sec:beta_equilibrium}.  In Ch.~\ref{sec:bulk_viscosity}, we calculated the energy dissipation due to bulk viscosity in the merger, and in Ch.~\ref{sec:axions}, we examined the role that axions might play in thermal transport in mergers.

At zero temperature, the beta equilibrium condition is known to be $\mu_n = \mu_p + \mu_e$, but as temperature increases above about 1 MeV, a correction to this beta equilibrium condition appears.  The rates of the Urca processes that determine beta equilibrium begin to differ due to the kinematic impact of the neutrino, which becomes increasingly relevant as temperature increases.  We found the correction to the beta equilibrium condition as a function of temperature and density for conditions where nuclear matter is transparent to neutrinos.  This modification of the beta equilibrium condition slightly changes the equation of state.  There is no modification for neutrino-trapped nuclear matter.  

We calculated the bulk-viscous dissipation time for adiabatic density oscillations in neutrino-transparent nuclear matter.  For density oscillations with frequency 1 kHz, we found that in certain thermodynamic conditions - densities from 0.5-2 $n_0$ and temperatures from 2-4 MeV, significant energy dissipation from the oscillations occurs in 5-20 ms, which makes bulk viscosity relevant for neutron star mergers.  The exact thermodynamic conditions for low dissipation time depend on the nuclear equation of state.  Higher frequency oscillations, which likely occur post-merger, damp even more quickly.  We considered only small oscillations, but in the future we want to study the large oscillations that occur right after the two stars touch (see Fig.~\ref{fig:density_oscillations}).

Energy transport in neutron stars and mergers is done by particles with long mean free paths, with neutrinos being by far the most likely candidate.  We considered the possibility of energy transport due to axions, calculating their mean free path and determining that they are not trapped in any part of a neutron star merger.  We calculated the cooling time of a fluid element radiating axions and found that under current constraints on the axion-nucleon coupling, axions could cool certain regions of a merger in under 10 milliseconds.  Importantly, this could occur for fluid elements at temperatures where neutrinos are trapped, and rapid cooling would not ordinarily be expected.

In this thesis we have assumed that the matter in mergers is nuclear matter, that is, it consists only of neutrons, protons, electrons, and in certain regions, neutrinos.  We also considered the possibility of trapped axions, though this situation is unlikely to occur in mergers.  There is still much uncertainty in the composition of nuclear matter, as the nuclear equation of state is not well known except around nuclear saturation density.  A natural next step would be to consider other forms of matter in mergers.  One option is to include muons and other flavors of neutrinos.  Another would be to consider a phase transition from nuclear to quark matter, and to see what influence the presence of quarks will have on the merger.  As discussed in Sec.~\ref{sec:exotic_phases}, there are many possible ways this phase transition could occur, each of which would surely have different physical consequences in mergers.  In addition, the presence of meson condensates \cite{Pethick:2015jma} like pions or kaons could have important consequences. 
The equation of state $\varepsilon = \varepsilon(P)$ can inform us to a certain extent about the particle content, for example, if there is a strong first-order phase transition, but the possibility exists of two different types of matter having indistinguishable equations of state \cite{Alford:2004pf}.  We believe that understanding transport in neutron star mergers offers a great opportunity to determine the particle content of nuclear matter.  Transport properties depend on particle interaction rates, densities of states, and particle statistics, providing an increased number of signatures of the presence of certain particles.  The task of identifying the consequences of different forms of matter in mergers has already begun \cite{Fore:2019wib,Hanauske:2019qgs,Most:2019onn,Chesler:2019osn,Malfatti:2019tpg}.  
\appendix
\chapter{Thermodynamics in grand canonical ensemble}
\pagestyle{myheadings}
\label{sec:GCE_thermo}
For much of this thesis we work in grand canonical ensemble, which considers a system hooked up to a large reservoir.  The system and the reservoir are in equilibrium with respect to energy and particle number.  This means that they have the same temperature and chemical potential (we assume one species of particle) \cite{swendsen2020introduction}.  The point of this setup is that we can consider a system with external control parameters temperature and chemical potential.  These are the axes of the QCD phase diagram (Fig.~\ref{fig:qcd_phase_diagram}), for example.  

We also want to work in terms of intensive quantities, which do not depend on the system size, so instead of considering the energy $E$, the entropy $S$, the particle number $N$, we consider the energy density $\varepsilon$, the entropy density $s$, and the particle number density $n$.

The first law of thermodynamics for a one-component system is
\begin{equation}
    \mathop{dE} = -P\mathop{dV} + T\mathop{dS} + \mu\mathop{dN}.
\end{equation}
We write this law in terms of intensive quantities by turning writing $E = \varepsilon V$, $S = sV$, and $N=nV$, and then using product rule to expand the differentials.  Then we see that the $\mathop{dV}$ differential is multiplied by $\varepsilon+P-Ts-\mu n$, which equals 0 because it is nothing by the Euler equation in thermodynamics (see Eq. 13.29 in \cite{swendsen2020introduction}), which is valid for extensive systems\footnote{Extensive systems are homogeneous, for example, there are no surface effects.  When we describe systems thermodynamically, we usually assume that our description is for the bulk of the material, and that surface effects are negligible, allowing us to take advantage of extensivity.}.  Thus, the volume differential vanishes and we are left with
\begin{equation}
    \mathop{d\varepsilon} = T\mathop{ds}+\mu\mathop{dn}.
    \label{eq:1st_law_intensive}
\end{equation}
In this formulation, entropy density and number density are the control parameters, so we must do a Legendre transformation \cite{swendsen2020introduction} to make $T$ and $\mu$ the control parameters.  This is accomplished by adding and subtracting $s\mathop{dT}$ and $n\mathop{d\mu}$ to the right hand side of Eq.~(\ref{eq:1st_law_intensive}), yielding
\begin{align*}
    \mathop{d\varepsilon} &= \mathop{d(Ts)}+\mathop{d(\mu n)}-s\mathop{dT}-n\mathop{d\mu},\\
    \mathop{d(\varepsilon-Ts-\mu n)} &= -s\mathop{dT}-n\mathop{d\mu}.
\end{align*}
From Euler's equation, the left hand side is just $-\mathop{dP}$, and so we have derived the first law of thermodynamics in grand canonical ensemble
\begin{equation}
    \mathop{dP} = s\mathop{dT}+n\mathop{d\mu},
\end{equation}
and it is now evident that $T$ and $\mu$ are our control parameters.  It is also apparent that
\begin{equation}
    s = \frac{\partial P}{\partial T}\bigg\vert_{\mu} \qquad \text{and} \qquad n = \frac{\partial P}{\partial \mu}\bigg\vert_T.
\end{equation}
\chapter{Free Fermi gas}
\pagestyle{myheadings}
\label{sec:free_fermi_gas}
The simplest way to model a uniform phase of (fermionic) matter is as a free Fermi gas, which consists of free particles which obey the Pauli exclusion principle.  We will assume the particles have energy dispersion relation $E(p) = \sqrt{p^2+m^2}$ and that two particles can occupy each momentum state because the fermions have spin-1/2.  The occupation probability of a specific energy state is given by the Fermi-Dirac distribution.  

The number density of gas is found by adding up $1$ for each momentum state weighted by the Fermi-Dirac distribution, with a factor of 2 because of spin degeneracy
\begin{equation}
    n = 2\int\frac{\mathop{d^3k}}{(2\pi)^3}\left(1+e^{\beta(\sqrt{k^2+m^2}-\mu)}\right)^{-1} = \frac{1}{\pi^2}\int_0^{\infty}\mathop{dk}k^2\left(1+e^{\beta(\sqrt{k^2+m^2}-\mu)}\right)^{-1}.
\end{equation}
The energy density is similar, except that for each momentum state, the energy of that momentum state $\sqrt{k^2+m^2}$ is counted
\begin{equation}
    \varepsilon = 2\int\frac{\mathop{d^3k}}{(2\pi)^3}\sqrt{k^2+m^2}\left(1+e^{\beta(\sqrt{k^2+m^2}-\mu)}\right)^{-1} = \frac{1}{\pi^2}\int_0^{\infty}\mathop{dk}k^2\sqrt{k^2+m^2}\left(1+e^{\beta(\sqrt{k^2+m^2}-\mu)}\right)^{-1}.
\end{equation}
The pressure is
\begin{equation}
    P = \frac{T}{V}\ln{Z_{GC}},
\end{equation}
where $Z_{GC}$ is the grand canonical partition function \cite{swendsen2020introduction}.  We find that
\begin{align}
    P = \frac{T}{V}\ln{Z_{GC}} &= \frac{T}{V}\left[2V\int\frac{\mathop{d^3k}}{(2\pi)^3}\ln{\left(1+e^{-\beta(\sqrt{k^2+m^2}-\mu)}\right)}\right]\\
    &= \frac{T}{\pi^2}\int_0^{\infty}\mathop{dk}k^2\ln{\left(1+e^{-\beta(\sqrt{k^2+m^2}-\mu)}\right)}.\nonumber
\end{align}
\chapter{Direct Urca matrix element}
\pagestyle{myheadings}
\label{sec:durca_M_derivation}
Using Feynman rules from, for example, \cite{Griffiths:2008zz}, the matrix element for neutron decay $n\rightarrow p + e^- + \bar{\nu}$ is
\begin{align}
    \sum_{\text{spins}}\vert\mathcal{M}\vert^2 &= 32G_F^2\cos^2{\theta_c}\big[(1+g_A)^2(p_p\cdot p_e)(p_n\cdot p_{\nu}) + (g_A^2-1)m^2(p_e\cdot p_{\nu}) \nonumber \\
    &+ (g_A-1)^2(p_n\cdot p_e)(p_p\cdot p_{\nu}) \big].
\end{align}
In the nonrelativistic limit, $E_n\approx m$ and $E_p\approx m$, and the nucleon momentum $\vert \mathbf{p}\vert \ll m$.  Applying these approximations to the matrix element yields
\begin{equation}
   \sum_{\text{spins}}\vert\mathcal{M}\vert^2 \approx  32G_F^2\cos^2{\theta_c}\big[(1+3g_A^2)m^2E_eE_{\nu} + (1-g_A^2)m^2\mathbf{p_e}\cdot\mathbf{p_{\nu}}\big].
\end{equation}
In calculation of the Urca rate, the quantity that appears under the integral is \newline $ \sum_{\text{spins}}\vert\mathcal{M}\vert^2/(2E_n2E_p2E_e2E_{\nu})$, which is the nonrelativistic transition amplitude in Fermi's Golden rule.  In the non-relativstic approximation, this quantity becomes
\begin{equation}
    \frac{\sum_{\text{spins}}\vert\mathcal{M}\vert^2}{2E_n2E_p2E_e2E_{\nu}} = 2G_F^2\cos^2{\theta_c}\left[1+3g_A^2+(1-g_A^2)\frac{\mathbf{p_e}\cdot\mathbf{p_{\nu}}}{E_eE_{\nu}}\right],
\end{equation}
which is the expression used in \cite{Yakovlev:2000jp} and hence in our papers \cite{Alford:2018lhf,Alford:2019qtm}.  Thanks to crossing symmetry \cite{Halzen:1984mc}, this is also the matrix element for electron capture $e^- + p \rightarrow n + \nu$.
\chapter{Modified Urca rates when $\mu_n \neq \mu_p + \mu_e$}
\pagestyle{myheadings}
\label{sec:out_of_equil_mU}
We present here the Fermi-surface approximation of the modified Urca rates, when allowed to deviate from the low-temperature beta equilibrium criterion (\ref{eq:beq_nu_trans}) by an amount $\xi \equiv \left(\mu_n-\mu_p-\mu_e\right)/T$.  In the following expressions for the rates the non-equilibrium behavior is encapsulated in the function
\begin{align}
F(\xi) &= -\left(\xi^4+10\pi^2\xi^2+9\pi^4\right)\text{Li}_3(-e^{\xi}) +12\xi\left(\xi^2+5\pi^2\right)\text{Li}_4(-e^{\xi})\nonumber\\ &-24\left(3\xi^2+5\pi^2\right)\text{Li}_5(-e^{\xi}) +240\xi \text{Li}_6(-e^{\xi})-360 \text{Li}_7(-e^{\xi}),
\end{align}
where $\text{Li}_n$ is the Polylogarithm function of order $n$ \cite{olver2010nist}.  We note that $F(0) \approx 2300.$
The rate of modified Urca neutron decay with a neutron spectator is
\begin{equation}
\Gamma_{mU, nd (n)}(\xi) = \dfrac{7}{64\pi^9} G^2 g_A^2 f^4 \dfrac{m_n^3m_p}{m_{\pi}^4} \dfrac{p_{F_n}^4p_{F_p}}{\left(p_{F_n}^2+m_{\pi}^2\right)^2} F(\xi) \vartheta_n T^7
\end{equation}
where $\vartheta_n$ is defined as in Eq.~(\ref{eq:mUrca_n}).  The rate of modified Urca electron capture with a neutron spectator is 
\begin{equation}
\Gamma_{mU, ec (n)}(\xi) = \Gamma_{mU, nd (n)}(-\xi),
\end{equation}
 and so the two neutron-spectator modified Urca rates agree in low-temperature beta equilibrium (\ref{eq:beq_nu_trans}).
The Fermi surface approximation of the modified Urca neutron decay process with a proton spectator is
\begin{equation}
\Gamma_{mU, nd (p)}(\xi) = \dfrac{1}{64\pi^9}G^2g_A^2f^4 \dfrac{m_nm_p^3}{m_{\pi}^4} \dfrac{p_{F_n}\left(p_{F_n}-p_{F_p}\right)^4}{\left(\left(p_{F_n}-p_{F_p}\right)^2+m_{\pi}^2\right)^2}F(\xi)\vartheta_p T^7
\end{equation}
with $\vartheta_p$ defined as in Eq.~(\ref{eq:mUrca_p}), and the modified Urca electron capture rate with a proton spectator is
\begin{equation}
\Gamma_{mU, ec (p)}(\xi) = \Gamma_{mU, nd (p)}(-\xi),
\end{equation}
where again both modified Urca rates with a proton spectator agree in low-temperature beta equilibrium (\ref{eq:beq_nu_trans}).
\chapter{Exact direct Urca rate integral}
\pagestyle{myheadings}
\label{sec:rate-integral}

The neutron decay rate given by a twelve dimensional integral in Eq.~(\ref{eq:ndecay}) can be reduced, without approximation, to a three dimensional integral.  Integrating over neutrino three-momentum, we have
\begin{align}
\Gamma_{n} &=\dfrac{G^2}{128\pi^8} \int \mathop{d^3p_n}\mathop{d^3p_p}\mathop{d^3p_e}f_n\left(1-f_p\right)\left(1-f_e\right)\delta(q-\vert \mathbf{p}_n-\mathbf{p}_p-\mathbf{p}_e\vert )\nonumber\\
&\times\left( 1+3g_A^2+\left(1-g_A^2\right)\hat{\mathbf{p}}_e \cdot \dfrac{\mathbf{p}_n-\mathbf{p}_p-\mathbf{p}_e}{\vert \mathbf{p}_n-\mathbf{p}_p-\mathbf{p}_e\vert}\right),
\end{align}
where we define $q \equiv E_n - E_p - E_e$, and the ``hat'' denotes a unit vector.
We adopt spherical coordinates for the momentum of each of the three particles.  We have the freedom to choose the coordinates such that the neutron momentum lies along the $z$ axis, and the proton momentum lies in the same plane as the neutron momentum, and so the momentum unit vectors are written as
\begin{align}
\hat{\mathbf{p}}_n &= \left(0,0,1\right)\\
\hat{\mathbf{p}}_p &= \left(\sqrt{1-z_p^2},0,z_p\right)\\
\hat{\mathbf{p}}_e &= \left(\sqrt{1-z_e^2}\cos{\phi},\sqrt{1-z_e^2}\sin{\phi},z_e\right),
\end{align}
where $z_p$ and $z_e$ are cosines of the polar angles of the proton and electron momenta and as such, take values from $-1$ to $1$.  The azimuthal angle $\phi$ of the electron with respect to the plane formed by the proton and neutron momenta ranges from $0$ to $2\pi$.

This choice of coordinates allows us to integrate over the three trivial angles, giving a factor of $8\pi^2$.  The rate integral can now be written as
\begin{equation}
\Gamma_{n} = \dfrac{G^2}{16\pi^6}\int_0^{\infty} \mathop{dp_n}\mathop{dp_p}\mathop{dp_e} p_n^2p_p^2p_e^2  f_n\left(1-f_p\right)\left(1-f_e\right)  I(p_n,p_p,p_e),
\end{equation}
where 
\begin{align}
&I = \int_{-1}^1\mathop{dz_e}\int_{-1}^1\mathop{dz_p}\int_0^{2\pi}\mathop{d\phi}\\
&\times \delta\Bigg(q-\sqrt{p_n^2+p_p^2+p_e^2+2p_pp_e\sqrt{1-z_p^2}\sqrt{1-z_e^2}\cos{\phi}+2p_pp_ez_pz_e-2p_np_pz_p-2p_np_ez_e   }\Bigg)\nonumber\\
&\times \left( 1+3g_A^2+\left(1-g_A^2\right)\dfrac{p_nz_e-p_pz_pz_e-p_e-p_p\sqrt{1-z_p^2}\sqrt{1-z_e^2}\cos{\phi}}{\sqrt{ p_n^2+p_p^2+p_e^2+2p_pp_e\sqrt{1-z_p^2}\sqrt{1-z_e^2}\cos{\phi}+2p_pp_ez_pz_e-2p_np_pz_p-2p_np_ez_e  }} \right).\nonumber
\end{align}
We do the $\phi$ integral first, using the delta function.  Clearly we require $q>0$, because if $q$ is negative, the argument of the delta function could never be zero and thus the integral would be zero.  For $q>0$, the delta function argument vanishes for either zero or two values of $\phi$ between $0$ and $2\pi$.  We find that
\begin{align}
I &= 4\vert q\vert\int_{-1}^1\mathop{dz_p}\left(1+3g_A^2+\dfrac{1-g_A^2}{2p_eq}\left(p_n^2+p_p^2-p_e^2-q^2-2p_np_pz_p\right)\right)\Theta(q)\\
&\times \int_{-1}^{1}\mathop{dz_e} \dfrac{\Theta(B)}{\sqrt{ 4p_p^2p_e^2\left(1-z_p^2\right)\left(1-z_e^2\right)-\left(q^2-p_n^2-p_p^2-p_e^2-2p_pp_ez_pz_e+2p_np_pz_p+2p_np_ez_e\right)^2 }},\nonumber
\end{align}
where the step function $\Theta(B)$ enforces $B>0$, where 
\begin{align}
&B = 2p_pp_e\sqrt{1-z_p^2}\sqrt{1-z_e^2}\\
&-\vert q^2-p_n^2-p_p^2-p_e^2-2p_pp_ez_pz_e+2p_np_pz_p+2p_np_ez_e\vert.\nonumber
\end{align}
This is the condition for there to be two, not zero, values of $\phi$ in the integration range which make the argument of the delta function vanish and thus contribute to the integral.

We now evaluate the $z_e$ integral, noting that the step function $\Theta(B)$ adjusts the range of integration.  
Only if $C>0$, where
\begin{equation}
C = 2p_e\vert q\vert-\vert p_e^2+q^2-p_n^2-p_p^2+2p_np_pz_p\vert,
\end{equation}
is the step function nonzero for any range of $z_e$ in the interval $[-1,1]$, in which case the step function is nonzero only for $z_e^- < z_e < z_e^+$, where $z_e^{\pm}$ lie inside the interval $[-1,1]$, and thus $z_e^{\pm}$ become the new integration bounds.  Doing the $z_e$ integral, we find
\begin{equation}
I = \dfrac{2\pi \vert q\vert}{p_e}\Theta(q)\int_{-1}^1\mathop{dz_p}\Theta(C)\left( \dfrac{1+3g_A^2+\dfrac{1-g_A^2}{2p_eq}\left(p_n^2+p_p^2-p_e^2-q^2-2p_np_pz_p\right)}{\sqrt{p_n^2+p_p^2-2p_np_pz_p}}\right).
\end{equation}

The step function $\Theta(C)$ creates a restriction on the bounds of $z_p$ and so the actual range of integration over $z_p$ is the intersection of the intervals $[-1,1]$ and $[z_p^-,z_p^+]$, where $z_p^{\pm} = (p_n^2+p_p^2-p_e^2-q^2 \pm 2p_e\vert q \vert )/(2p_n p_p)$, and so the range of integration will depend on the values of  $\{p_n,p_p,p_e\}$.  Evaluating the integral over $z_p$ with the bounds $y^+$ and $y^-$, we have
\begin{equation}
I(p_n,p_p,p_e) =  \frac{2\pi\vert q\vert}{p_np_pp_e}\Theta(q) J(p_n,p_p,p_e)
\end{equation}
where
\begin{equation}
\label{eq:Jdef}
J(p_n,p_p,p_e) = \bigg[ \Big(1+3g_A^2- (1-g_A^2)\frac{p_e^2+q^2}{2p_eq}\Big)y^{1/2} +\frac{1-g_A^2}{6p_eq}y^{3/2}  \bigg] \Bigg\rvert_{y=y^-}^{y=y^+}
\end{equation}
with
\begin{equation}
\label{eq:upper}
y^+ =
 \begin{cases}
(p_e+\vert q\vert)^2 & \text{ if } -1<z_p^-<1<z_p^+ \\
(p_e+\vert q\vert)^2 & \text{ if } -1<z_p^-<z_p^+<1 \\
(p_n+p_p)^2 & \text{ if  }  z_p^-<-1<1<z_p^+\\
(p_n+p_p)^2 &  \text{ if  } z_p^-<-1<z_p^+<1
\end{cases}
\end{equation}
and
\begin{equation}
\label{eq:lower}
y^- = 
\begin{cases}
(p_n-p_p)^2 & \text{ if } -1<z_p^-<1<z_p^+\\
(p_e-\vert q \vert)^2 & \text{ if } -1 < z_p^- < z_p^+<1\\
(p_n-p_p)^2 & \text{ if } z_p^- < -1 < 1 < z_p^+\\
(p_e-\vert q\vert)^2 & \text{ if } z_p^- < -1 < z_p^+ < 1.
\end{cases}
\end{equation}

Thus, the direct Urca neutron decay rate is
\begin{equation}
\label{eq:final_urca_rate}
\Gamma_n = \frac{G^2}{8\pi^5}\int_0^{\infty} \mathop{dp_n}\mathop{dp_p}\mathop{dp_e} p_np_pp_e\vert q\vert \Theta(q) f_n\left(1-f_p\right)\left(1-f_e\right)  J(p_n,p_p,p_e)
\end{equation}
with $J(p_n,p_p,p_e)$ as defined in Eqs.~(\ref{eq:Jdef}), (\ref{eq:upper}), and (\ref{eq:lower}).

The direct Urca electron capture rate integral is identical, except for $f_n\left(1-f_p\right)\left(1-f_e\right)$ is replaced by $\left(1-f_n\right)f_pf_e$ and, because the neutrino is now on the same side of the reaction as the neutron, instead of with the electron and proton, $\Theta(q)$ is replaced by $\Theta(-q)$.

 The remaining three dimensional integral in Eq.~(\ref{eq:final_urca_rate}) can be done numerically (we used Mathematica's Monte Carlo integration routine), giving the direct Urca rate results shown in Fig \ref{fig:urca_rates}.  Mathematica's estimated error on the numerical integrals is under twenty percent, and repeated evaluation of the integrals leads to results within ten percent of their average. 
\chapter{A Maxwell relation for bulk viscosity}
\pagestyle{myheadings}
\label{sec:maxwell}
We begin with the first law of thermodynamics for $npe$ matter
\begin{equation}
    \mathop{dE} = -P\mathop{dV}+T\mathop{dS}+\mu_n\mathop{dN_n}+\mu_p\mathop{dN_p}+\mu_e\mathop{dN_e}.
\end{equation}
We normalize the extensive quantities by the baryon number, where $\sigma = S/N_B$, $x_i = N_i / N_B$,  and $V = N_B / n_B$, and the first law becomes
\begin{equation}
    \mathop{d\left(\frac{\varepsilon}{n_B}\right)} = \frac{P}{n_B^2}\mathop{dn_B} + T\mathop{d\sigma} + \mu_n\mathop{dx_n}+\mu_p\mathop{dx_p}+\mu_e\mathop{dx_e}.
\end{equation}
Because of charge neutrality, $n_p = n_e$, so $\mathop{dx_p}=\mathop{dx_e}$, and the Urca processes indicate that $\mathop{dx_n}=-\mathop{dx_p}$.  Turning the particle fractions into proton fractions and defining $\mu_{\Delta} = \mu_n-\mu_p-\mu_e$, the first law becomes
\begin{equation}
    \mathop{d\left(\frac{\varepsilon}{n_B}\right)} = \frac{P}{n_B^2}\mathop{dn_B} + T\mathop{d\sigma} - \mu_{\Delta}\mathop{dx_p}.
\end{equation}
One of the three Maxwell relations \cite{swendsen2020introduction} that one can derive from this expression is Eq.~(\ref{eq:maxwell}).
\chapter{Adiabatic and isothermal oscillations}
\pagestyle{myheadings}
\label{app:adiabatic}

Most previous works use the isothermal susceptibilities 
\begin{align}
 B_T &= -\frac{1}{n_B} \frac{\partial\mu_{\Delta}}{\partial x_p} \bigg\rvert_{n_B,T} \ , \\
 C_T &= n_B \frac{\partial \mu_{\Delta}}{\partial n_B}\bigg\rvert_{x_p,T} \ ,
\end{align}
often only considering the zero temperature case \cite{Yakovlev:2018jia,Alford:2010gw,2000A&A...357.1157H,Haensel:1992zz,PhysRevD.39.3804}.
As discussed in Sec.~\ref{sec:bv}, because thermal equilibration is so slow in neutrino-transparent nuclear matter in merger conditions, we must use the adiabatic susceptibilities
\begin{align}
 B &= -\frac{1}{n_B} \frac{\partial\mu_{\Delta}}{\partial x_p} \bigg\rvert_{n_B,s/n_B} \ , \\
 C &= n_B \frac{\partial \mu_{\Delta}}{\partial n_B}\bigg\rvert_{x_p,s/n_B} \ .
\end{align}
Note that at zero temperature, adiabatic and isothermal quantities become equivalent \cite{schroederintroduction,swendsen2020introduction}.  
Often, it is convenient to work with thermodynamic derivatives at constant temperature $T$, or baryon density $n_B$, or proton fraction $x_p$.  In particular, these three variables are the degrees of freedom in the CompOSE database of EoSs \cite{CompOSE}.  Using a Jacobian coordinate transformation \cite{greiner2012thermodynamics,swendsen2020introduction}, we can relate adiabatic derivatives (derivatives at constant entropy per baryon) to isothermal derivatives.  The adiabatic susceptibility derivatives are related to the isothermal susceptibility
derivatives by
\begin{align}
    \frac{\partial \mu_{\Delta}}{\partial n_B}\bigg \vert_{s/n_B,x_p} &= \frac{\partial \mu_{\Delta}}{\partial n_B}\bigg \vert_{T,x_p} - \frac{\frac{\partial (s/n_B)}{\partial n_B}\big \vert_{T,x_p}\frac{\partial \mu_{\Delta}}{\partial T}\big \vert_{n_B,x_p}}{\frac{\partial (s/n_B)}{\partial T}\big \vert_{n_B,x_p}}\\
    \frac{\partial \mu_{\Delta}}{\partial x_p}\bigg \vert_{s/n_B,n_B} &= \frac{\partial \mu_{\Delta}}{\partial x_p}\bigg \vert_{T,n_B} - \frac{\frac{\partial (s/n_B)}{\partial x_p}\big \vert_{T,n_B}\frac{\partial \mu_{\Delta}}{\partial T}\big \vert_{n_B,x_p}}{\frac{\partial (s/n_B)}{\partial T}\big \vert_{n_B,x_p}}.
\end{align}

The isothermal compressibility is 
\begin{equation}
    \kappa_T^{-1} = n_B \frac{\partial P}{\partial n_B}\bigg\vert_{x_p,T}
\end{equation}
and the adiabatic compressibility is given by
\begin{equation}
    \kappa_S^{-1} = n_B \frac{\partial P}{\partial n_B}\bigg\vert_{x_p,s/n_B}.
\end{equation}
The adiabatic derivative can be obtained from the isothermal derivative by
\begin{equation}
     \frac{\partial P}{\partial n_B}\bigg \vert_{s/n_B,x_p} = \frac{\partial P}{\partial n_B}\bigg \vert_{T,x_p} - \frac{\frac{\partial (s/n_B)}{\partial n_B}\big \vert_{T,x_p}\frac{\partial P}{\partial T}\big \vert_{n_B,x_p}}{\frac{\partial (s/n_B)}{\partial T}\big \vert_{n_B,x_p}}.
\end{equation}

Above $n_0$, the adiabatic and isothermal derivatives are within 25\% of each other for all temperatures considered here. For nuclear matter that is below $n_0$ with $T> 5\,\text{MeV}$, there are noticeable differences between the isothermal and adiabatic susceptibility $C$ and the compressibility $\kappa$.  The susceptibility $B$ is not sensitive to differences between adiabaticity and isothermality (the differences are below 10\%).    

Below $n_0$, the adiabatic $C$ is greater than the isothermal $C$ by as much as a factor of 2.5 (DD2) or 5.5 (IUF).  These large differences are at temperatures above 5\,MeV.  Thus, the adiabatic $C^2/B$ is larger than the isothermal version by factors of up to 6 (DD2) or 30 (IUF).  However, these large differences occur at low densities ($\approx 0.5n_0$) and high temperatures ($T\approx 10\,\text{MeV}$) where the bulk viscosity is small anyway, since the equilibration rate $\gamma$ is much faster than a 1 kHz density oscillation.  In the regions where bulk viscosity is large, the difference between adiabatic and isothermal susceptibilities is at most a factor of 2 in the quantity $C^2/B$.

At the densities and temperatures where bulk viscosity is large, the isothermal compressibility is at most 20\% larger than the adiabatic compressibility, which means that adiabatic density oscillations would lose energy slightly more slowly than isothermal density oscillations.  At densities below $n_0$ and temperatures above 5\,MeV, the isothermal compressibility can be up to 40\% (DD2) or 80\% (IUF) larger than the adiabatic value, but the bulk viscosity is too small for fluid elements under these conditions for this to matter.
\chapter{Axion emissivity integrals}
\label{sec:H}
\pagestyle{myheadings}
Below, we detail a series of approximations for the axion emissivity, Eq.~(\ref{eq:emissivity_integral}).
\section{Relativistic, constant-matrix-element phase space integration}
\label{sec:Q_rel_PS}
The phase space integral Eq.~(\ref{eq:emissivity_integral}) can be done if we make the approximation of a momentum-independent matrix element [Eq.~(\ref{eq:cst_matrix})].  The resulting emissivity will be valid at arbitrary neutron degeneracy and arbitrary degree of relativistic nature of the neutrons.  
In the constant-matrix element approximation the axion emissivity is
\begin{align}
    Q &= \left(1-\frac{\beta}{3}\right)\frac{f^4m_n^4G_{an}^2}{256\pi^{11}m_{\pi}^4}\left(1+\frac{m_{\pi}^2}{k_{\text{typ}}^2}\right)^{-2}\\
    &\times \int \mathop{d^3p_1}\mathop{d^3p_2}\mathop{d^3p_3}\mathop{d^3p_4}\mathop{d^3\omega}\delta^4(p_1+p_2-p_3-p_4-\omega)\frac{f_1f_2(1-f_3)(1-f_4)}{E_1^*E_2^*E_3^*E_4^*}\nonumber.
\end{align}

The zeroth component of the delta function can be written as $\delta(E_1^*+E_2^*-E_3^*-E_4^*-\omega)$ since the neutron mean fields $U_n$ cancel out.
This integral can be broken up into 2 subsystems $A$ and $B$ which exchange some 4-momentum $q$, which is integrated over (a similar approach was used by \cite{Kaminker:2016ayg} in a different context).  Thus,
\begin{equation}
    Q = \left(1-\frac{\beta}{3}\right)\frac{f^4m_n^4G_{an}^2}{256\pi^{11}m_{\pi}^4} \left(1+\frac{m_{\pi}^2}{k_{\text{typ}}^2}\right)^{-2}\int \mathop{d^4q} A(q_0,q) B(q_0,q)
\end{equation}
where 
\begin{equation}
    A(q_0,q) = \int \mathop{d^3p_1}\mathop{d^3p_2}\delta^4(p_1+p_2-q)\frac{f_1f_2}{E_1^*E_2^*}
\end{equation}
and
\begin{equation}
    B(q_0,q) = \int \mathop{d^3p_3}\mathop{d^3p_4}\mathop{d^3\omega}\delta^4(q-p_3-p_4-\omega)\frac{(1-f_3)(1-f_4)}{E_3^*E_4^*}.
\end{equation}
Then we split $A$ and $B$ up into subsystem with 4-momentum transfers $k$ and $l$
\begin{align}
    A(q_0,q) &= \int\mathop{d^4k}I_1(k_0,k)I_2(k_0,k)\label{eq:A_integral}\\
    B(q_0,q) &= \int\mathop{d^4l}I_3(l_0,l)I_4(l_0,l)\label{eq:B_integral}
\end{align}
where we define and compute
\begin{align}
    I_1 \equiv \int\mathop{d^3p_1}\delta^4(p_1+k)\frac{f_1}{E_1^*} &= \frac{\delta(k_0+\sqrt{\mathbf{k}^2+m_*^2})}{\sqrt{\mathbf{k}^2+m_*^2}(1+e^{(\sqrt{\mathbf{k}^2+m_*^2}-\mu_n^*)/T})}\\
    I_2 \equiv \int\mathop{d^3p_2}\delta^4(p_2-q-k)\frac{f_2}{E_2^*} &= \frac{\delta(\sqrt{m_*^2+(\mathbf{q}+\mathbf{k})^2}-q_0-k_0)}{\sqrt{m_*^2+(\mathbf{q}+\mathbf{k})^2}(1+e^{(\sqrt{m_*^2+(\mathbf{q}+\mathbf{k})^2}-\mu_n^*)/T})}\\
    I_3 \equiv \int\mathop{d^3p_3}\delta^4(q-p_3-l)\frac{(1-f_3)}{E_3^*} &= \frac{\delta(q_0-l_0-\sqrt{m_*^2+(\mathbf{q}-\mathbf{l})^2})}{\sqrt{m_*^2+(\mathbf{q}-\mathbf{l})^2}(1+e^{-(\sqrt{m_*^2+(\mathbf{q}-\mathbf{l})^2)}-\mu_n^*)/T})}\\
    I_4 \equiv \int\mathop{d^3p_4}\mathop{d^3\omega}\delta^4(l-p_4-\omega)\frac{(1-f_4)}{E_4^*} &= \frac{2\pi}{l}\int_{\omega_-}^{\omega_+}\mathop{d\omega}\frac{\omega}{1+e^{-(l_0-\omega-\mu_n^*)/T}}\theta(l_0^2-l^2-m_*^2),
\end{align}
where
\begin{align}
    \omega_- &= \frac{l_0^2-l^2-m_*^2}{2(l_0+l)}\\
    \omega_+ &= \frac{l_0^2-l^2-m_*^2}{2(l_0-l)}.
\end{align}
We can now calculate $A(q)$ and $B(q)$ using Eq.~(\ref{eq:A_integral}) and (\ref{eq:B_integral}).  We find
\begin{align}
    A(q_0,q) &= \frac{2\pi}{q}\int_0^{\sqrt{q_0^2-m_*^2}}\mathop{dk}\frac{k(q_0-\sqrt{k^2+m_*^2})}{\sqrt{k^2+m_*^2}\sqrt{k^2+m_*^2+q_0^2-2q_0\sqrt{k^2+m_*^2}}}\\
    &\times \frac{\theta(2kq-\vert q_0^2-q^2-2q_0\sqrt{k^2+m_*^2}\vert)}{(1+e^{(\sqrt{k^2+m_*^2}-\mu_n^*)/T})(1+e^{(\sqrt{k^2+m_*^2+q_0^2-2q_0\sqrt{k^2+m_*^2}}-\mu_n^*)/T})}\nonumber
\end{align}
and
\begin{align}
    B(q_0,q) &= \frac{4\pi^2}{q}\int_m^{q_0}\mathop{dl_0}\int_0^{\sqrt{l_0^2-m_*^2}}\mathop{dl}\frac{\theta(2ql-\vert m_*^2+q^2+l^2-q_0^2-l_0^2+2q_0l_0\vert)}{1+e^{-(q_0-l_0-\mu_n^*)/T}}\nonumber\\
    &\times\int_{\omega_-}^{\omega_+}\mathop{d\omega}\frac{\omega}{1+e^{-(l_0-\omega-\mu_n^*)/T}}.
\end{align}
Thus, the full expression for the emissivity is
\begin{align}
    Q &= \left(1-\frac{\beta}{3}\right) \frac{f^4m_n^4G_{an}^2}{8\pi^7m_{\pi}^4}\left(1+\frac{m_{\pi}^2}{k_{\text{typ}}^2}\right)^{-2}\int_{m_*}^{\infty}\mathop{dq_0}\int_0^{\infty}\mathop{dq}\int_0^{\sqrt{q_0^2-m_*^2}}\mathop{dk}\int_{m_*}^{q_0}\mathop{dl_0}\int_0^{\sqrt{l_0^2-m_*^2}}\mathop{dl}\int_{\omega_-(l_0,l)}^{\omega_+(l_0,l)}\mathop{d\omega}\nonumber\\
    &\times k\omega(q_0-\sqrt{k^2+m_*^2})\frac{\theta(2kq-\vert q_0^2-q^2-2q_0\sqrt{k^2+m_*^2}\vert)\theta(2ql-\vert m_*^2+q^2+l^2-q_0^2-l_0^2+2q_0l_0\vert)}{\sqrt{k^2+m_*^2}\sqrt{k^2+m_*^2+q_0^2-2q_0\sqrt{k^2+m_*^2}}}    \label{eq:exact_emissivity}\\
    &\times \frac{1}{(1+e^{(\sqrt{k^2+m_*^2}-\mu_n^*)/T})(1+e^{(\sqrt{k^2+m_*^2+q_0^2-2q_0\sqrt{k^2+m_*^2}}-\mu_n^*)/T})(1+e^{-(q_0-l_0-\mu_n^*)/T})(1+e^{-(l_0-\omega-\mu_n^*)/T})}.\nonumber
\end{align}
This six-dimensional integral can be done numerically in Mathematica.
\section{Non-relativistic phase space integration}
\label{sec:Q_nonrel_PS}
The axion emissivity [Eq.~(\ref{eq:emissivity_integral})] can be computed assuming non-relativistic neutrons.  In this case, it is possible to keep the full momentum-dependence of the matrix element [Eq.~(\ref{eq:matrix_element})].  However, the axion 3-momentum is neglected in the 3-momentum conserving delta function.  Calculations similar to this have been done in the literature, mostly for non-degenerate nucleons \cite{PhysRevD.52.1780,Giannotti:2005tn,Stoica:2009zh,Dent:2012mx,Turner:1987by}.  The full calculation of the axion emissivity with non-relativistic neutrons and at arbitrary degeneracy has been done recently by \cite{Carenza:2019pxu}.  Here we will apply this calculation to neutrons described by the NL$\rho$ EoS, that is, neutrons with dispersion relation given by Eq.~(\ref{eq:En}).  We will do the calculation for arbitrary degeneracy, using Fermi-Dirac factors instead of a Maxwell-Boltzmann distribution.  In the nonrelativistic approximation, $E-\mu_n = E^* - \mu_n^* \approx m^*+p^2/(2m^*)-\mu^* = p^2/(2m^*)-(\mu^*-m^*) \equiv p^2/(2m^*)-\hat{\mu}$, where $\hat{\mu} \equiv \mu^*-m^*$ is the non-relativistic definition of the chemical potential.  

Starting with Eq.~(\ref{eq:emissivity_integral}), we neglect the axion three-momentum in the three-dimensional delta function and we set the neutron energy $E^* = m^*$, in the factor $(2^5E_1E_2E_3E_4\omega)^{-1}$ as is conventional \cite{Dent:2012mx}, as it often simplifies the integration.  Converting the integral over the axion momentum to spherical coordinates and doing the trivial integral over the axion momentum solid angle, we obtain
\begin{align}
    &Q_a = \frac{1}{96\pi^{10}}\frac{f^4 m_n^4 G_{an}^2}{m_{\pi}^4m_*^3}\int \mathop{d^3p_1}\mathop{d^3p_2}\mathop{d^3p_3}\mathop{d^3p_4}\int_0^{\infty}\mathop{d\omega}\omega^2\delta(p_1^2+p_2^2-p_3^2-p_4^2-2m_*\omega)\\
    &\times \delta^3(\mathbf{p}_1+\mathbf{p}_2-\mathbf{p}_3-\mathbf{p}_4)f_1f_2(1-f_3)(1-f_4)\left[ \frac{\mathbf{k}^4}{\left(\mathbf{k}^2+m_{\pi}^2\right)^2}+\frac{\mathbf{l}^4}{\left(\mathbf{l}^2+m_{\pi}^2\right)^2}+\frac{\mathbf{k}^2\mathbf{l}^2-3\left(\mathbf{k}\cdot\mathbf{l}\right)^2}{\left(\mathbf{k}^2+m_{\pi}^2\right)\left(\mathbf{l}^2+m_{\pi}^2\right)}    \right].\nonumber
\end{align}
Now we define a new coordinate system, $\mathbf{p}_+=(\mathbf{p}_1+\mathbf{p}_2)/2$, $\mathbf{p}_-=(\mathbf{p}_1-\mathbf{p}_2)/2$, $\mathbf{a}=\mathbf{p}_3-\mathbf{p}_+$, $\mathbf{b}=\mathbf{p}_4-\mathbf{p}_+$,
which has Jacobian $\mathop{d^3p}_1\mathop{d^3p}_2\mathop{d^3p}_3\mathop{d^3p}_4=8\mathop{d^3p}_+\mathop{d^3p}_-\mathop{d^3a}\mathop{d^3b}$.  The three-dimensional delta function becomes $\delta^3(\mathbf{a}+\mathbf{b})$, so we integrate over the three-momentum $\mathbf{b}$ and then over the axion energy $\omega$, using the delta function (which became $\delta(\mathbf{p}_-^2-\mathbf{a}^2-m_*\omega)$).  

We are now left with an integral over the three-vectors $\mathbf{p}_+$, $\mathbf{p}_-$, and $\mathbf{a}$, so we choose a coordinate system where $\mathbf{a} = a(0,0,1)$, $\mathbf{p}_- = p_-(\sqrt{1-r^2},0,r)$, and \newline $\mathbf{p}_+ = p_+(\sqrt{1-s^2}\cos{\phi},\sqrt{1-s^2}\sin{\phi},s)$.  Now, $\mathbf{k}^2=p_-^2+a^2-2p_-ar$, $\mathbf{l}^2= p_-^2+a^2+2p_-ar$, $\mathbf{k}\cdot\mathbf{l}=p_-^2-a^2$, $\mathbf{p}_1^2 = p_+^2+p_-^2+2p_+p_-(\sqrt{1-r^2}\sqrt{1-s^2}\cos{\phi}+rs)$, $\mathbf{p}_2^2=p_+^2+p_-^2-2p_+p_-(\sqrt{1-r^2}\sqrt{1-s^2}\cos{\phi}+rs)$, $\mathbf{p}_3^2=p_+^2+a^2+2p_+as$, and $\mathbf{p}_4^2=p_+^2+a^2-2p_+as$.

Now we can integrate over the three trivial angles, giving a factor of $8\pi^2$, leaving us with a six-dimensional integral that we simplify with the coordinate transformations $u = p_+^2/(2m_*T)$, $v=p_-^2/(2m_*T)$, $w=a^2/(2m_*T)$, and we define $\hat{y}=\hat{\mu}/T$.  Thus the key variables in the emissivity expression become
\begin{align}
    \mathbf{k}^2 &= 2m_*T(v+w-2\sqrt{vw}r)\nonumber\\
    \mathbf{l}^2 &= 2m_*T(v+w+2\sqrt{vw}r)\nonumber\\
    \mathbf{k}\cdot\mathbf{l}&=2m_*T(v-w)\nonumber\\
    \beta (E_1-\mu_n) &= -\hat{y}+u+v+2\sqrt{uv}(\sqrt{1-r^2}\sqrt{1-s^2}\cos{\phi}+rs)\\
    \beta (E_2-\mu_n) &= -\hat{y}+u+v-2\sqrt{uv}(\sqrt{1-r^2}\sqrt{1-s^2}\cos{\phi}+rs)\nonumber\\
    \beta (E_3-\mu_n) &= -\hat{y}+u+w+2\sqrt{uw}s\nonumber\\
    \beta (E_4-\mu_n) &= -\hat{y}+u+w-2\sqrt{uw}s.\nonumber
\end{align}
and the axion emissivity is
\begin{align}
    Q &= \frac{32\sqrt{2}}{3\pi^8}\frac{f^4m_n^4G_{an}^2}{m_{\pi}^4}m_*^{1/2}T^{6.5}\int_0^{\infty}\mathop{du}\mathop{dv}\int_0^v\mathop{dw}\int_{-1}^1\mathop{dr}\mathop{ds}\int_0^{2\pi}\mathop{d\phi}u^{1/2}v^{3/2}w^{3/2}(v-w)^2\nonumber\\
    &\times \frac{\left(\alpha^4 \left(r^2+3\right)-6 \alpha^2 \left(r^2-1\right) (v+w)-3 \left(r^2-1\right) \left(2 \left(1-2 r^2\right) v w+v^2+w^2\right)\right)}{\left(2 w \left(\alpha^2-2 r^2
   v+v\right)+\left(\alpha^2+v\right)^2+w^2\right)^2}\nonumber\\
   &\times \left((1+e^{\beta(E_1-\mu_n)})(1+e^{\beta(E_2-\mu_n)})(1+e^{-\beta(E_3-\mu_n)})(1+e^{-\beta(E_4-\mu_n)})\right)^{-1}\label{eq:Q_NR_PS}
\end{align}
where $\alpha = m_{\pi}/\sqrt{2m_*T}$.
This integral can be done numerically.
\section{Fermi surface approximation and its improvement}
\label{sec:Q_FS_derivation}
In nuclear matter, to good approximation the dominant contribution to a process involving degenerate fermions will be from those fermions near their Fermi surface.  Since calculations of the full phase space integral are often not possible, and almost always result in an integral that must be done numerically, the calculations of the mean free path, rate, and emissivity are almost always done via the Fermi surface approximation, which we describe below.  

In the Fermi surface approximation, the phase space integral (like in Eq.~(\ref{eq:MFP_integral}) and (\ref{eq:emissivity_integral})) is converted into spherical coordinates for each momentum-three vector and then broken up (termed ``phase space decomposition'' \cite{Shapiro:1983du}) into an angular integral $A$ and an integral $J$ over momentum magnitudes (equivalently, energies).  In the angular integral, the fermion momentum magnitudes are set equal to their respective Fermi momenta, while in the energy integral, the energies are integrated over, consistent with thermal blurring of the Fermi surface (although inconsistent with the momenta magnitudes in the angular integral, which were not allowed to vary above or below the Fermi surface).   

The emissivity due to axion emission via $n+n\rightarrow n+n+a$ is given by Eq.~(\ref{eq:emissivity_integral}).  We again neglect the axion 3-momentum in the momentum-conserving delta function and then multiply by one in the form
\begin{align}
    1 &= \int_0^{\infty} \mathop{dp_1}\mathop{dp_2}\mathop{dp_3}\mathop{dp_4} \delta(p_1-p_{Fn})\delta(p_2-p_{Fn})\delta(p_3-p_{Fn})\delta(p_4-p_{Fn})    \label{eq:cleverone}\\
    &=\frac{1}{p_{Fn}^4}\int_{m_*+U_n}^{\infty} \mathop{dE_1}\mathop{dE_2}\mathop{dE_3}\mathop{dE_4}E_1^*E_2^*E_3^*E_4^*\delta(p_1-p_{Fn})\delta(p_2-p_{Fn})\delta(p_3-p_{Fn})\delta(p_4-p_{Fn}),\nonumber
\end{align}
where we have used $\mathop{dE} = (p/E^*) \mathop{dp}$ as can be seen from the neutron dispersion relation Eq.~(\ref{eq:En}).  
We perform phase space decomposition, obtaining
\begin{equation}
    Q_{FS} = \frac{1}{768\pi^{11}}\frac{G_{an}^2f^4m_n^4}{m_{\pi}^4p_{Fn}^4}A(p_{Fn})J_2(T,\hat{y}).  
\end{equation}
The angular integral $A$ is 
\begin{align}
A(p_{Fn}) &= \int \mathop{d^3p_1}\mathop{d^3p_2}\mathop{d^3p_3}\mathop{d^3p_4} \delta(p_1-p_{Fn})\delta(p_2-p_{Fn})\delta(p_3-p_{Fn})\delta(p_4-p_{Fn})\delta^3(\mathbf{p}_1+\mathbf{p}_2-\mathbf{p}_3-\mathbf{p}_4) \nonumber\\
&\times\left( \frac{\mathbf{k}^4}{\left(\mathbf{k}^2+m_{\pi}^2\right)^2}+\frac{\mathbf{l}^4}{\left(\mathbf{l}^2+m_{\pi}^2\right)^2}+\frac{\mathbf{k}^2\mathbf{l}^2\left(1-3\left(\hat{\mathbf{k}}\cdot \hat{\mathbf{l}}\right)^2\right)}{\left(\mathbf{k}^2+m_{\pi}^2\right)\left(\mathbf{l}^2+m_{\pi}^2\right)}    \right)= 32\pi^3 p_{Fn}^5 F(y),
\label{eq:ang_expression}
\end{align}
where $F(y)$ is given in Eq.~(\ref{eq:Fofy}).  The energy integral is 
\begin{equation}
    J_2(T) = \int\mathop{d^3\omega} \int_{m_*+U_n}^{\infty} \mathop{dE_1}\mathop{dE_2}\mathop{dE_3}\mathop{dE_4}\delta(E_1+E_2-E_3-E_4-\omega)f_1f_2(1-f_3)(1-f_4).
    \label{eq:J2_original}
\end{equation}
The energy integral is evaluated by changing to dimensionless variables centered at the Fermi energy $x_i = (E_i - \mu_n)/T$, and then to variables $u=x_1+x_2$ and $v=x_1-x_2$ and so Eq.~(\ref{eq:J2_original}) becomes
\begin{align}
   J_2(T,\hat{y}) &\equiv T^3\int\mathop{d^3\omega}\int_{-\hat{y}}^{\infty}\mathop{dx_1}\mathop{dx_2}\mathop{dx_3}\mathop{dx_4}\frac{\delta(x_1+x_2-x_3-x_4-\omega/T)}{(1+e^{x_1})(1+e^{x_2})(1+e^{-x_3})(1+e^{-x_4})}\nonumber\\
    &= 8\pi T^6 \int_{-\hat{y}}^{\infty}\mathop{dx_1}\mathop{dx_2}\int_0^{x_1+x_2+2\hat{y}}\mathop{dz}z^2\frac{\ln{\left(\frac{\cosh{((x_1+x_2-z+\hat{y})/2)}}{\cosh{(\hat{y}/2)}}\right)}}{(1+e^{x_1})(1+e^{x_2})(1-e^{z-x_1-x_2})}
    \label{eq:J2}\nonumber\\
    &= 16\pi T^6\int_{-2\hat{y}}^{\infty}\mathop{du}\frac{1}{1-e^u}\ln{\left\{\frac{\cosh{(\hat{y}/2)}}{\cosh{\left[(u+\hat{y})/2\right]}}\right\}}\int_0^{u+2\hat{y}}\mathop{dw}\frac{w^2}{1-e^{w-u}}\ln{\left\{\frac{\cosh{\left[(u+\hat{y}-w)/2\right]}}{\cosh{(\hat{y}/2)}}\right\}}\nonumber\\
    &= 16\pi T^6K_2(\hat{y})
\end{align}
where $\hat{y} = (\mu_n^*-m_*)/T$ and $K_2(\hat{y})$ is given in Eq.~(\ref{K2}).  For strongly degenerate nuclear matter ($\hat{y}\rightarrow\infty$), 
\begin{equation}
    K_2(\hat{y}\rightarrow\infty) = \frac{31}{1890}\pi^6.
\end{equation}

\section{Comparison of emissivity expressions}
\begin{figure*}[t!]
\begin{minipage}[t]{0.5\linewidth}
\includegraphics[width=.95\linewidth]{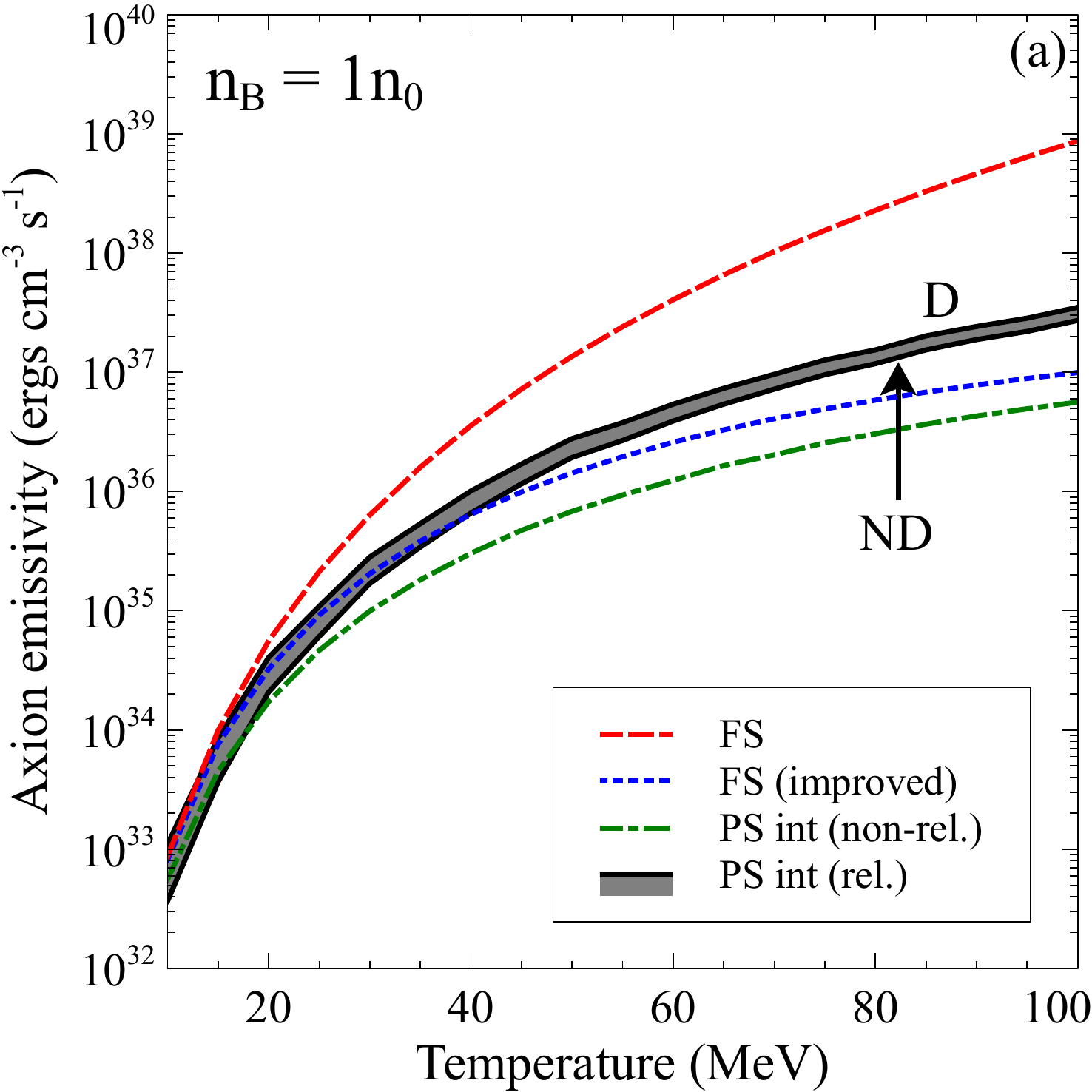}
\end{minipage}\hfill%
\begin{minipage}[t]{0.5\linewidth}
\includegraphics[width=.95\linewidth]{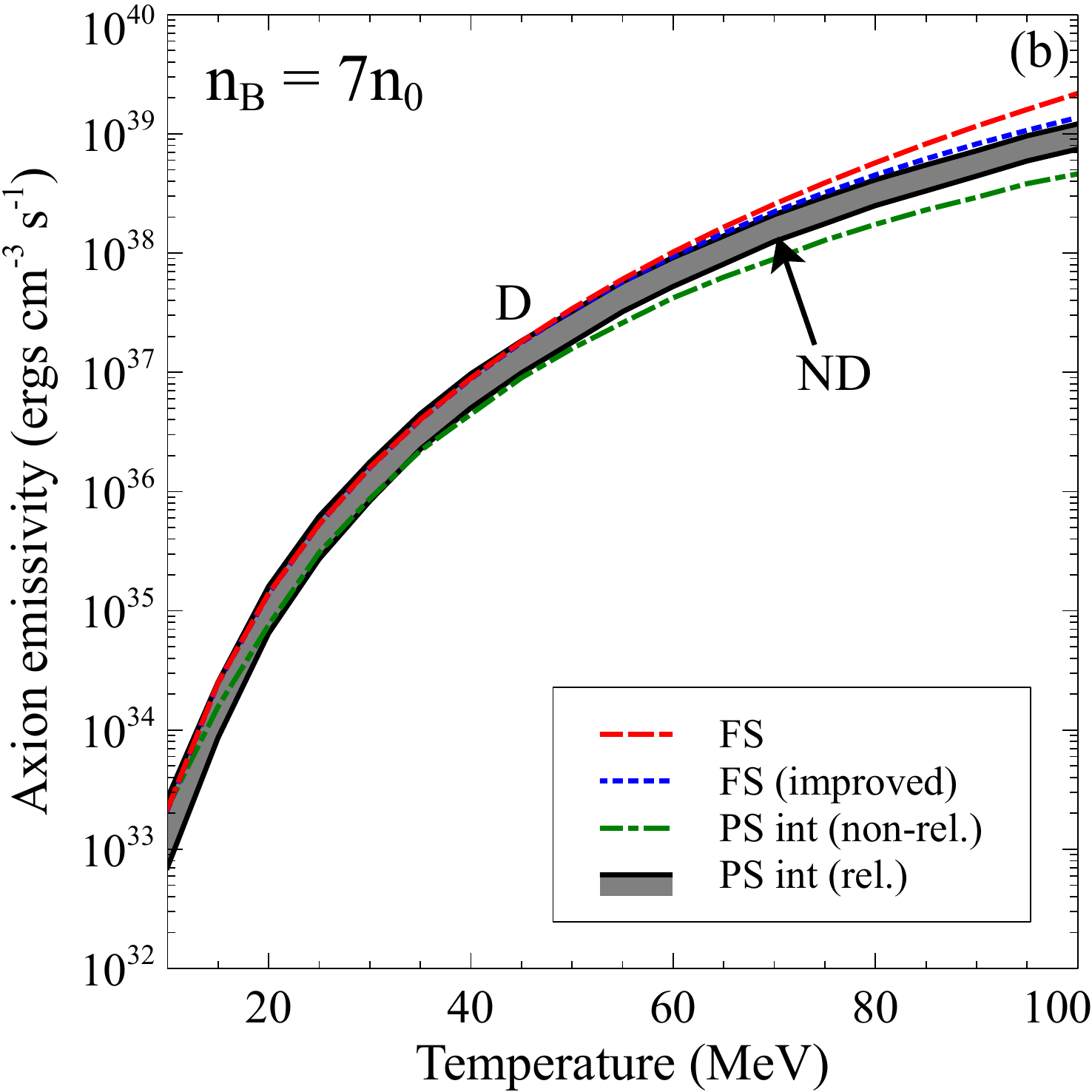}
\end{minipage}%
\caption{Comparison of several approximations of the axion emissivity, including the FS approximation [Eq.~(\ref{eq:Q_FS})] (red, dashed), our improvement to the FS approximation [Eq.~(\ref{eq:Q_new})] (blue, dotted), the non-relativistic phase space integral [Eq.~(\ref{eq:Q_NR_PS})] (green, dash-dotted), and the fully relativistic phase space integral (with constant matrix element) [Eq.~(\ref{eq:exact_emissivity})] (black, solid) at densities of $1n_0$ (a) and $7n_0$ (b).}
\label{fig:compare_Q}
\end{figure*}

In Fig.~\ref{fig:compare_Q} we compare various approximations of the axion emissivity, namely, the FS approximation [Eq.~(\ref{eq:Q_FS})], our improvement to the FS approximation [Eq.~(\ref{eq:Q_new})], the non-relativistic phase space integral [Eq.~(\ref{eq:Q_NR_PS})], and the fully relativistic phase space integral\footnote{In the fully relativistic phase space integral, we choose two values of $k_{\text{typ}}$: $k_{\text{typ}}^2=3m_*T$ and also $k_{\text{typ}}^2=p_{Fn}^2$ as upper and lower bounds, marked ``ND'' and ``D'' respectively in Fig.~\ref{fig:compare_Q}.} (with constant matrix element) [Eq.~(\ref{eq:exact_emissivity})].  We see that at $1n_0$, at low temperature the approximations all agree.  Here the neutrons are degenerate, which explains the success of the FS approximation, and the neutrons are indeed nonrelativistic because their Fermi momentum is not yet large, which explains the success of the nonrelativistic approximation.  As the temperature increases, the neutrons become non-degenerate and so the FS approximation fails dramatically.  The improved treatment of the FS approximation does better than the original, but still fails at temperatures above 40-50 MeV.  The NR approximation fails at high temperature because it neglects the axion momentum in the 3d delta function (at $T=100 \text{ MeV}$, the axion spectrum peaks at $\omega = 300 \text{ MeV}$, which is not negligible compared to the neutron Fermi momentum of 320 MeV).  At $7n_0$, the neutrons are always degenerate, and thus the Fermi surface approximation and its improvement match the full phase space integral quite well.  The non-relativistic phase space integral doesn't match as well because the approximation $E^* \approx m_*$ in the denominator of the phase space integral becomes a very poor approximation as the effective mass dwindles to around 250 MeV at this high density.

\chapter{Axion mean free path integrals}
\label{sec:I}
\pagestyle{myheadings}
Below, we detail a series of approximations for the axion mean free path, Eq.~(\ref{eq:MFP_integral}).
\section{Relativistic, constant-matrix-element phase space integration}
\label{sec:rel_PS_MFP}
Following the same procedure as for the emissivity full phase space integral, we can do the full phase space integration of the axion mean free path (with the same approximation of a constant matrix element) and we obtain
\begin{align}
    &\lambda^{-1} = \left(1-\frac{\beta}{3}\right) \frac{f^4 m_n^4 G_{an}^2}{4\pi^5 m_{\pi}^4\omega} \left(1+\frac{m_{\pi}^2}{k_{\text{typ}}^2}\right)^{-2}\int_{m_*}^{\infty}\mathop{dq_0}\int_0^{\infty}\mathop{dq}\int_{-1}^1\mathop{dr}\int_0^{\sqrt{(q_0+\omega)^2-m_*^2}}\mathop{dx}\int_0^{\sqrt{q_0^2-m_*^2}}\mathop{dk}\nonumber\\
    &\times \frac{kqx(q_0-\sqrt{k^2+m_*^2})\theta(2kq-\vert q_0^2-q^2-2q_0\sqrt{k^2+m_*^2} \vert)}{\sqrt{x^2+m_*^2}\sqrt{k^2+m_*^2}\sqrt{\omega^2+q^2+2\omega qr}\sqrt{k^2+m_*^2+q_0^2-2q_0\sqrt{k^2+m_*^2}}(1+e^{(\sqrt{k^2+m_*^2}-\mu_n^*)/T})}\label{eq:MFPexactanswer}\\
    &\times \frac{\theta(2x\sqrt{\omega^2+q^2+2\omega qr}-\vert q^2+2\omega qr-q_0^2-2q_0\omega +2(q_0+\omega)\sqrt{x^2+m_*^2} \vert)}{(1+e^{(\sqrt{k^2+m_*^2+q_0^2-2q_0\sqrt{k^2+m_*^2}}-\mu_n^*)/T})(1+e^{-(\sqrt{x^2+m_*^2}-\mu_n^*)/T})(1+e^{-(\omega + q_0 - \sqrt{x^2+m_*^2}-\mu_n^*)/T})}.\nonumber
\end{align}
\section{Non-relativistic phase space integration}
\label{sec:MFP_nonrel_PS}
The axion mean free path [Eq.~(\ref{eq:MFP_integral})] can be computed assuming non-relativistic neutrons.  A similar calculation (for nondegenerate neutrons) has been done in \cite{PhysRevD.42.3297,Giannotti:2005tn} and was extended to arbitrary degeneracy in \cite{Carenza:2019pxu}.  We start with Eq.~(\ref{eq:MFP_integral}), keeping the matrix element as momentum-dependent and thus inside the integral.  We use the nonrelativistic (quadratic) approximation for the neutron dispersion relations except in the four energy denominators where $E^* = m_*$, just as in the emissivity calculation.  We obtain
\begin{align}
    \lambda_a^{-1} &= \frac{1}{48\pi^{8}}\frac{f^4 m_n^4 G_{an}^2}{m_{\pi}^4\omega m_*^3}\int \mathop{d^3p_1}\mathop{d^3p_2}\mathop{d^3p_3}\mathop{d^3p_4}\delta(p_1^2+p_2^2-p_3^2-p_4^2+2m_*\omega)\delta^3(\mathbf{p}_1+\mathbf{p}_2-\mathbf{p}_3-\mathbf{p}_4)\nonumber\\
    &\times f_1f_2(1-f_3)(1-f_4)\left[ \frac{\mathbf{k}^4}{\left(\mathbf{k}^2+m_{\pi}^2\right)^2}+\frac{\mathbf{l}^4}{\left(\mathbf{l}^2+m_{\pi}^2\right)^2}+\frac{\mathbf{k}^2\mathbf{l}^2-3\left(\mathbf{k}\cdot\mathbf{l}\right)^2}{\left(\mathbf{k}^2+m_{\pi}^2\right)\left(\mathbf{l}^2+m_{\pi}^2\right)}    \right].
\end{align}
Just as in the emissivity calculation, we transform coordinates to $\{\mathbf{p}_+,\mathbf{p}_-,\mathbf{a},\mathbf{b}\}$, picking up a factor of 8 from the Jacobian, and then we integrate over $\mathbf{b}$, using the three-dimensional delta function $\delta^3(\mathbf{a}+\mathbf{b})$.  We choose the same coordinate system as in the emissivity calculation (Appendix \ref{sec:Q_nonrel_PS}), aligning the ``z'' axis along the $\mathbf{a}$ three-momentum vector.  We integrate over the 3 trivial angles and then use the energy delta function to integrate over $a$.  Then we create nondimensional variables $u=p_+^2/(2m_*T)$ and $v=p_-^2/(2m_*T)$, and define $\hat{y}=\hat{\mu}/T=(\mu^*-m_*)/T$ as in the emissivity calculation.  We also define $\gamma = \omega/(2T)$.
\begin{align}
    \mathbf{k}^2 &= 2m_*T(2v+\gamma-2\sqrt{v}\sqrt{v+\gamma}r)\nonumber\\
    \mathbf{l}^2 &= 2m_*T(2v+\gamma+2\sqrt{v}\sqrt{v+\gamma}r)\nonumber\\
    \mathbf{k}\cdot\mathbf{l} &= -m_*\omega\nonumber\\
    \beta(E_1-\mu_n) &= -\hat{y}+u+v+2\sqrt{uv}(\sqrt{1-r^2}\sqrt{1-s^2}\cos{\phi}+rs)\\
    \beta(E_2-\mu_n) &= -\hat{y}+u+v-2\sqrt{uv}(\sqrt{1-r^2}\sqrt{1-s^2}\cos{\phi}+rs)\nonumber\\
    \beta(E_3-\mu_n) &= -\hat{y}+u+v+\gamma + 2\sqrt{u}\sqrt{v+\gamma}s\nonumber\\
    \beta(E_4-\mu_n) &= -\hat{y}+u+v+\gamma-2\sqrt{u}\sqrt{v+\gamma}s\nonumber.
\end{align}
We find that
\begin{align}
    \lambda^{-1} &= \frac{8\sqrt{2}}{3\pi^6}\frac{f^4m_n^4G_{an}^2}{m_{\pi}^4\omega}m_*^{1/2}T^{3.5}\int_0^{\infty}\mathop{du}\mathop{dv}\int_{-1}^1\mathop{dr}\mathop{ds}\int_0^{2\pi}\mathop{d\phi}u^{1/2}v^{3/2}(v+\gamma)^{3/2}\label{eq:MFP_NR_PS}\\
    &\times \frac{\alpha ^4 \left(r^2+3\right)-6 \alpha ^2 \left(r^2-1\right) (\gamma +2 v)+3 \left(r^2-1\right) \left(-\gamma ^2+4 \gamma  \left(r^2-1\right) v+4 \left(r^2-1\right) v^2\right)}{\left(4 v \left(\alpha
   ^2-\gamma  r^2+\gamma \right)+\left(\alpha ^2+\gamma \right)^2-4 \left(r^2-1\right) v^2\right)^2}\nonumber\\
   &\left((1+e^{\beta(E_1-\mu_n)})(1+e^{\beta(E_2-\mu_n)})(1+e^{-\beta(E_3-\mu_n)})(1+e^{-\beta(E_4-\mu_n)})\right)^{-1}\nonumber
\end{align}
\section{Fermi surface approximation and its improvement}
\label{sec:mfp_FS_calculation}
The mean free path of an axion due to the process $n+n+a\rightarrow n+n$ is given by Eq.~(\ref{eq:MFP_integral}).  We neglect the 3-momentum of the axion in the momentum-conserving delta function, and then multiply by one in the form Eq.~(\ref{eq:cleverone}).  Then we perform phase space decomposition, splitting the integral into integral expressions $A$ and $J_1$
\begin{equation}
    \lambda^{-1} = \frac{f^4G_{an}^2m_n^4}{96\pi^8\omega m_{\pi}^4 p_{Fn}^4}A(p_{Fn})J_1(\omega,T,\hat{y}),
\end{equation}
The angular integral is the same as for the emissivity calculation (see Eq.~(\ref{eq:ang_expression})) and the energy integral is
\begin{equation}
    J_1(\omega,T,\hat{y}) = \int_{m_*+U_n}^{\infty} \mathop{dE_1}\mathop{dE_2}\mathop{dE_3}\mathop{dE_4}\delta(E_1+E_2-E_3-E_4+\omega)f_1f_2(1-f_3)(1-f_4).
    \label{eq:J1_original}
\end{equation}
The energy integral is evaluated by changing to dimensionless variables centered at the Fermi energy $x_i = (E_i - \mu_n)/T$, and then to variables $u=x_1+x_2$ and $v=x_1-x_2$ and so Eq.~(\ref{eq:J1_original}) becomes
\begin{align}
    J_1(\omega,T,\hat{y}) &\equiv T^3 \int_{-\hat{y}}^{\infty}\mathop{dx_1}\mathop{dx_2}\mathop{dx_3}\mathop{dx_4}\frac{\delta(x_1+x_2-x_3-x_4+\omega/T)}{(1+e^{x_1})(1+e^{x_2})(1+e^{-x_3})(1+e^{-x_4})}\nonumber\\
    &= 2T^3\int_{-\hat{y}}^{\infty}\mathop{dx_1}{dx_2}\frac{\ln{\left(\frac{\cosh{((x_1+x_2+\omega/T+\hat{y})/2})}{\cosh{(\hat{y}/2)}}\right)}}{(1+e^{x_1})(1+e^{x_2})(1-e^{-x_1-x_2-\omega/T})}\nonumber\\
    &= 4T^3 K_1(\hat{y},\omega/T),
\end{align}
where $\hat{y} = (\mu_n^*-m_*)/T$ and $K_1(\hat{y},\omega/T)$ is given in Eq.~(\ref{eq:K1}).
In the literature, it is standard to consider strongly degenerate nuclear matter where $\hat{y}\rightarrow\infty$ and so
\begin{equation}
    K_1(\hat{y}\rightarrow\infty,\omega/T) = \left(\frac{\omega/T}{24}\right)\frac{(\omega/T)^2+4\pi^2}{1-e^{-\omega/T}},
\end{equation}
and we arrive at the formula for the axion mean free path in strongly degenerate nuclear matter, Eq.~(\ref{eq:lambdaFS}).  In semi-degenerate matter, the improved FS approximation yields Eq.~(\ref{eq:lambda_new}) for the axion mean free path.
\section{Comparison of MFP expressions}
\begin{figure*}[t!]
\begin{minipage}[t]{0.5\linewidth}
\includegraphics[width=.95\linewidth]{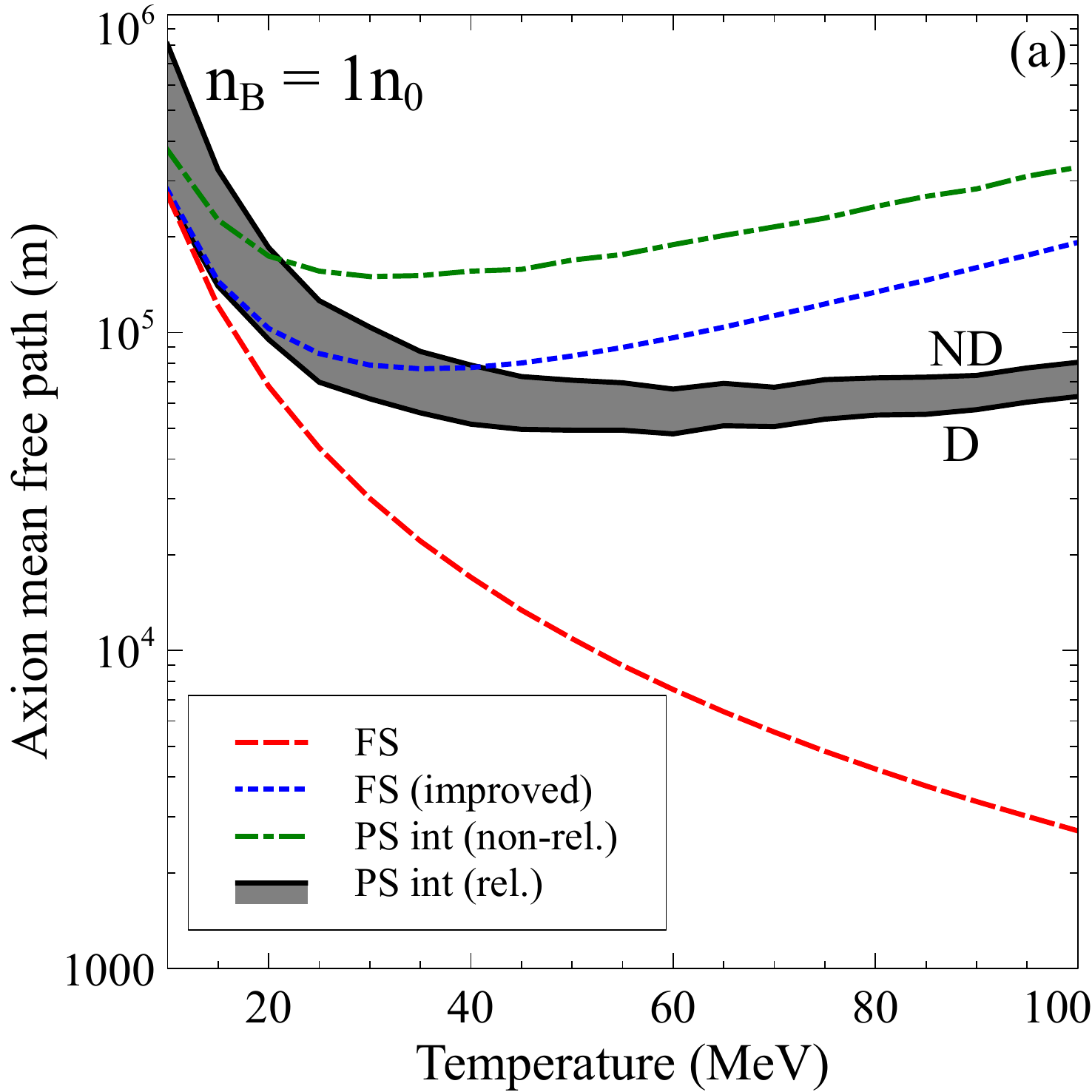}
\end{minipage}\hfill%
\begin{minipage}[t]{0.5\linewidth}
\includegraphics[width=.95\linewidth]{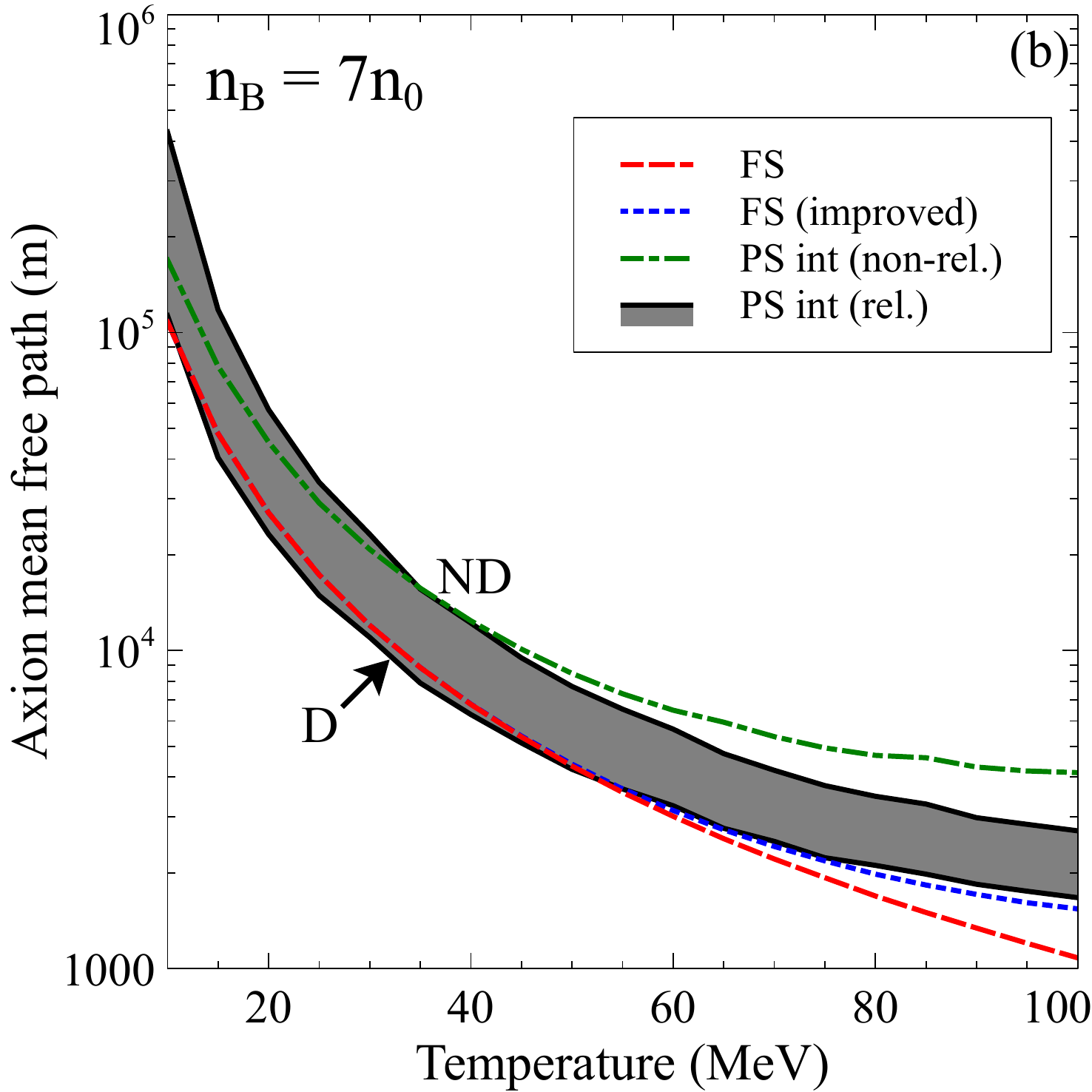}
\end{minipage}%
\caption{Comparison of the various approximations of the axion MFP including the FS approximation [Eq.~(\ref{eq:lambdaFS})] (red, dashed), our improvement to the FS approximation [Eq.~(\ref{eq:lambda_new})] (blue, dotted), the non-relativistic phase space integral [Eq.~(\ref{eq:MFP_NR_PS})] (green, dash-dotted), and the fully relativistic phase space integral (with constant matrix element) [Eq.~(\ref{eq:MFPexactanswer})] (black, solid) at densities of $1n_0$ (a) and $7n_0$ (b).}
\label{fig:MFP_comparison}
\end{figure*}
In Fig.~\ref{fig:MFP_comparison} we compare different approximations for the axion mean free path, namely, the FS approximation [Eq.~(\ref{eq:lambdaFS})], our improvement to the FS approximation [Eq.~(\ref{eq:lambda_new})], the non-relativistic phase space integral [Eq.~(\ref{eq:MFP_NR_PS})], and the fully relativistic phase space integral\footnote{In the fully relativistic phase space integral, we choose two values of $k_{\text{typ}}$: $k_{\text{typ}}^2=3m_*T$ and also $k_{\text{typ}}^2=p_{Fn}^2$ as upper and lower bounds, marked ``ND'' and ``D'' respectively in Fig.~\ref{fig:MFP_comparison}.} (with constant matrix element) [Eq.~(\ref{eq:MFPexactanswer})]. 

 We see in Fig.~\ref{fig:MFP_comparison} that at $1n_0$, at low temperature the approximations all agree (though the nonrelativistic phase space integral deviates slightly from the rest, because of the approximation in the energy denominators $E^*\approx m_*$, which shows up more in this figure than in Fig.~\ref{fig:compare_Q} because of the difference in y-axis scales).  At saturation density and low temperature, neutrons are degenerate, which explains the success of the FS approximation, and the neutrons are indeed nonrelativistic because their Fermi momentum is not yet large, which explains the success of the nonrelativistic approximation.  As the temperature increases, the neutrons become non-degenerate and so the FS approximation fails dramatically.  The improved treatment of the FS approximation does better than the original, but still fails at temperatures above about 40 MeV.  The NR approximation fails at high temperature because it neglects the axion momentum in the 3d delta function (again, at $T=100 \text{ MeV}$, the axion spectrum peaks at $\omega = 300 \text{ MeV}$, which is not negligible compared the neutron Fermi momentum of 320 MeV).  At $7n_0$, the neutrons are always degenerate, and thus the Fermi surface approximation and its improvement match the full phase space integral quite well.  Again, the non-relativistic phase space integral doesn't match as well because the approximation $E^* \approx m_*$ in the denominator of the phase space integral becomes a very poor approximation as the effective mass dwindles to around 250 MeV.
\addtocontents{toc}{\protect\vspace{1ex}}
\addcontentsline{toc}{chapter}{References}
\addtocontents{toc}{\protect\vspace{1ex}}
\bibliographystyle{JHEP}
\bibliography{main}

\end{document}